\documentclass[11pt]{article}
\newif\iffigs\figstrue

\usepackage{epsfig,latexsym}
\usepackage{amsmath}
\usepackage{verbatim}
\usepackage{mathrsfs}
\usepackage{amssymb}

\DeclareMathAlphabet{\mathpzc}{OT1}{pzc}{m}{it}

 \csname
@addtoreset\endcsname{equation}{section}

\def\gz0{\gamma^{0}}

\def\ft#1#2{{\textstyle{{\scriptstyle #1}
\over {\scriptstyle #2}}}}

\def\vf{\varphi}

\def\be{\begin{equation}}
\def\ee{\end{equation}}
\def\bea{\begin{eqnarray}}
\def\eea{\end{eqnarray}}
\def\ba{\begin{array}}
\def\ea{\end{array}}
\def\bec{\begin{center}}
\def\ec{\end{center}}
\def\ba{\begin{align}}
\def\ena{\end{align}}
\def\ft{\footnote}

\def\12{\frac{1}{2}}

\newcommand{\so}{\mathfrak{so}}

\newcommand{\uu}{\mathfrak{u}}

\newcommand{\slal}{\mathfrak{sl}}

\newcommand{\SO}{\mathop{\rm SO}}

\newcommand{\U}{\mathop{\rm {}U}}

\begin{document}
\begin{titlepage}
\begin{center}
\vskip 0.2cm
\vskip 0.2cm
{\Large\sc Integrable Scalar Cosmologies }\vskip 4pt
{\sc II. Can they fit into Gauged Extended Supergavity or be encoded in $\mathcal{N}$=1 superpotentials?}\\[1cm]
{\sc P.~Fr\'e${}^{\; a}$\footnote{Prof. Fr\'e is presently fulfilling the duties of Scientific Counselor of the Italian Embassy in the Russian Federation, Denezhnij pereulok, 5, 121002 Moscow, Russia.},  A.S.~Sorin$^{\; b}$ and M. Trigiante$^{\; c}$}\\[10pt]
{${}^a$\sl\small Dipartimento di Fisica, Universit\'a di Torino\\INFN -- Sezione di Torino \\
via P. Giuria 1, \ 10125 Torino \ ITALY \\}\emph{e-mail:} \quad {\small {\tt
fre@to.infn.it}}\\
\vspace{5pt}
{{\em $^{b}$\sl\small Bogoliubov Laboratory of Theoretical Physics}}\\
{{\em  {\tt and} Veksler and Baldin Laboratory of High Energy Physics,}}\\
{{\em Joint Institute for Nuclear Research,}}\\
{\em 141980 Dubna, Moscow Region, Russia}~\quad\\
\emph{e-mail:}\quad {\small {\tt sorin@theor.jinr.ru}}
\quad \\
\vspace{5pt}
{{\em $^c$\sl\small  Dipartimento di Fisica Politecnico di Torino,}}\\
{\em C.so Duca degli Abruzzi, 24, I-10129 Torino, Italy}~\quad\\
\emph{e-mail:}\quad {\small {\tt mario.trigiante@gmail.com}}
\quad \vspace{8pt}
\vspace{15pt}
\begin{abstract}
The question whether the integrable one-field cosmologies classified in a previous paper by Fr\'e, Sagnotti and Sorin can be embedded as consistent one-field truncations into Extended  Gauged Supergravity or in $\mathcal{N}=1$ supergravity gauged  by a superpotential without the use of $D$-terms is addressed in this paper. The answer is that such an embedding  is very difficult and rare but not impossible. Indeed we were able to find two examples of integrable models embedded in  Supergravity in this way. Both examples are fitted into $\mathcal{N}=1$ Supergravity by means of a very specific and interesting choice of the superpotential $W(z)$. The question whether there are examples of such an embedding in extended Gauged Supergravity remains open. In the present paper, relying on the embedding tensor formalism we classified all gaugings of the $\mathcal{N}=2$ STU model, confirming, in the absence on hypermultiplets, the uniqueness of the stable de Sitter vacuum found several years ago by Fr\'e, Trigiante and Van Proeyen and excluding the embedding of any integrable cosmological model.
A detailed analysis of the space of exact  solutions of the first Supergravity--embedded  integrable cosmological model revealed several new features worth an in depth consideration. When the scalar potential has an extremum at a negative value, the universe necessarily collapses into a Big Crunch notwithstanding its spatial flatness. The causal structure of these universes is quite different from that of the closed, positive curved, universe: indeed in this case the particle and event horizons do not coincide and develop complicated patterns. The cosmological consequences of this unexpected mechanism deserve careful consideration.
\end{abstract}
\end{center}
\end{titlepage}
\tableofcontents
\noindent {}
\newpage
\section{\sc Introduction}
In a recent paper \cite{primopapero} some of us have addressed the question of classifying integrable one-field cosmological models based on a slightly generalized ansatz for the spatially flat metric,
\begin{equation}\label{piatttosa}
        ds^2 \ = \ - \ e^{\,2\,{\cal B}(t)} \, \mathrm{d}t^2 \ + \ a^2(t) \ \mathrm{d}\mathbf{x}\cdot \mathrm{d}\mathbf{x}
        \ ,
      \end{equation}
and on a suitable choice of a potential $V(\phi)$ for the unique scalar field $\phi$, whose kinetic term is supposed to be canonical:
\begin{equation}\label{kin}
    \mathcal{L}_{kin}(\phi) \, = \, \frac{1}{2} \partial_\mu\phi \, \partial^\mu \phi \, \sqrt{-g}\, .
\end{equation}
The suitable potential functions $V(\phi)$ that lead to exactly integrable  Maxwell Einstein field equations were searched within the family of linear combinations of exponential functions $\exp \beta \phi$ or rational functions thereof. The motivations for such a choice were provided both with String and Supergravity arguments and a rather remarkable bestiary of exact cosmological solutions was uncovered, endowed with quite interesting mathematical properties. Some of these solutions have also some appeal as candidate models of the inflationary scenario, capable of explaining the structure of the primordial power spectrum.
\par
In \cite{primopapero} the classical Friedman equations
\begin{eqnarray}
 && H^2 \ = \  \frac{1}{3} \ \dot{\phi}^2 \, + \, \frac{2}{3} \ V(\phi)  \ , \nonumber\\
 &&  \dot{H}  \ = \ - \, \dot{\phi}^2   \ , \nonumber\\
 && \ddot{\phi} \,+ \,  3 \, H \, \dot{\phi} \, + \, V^{\,\prime}  \ = \ 0 \ , \label{fridmano}
\end{eqnarray}
where
\be
a(t) \ = \ e^{\, A(t)} \ , \qquad H(t) \, \equiv \, \frac{\dot{a}(t)}{a(t)} \ = \ \dot{A}(t) \,
\ee
are respectively the scale factor and the Hubble function, were revisited  in the more general gauge  with ${\cal B}\ne 0$ which allows for the construction of exact solutions, whenever the effective two dimensional dynamical system lying behind
eq.s (\ref{fridmano}) can be mapped, by means of a suitable canonical transformation,  into an integrable dynamical model endowed with two hamiltonians in involution. Such procedure produced the bestiary constructed and analyzed in \cite{primopapero}.
\par
After the change of perspective produced by the recent series of papers \cite{johndimitri},\cite{Ketov:2010qz},\cite{Ketov:2012jt},\cite{Kallosh:2013hoa},\cite{Kallosh:2013lkr},\cite{Farakos:2013cqa},
\cite{minimalsergioKLP},\cite{primosashapietro},\cite{Ferrara:2013wka},\cite{Ferrara:2013kca},\cite{Ferrara:2013pla} and in particular after
\cite{minimalsergioKLP},\cite{primosashapietro}, we know that all positive definite members of the above mentioned bestiary can be embedded into $\mathcal{N}=1$ supergravity as $D$-terms produced by the gauging of an axial symmetry, provided the K\"ahler manifold to which we assign the Wess-Zumino multiplet of the inflaton is consistent with the chosen potential $V(\phi)$, namely it has an axial symmetric K\"ahler potential defined  in a precise way by $V(\phi)$. In \cite{secondosashapietro} which is published at the same time as the present paper, two of us have analysed the mathematical algorithm lying behind  this embedding mechanism which we have named the $D$-map. In the same paper a possible path toward the microscopic interpretation of the peculiar axial symmetric K\"ahler manifolds requested by the $D$-type supergravity embedding of the integrable potentials is proposed and discussed.  Such a microscopic interpretation is obligatory in order  to give a sound physical meaning to the supergravity embedding.
\par
The main theme that we are going to debate in  the present paper is instead the following. Can  integrable cosmologies be embedded into gauged extended supergravity or in $\mathcal{N}=1$ supergravity gauged by the F-terms that are produced by the choice of some suitable superpotential $W(z)$? In the present paper the choice of the K\"ahler geometry for the inflaton will not depend on the potential $V(\phi)$. The inflaton Wess-Zumino multiplet will always be assigned to a constant curvature K\"ahler manifold as it is the case in compactifications on torii, orbifolds  or orientifolds.
\par
Having clarified this fundamental distinction between the complementary approaches of the present paper and of the parallel paper \cite{secondosashapietro}, we continue our discussion of the Friedman system (\ref{fridmano}).
Referring to the classical cosmic time formulation (\ref{fridmano}) of Friedman equations and to the very instructive hydrodynamical picture, we recall that the energy density and the pressure of the fluid describing the scalar matter can be identified with the two combinations
 \begin{eqnarray}
&& \rho  \ = \  \frac{1}{4} \ \dot{\phi}^2 \, + \, \frac{1}{2}\ V(\phi)\ , \nonumber\\
&&   p  \ = \ \frac{1}{4} \ \dot{\phi}^2 \, - \, \frac{1}{2} \ V(\phi)\ ,\label{patatefritte}
 \end{eqnarray}
since, in this fashion, the first of eqs.~\eqref{fridmano} translates into the familiar link between the Hubble constant and the energy density of the Universe,
\begin{equation}\label{gordilatinus}
    H^2 \, = \, \frac{4}{3} \ \rho \ .
\end{equation}
A standard result in General Relativity (see for instance \cite{pietroGR}) is that for a fluid whose equation of state is
\begin{equation}\label{equatastata}
    p\, = \, w \, \rho \qquad  \quad  w\, \in \,\mathbb{R}
\end{equation}
the relation between the energy density and the scale factor takes the form
\begin{equation}\label{forense2}
    \frac{\rho}{\rho_0} \, = \, \left(\frac{a_0}{a} \right)^{3(1+w)} \ ,
\end{equation}
where $\rho_0$ and $a_0$ are their values at some reference time $t_0$.
Combining eq.~(\ref{equatastata}) with the first of eqs.~(\ref{fridmano}) one can then deduce that
\begin{equation}\label{andamentus}
    a(t) \, \sim \, \left(t-t_i\right)^{\frac{2}{3 (w+1)}} \ ,
\end{equation}
where $t_i$ is an initial cosmic time. All values $-1\leq w \leq 1$ can be encompassed by eqs.~\eqref{patatefritte}, including the two particularly important cases of a dust--filled Universe, for which $w=0$ and $a(t) \, \sim \, \left(t-t_i\right)^{\frac{2}{3}}$, and of a radiation--filled Universe, for which $w=\frac{1}{3}$ and $a(t) \, \sim \, \left(t-t_i\right)^{\frac{1}{2}}$. Moreover, when the potential energy $V(\phi)$ becomes negligible with respect to the kinetic energy in eqs.~\eqref{patatefritte}, $w \approx 1$. On the other hand, when the potential energy $V(\phi)$ dominates $w\approx-1$, and eq.~(\ref{forense2}) implies that the energy density is approximately constant (vacuum energy) $\rho \, = \, \rho_0$. The behavior of the scale factor is then exponential, since the Hubble function is also a constant $H_0$ on account of eq.~\eqref{gordilatinus}, and therefore
      $$ a(t) \, \sim \, \exp\left [ H_0 \, t \right ] \quad ; \quad H_0 \, = \, \sqrt{\frac{4}{3}\ \rho_0}$$
The actual solutions of the bestiary described in \cite{primopapero} correspond to complicated equations of state whose index $w$  varies in time. Nonetheless they are qualitatively akin, at different epochs, to these simple types of behavior.
\par
As we just stressed, the next question that constitutes the main issue of the present paper  is whether the integrable potentials classified in \cite{primopapero} play a role in consistent one--field truncations of \emph{four--dimensional} gauged Supergravity.
A striking and fascinating feature of Supergravity is in fact that its scalar potentials are not completely free. Rather, they emerge from a well defined gauging procedure that becomes more and more restrictive as the number $\mathcal{N}$ of supercharges increases, so that the link between the integrable cosmologies of \cite{primopapero} and this structure is clearly of interest.
\par
The first encouraging observation was already mentioned: in all integrable models found in \cite{primopapero}  the potential ${V}(\phi)$ is a polynomial or rational function  of exponentials  $\exp[ \beta \, \phi]$ of a field $\phi$ whose kinetic term is canonical.  If we discard the rational cases and we retain only the polynomial ones that are the majority, this feature connects naturally such cosmological models  to Gauged Supergravity with scalar fields belonging to \emph{non compact, symmetric coset manifolds} $\mathrm{G/H}$. This wide class encompasses not only all $\mathcal{N}>2$ theories, but also some $\mathcal{N} \le 2$ models that are frequently considered in connection with Cosmology, Black Holes, Compactifications and other issues.
%
Since the coset manifolds $\mathrm{G/H}$ relevant to supergravity are characterized by a numerator group $\mathrm{G}$ that is a non-compact semi-simple group, in these models one can always resort to a \textit{solvable parameterization} of the scalar manifold \cite{SUGRA_solvable}, so that the scalar fields fall into two classes:
\begin{enumerate}
  \item The \textit{Cartan fields} $\mathfrak{h}^i$ associated with the Cartan generators of the Lie algebra $\mathbb{G}$, whose number equals the rank $r$ of $\mathrm{G/H}$. For instance, in models associated with toroidal or orbifold compactifications, fields of this type are generically interpreted as radii of the underlying multi--tori.
   \item The \textit{axion fields} $b^I$ associated with the roots of the Lie algebra $\mathbb{G}$.
\end{enumerate}
The kinetic terms of Cartan scalars have the canonical form
\be
\sum_i^r\frac{\alpha_i^2}{2}\ \partial_\mu \mathfrak{h}^i \, \partial^\mu \mathfrak{}h^i \ ,
\ee
up to constant coefficients, while for the axion scalars entering solvable coset representatives, the $\alpha_i^2$ factors leave way to exponential functions $\exp[ \beta_i \, \mathfrak{h}^i]$ of Cartan fields. The scalar potentials of gauged Supergravity are polynomial functions of the coset representatives, so that after the truncation to Cartan sectors, setting the axions to constant values, one is led naturally to combinations of exponentials of the type encountered in \cite{primopapero}. Yet the devil lies in the details, since the integrable potentials do result from exponential functions $\exp[ \beta \, \mathfrak{h}] $, but with rigidly fixed ratios between the $\beta_i$ entering the exponents and the $\alpha_i$ entering the kinetic terms. The candidate potentials are displayed in tables \ref{tab:families} and \ref{Sporadic} following the notations and the nomenclature of \cite{primopapero}.
\begin{table}[ht!]
\centering
\begin{tabular}{|lc|}
\hline
\null & Potential function \\
\hline
\null&\null\\
$I_1$ & $\! C_{11} \, e^{\,\vf} \, + \, 2\, C_{12} \, + \, C_{22} \, e^{\, - \vf}$  \\
\null&\null\\
$I_2$ & $\! C_1 \, e^{\,2\,\gamma \,\varphi}\, +\, C_2e^{\,(\gamma+1)\, \varphi} $  \\
\null&\null\\
$I_3$ & $\! C_1 \, e^{\, 2\, \varphi} \ + \ C_2$    \\
\null&\null\\
$I_7$ & $\! C_1 \, \Big(\cosh\,\gamma\,\varphi \Big)^{\frac{2}{\gamma} \, - \, 2}\, + \, C_2 \Big( \sinh\,\gamma\,\varphi \Big)^{\frac{2}{\gamma} \, - \, 2}$  \\
\null&\null\\
$I_8$ & $\! C_1 \left(\cosh [2 \, \gamma \, \varphi] \right)^{\frac {1}{\gamma} -1}\,\cos\left[\left(\frac {1}{\gamma} -1\right)\, \arccos\left(\tanh[2\,\gamma\, \varphi]\,+\,C_2\right)\right]$ \\
\null&\null\\
$I_9$ & $\! C_1 \ e^{2\,\gamma\,\varphi} \  + \ C_2 \
e^{\frac{2}{\gamma}\,\varphi} $ \\
\null&\null\\
\hline
\end{tabular}
\caption{The families of integrable potential classified in \cite{primopapero} (and further extended in \cite{secondosashapietro})  that, being pure linear combinations of exponentials, might have a chance to be fitted into Gauged Supergravity are those corresponding to the numbers $I_1$,$I_2$,$I_3$,$I_7$,$I_8$ (if $\gamma = \frac{1}{n}$ with $n\in \mathbb{Z}$) and $I_9$. In all   cases  $C_i$ should be real parameters and $\gamma \in \mathbb{Q}$ should just be a rational number.}
\label{tab:families}
\end{table}
As a result, the possible role of integrable potentials in gauged supergravity theories is not evident a priori, and actually, the required ratios are quite difficult to be obtained. Notwithstanding these difficulties we were able to identify a pair of examples, showing that although rare, supergravity integrable cosmological models based on $\mathrm{G/H}$ scalar manifolds\footnote{The main consequence of the $D$-embedding of integrable potentials discussed in the parallel paper \cite{secondosashapietro} is that the K\"ahler manifold hosting the inflaton is not a constant curvature coset manifold $\mathrm{G/H}$.} do exist and might provide a very useful testing ground where exact calculations can be performed \textit{ab initio} to the very end.
\begin{table}[ht!]
\centering
\begin{tabular}{|lc|}
\hline
\null & \null \\
\null & Sporadic Integrable Potentials \\
\null & \null \\
 \null & $\begin{array}{lcr}\mathcal{V}_{Ia}(\varphi)
   & = & \frac{\lambda}{4} \left[(a+b)
   \cosh\left(\frac{6}{5}\varphi\right)+(3 a-b)
   \cosh\left(\frac{2}{5}\varphi\right)\right] \end{array}$ \\
   \null & \null \\
   \null & $\begin{array}{lcr}\mathcal{V}_{Ib}(\varphi) & = & \frac{\lambda}{4} \left[(a+b)
   \sinh\left(\frac{6}{5}\varphi\right)-(3 a-b)
   \sinh\left(\frac{2}{5}\varphi\right)\right] \end{array} $\\
   \null & \null \\
   where & $ \left\{a,b\right\} \, = \, \left\{
\begin{array}{cc}
 1  & -3  \\
 1  & -\frac{1}{2} \\
 1  & -\frac{3}{16}
\end{array}
\right\} $\\
\null & \null \\
\hline
\null & \null \\
\null & $\begin{array}{lcr}
\mathcal{V}_{II}(\varphi)
   & = & \frac{\lambda}{8} \left[3 a+3 b- c+4( a- b)
   \cosh\left(\frac{2}{3}\varphi \right)+(a+b+c) \cosh\left(\frac{4}{3}\varphi \right)\right]\nonumber\ ,
\end{array}$\\
\null & \null \\
where  & $\left\{a,b,c\right\} \, = \, \left\{
\begin{array}{ccc}
 1  & 1  & -2
   \\
 1  & 1  & -6
   \\
 1  & 8  & -6
   \\
 1  & 16  & -12
    \\
 1  & \frac{1}{8} &
   -\frac{3}{4} \\
 1  & \frac{1}{16} &
   -\frac{3}{4}
\end{array}
\right\} $\\
\null & \null \\
\hline
\null & \null \\
\null & $\begin{array}{lcr}
\mathcal{V}_{IIIa}(\varphi) & = & \frac{\lambda}{16} \left[\left(1-\frac{1}{3
   \sqrt{3}}\right) e^{-6 \varphi
   /5}+\left(7+\frac{1}{\sqrt{3}}\right)
   e^{-2 \varphi
   /5} \right. \\
   && \left. +\left(7-\frac{1}{\sqrt{3}}\right)
   e^{2 \varphi /5}+\left(1+\frac{1}{3
   \sqrt{3}}\right) e^{6 \varphi
   /5}\right]\ .
   \end{array}$\\
\null&\null\\
\hline
\null & \null \\
\null &$\begin{array}{lcr}
  \mathcal{V}_{IIIb}(\varphi) &=& \frac{\lambda}{16} \left[\left(2-18
   \sqrt{3}\right) e^{-6 \varphi
   /5}+\left(6+30 \sqrt{3}\right) e^{-2
   \varphi /5}\right.\\
   &&\left. +\left(6-30
   \sqrt{3}\right) e^{2 \varphi
   /5}+\left(2+18 \sqrt{3}\right) e^{6
   \varphi /5}\right]
   \end{array}
 $ \\
 \null&\null\\
\hline
\end{tabular}
\caption{In this table of the sporadic integrable potentials classified in \cite{primopapero} we retain only those that being pure linear combinations of exponentials have an a priori possibility of being realized in some truncation of Gauged Supergravity models}
\label{Sporadic}
\end{table}
\section{\sc The set up for comparison with Supergravity}\label{sec:supergravity}
In this paper we focus on $D=4$ supergravity models.
In order to compare the effective dynamical model considered in \cite{primopapero} with the possible one-field truncations of supergravity,
it is convenient to adopt a slightly different starting point which touches upon some of the fundamental features of all supersymmetric extension of gravity. As we have already mentioned, differently from non supersymmetric theories, where the kinetic and potential terms of the scalar fields are uncorrelated and disposable at will, the fascination of \textit{sugras} is precisely that a close relation between these two terms here exists  and is mandatory. Indeed the potential is just created by means of the gauging procedure.  The explicit formulae for the potential always involve the metric of the target manifold which, on the other hand,  determines the scalar field kinetic terms. Thus, in one-field truncations, the form of the kinetic term cannot be normalized at will but comes out differently, depending on the considered model and on the chosen truncation. A sufficiently ductile Lagrangian that encodes the various sugra-truncations discussed in this paper is the following one:
 \begin{equation}\label{unopratino}
    {\cal L}_{eff} \, = \, \mbox{const} \, \times \, e^{\,{3 \, A} \ - \ {\cal B} } \ \left[ \ - \, \frac{3}{2} \dot{ A}^{\,2} \ + \ \frac{q}{2} \, \dot{\mathfrak{h}}^2 \ - \  e^{\,2\,{\cal B}} \  {{V}}(\mathfrak{h}) \right]
\end{equation}
The field $\mathfrak{h}$ is a residual dilaton field after all the the other dilatonic and axionic fields have been fixed to their stationary values and $q$ is a parameter, usually integer, that depends both on the chosen supergravity model and on the chosen truncation. The correspondence with the set up of \cite{primopapero} is simple: $\phi=\sqrt{q} \, \mathfrak{h}$.
Hence altogether the transformation formulae that correlate the general discussion of this paper with the bestiary of supergravity potentials, found in \cite{primopapero} and displayed in tables \ref{tab:families} and \ref{Sporadic} are the following ones:
\begin{eqnarray}
\label{babushka}
  \dot{\cal A}(t) &=& 3 \, H(t) \, = \, 3 \, \frac{\mathrm{d}}{\mathrm{d}t}  \,  \log [a(t)]\nonumber\\
  &\updownarrow & \nonumber\\
  {\cal A}(t) & = & 3 A(t) \nonumber\\
   {\cal B}(t) & = & {\cal B}(t)\nonumber\\
  \varphi &=& \sqrt{3 q} \, \mathfrak{h} \nonumber\\
  \mathcal{V}(\varphi) &=& 3 \, V(\mathfrak{h}) \, = \, 3 \,V(\frac{\varphi}{\sqrt{3q}})
\end{eqnarray}
We will consider examples of $\mathcal{N}=2$ and $\mathcal{N}=1$ models trying to spot the crucial points that make it unexpectedly difficult to fit integrable cosmological models into the well established framework of \textit{gauged supergravities}. Difficult but not impossible since we were able to identify at least one integrable $\mathcal{N}=1$ supergravity  model based on the coupling of a single Wess-Zumino multiplet endowed with a very specific superpotential.
While postponing to a further paper the classification of all the gaugings of the $\mathcal{N}=2$ models based on symmetric spaces \cite{ToineCremmerOld} (see table \ref{homomodels}) and the analysis of their one-field truncation in the quest of possible matching with
\begin{table}[ht!]
\begin{center}
{\small
\begin{tabular}{||c|c||c||}
  \hline
   coset &coset &    susy\\
   D=4 & D=3 &   \\
  \hline
\null & \null &\null \\
 $ \frac{\mathrm{SU(1,1)}}{\mathrm{U(1)}}$ & $ \frac{\mathrm{G_{2(2)}}}{\mathrm{SU(2)\times SU(2)}}$ & $\mathcal{N}=2$ \\
\null & \null & n=1 \\
\hline
\null & \null &\null \\
  $ \frac{\mathrm{Sp(6,R)}}{\mathrm{SU(3)\times  U(1)}}$ & $ \frac{\mathrm{F_{4(4)}}}{\mathrm{USp(6)\times SU(2)}}$  & $\mathcal{N}=2$ \\
 \null & \null & $n=6$ \\
 \null & \null &\null \\
\hline
\null & \null &\null \\
 $ \frac{\mathrm{SU(3,3)}}{\mathrm{SU(3)\times SU(3) \times U(1)}}$ & $ \frac{\mathrm{E_{6(2)}}}{\mathrm{SU(6)\times SU(2)}}$    & $\mathcal{N}=2$ \\
 \null & \null & $n=9$ \\
\null & \null &\null \\
\hline
\null & \null &\null \\
 $ \frac{\mathrm{SO^\star(12)}}{\mathrm{SU(6)\times U(1)}}$ & $ \frac{\mathrm{E_{7(-5)}}}{\mathrm{SO(12)\times SU(2)}}$  & $\mathcal{N}=2$ \\
\null & \null & n=15 \\
\null & \null &\null \\
\hline
\null & \null &\null \\
$ \frac{\mathrm{E_{7(-25)}}}{\mathrm{E_{6(-78)} \times U(1)}}$ & $ \frac{\mathrm{E_{8(-24)}}}{\mathrm{E_{7(-133)}\times SU(2)}}$    & $\mathcal{N}=2$ \\
\null & \null & $n=27$ \\
\hline
\null & \null &\null \\
 $ \frac{\mathrm{SL(2,\mathbb{R})}}{\mathrm{SO(2)}}\times\frac{\mathrm{SO(2,2+p)}}{\mathrm{SO(2)\times SO(2+p)}}$ & $ \frac{\mathrm{SO(4,4+p)}}{\mathrm{SO(4)\times SO(4+p)}}$   & $\mathcal{N}=2$ \\
  \null & \null & n=3+p \\
\hline
\null & \null &\null \\
$ \frac{\mathrm{SU(p+1,1)}}{\mathrm{SU(p+1)\times U(1)}}$ & $ \frac{\mathrm{SU(p+2,2)}}{\mathrm{SU(p+2)\times SU(2)}}$    & $\mathcal{N}=2$\\
\null & \null &\null \\
\hline
\end{tabular}
}
\caption{List of special K\"ahler homogeneous spaces in $D=4$ with their $D=3$ enlarged counterparts, obtained through Kaluza-Klein reduction. The number $n$ denotes the number of vector multiplets. The total number of vector fields is therefore $n_V \, =\, n+1$. \label{homomodels}}
\end{center}
\end{table}
the integrable potentials, in the present paper we will consider  in some detail another possible point of view. It was named the \textit{minimalist approach} in the conclusions of \cite{primopapero}. Possibly no physically relevant cosmological model extracted from Gauged Supergravity is integrable, yet the solution of its field equations might be effectively simulated in their essential behavior by the exact solution of a neighboring integrable model. Relying on the classification of fixed points presented in \cite{primopapero} we advocate that if there is a one parameter family of potentials that includes both an integrable case and a case derived from supergravity and if the fixed point type is the same for the integrable case and for the supergravity case, then the integrable model provides a viable substitute of the physical one and its solutions provide good approximations of the physical ones accessible only with numerical evaluations. We will illustrate this viewpoint  with the a detailed analysis of one particularly relevant case.
\par
The obvious limitation of this approach is the absence of an algorithm to evaluate the error that separates  the unknown physical solution from its integrable model clone. Yet a posteriori numerical experiments show that is error is rather small and that all essential features of the physical solution are captured by the solutions of the appropriate integrable member of the same family.
\par
Certainly it would be very much rewarding if other integrable potentials could be derived from specific truncations of specially chosen supergravity gaugings. If such a case is realized  the particular choice of parameters that leads to integrability would certainly encode some profound physical significance.
\section{\sc  $\mathcal{N}=2$ Models and Stable de Sitter Vacua}
\label{STUgauginghi}
An issue of high relevance for a theoretical explanation of current cosmological data is the
construction of \textit{stable de Sitter string vacua} that break all supersymmetries \cite{kklt}, a question that is actually formulated at the level of the low--energy $\mathcal{N}$--extended Supergravity. As recently reviewed in \cite{scrucca}, for $\mathcal{N} > 2$ no stable de Sitter vacua have ever been found and do not seem to be possible. In $\mathcal{N}=1$ Supergravity coupled only to chiral multiplets, stability criteria can be formulated in terms of sectional curvatures of the underlying K\"ahler manifold that are quite involved, so that their general solution has not been worked out to date.
\par
In $\mathcal{N}=2$ Supergravity stable de Sitter vacua have been obtained, until very recently, only in a unique class of models \cite{mapietoine} (later generalized in \cite{Roest:2009tt})\footnote{For a recent construction of meta-stable de Sitter vacua in abelian gaugings of $\mathcal{N}=2$ supergravity, see \cite{Catino:2013syn}.} and, as stressed there, the mechanism that generates a scalar potential with the desired properties results from three equally essential ingredients:
\begin{enumerate}
  \item The gauging of a \textit{non-compact, non-abelian group} that in the models that were considered is $\so(2,1)$.
  \item The introduction of Fayet Iliopoulos terms corresponding to the gauging of compact $\uu(1)$ factors.
  \item The introduction of a Wagemans-de Roo angle that within special K\"ahler geometry rotates the directions associated to the non-compact gauge group with respect to those associated with the compact one.
\end{enumerate}
The class of models constructed in \cite{mapietoine} relies on the coupling of vector multiplets to Supergravity as dictated by the special K{\"a}hler manifold
\begin{eqnarray}
\mathcal{SK}_n & = & \mathcal{ST}[2,n] \, \equiv \,
\frac{\mathrm{SU(1,1)}}{\mathrm{U(1)}} \, \times \,
\frac{\mathrm{SO(2,n)}}{\mathrm{SO(2)\times SO(n)}}\ , \label{spiffero}
\end{eqnarray}
which accommodates the scalar fields and governs the entire structure of the Lagrangian. There are two interesting special cases: for $n=1$ one obtains the $\mathrm{ST}$ model, which describes two vector multiplets, while for $n=2$ one obtains the $\mathrm{STU}$-model, which constitutes the core of most supergravity theories and is thus ubiquitous in the study of string compactifications at low energies. In this case, due to accidental Lie algebra automorphisms, the scalar manifold factorizes, since
\begin{equation}\label{vorticino}
    \mathcal{ST}[2,2] \, \equiv \,\frac{\mathrm{SU(1,1)}}{\mathrm{U(1)}} \, \times \,\frac{\mathrm{SU(1,1)}}{\mathrm{U(1)}} \, \times \,\frac{\mathrm{SU(1,1)}}{\mathrm{U(1)}}
\end{equation}
Starting from the Lagrangian of ungauged $\mathcal{N}=2$ Supergravity based on this special K\"ahler geometry, the scalar potential is generated gauging a subgroup $\mathrm{G_{gauge}} \subset \mathrm{SU(1,1)} \times \mathrm{SO(2,n)}$.
The three models explicitly constructed in \cite{mapietoine}, whose scalar potential admits stable de Sitter extrema,  are
\begin{itemize}
  \item The $\mathrm{STU}$ model with 3 vector multiplets, in the manifold $\mathcal{ST}[2,2]$,
  which, together with the graviphoton, are gauging
  $\SO(2,1)\times \U(1)$, with a Fayet--Iliopoulos term for the $\U(1)$ factor.
  \item a model with 5 vector multiplets, in the manifold $\mathcal{ST}[2,4]$,
  which, together with the graviphoton, are gauging
  $\SO(2,1)\times \SO(3)$, with a Fayet--Iliopoulos term for the
  $\SO(3)$; and
  \item the last model extended with 2 hypermultiplets with 8 real
  scalars in the coset $\frac{\SO(4,2)}{\SO(4)\times \SO(2)}$.
\end{itemize}
The choice of the hypermultiplet sector for the third model  is possible since
the coset $\frac{\SO(4,2)}{\SO(4)\times \SO(2)}$ can be viewed as a
factor in the special K{\"a}hler manifold $\mathcal{ST}[2,4]$, or alternatively as a
quaternionic-K{\"a}hler manifold by itself. The scalar potentials of the three models are qualitatively very similar, while the key ingredient behind the emergence of de Sitter extrema is the introduction of a non--trivial Wagemans--de Roo angle. For this reason we shall analyze only the first and simplest of these three models.

The explicit form of the scalar potential obtained in this gauging can be illustrated by introducing
a parametrization of the scalar sector according to Special Geometry, and symplectic sections are the main ingredient to this effect. In
the notation of \cite{Andrianopoli:1997cm}, the holomorphic
section reads
\begin{equation}
  \Omega= \left( \begin{array}{c}
  X^\Lambda \\
  F_\Sigma
\end{array}\right),
\label{Omegabig}
\end{equation}
where
\begin{eqnarray}
X^\Lambda(S,y) & = & \left( \begin{array}{c}
  \frac{1}{2} \, \left( 1+y^2\right)  \\
  \frac{1}{2} \, {\rm i} \, (1-y^2) \\
  y^a
\end{array}\right)  \qquad ;\quad a=1,\dots , n\,, \quad \nonumber\\
F_\Lambda(S,y) & = & \left( \begin{array}{c}
  \frac{1}{2} \, S \,  \left( 1+y^2\right)   \\
  \frac{1}{2} \, {\rm i} \, S \,  (1-y^2)  \\
  - S \, y^a
\end{array}\right) \quad ;\quad y^2 = \sum_{a=1}^{n} (y^a)^2\ ,
\label{symsecso21}
\end{eqnarray}
The complex $y^a$ fields are
Calabi--Vesentini coordinates for the homogeneous manifold
$\frac{\mathrm{SO(2,n)}}{\mathrm{SO(2)\times SO(n)}}$, while the complex
field $S$ parameterizes the homogeneous space
$\frac{\mathrm{SU(1,1)}}{\mathrm{U(1)}}$, which is identified with the
complex upper half-plane. With these conventions, the positivity domain of our Lagrangian is
\begin{equation}
  \mbox{Im} \, S \, > \, 0\,.
\label{positdom}
\end{equation}
The  K{\"a}hler potential is by definition
\begin{equation}
{\cal K}\,  = \,  -\mbox{log}\left (- {\rm i}\langle \Omega \,
 \vert \, \bar \Omega
\rangle \right )\, =\, -\mbox{log}\left [- {\rm i} \left ({\bar X}^\Lambda
F_\Lambda - {\bar F}_\Sigma X^\Sigma \right ) \right ] \ , \label{specpot}
\end{equation}
so that in this example the K{\"a}hler potential and the K{\"a}hler metric read
\begin{eqnarray}
\mathcal{K} & = & \mathcal{K}_1 + \mathcal{K}_2 \,,\nonumber\\
\mathcal{K}_1  & = & -\log \, \left[ - {\rm i} \, \left(
S-\overline{S}\right) \right],\qquad
 \mathcal{K}_2 = -\log \left[ \frac{1}{2}
\, \left( 1-2\overline{y}^a \, y^a + | y^a y^a|^2
\right) \right],
\nonumber\\
g_{S\overline{S}}&= & \frac{1}{(2\mbox{Im} S)^2}\,,\qquad \,\qquad \, \quad
g_{a \bar b}= \,\frac{\partial }{\partial
y^a}\,\frac{\partial }{\partial \bar {y}^b}\, \mathcal{K}_2\,.
\label{kalermetr}
\end{eqnarray}
%
\subsection{\sc  de Roo -- Wagemans angles}
\label{stabdesitter}
As we have stressed, in the construction of \cite{mapietoine}, the \textit{de
Roo--Wagemans angles} are essential ingredients for the existence of de Sitter extrema. They were originally introduced \cite{deRoo:1985jh,Wagemans:1990mv}  in ${\cal N}=4$ supergravity with semisimple gaugings to characterize the relative embeddings of each simple factor  $\mathrm{G_k}$ of the gauge group inside
$\mathrm{Sp(2({\bf n}+2),\mathbb{R})}$, performing a \textit{symplectic rotation} on the
holomorphic section of the manifold prior to gauging. Different choices
of the angles yield inequivalent gauged models with different properties. For $n=2$, with ${\rm SO(2,1)\times U(1)}$ gauging, there is just one  de Roo-Wagemans' angle and the corresponding rotation matrix reads
\be
\mathcal{R} \ = \ \left(\begin{array}{cc} A & B\cr -B & A \end{array}\right)\ ,\ee
where
\be
 A \ = \ \left(\begin{array}{cc} {\mathbf 1}_{3\times 3} &0 \\
                         0  & \cos{(\theta)} \end{array} \right) \ , \qquad
 B \ = \ \left(\begin{array}{cc} \mathbf{0}_{3\times 3} & 0\\
                         0  & \sin{(\theta)} \end{array} \right) \ .
                           \label{dRWagangles}
\ee
The symplectic section is rotated as
\begin{eqnarray}
\Omega & \rightarrow & \Omega_R \, \equiv \, \mathcal{R}\cdot \Omega\ ,
\end{eqnarray}
while the K{\"a}hler potential is clearly left invariant by the
transformation. The de Roo--Wagemans' angle appears explicitly in the scalar potential, which is determined by the symplectic section $\Omega_R$ and by
\be
V_R
\equiv \exp \left[\mathcal{ K}\right] \, \Omega_R
\ee
and reads~\cite{mapietoine}
\begin{equation}
\mathcal{V}_{\SO(2,1)\times \U(1)}=\mathcal{V}_3+\mathcal{V}_1=
\frac{1}{2\mbox{Im} S} \,\left(e_1{}^2 |\cos \theta -S\,\sin \theta|^2 +
      e_0{}^2\,\frac{{P_2^+}(y)}{{P_2^-}(y)}
      \right) \ ,
\label{Potentabel}
\end{equation}
where $P_2^\pm  (y) $ are polynomial functions in the Calabi--Vesentini
variables of holomorphic degree specified by their index,
\begin{equation}
P_2^\pm (y)  =  1 - 2\,y_0\,\overline{y}_0\pm
  2\,y_1\,{\overline{y}_1} +y^2\bar y^2\ ,
\label{polinabel}
\end{equation}
while $e_{0,1}$ are the coupling constants for the $\so(2,1)$ and $\uu(1)$ gauge algebras.

In order to study the properties of this potential one has to perform a coordinate transformation from the Calabi--Vesentini coordinates to the standard ones that provide a \textit{solvable parametrization} of the three
Lobachevsky--Poincar\'e planes displayed in eq.~(\ref{vorticino}).
With some care such a transformation can be worked out and reads
\begin{eqnarray}
  y_1 &=& -\frac{{\rm i} \left({\rm i} b_1 \left({\rm i}
   b_2+e^{h_2}\right)+e^{h_1+h_2}+{\rm i} e^{h_1}
   b_2-1\right)}{\left({\rm i} b_1+e^{h_1}+1\right) \left({\rm i}
   b_2+e^{h_2}+1\right)}\nonumber \\
  y_2 &=& \frac{{\rm i} b_1+e^{h_1}-e^{h_2}-{\rm i} b_2}{\left({\rm i}
   b_1+e^{h_1}+1\right) \left({\rm i} b_2+e^{h_2}+1\right)}\nonumber \\
  S &=& {\rm i}  e^h \, + \, b \ . \label{transformus}
\end{eqnarray}
After this coordinate change, the complete K\"ahler potential becomes
\begin{equation}\label{nuovoK}
 \mathcal{K} \, = \,    -\log \left(-\frac{16 b
   e^{h_1+h_2}}{\left(\left(1+e^{h_1}\right)^2+b_1^2\right)
   \left(\left(1+e^{h_2}\right)^2+b_2^2\right)}\right)
\end{equation}
so that the K\"ahler metric is
\begin{equation}\label{rillu}
    ds_{K}^2 \, = \, \frac{1}{4} \left(e^{-2 h} {db}^2+{dh}^2+e^{-2
   h_1} {db}_1^2+e^{-2 h_2}
   {db}_2^2+{dh}_1^2+{dh}_2^2\right) \ ,
\end{equation}
while in the new coordinates the scalar potential takes the form
\begin{eqnarray}\label{potentissimo}
    V & = & \, - \, \frac{1}{8} e^{-h-h_1-h_2} \left[2 e^{h_1+h_2}
   \left(-b^2+2 \sin (2 \theta ) b-e^{2
   h}+\left(b^2+e^{2 h}-1\right) \cos (2 \theta
   )-1\right)
   e_1^2 \nonumber \right.\\
   && \left.-\left(\left(e^{h_1}+e^{h_2}\right)^2+b_1^2+b_2^
   2-2 b_1 b_2\right) e_0^2\right] \ .
\end{eqnarray}

Let us now turn to exploring consistent truncation patterns to one--field models with standard kinetic terms for the residual scalars. To this effect, one can verify that the constant values
\begin{equation}\label{assioni}
    \left \{b \, , \, b_1 \, , \, b_2 \right \} \, \Rightarrow \, \vec{b}_0 \, \equiv \, \left\{-\frac{\sin (2 \theta )}{\cos (2 \theta )-1} \,  , \,  \kappa \, ,  \, \kappa \right \}
\end{equation}
result in the vanishing of the derivatives of the potential with respect to the three axions, identically in the remaining fields, so that one can safely introduce these values (\ref{assioni}) in the potential to arrive at the reduced form
\begin{equation}\label{barpot}
    \bar{V} \, = \, V|_{\vec{b}=\vec{b}_0} \, = \, \frac{1}{4} e^{-h} e_0^2+\frac{1}{8} e^{-h+h_1-h_2}
   e_0^2+\frac{1}{8} e^{-h-h_1+h_2} e_0^2+\frac{1}{2}
   e^h \sin ^2(\theta ) e_1^2 \ .
\end{equation}
The last step of the reduction is performed setting the two fields $h_{1,2}$ to a common constant value:
\begin{equation}\label{fixingh12}
    h_{1,2}\, = \, \ell
\end{equation}
Indeed, it can be simply verified that for these values the derivatives of $\bar V$ with respect to $h_{1,2}$
vanish identically. Finally, redefining the field $h$ by means of the constant shift
\begin{equation}\label{rulito}
    h \, = \, \mathfrak{h} \, + \, \log \left(\frac{\csc (\theta ) e_0}{e_1}\right)
\end{equation}
the one-field potential becomes
\begin{equation}\label{finalpotential}
    V(\mathfrak{h}) \, = \, \underbrace{\,\sin(\theta ) e_0 e_1}_{\bar{\mu}} \, \cosh \, \mathfrak{h}
\end{equation}
This information suffices to determine the corresponding dynamical system.
We start from the general form of the $\mathcal{N}=2$ supergravity action truncated to the scalar sector which is the following:
\begin{eqnarray}
\mathcal{S}^{\mathcal{N}=2} & = & \int d^4x \, \, \mathcal{L}^{\mathcal{N}=2}_{SUGRA} \nonumber\\
    \mathcal{L}^{\mathcal{N}=2}_{SUGRA} & = & \sqrt{-g} \, \left[ R[g] \, +  2\, g^{SK}_{ij^\star} \, \partial_\mu z^i \, \partial^\mu {\bar z}^{j^\star} \, - 2\, V(z,{\bar z}) \,\right ] \label{n2sugra}
\end{eqnarray}
where $g^{SK}_{ij^\star}$ is the special K\"ahler metric of the target manifold and $V(z,{\bar z})$ is the potential that we have been discussing. Reduced to the residual  dynamical field content, after fixing the other fields to their extremal values, the above action becomes:
\begin{eqnarray}
\mathcal{S}^{\mathcal{N}=2} & = & \int d^4x \, \sqrt{-g} \, \left( \mathcal{R}[g] \, + \, \frac{1}{2} \, \partial_\mu \mathfrak{h} \,   \partial^\mu \mathfrak{h} \, - \, \mu^2 \,  \cosh\left[ h\right] \, \right ) \label{n2sugrareduzia}
\end{eqnarray}
where we have redefined $\mu^2 \, = \, 2 \, {\bar{\mu}}^2$
Hence the effective one-field dynamical system is described by the following Lagrangian
\begin{equation}\label{lagruccona}
    \mathcal{L}_{eff} \, = \, \exp[3 A \, - \, \mathcal{B}] \,  \left(\frac{1}{2} \, \dot{\mathfrak{h}}^2 \, - \, \frac{3}{2} \, \dot{A}^2 \, - \, \exp[2 \mathcal{B}] \, \underbrace{\mu^2 \,\cosh [ \mathfrak{h} ]}_{V(\mathfrak{h})}\right )
\end{equation}
which agrees with the general form (\ref{unopratino}), introduced above.
\par
In light of this, the effective dynamical model of the gauged $\mathrm{STU}$ model would be integrable if the potential
\begin{equation}\label{signorina}
 \mathcal{V}(\varphi) \, = \, 3 \, \mu^2 \,    \cosh\left[\frac{1}{\sqrt{3}}\, \varphi \right ]
\end{equation}
could be identified with any of the integrable potentials listed in tables \ref{tab:families} and \ref{Sporadic}. We show below that this is not the case. Nonetheless, the results of \cite{primopapero} provide a  qualitative information on the behavior of the solutions of this supergravity model. As a special case, one can simply retrieve the de Sitter vacuum from this formulation in terms of a dynamical system. Choosing the gauge $\mathcal{B}=0$, the field equations associated with the Lagrangian \eqref{lagruccona} are solved by setting $\mathfrak{h}\, = \, 0$, which corresponds to the extremum of the potential, and
\begin{equation}\label{rutilant}
    A(t) \, = \, \exp \left[ H_0 \, t \right ] \quad ; \quad  H_0 \, = \, \sqrt{ \frac{2}{3} \mu^2} \, = \, \sqrt{  \frac{4}{3} \sin (\theta ) e_0 e_1} \ ,
\end{equation}
which corresponds to the eternal exponential expansion of de-Sitter space. This solution is an attractor for all the other solutions as shown in \cite{primopapero}.
\par
In order to answer the question  whether the Lagrangian (\ref{lagruccona}) defines an integrable system, so that its general solutions can be written down in analytic form, it is useful to reformulate our question in slightly more general terms, observing that the Lagrangian under consideration belongs to the family
\begin{equation}\label{lagrucconata}
    \mathcal{L}_{cosh} \, = \, \exp[3 A \, - \, \mathcal{B}] \,  \left(\frac{q}{2} \, \dot{\mathfrak{h}}^2 \, - \, \frac{3}{2} \, \dot{A}^2 \, - \, \exp[2 \mathcal{B}] \, \mu^2 \,\cosh [ p \,\mathfrak{h} ]\right )
\end{equation}
that depends on two parameters $q$ and $ p$. Comparing with the list of integrable models one can see that there are just two integrable cases corresponding to the choices
\begin{equation}\label{integratus}
    \frac{p}{\sqrt{3\, q}} \, = \, 1 \quad ; \quad \frac{p}{\sqrt{3\, q}}\, = \, \frac{2}{3}
\end{equation}
The first case $\frac{p}{\sqrt{3\, q}} \, = \, 1$, corresponding to the potential $\mathcal{V}(\varphi)\sim \cosh[\varphi]$ can be mapped into three different integrable series among those displayed in table \ref{tab:families}. The first embedding is into the series $I_1$ by choosing $C_{11}=C_{22} \ne 0$ and $C_{12}=0$. The second embedding is into series $I_2$, by choosing  $\gamma \, = \,\frac{1}{2}$ and   $C_1=C_2$. The third embedding is into model $I_7$, by choosing once again $\gamma \, = \,\frac{1}{2}$ and   $C_1=C_2$. The second case $\frac{p}{\sqrt{3\, q}}\, = \, \frac{2}{3}$ corresponding to the potential $\mathcal{V}(\varphi)\sim \cosh[\frac{2}{3}\varphi]$ can be mapped into  series $I_2$ of table \ref{tab:families}, by choosing $\gamma \, = \, - \,\frac{1}{3}$ and $C_1=C_2$. It can also me mapped into the series $I_7$ by choosing $\gamma \, = \, \frac{1}{3}$ and $C_1= - C_2$. Unfortunately, none of these solutions correspond to the Lagrangian (\ref{lagruccona}), where
\begin{equation}\label{ginopaoli}
    p \, = \,  1 \quad ; \quad q \, = \, 1
\end{equation}
so that the one--field cosmology emerging from the non--compact non--abelian $\so(2,1)$ gauging of the $\mathrm{STU}$ model is indeed not integrable! This analysis  emphasizes that embedding an integrable model into the gauging of an extended supergravity theory is a difficult task.
\par
In  section \ref{zerlina} we will consider in more detail the $Cosh$-model defined by eq.~(\ref{lagrucconata}). There we will show that it can be reduced to a normal form depending only on one parameter that we name the index:
\begin{equation}\label{indexomega}
    \omega \, = \, \frac{p}{\sqrt{q}}
\end{equation}
and we will compare its behavior for various values of the index $\omega$. The  two integrable cases mentioned above correspond respectively to the following critical indices,
 \begin{equation}\label{criticalindices}
    \omega_c^f \, =\, \sqrt{3} \quad , \quad \omega_c^n \, = \, \frac{2}{\sqrt{3}}
 \end{equation}
 The first critical index has been denoted with the superscript $f$ since, in the language of \cite{primopapero} the fixed point of the corresponding dynamical system is of \textit{focus} type. Similarly, the second critical index has been given the superscript $n$ since the fixed point of the corresponding dynamical system is of the \textit{node} type.
In these two cases we are able to integrate the field equations explicitly. For the other values of $\omega$ we are confined to numerical integration. Such a numerical study reveals that when the initial conditions are identical, the   solutions of the non integrable models have a behavior very similar to that of the exact solutions of the integrable model, as long as the type of fixed point defined by the extremum of the potential is the same. Hence the the behavior of the one-field cosmology emerging from the  $\so(2,1)$ gauging of the $\mathrm{STU}$ model can be approximated by the exact analytic solutions of the $cosh$-model with index $\omega_c^n$.
\par
It remains a fact that the value of $\omega$ selected by the Gauged Supergravity model is $\omega = 1$ rather than the integrable one, a conclusion that will be reinforced by a study of Fayet--Iliopoulos gaugings in the $S^3$ model~\cite{FayetIlio}.
\par
Considering instead the  integrable series $I_2$ of table \ref{tab:families}, we will show in section \ref{integsusymodel} that there is just one case
there that can be fitted into a Gauged Supergravity model. It corresponds to the value $\gamma\, = \, \frac{2}{3}$, which can be realized in $\mathcal{N}=1$ supergravity by an acceptable and well defined superpotential. After a wide inspection, this seems to be one of the very few integrable supersymmetric models so far available. A second one will be identified in section \ref{fluxscan}. As we shall emphasize, the superpotential underlying both instances of supersymmetric integrable models  is strictly $\mathcal{N}=1$ and does not arise from a
Fayet--Iliopoulos gauging of a corresponding $\mathcal{N}=2$ model.
\subsection{\sc  Behavior of the solutions in the $\mathcal{N}=2$ $STU$ model with $\so(1,2)$-gauging}

Although the $\mathcal{N}=2$ model that we have been considering is not integrable, its Friedman equations can be integrated numerically providing a qualitative understanding of the nature of the solutions
\begin{figure}[!hbt]
\begin{center}
\iffigs
 \includegraphics[height=40mm]{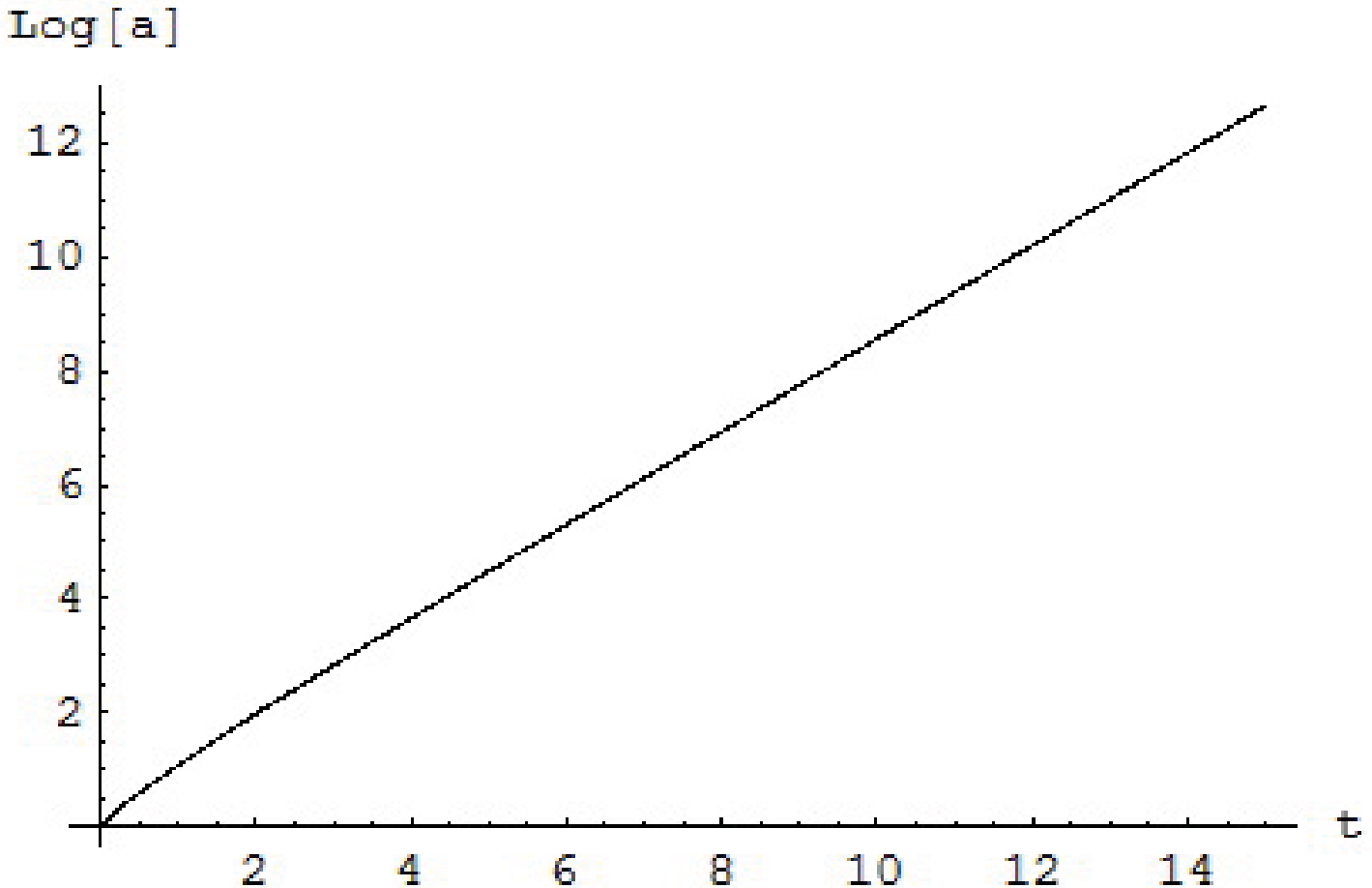}
 \includegraphics[height=40mm]{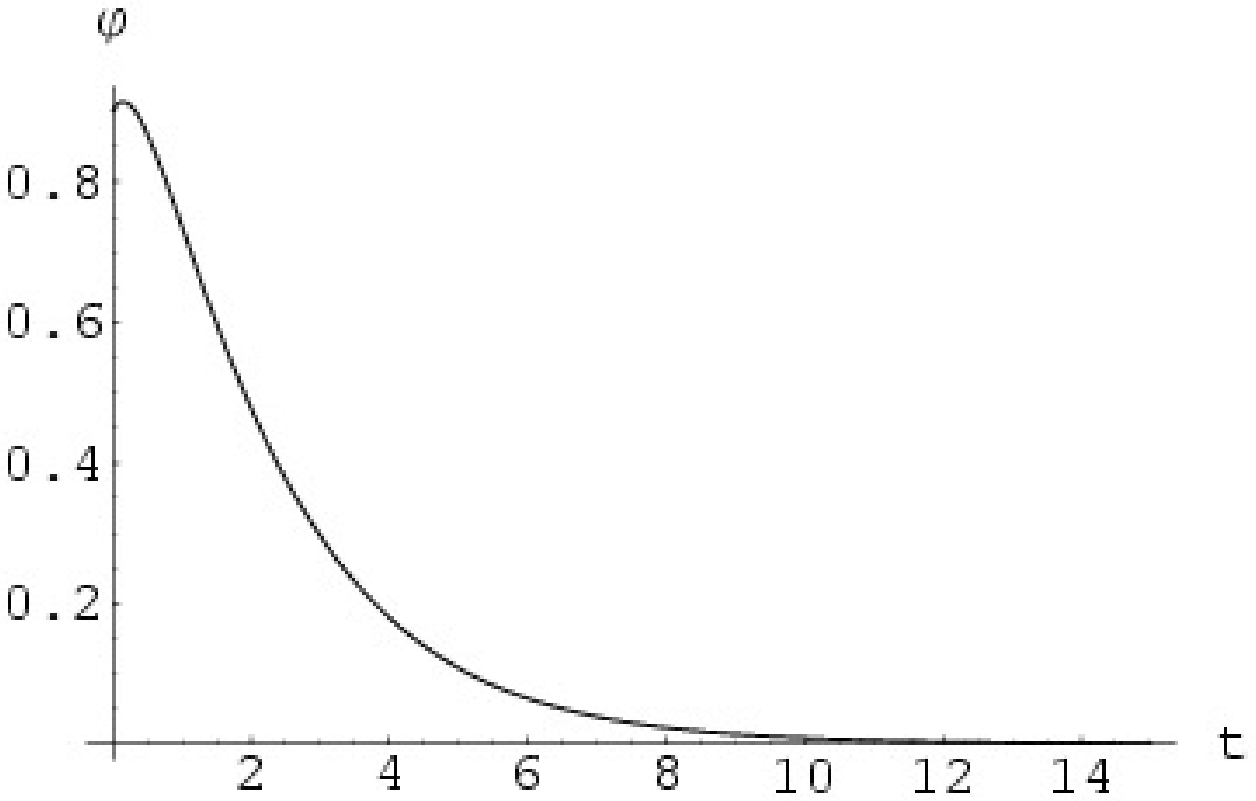}
\else
\end{center}
 \fi
\caption{\it
The numerical solution shows that the de Sitter solution is indeed an attractor.}
\label{soluziapiat}
 \iffigs
 \hskip 1cm \unitlength=1.1mm
 \end{center}
  \fi
\end{figure}

In fig.~\ref{soluziapiat} we show the behavior of both the scale factor and the scalar field for  any regular initial conditions. The plot clearly shows that the de Sitter solution corresponding to an indefinite exponential expansion is an attractor as predicted by the fixed point analysis of the differential system. Indeed  the numerical integration reveals a slow--roll phase that works in reversed order with respect to the standard inflationary scenario \cite{inflation}. When the scalar field is high up and descends rapidly, the expansion of the Universe proceeds rather slowly, then the scalar field reaches the bottom of the potential and rolls slowly toward its minimum, while the Universe expands exponentially becoming asymptotically de Sitter.
\subsection{\sc A More Systematic Approach: The Embbedding Tensor Formalism}
In the previous section we have reviewed and analyzed, from a cosmological perspective, a special class of  $\mathcal{N}=2$ models which exhibit stable de Sitter vacua. A complete analysis of one-field cosmological models emerging from $\mathcal{N}=2$ supergravities (or even $\mathcal{N}>2$ theories) is a considerably more ambitious project, which requires a systematic study of the possible gaugings of extended supergravities. A precious tool in this respect is the embedding tensor formulation of gauged extended supergravities \cite{embeddingtensor}. This approach consists in writing the gauged theory as a deformation of an ungauged one (with the same field content and supersymemtry) in which the additional terms in the Lagrangian  (minimal couplings, fermion mass terms and scalar potential) and in the supersymmetry transformation laws, which are needed in order to make the theory locally invariant with respect to the chosen gauge group $\mathcal{G}$ while keeping the original  supersymmetry unbroken, are all expressed in terms of a single matrix of coupling constants (the embedding tensor) which can be described as a covariant tensor with respect to the global symmetry group ${\rm G}$ of the original ungauged model. If we denote by $\{t_\alpha\}$ the generators of the Lie algebra $\mathfrak{g}$ of ${\rm G}$ and by $X_\Lambda$ the gauge generators, to be gauged by the vector fields $A^\Lambda_\mu$ of the model,
since the gauge group must be contained in ${\rm G}$,   $X_\Lambda$ must be a linear combination of the $t_\alpha$:
\begin{equation}
X_\Lambda=\Theta_\Lambda{}^\alpha\,t_\alpha\,.
\end{equation}
The matrix $\Theta_\Lambda{}^\alpha$ is the embedding tensor and defines all the information about the embedding of the gauge algebra inside $\mathfrak{g}$. A formulation of the gauging which is independent of the  symplectic frame of the original ungauged theory, was given in \cite{magnetic} and extends the definition of the embedding tensor by including, besides the electric components defined above, also  magnetic ones:
\begin{equation}
\Theta_M{}^\alpha=\{\Theta_\Lambda{}^\alpha,\,\Theta^{\Lambda\,\alpha}\}\,.
\end{equation}
The index $M$ is now associated with the symplectic duality-representation ${\bf W}$ of ${\rm G}$ in which the electric field strengths and their  magnetic duals transform, so that $\Theta_M{}^\alpha$ formally belongs to the product ${\bf W}\otimes {\bf Adj}({\rm G})$, namely is a ${\rm G}$-covariant tensor. Since all the deformations of the original ungauged action, implied by the gauging procedure, are written in terms of $\Theta_M{}^\alpha$ in a ${\rm G}$-covariant way, the gauged  equations of motion and the  Bianchi identities formally retain the original global ${\rm G}$-invariance, provided  $\Theta_M{}^\alpha$ is transformed as well. Since, however, the action of ${\rm G}$, at the gauged level, affects the coupling constants of the theory,  encoded in the embedding tensor, it should be viewed as an equivalence between theories rather than a symmetry, and gauged models whose embedding tensors are related by ${\rm G}$-transformations, share the same physics. Thus gauged extended supergravities obtained from the  same ungauged model, can be classified in universality classes defined by the orbits of the embedding tensor under the action of    ${\rm G}$. Classifying such classes is  a rather non trivial task. In simple models like the STU one, this can be done thoroughly. In the following we perform this analysis and analyze the possible one-field cosmological models  for each class, leaving its extension to more general $\mathcal{N}=2$ gauged models to a future investigation \cite{FayetIlio}.\par
To set the stage, let us consider an $\mathcal{N}=2$ theory with $n_V$ vector fields and a global symmetry group of the form:
\begin{equation}
\mathrm{G}=\mathrm{U}_{SK}\times \mathrm{G}_{QK}\,,
\end{equation}
where $\mathrm{U}_{SK},\, \mathrm{G}_{QK}$ are the isometry groups of the Special K\"ahler and Quaternionic K\"ahler manifolds (in the absence of hypermultiplets $\mathrm{G}_{QK}={\rm SO}(3)$).
Let $\mathfrak{g},\,\mathfrak{g}_{SK},\, \mathfrak{g}_{QK}$ denote the Lie algebras of $\mathrm{G},\,\mathrm{U}_{SK},\, \mathrm{G}_{QK}$ and $\{t_\alpha\},\,\{t_A\},\,\{t_a\}$, $\alpha=1,\dots, {\rm dim}(\mathrm{G}),\,A=1,\dots, {\rm dim}(\mathrm{U}_{SK}),\,a=1,\dots, {\rm dim}(\mathrm{G}_{QK})$, a set of corresponding bases.
Only the group $\mathrm{U}_{SK}$ has a symplectic duality action on the $2n_V$-dimensional  vector $\mathbb{F}^M_{\mu\nu}$ , $M=\,\dots, 2n_V$, consisting of the electric field strengths and their duals
\begin{equation}\label{symplettone}
    \mathbb{F}^M_{\mu\nu} \, \equiv \, \left( \begin{array}{c}
                                                F^\Lambda_{\mu\nu} \\
                                                G_{\Lambda\,\mu\nu}
                                              \end{array}
    \right )
\end{equation}
namely:
\begin{equation}
\forall u\in \mathrm{U}_{SK}\,\,:\,\,\,\mathbb{F}^M_{\mu\nu}\rightarrow \mathbb{F}^{'M}_{\mu\nu}=u^M{}_N\,\mathbb{F}^N_{\mu\nu}\,.
\end{equation}
We have denoted by ${\bf W}$ the corresponding $2n_V$-dimensional, symplectic representation of $\mathrm{U}_{SK}$.
\par
For  reader's convenience we summarize the index conventions in the following table
{\small\begin{center}
\begin{tabular}{|l||c|c|c|c|}
  \hline
  groups and & $\mathrm{G}$ & $\mathrm{U}_{SK}$ & $\mathrm{G}_{QK}$ & $\mathbf{W}$-rep  \\
  represent. & \null & \null & \null  &  of $\mathrm{U}_{SK}$\\
  \hline
  Lie algebras & $\mathfrak{g}$ & $\mathfrak{g}_{SK}$ & $\mathfrak{g}_{QK}$ & $\mathbf{W}$-rep \\
  \hline
  action & global & on vector  & on  & on elect/magn.   \\
   \null & \null &  multiplets & hypermultiplets & $\mathbb{F}^M_{\mu\nu}$  \\
  \hline
  generators & $t_\alpha$ & $t_A$ & $t_a$  & $t_{AM}^{\phantom{AM}N}$ \\
  \hline
  range & $\alpha=1,\dots, {\rm dim}(\mathrm{G})$ & $A=1,\dots, {\rm dim}\mathrm{U}_{SK}$ & $a=1,\dots, {\rm dim}\mathrm{G}_{QK}$ & $M=1,\dots, 2 n_V$\\
  \null & \null & \null & \null & $\Lambda = 1,\dots, n_V$\\
  \hline
\end{tabular}
\end{center}
}
\par
The embedding tensor has the general form:
\begin{equation}
\{\Theta_M{}^\alpha\}=\{\Theta_M{}^A,\,\Theta_M{}^a\}\,,
\end{equation}
and defines the embedding of the gauge algebra $\mathfrak{g}_{gauge}=\{X_M\}$ inside $\mathfrak{g}$:
\begin{equation}
\label{Xcombo}
X_M=\Theta_M{}^A\,t_A+\Theta_M{}^a\,t_a\,.
\end{equation}
In the absence of hypermultiplets, the  components $\Theta_M{}^a$, $a=1,2,3$ running over the adjoint representation of the global symmetry ${\rm SO}(3)$, are the Fayet-Iliopoulos terms. The generators $t_A$ of $\mathfrak{g}_{SK}$ have a non-trivial ${\bf W}$-representation: $t_A=(t_{A M}{}^N)$ while the generators $t_a$ do not.
Thus we can define the following tensor:
\begin{equation}
\label{Xtensoro}
X_{MN}{}^P=\Theta_M{}^A\,t_{A N}{}^P\,\,;\,\,\,\,X_{MNP}=X_{MN}{}^Q\,\mathbb{C}_{QP}\,,
\end{equation}
where $\mathbb{C}$ is the $2n_V\times 2n_V$ skew-symmetric, invariant ${\rm Sp}(2n_V,\mathbb{R})$-matrix.
\par
 Gauge-invariance and supersymmetry of the action impose linear and quadratic constraints on $\Theta$:
 \begin{itemize}
 \item The {\bf linear constraints} are:
 \begin{equation}
 X_{MN}{}^M=0\,\,;\,\,\,X_{(MNP)}=0\,.\label{1constr}
 \end{equation}
 \item The {\bf quadratic constraints} originate from the condition that $X_M$ close an algebra inside $\mathfrak{g}$ with structure constants given in terms of $X_{MN}{}^P$, and from the condition that the symplectic vectors $\Theta_M{}^\alpha$, labeled by $\alpha$, be \emph{mutually local}:
     \begin{align}
     [X_M,\,X_N]&=-X_{MN}{}^P\,X_P\,,\label{2constr1}\\
     \mathbb{C}^{MN}\Theta_M{}^\alpha \Theta_N{}^\beta&=0\,\,\Leftrightarrow \,\,\,\Theta^{\Lambda\,[\alpha} \Theta_\Lambda{}^{\beta]}=0\label{2constr2}\,.
     \end{align}
     the former can be rewritten as the  following set of two equations:
     \begin{align}
     \Theta_M{}^A\Theta_N{}^B\,f_{AB}{}^C+ \Theta_M{}^A\,t_{A N}{}^P\,\Theta_P{}^C&=0\,,\label{2constr11}\\
       \Theta_M{}^a\Theta_N{}^b\,f_{ab}{}^c+ \Theta_M{}^A\,t_{A N}{}^P\,\Theta_P{}^c&=0\,,\label{2constr12}
     \end{align}
     where $f_{AB}{}^C,\,f_{ab}{}^c$ are the structure constants of $\mathfrak{g}_{SK}$ and $\mathfrak{g}_{QK}$, respectively.
     It can be shown that eq. s (\ref{1constr}) and (\ref{2constr11}) imply $\Theta^{\Lambda\,[A} \Theta_\Lambda{}^{B]}=0$, which is the part of (\ref{2constr2}) corresponding to $\alpha=A,\,\beta=B$.
 \end{itemize}
Let us now denote by $k_A^i,\,k_A^{\bar{\imath}}$ the Killing vectors on the Special K\"ahler manifold corresponding to the isometry generator $t_A$, $k_a^u$ the Killing vectors on the Quaternionic K\"ahler manifold corresponding to the isometry generator $t_a$, and $\mathcal{P}_a^x$, $x=1,2,3$, the corresponding momentum map (note that these quantities are, by definition, associated only with the geometry of the scalar manifold and thus independent of the gauging). The scalar potential can be written in the following way \cite{Andrianopoli:1997cm}
\begin{align}
\mathcal{V}&=\mathcal{V}_{hyperino}+\mathcal{V}_{gaugino,1}+\mathcal{V}_{gaugino,2}+\mathcal{V}_{gravitino}\,,\nonumber\\
\mathcal{V}_{hyperino}&= 4\,\overline{V}^M\,V^N\,\Theta_M{}^a\Theta_N{}^b\,k_a^u\,k_b^v\,h_{uv}\,,\nonumber\\
\mathcal{V}_{gaugino,1}&= \overline{V}^M\,V^N\,\Theta_M{}^A\Theta_N{}^B\,k_A^i\,k_B^{\bar{\jmath}}\,g_{i\bar{\jmath}}\,,\nonumber\\
\mathcal{V}_{gaugino,2}&= g^{i\bar{\jmath}}\,D_iV^M\,D_{\bar{\jmath}}\overline{V}^N\,\Theta_M{}^a\Theta_N{}^b\,\mathcal{P}_a^x\,\mathcal{P}_b^x\,,\nonumber\\
\mathcal{V}_{gravitino}&=-3\,\overline{V}^M\,V^N\,\Theta_M{}^a\Theta_N{}^b\,\mathcal{P}_a^x\,\mathcal{P}_b^x\,,
\end{align}
where $V^M$ is the covariantly holomorphic symplectic section of the Special K\"ahler manifold under consideration: $(V^M)=(L^\Lambda,\,M_\Lambda)$.
\subsection{\sc Scan of the STU Model Gaugings and Their Duality  Orbits}\label{gaustu}
Consider the STU model with no hypermultiplets. This corresponds to the sixth item in table \ref{homomodels} for $p=0$.
The global symmetry group is $ \mathrm{G} \, = \, {\rm SL}(2,\mathbb{R})^3\times {\rm SO}(3)$, the latter
factor being the form of  $\mathrm{G}_{QK}$ in the absence of hypermultiplets. It  is
relevant to our discussion only in the case we want to add FI terms,
i.e. when we introduce non vanishing  components $\Theta_M{}^a$, $a=1,2,3$.
\par
The symplectic
$\mathbf{W}$-representation of the electric-magnetic charges is the
$(\frac{1}{2},\frac{1}{2},\frac{1}{2})$ of $\mathrm{U}_{SK}={\rm
SL}(2,\mathbb{R})^3$. Let us use the indices $i,j,k=1,2$ to label
the fundamental   representation of ${\rm SL}(2,\mathbb{R})$.
As $\mathfrak{sl}(2)$-generators in this  spinor representation of $\so(2,1) \sim \slal(2)$, we make the following choice:
$\{s_x\}=\{\sigma_1,i\,\sigma_2,\,\sigma_3\}$, $\sigma_x$ being the
Pauli matrices. The index $M$ can be written as $M=(i_1,i_2,i_3)$
and the embedding tensor $\Theta_M{}^A$ takes the following form:
\begin{equation}
\Theta_M{}^A=\{\Theta_{(i_1,i_2,i_3)}{}^{x_1},\,\Theta_{(i_1,i_2,i_3)}{}^{x_2},\,\Theta_{(i_1,i_2,i_3)}{}^{x_3}\}\,\,\in\,\,\, \left(\frac{1}{2},\frac{1}{2},\frac{1}{2}\right)\times [(1,0,0)+(0,1,0)+(0,0,1)]\,,
\end{equation}
where $x_i$ run over the adjoint (vector)-representations of the three $\mathfrak{sl}(2)$ algebras.
 Since:
\begin{equation}
\left(\frac{1}{2},\frac{1}{2},\frac{1}{2}\right)\times [(1,0,0)+(0,1,0)+(0,0,1)]=3\times
 \left(\frac{1}{2},\frac{1}{2},\frac{1}{2}\right)+\left(\frac{3}{2},\frac{1}{2},\frac{1}{2}\right)+\left(\frac{1}{2},\frac{3}{2},\frac{1}{2}\right)+\left(\frac{1}{2},\frac{1}{2},\frac{3}{2}\right)\,,
\end{equation}
each component of the embedding tensor can be split into its $\left(\frac{1}{2},\frac{1}{2},\frac{1}{2}\right)$ and
$\left(\frac{3}{2},\frac{1}{2},\frac{1}{2}\right)$ irreducible parts as follows:
\begin{align}
\Theta_{(i_1,i_2,i_3)}{}^{x_1}&=(s^{x_1})_{i_1}{}^j\,\xi^{(1)}_{j\,i_2\,i_3}+\Xi_{i_1,i_2,i_3}{}^{x_1}\,,\nonumber\\
\Theta_{(i_1,i_2,i_3)}{}^{x_2}&=(s^{x_2})_{i_2}{}^j\,\xi^{(1)}_{i_1\,j\,i_3}+\Xi_{i_1,i_2,i_3}{}^{x_2}\,,\nonumber\\
\Theta_{(i_1,i_2,i_3)}{}^{x_3}&=(s^{x_3})_{i_3}{}^j\,\xi^{(1)}_{i_1\,i_2\,j}+\Xi_{i_1,i_2,i_3}{}^{x_3}\,,\nonumber\\
\end{align}
The irreducible $\left(\frac{3}{2},\frac{1}{2},\frac{1}{2}\right)$tensors $ \Xi_{i_1,i_2,i_3}{}^{x_i}$  are defined by the the vanishing of the approriate gamma-trace, namely:
\begin{equation}
 \Xi_{j,i_2,i_3}{}^{x_1} (s_{x_1})_{i_1}{}^j=\Xi_{i_1,j,i_3}{}^{x_2} (s_{x_2})_{i_2}{}^j=\Xi_{i_1,i_2,j}{}^{x_3} (s_{x_3})_{i_3}{}^j=0\,.
 \label{purlo}
\end{equation}
 Let us now define embedded  gauge generators $X_{MN}{}^P$:
 \begin{align}
 X_{(i_1,i_2,i_3),(j_1,j_2,j_3)}{}^{(k_1,k_2,k_3)}&=\Theta_{(i_1,i_2,i_3)}{}^{x_1}\,(s_{x_1})_{j_1}{}^{k_1}\delta_{j_2}^{k_2}\delta_{j_3}^{k_3}+
 \Theta_{(i_1,i_2,i_3)}{}^{x_2}\,(s_{x_2})_{j_2}{}^{k_2}\delta_{j_1}^{k_1}\delta_{j_3}^{k_3}+\nonumber\\
 &+\Theta_{(i_1,i_2,i_3)}{}^{x_3}\,(s_{x_3})_{j_3}{}^{k_3}\delta_{j_2}^{k_2}\delta_{j_1}^{k_1}\,.
 \label{fischietto}
 \end{align}
The linear  constraints (\ref{1constr}) become:
 \begin{align}
& X_{(i_1,i_2,i_3),(j_1,j_2,j_3)}{}^{(i_1,i_2,i_3)}=0\,\,\Rightarrow\,\,\,\,\xi^{(1)}_{i_1\,i_2\,i_3}+
 \xi^{(2)}_{i_1\,i_2\,i_3}+\xi^{(3)}_{i_1\,i_2\,i_3}=0\,,\nonumber\\
&X_{(i_1,i_2,i_3),(j_1,j_2,j_3),(k_1,k_2,k_3)}+X_{(k_1,k_2,k_3),(j_1,j_2,j_3),(i_1,i_2,i_3)}+
X_{(j_1,j_2,j_3),(i_1,i_2,i_3),(k_1,k_2,k_3)}=0\,\Rightarrow\nonumber\\&\Rightarrow\,\, \Xi_{i_1,i_2,i_3}{}^{x_i}=0\,.
 \end{align}
This corresponds to the elimination of the $\left(\frac{3}{2},\frac{1}{2},\frac{1}{2}\right)$ representation leaving us only with three tensors in the $\left(\frac{1}{2},\frac{1}{2},\frac{1}{2}\right)$ representation.
Explicitly the linearly constrained embedding tensor reads as follows:
 \begin{align}
\Theta_{(i_1,i_2,i_3)}{}^{A}&=\{(s^{x_1})_{i_1}{}^j\,\xi^{(1)}_{j\,i_2\,i_3},\,(s^{x_2})_{i_2}{}^j\,\xi^{(2)}_{i_1\,j\,i_3},\,
(s^{x_3})_{i_3}{}^j\,\xi^{(3)}_{i_1\,i_2\,j}\}\,,
\end{align}
and the additional linear constraint reduces further the independent tensors in the $\left(\frac{1}{2},\frac{1}{2},\frac{1}{2}\right)$ to two, since we get condition
\begin{equation}
\xi^{(1)}_{i_1\,i_2\,i_3}+
 \xi^{(2)}_{i_1\,i_2\,i_3}+\xi^{(3)}_{i_1\,i_2\,i_3}=0
\end{equation}
The quadratic condition for $\Theta_M{}^A$ has the form (\ref{2constr2}), which applied to our solution of the linear constraint takes the following appearance:
\begin{equation}
 \epsilon^{i_1 j_1}\epsilon^{i_2 j_2}\epsilon^{i_3 j_3}\,\Theta_{(i_1,i_2,i_3)}{}^{A}\,\Theta_{(j_1,j_2,j_3)}{}^{B}=0\,.
\end{equation}
 By means of a MATHEMATICA computer code we were able to find 36 solutions to this equation, all of which corresponding to non-semisimple gauge groups. We do not display them here, since, in section \ref{noFIgauge} we show how to classify the orbits into which such solutions are organized and it will be sufficient to consider only one representative for each orbit.
\subsubsection{\sc The special Geometry of the STU model}
 In equation (\ref{transformus}) we derived the transformation from the Calabi--Vesentini coordinates $\{S,y_{1,2}\}$ to a triplet of complex coordinates $z_{1,2,3}$ parameterizing the three identical copies of the coset manifold $\frac{\mathrm{SL(2,\mathbb{R})}}{\mathrm{SO(2)}}$ which compose this special instance of special K\"ahler manifold. Indeed setting:
 \begin{equation}\label{rinomo}
   {\rm i} e^{h} + b \, \equiv \, z_1 \quad ; \quad {\rm i} e^{h_1} + b_1 \, \equiv \, z_2 \quad ; \quad {\rm i} e^{h_2} + b_2 \, \equiv \, z_3
 \end{equation}
 the transformation (\ref{transformus}) can be  rewritten as follows:
\begin{equation}\label{transformer}
\begin{array}{lll}
 S & =  & z_1 \\
 y_1 & =  & -\frac{{\rm i}
   \left(z_2
   z_3+1\right)}{\left(z_2+{\rm i}\right) \left(z_3+{\rm i}\right)}
   \\
 y_2 & =  & \frac{{\rm i}
   \left(z_2-z_3\right)}{\left (z_2+{\rm i}\right) \left(z_3+{\rm i}\right)}
\end{array}
\end{equation}
In the sequel we will adopt the symmetric renaming of variables
\begin{equation}\label{bongobongo}
    z_i \, = \, {\rm i} e^{\mathfrak{h}_i} + \mathfrak{b}_i
\end{equation}
Applying the transformation (\ref{transformer}) of its arguments to the Calabi-Vesentini holomorphic section $\Omega_{CV} $, we find:
\begin{eqnarray}
  \Omega_{CV}(z) &=& \mathfrak{f}(z) \,\left(
\begin{array}{l}
\frac{z_2}{\sqrt{2}}+\frac{
   z_3}{\sqrt{2}} \\
 \frac{z_2
   z_3}{\sqrt{2}}-\frac{1}{\sqrt{2}} \\
 -\frac{z_2
   z_3}{\sqrt{2}}-\frac{1}{\sqrt{2}} \\
\frac{z_2}{\sqrt{2}}-\frac{
   z_3}{\sqrt{2}} \\
 \frac{z_1
   z_2}{\sqrt{2}}+\frac{z_1
   z_3}{\sqrt{2}} \\
 \frac{z_1 z_2
   z_3}{\sqrt{2}}-\frac{z_1}{\sqrt{2}} \\
 \frac{z_2 z_3
   z_1}{\sqrt{2}}+\frac{z_1}{\sqrt{2}} \\
 \frac{z_1
   z_3}{\sqrt{2}}-\frac{z_1
   z_2}{\sqrt{2}}
\end{array}
\right)\\
  \mathfrak{f}(z) &=& \frac{{\rm i} \, \sqrt{2}}{\left(z_2+{\rm i}\right)
   \left(z_3+{\rm i}\right)}
\end{eqnarray}
As it is well known the overall holomorphic factor $\mathfrak{f}(z)$ in front of the section has no consequences on the determination of the K\"ahler metric and  simply it adds the real part of a holomorphic function to the K\"ahler potential. Similarly, at the level of ungauged supergravity, the symplectic frame plays no role on the Lagrangian and we are free to perform any desired symplectic rotation on the section, the preserved symplectic metric being the following one:
\begin{equation}\label{CCmatruzza}
    \mathbb{C} \, = \, \left(
\begin{array}{llllllll}
 0 & 0 & 0 & 0 & 1 & 0 & 0 & 0
   \\
 0 & 0 & 0 & 0 & 0 & 1 & 0 & 0
   \\
 0 & 0 & 0 & 0 & 0 & 0 & 1 & 0
   \\
 0 & 0 & 0 & 0 & 0 & 0 & 0 & 1
   \\
 -1 & 0 & 0 & 0 & 0 & 0 & 0 &
   0 \\
 0 & -1 & 0 & 0 & 0 & 0 & 0 &
   0 \\
 0 & 0 & -1 & 0 & 0 & 0 & 0 &
   0 \\
 0 & 0 & 0 & -1 & 0 & 0 & 0 &
   0
\end{array}
\right)
\end{equation}
On the other hand at the level of \textit{gauge supergravity} the choice of the symplectic frame is physically relevant. The Calabi--Vesentini frame is that one where the $\mathrm{SO(2,2)}$ isometries of the manifold are all linearly realized on the electric vector field strengths, while the $\mathrm{SL(2,\mathbb{R})}$ factor acts as a group of electric/magnetic duality transformations. For this reason the CV frame was chosen in paper \cite{mapietoine}, since in such a frame it was easy to single out the non-compact gauge group $\mathrm{SO(2,1)}$. On the other the so named special coordinate frame which admits a description in terms of a prepotential, is that one where the three group factors $\mathrm{SL(2,\mathbb{R})}$ are all on the same footing and the $\mathbf{W}$-representation is identified as the $(2,2,2)\sim \left(\frac{1}{2},\frac{1}{2},\frac{1}{2}\right)$.
\par
The philosophy underlying the \textit{embedding tensor approach} to gaugings is that the embedding tensor already contains all possible symplectic frame choices, since it transforms as a good tensor under the symplectic group. Hence we can choose any preferred symplectic frame to start with.
\par
In view of these considerations we introduce the following symplectic matrix:
\begin{eqnarray}\label{symgroupel}
    \mathcal{S}& = & \left(
\begin{array}{llllllll}
 0 & 0 & \frac{1}{\sqrt{2}} &
   \frac{1}{\sqrt{2}} & 0 & 0
   & 0 & 0 \\
 -\frac{1}{\sqrt{2}} & 0 & 0 &
   0 & 0 & \frac{1}{\sqrt{2}}
   & 0 & 0 \\
 -\frac{1}{\sqrt{2}} & 0 & 0 &
   0 & 0 & -\frac{1}{\sqrt{2}}
   & 0 & 0 \\
 0 & 0 & \frac{1}{\sqrt{2}} &
   -\frac{1}{\sqrt{2}} & 0 & 0
   & 0 & 0 \\
 0 & 0 & 0 & 0 & 0 & 0 &
   \frac{1}{\sqrt{2}} &
   \frac{1}{\sqrt{2}} \\
 0 & -\frac{1}{\sqrt{2}} & 0 &
   0 & -\frac{1}{\sqrt{2}} & 0
   & 0 & 0 \\
 0 & \frac{1}{\sqrt{2}} & 0 &
   0 & -\frac{1}{\sqrt{2}} & 0
   & 0 & 0 \\
 0 & 0 & 0 & 0 & 0 & 0 &
   \frac{1}{\sqrt{2}} &
   -\frac{1}{\sqrt{2}}
\end{array}
\right)\label{symgroupebisl}\\
 \mathbb{C} &=& \mathcal{S}^T\, \mathbb{C} \, \mathcal{S}
\end{eqnarray}
and we introduce the symplectic section in the \textit{special coordinate frame} by setting
\begin{equation}\label{specialframe}
    \Omega_{SF}^M \, = \, \frac{1}{\mathfrak{f}(z)} \, \mathcal{S}^{-1}\, \Omega_{CV} \, = \, \left(\begin{array}{l}
 1 \\
 z_1 \\
 z_2 \\
 z_3 \\
 -z_1 z_2 z_3 \\
 z_2 z_3 \\
 z_1 z_3 \\
 z_1 z_2
\end{array}
\right) \, \equiv \,  \left(\begin{matrix}X^\Lambda (z)\cr
F_\Lambda(z)\end{matrix}\right)
\end{equation}
Note that this frame admits a prepotential description. Introducing the following holomorphic prepotential:
\begin{equation}
\mathcal{F}(z)=z^1 z^2 z^3=s t u\,,
\end{equation}
The  symplectic section $\Omega_{SF}$  can be written as follows:
\begin{equation}
\Omega_{SF}(z) =\left(1,\, \underbrace{z^m}_{m=1,2,3}, -\mathcal{F}(z), \underbrace{\frac{\partial  \mathcal{F}(z)}
{\partial z^m}}_{m=1,2,3} \right)\,,
\end{equation}
It is useful to work also with real fields $(\phi^r)=(\mathfrak{b}_m,\,\mathfrak{h}_m)$, defined  as in equation (\ref{bongobongo}).
The K\"ahler potential is expressed as follows:
\begin{equation}
\mathcal{K}(z^m,\bar{z})=-\log\left[-{\rm i}\,\Omega\mathbb{C}\bar{\Omega}\right])=
-\log\left[-{\rm i}\,(X^\Lambda \bar{F}_\Lambda-F_\Lambda
\bar{X}^\Lambda)\right]\,\label{Kom}
\end{equation}
and, in real coordinates we have:
\begin{equation}
e^{-\mathcal{K}}=8\, e^{\mathfrak{h} _1+\mathfrak{h} _2+\mathfrak{h} _3}\,.
\end{equation}
We also introduce the covariantly holomorphic symplectic section $V=e^{\mathcal{K}/2}\,\Omega_{SF}$ satisfying the condition:
\begin{equation}
\nabla_{\bar{a}}V\equiv
(\partial_{\bar{a}}-\frac{1}{2}\,\partial_{\bar{a}}\mathcal{K})V=0\,,
\label{covholo}
\end{equation}
and its covariant derivatives:
\begin{equation}
U_m=(U_m{}^M)\equiv
\nabla_{{m}}V=(\partial_m+\frac{1}{2}\,\partial_m\mathcal{K})V\,.
\end{equation}
The following properties hold:
\begin{align}
V\mathbb{C}\bar{V}&={\rm i}\,\,;\,\,\,U_m\mathbb{C}\bar{V}=\bar{U}_{\bar{m}}\mathbb{C}\bar{V}=0\,\,;\,\,\,
U_m\mathbb{C}\bar{U}_{\bar{n}}=-{\rm i}\,g_{m\bar{n}}\,.\label{props}
\end{align} If
$E_m{}^I$, $I=1,\dots, 3$,  is the complex vielbein matrix of the manifold,
$g_{m\bar{n}}=\sum_{I}E_m{}^I\bar{E}_{\bar{n}}{}^I$, and ${E}_I{}^m$
its inverse, we introduce the quantities $U_I\equiv E_I{}^m\,U_m$,
in terms of which the following $8\times 8$ matrix
$\hat{\mathbb{L}}_4=(\hat{\mathbb{L}}_4{}^M{}_N)$ is defined:
\begin{equation}
\hat{\mathbb{L}}_4(z,\bar{z})=\left(V,\overline{U}_{I},\,\overline{V},\,U_I\right)\mathcal{C}=\sqrt{2}\,\left({\rm Re}(V),\,{\rm
Re}(U_I),\,-{\rm Im}(V),{\rm Im}(U_I)\right)\,,
\end{equation}
where $\mathcal{C}$ is the Cayley matrix.
By virtue of eq.s (\ref{props}), the matrix $\hat{\mathbb{L}}_4$ is symplectic:
$\hat{\mathbb{L}}^T_4\mathbb{C}\hat{\mathbb{L}}_4=\mathbb{C}$.
\par
In order to find the coset representative $\mathbb{L}$ as an ${\rm Sp}(8,\mathbb{R})$ matrix in the solvable gauge, and the symplectic representation of the isometry generators $t_A$ in the special coordinate basis,
 we proceed as follows. We construct a symplectic matrix $\mathcal{L}$ which coincides with the identity at the origin where $\phi^r\equiv 0\,\Leftrightarrow\,\,\mathfrak{h}_m=\mathfrak{b}_m=0$:
 \begin{equation}
 \mathcal{L}(\phi^r)=\hat{\mathbb{L}}_4(\phi^r)\,\hat{\mathbb{L}}_4(\phi^r\equiv  0)^{-1}\,.
 \end{equation}
 The following property holds:
 \begin{equation}
 V(\phi^r)= \mathbb{L}(\phi^r)\,V(\phi^r\equiv  0)\,.
 \end{equation}
The matrix $ \mathbb{L}$ is the coset representative in the solvable gauge. To show this we compute the following generators:
\begin{equation}
{\bf h}_m=\left.\frac{\partial \mathbb{L}}{\partial
\mathfrak{h}_m}\right\vert_{\phi^r\equiv 0}\,\,;\,\,\,{\bf
a}_m=\left.\frac{\partial \mathbb{L}}{\partial
\mathfrak{b}_m}\right\vert_{\phi^r\equiv 0}\,.
\end{equation}
These generators close a solvable Lie algebra $Solv$ which is the Borel subalgebra of $\mathfrak{g}_{SK}$. The above construction is general and applies to any symmetric Special K\"ahler manifold. In our case $Solv=Solv_2^{(1)}\oplus Solv_2^{(2)}\oplus Solv_2^{(3)}$, where
\begin{equation}
Solv_2^{(m)}\equiv \{{\bf h}_m,\,{\bf
a}_m\}\,\,;\,\,\,[{\bf h}_m,\,{\bf
a}_n]=\delta_{mn}\,{\bf
a}_n\,\,;\,\,\,[Solv_2^{(m)},\,Solv_2^{(n)}]=0,.
\end{equation}
One can verify that
\begin{equation}
\mathbb{L}(\mathfrak{h}_m,\,\mathfrak{b}_m)=\mathbb{L}_{axion}(\mathfrak{b}_m)\,\mathbb{L}_{dilaton}(\mathfrak{h}_m)=
e^{\mathfrak{b}_m\,{\bf a}_m}\,e^{\mathfrak{h}_m\,{\bf h}_m}\,.\label{cosetrep}
\end{equation}
Each $\mathfrak{sl}(2)$ algebra is spanned by $\{{\bf h}_m,\,{\bf a}_m,\,{\bf a}^T_m\}$, $[{\bf a}_m,\,{\bf
a}_n^T]=2\,\delta_{mn}\,{\bf
h}_n$. The explicit matrix representations of these generators is:
\begin{align}
{\bf h}_1 &=\left(
\begin{array}{llllllll}
 -\frac{1}{2} & 0 & 0 & 0 & 0 & 0 & 0 & 0 \\
 0 & \frac{1}{2} & 0 & 0 & 0 & 0 & 0 & 0 \\
 0 & 0 & -\frac{1}{2} & 0 & 0 & 0 & 0 & 0 \\
 0 & 0 & 0 & -\frac{1}{2} & 0 & 0 & 0 & 0 \\
 0 & 0 & 0 & 0 & \frac{1}{2} & 0 & 0 & 0 \\
 0 & 0 & 0 & 0 & 0 & -\frac{1}{2} & 0 & 0 \\
 0 & 0 & 0 & 0 & 0 & 0 & \frac{1}{2} & 0 \\
 0 & 0 & 0 & 0 & 0 & 0 & 0 & \frac{1}{2}
\end{array}
\right)\,\,;\,\,\,{\bf a}_1=\left(
\begin{array}{llllllll}
 0 & 0 & 0 & 0 & 0 & 0 & 0 & 0 \\
 1 & 0 & 0 & 0 & 0 & 0 & 0 & 0 \\
 0 & 0 & 0 & 0 & 0 & 0 & 0 & 0 \\
 0 & 0 & 0 & 0 & 0 & 0 & 0 & 0 \\
 0 & 0 & 0 & 0 & 0 & -1 & 0 & 0 \\
 0 & 0 & 0 & 0 & 0 & 0 & 0 & 0 \\
 0 & 0 & 0 & 1 & 0 & 0 & 0 & 0 \\
 0 & 0 & 1 & 0 & 0 & 0 & 0 & 0
\end{array}
\right)\,,\nonumber\\
{\bf h}_2 &=\left(
\begin{array}{llllllll}
 - \frac{1}{2} & 0 & 0 & 0 & 0 & 0 & 0 & 0 \\
 0 & - \frac{1}{2} & 0 & 0 & 0 & 0 & 0 & 0 \\
 0 & 0 & \frac{1}{2} & 0 & 0 & 0 & 0 & 0 \\
 0 & 0 & 0 & - \frac{1}{2} & 0 & 0 & 0 & 0 \\
 0 & 0 & 0 & 0 & \frac{1}{2} & 0 & 0 & 0 \\
 0 & 0 & 0 & 0 & 0 & \frac{1}{2} & 0 & 0 \\
 0 & 0 & 0 & 0 & 0 & 0 & - \frac{1}{2} & 0 \\
 0 & 0 & 0 & 0 & 0 & 0 & 0 & \frac{1}{2}
\end{array}
\right)\,\,;\,\,\,{\bf a}_2=\left(
\begin{array}{llllllll}
 0 & 0 & 0 & 0 & 0 & 0 & 0 & 0 \\
 0 & 0 & 0 & 0 & 0 & 0 & 0 & 0 \\
 1 & 0 & 0 & 0 & 0 & 0 & 0 & 0 \\
 0 & 0 & 0 & 0 & 0 & 0 & 0 & 0 \\
 0 & 0 & 0 & 0 & 0 & 0 & - 1 & 0 \\
 0 & 0 & 0 & 1 & 0 & 0 & 0 & 0 \\
 0 & 0 & 0 & 0 & 0 & 0 & 0 & 0 \\
 0 & 1 & 0 & 0 & 0 & 0 & 0 & 0
\end{array}
\right)\,,\nonumber\\
{\bf h}_3 &=\left(
\begin{array}{llllllll}
 - \frac{1}{2} & 0 & 0 & 0 & 0 & 0 & 0 & 0 \\
 0 & - \frac{1}{2} & 0 & 0 & 0 & 0 & 0 & 0 \\
 0 & 0 & - \frac{1}{2} & 0 & 0 & 0 & 0 & 0 \\
 0 & 0 & 0 & \frac{1}{2} & 0 & 0 & 0 & 0 \\
 0 & 0 & 0 & 0 & \frac{1}{2} & 0 & 0 & 0 \\
 0 & 0 & 0 & 0 & 0 & \frac{1}{2} & 0 & 0 \\
 0 & 0 & 0 & 0 & 0 & 0 & \frac{1}{2} & 0 \\
 0 & 0 & 0 & 0 & 0 & 0 & 0 & - \frac{1}{2}
\end{array}
\right)\,\,;\,\,\,{\bf a}_3=\left(
\begin{array}{llllllll}
 0 & 0 & 0 & 0 & 0 & 0 & 0 & 0 \\
 0 & 0 & 0 & 0 & 0 & 0 & 0 & 0 \\
 0 & 0 & 0 & 0 & 0 & 0 & 0 & 0 \\
 1 & 0 & 0 & 0 & 0 & 0 & 0 & 0 \\
 0 & 0 & 0 & 0 & 0 & 0 & 0 & - 1 \\
 0 & 0 & 1 & 0 & 0 & 0 & 0 & 0 \\
 0 & 1 & 0 & 0 & 0 & 0 & 0 & 0 \\
 0 & 0 & 0 & 0 & 0 & 0 & 0 & 0
\end{array}
\right)\,.
\end{align}
The axionic and dilatonic parts of the coset representative in (\ref{cosetrep}) have the following matrix form:
\begin{align}
\mathbb{L}_{axion}(\mathfrak{b}_m)&=e^{\mathfrak{b}_m\,{\bf a}_m}=\left(
\begin{array}{llllllll}
 1 & 0 & 0 & 0 & 0 & 0 & 0 & 0 \\
 \mathfrak{b}_1 & 1 & 0 & 0 & 0 & 0 & 0 & 0 \\
 \mathfrak{b}_2 & 0 & 1 & 0 & 0 & 0 & 0 & 0 \\
 \mathfrak{b}_3 & 0 & 0 & 1 & 0 & 0 & 0 & 0 \\
 -\mathfrak{b}_1 \mathfrak{b}_2 \mathfrak{b}_3 & -\mathfrak{b}_2 \mathfrak{b}_3 & -\mathfrak{b}_1 \mathfrak{b}_3 & -\mathfrak{b}_1 \mathfrak{b}_2 & 1 & -\mathfrak{b}_1 & -\mathfrak{b}_2 & -\mathfrak{b}_3 \\
 \mathfrak{b}_2 \mathfrak{b}_3 & 0 & \mathfrak{b}_3 & \mathfrak{b}_2 & 0 & 1 & 0 & 0 \\
 \mathfrak{b}_1 \mathfrak{b}_3 & \mathfrak{b}_3 & 0 & \mathfrak{b}_1 & 0 & 0 & 1 & 0 \\
 \mathfrak{b}_1 \mathfrak{b}_2 & \mathfrak{b}_2 & \mathfrak{b}_1 & 0 & 0 & 0 & 0 & 1
\end{array}
\right)\,,\nonumber\\
\mathbb{L}_{dilaton}(\mathfrak{h}_m)&=e^{\mathfrak{h}_m\,{\bf h}_m}={\rm diag}\left(e^{-\frac{\mathfrak{h} _1}{2}-\frac{\mathfrak{h} _2}{2}-\frac{\mathfrak{h} _3}{2}},e^{\frac{\mathfrak{h} _1}{2}-\frac{\mathfrak{h} _2}{2}-\frac{\mathfrak{h}
   _3}{2}},e^{-\frac{\mathfrak{h} _1}{2}+\frac{\mathfrak{h} _2}{2}-\frac{\mathfrak{h} _3}{2}},e^{-\frac{\mathfrak{h} _1}{2}-\frac{\mathfrak{h} _2}{2}+\frac{\mathfrak{h}
   _3}{2}},\right.\nonumber\\&\left.e^{\frac{\mathfrak{h} _1}{2}+\frac{\mathfrak{h} _2}{2}+\frac{\mathfrak{h} _3}{2}},e^{-\frac{\mathfrak{h} _1}{2}+\frac{\mathfrak{h} _2}{2}+\frac{\mathfrak{h}
   _3}{2}},e^{\frac{\mathfrak{h} _1}{2}-\frac{\mathfrak{h} _2}{2}+\frac{\mathfrak{h} _3}{2}},e^{\frac{\mathfrak{h} _1}{2}+\frac{\mathfrak{h} _2}{2}-\frac{\mathfrak{h}
   _3}{2}}\right)\,,\nonumber\\
\end{align}
 To make contact with the discussion about the embedding tensor provided in the previous section, we define the transformation from the basis $(i_1,i_2,i_3)$ and the special coordinate symplectic frame. We start with an ordering of the independent components of a vector  $W_{i_1,i_2,i_3}$, which defines a symplectic basis to be dubbed ``old'':
 \begin{equation}
W^{old} =(W^{old}_M)=\left(W_{ 1,1,1 },W_{ 1,1,2 },W_{ 1,2,1 },W_{ 1,2,2 },W_{ 2,1,1 },W_{ 2,1,2 },W_{ 2,2,1 },W_{ 2,2,2 }
\right)\,.
 \end{equation}
 The new special coordinate basis is related to the old one by an orthogonal transformation  $\mathcal{O}$:
 \begin{align}
 W^{s.c.}_M&=\mathcal{O}_M{}^N\,W^{old}_M\,\,;\,\,\,\,
 \mathcal{O}=\frac{1}{2\sqrt{2}}\left(
\begin{array}{llllllll}
 1 & -1 & -1 & 1 & -1 & 1 & 1 & -1 \\
 -1 & 1 & 1 & -1 & -1 & 1 & 1 & -1 \\
 -1 & 1 & -1 & 1 & 1 & -1 & 1 & -1 \\
 -1 & -1 & 1 & 1 & 1 & 1 & -1 & -1 \\
 1 & 1 & 1 & 1 & 1 & 1 & 1 & 1 \\
 1 & 1 & 1 & 1 & -1 & -1 & -1 & -1 \\
 1 & 1 & -1 & -1 & 1 & 1 & -1 & -1 \\
 1 & -1 & 1 & -1 & 1 & -1 & 1 & -1
\end{array}
\right)\,.
 \end{align}
 The $\mathfrak{sl}(2)^3$ generators $t_A$ in the old basis read:
 \begin{align}
 (t_{x_1})_{j_1,j_2, j_3}{}^{k_1,k_2, k_3}&=(s_{x_1})_{j_1}{}^{k_1}\delta_{j_2}^{k_2}\delta_{j_3}^{k_3}\,,;\,\,
 (t_{x_2})_{j_1,j_2, j_3}{}^{k_1,k_2, k_3}=(s_{x_2})_{j_2}{}^{k_2}\delta_{j_1}^{k_1}\delta_{j_3}^{k_3}\,,;\nonumber\\
 (t_{x_3})_{j_1,j_2, j_3}{}^{k_1,k_2, k_3}&=(s_{x_3})_{j_3}{}^{k_3}\delta_{j_1}^{k_1}\delta_{j_2}^{k_2}\,,
 \end{align}
 In the new basis their representation is deduced from their relation to the $Solv$ generators and their transpose:
 \begin{equation}
 t_{1_m}=2\,\,{\bf h}_m\,\,;\,\,\,t_{2_m}= {\bf a}_m-{\bf a}_m^T\,\,;\,\,\,t_{3_m}= -{\bf a}_m-{\bf a}_m^T\,.
 \end{equation}
 The commutation relations among them read:
 \begin{equation}
 [t_{x_m},\,t_{y_n}]=-2\,\delta_{mn}\,\epsilon_{xy}{}^z\,t_{z_n}\,,
 \end{equation}
 where the adjoint index is raised with $\eta_{xy}={\rm diag}(+1,-1,+1)$.
 \paragraph{\sc The Killing Vectors}
 A standard procedure in coset geometry allows to compute the Killing vectors $\{k_A\}=\{k_{x_m}^r\,\frac{\partial}{\partial \phi^r}\}_{m=1,2,3}$:
 \begin{align}
 k_{1_m}&=-2 \,(\partial_{\mathfrak{h}_m}+\mathfrak{b}_m\,\partial_{\mathfrak{b}_m})\,\,;\,\,\,k_{2_m}=2 \mathfrak{b}_m\,\partial _{\mathfrak{h} _m} +\left(\mathfrak{b}_m^2-e^{2 \mathfrak{h} _m}+1\right)\,\partial _{\mathfrak{b}_m}\,,\nonumber\\
 k_{3_m}&=-2 \mathfrak{b}_m\,\partial _{\mathfrak{h} _m} - \left(\mathfrak{b}_m^2-e^{2 \mathfrak{h} _m}-1\right)\,\partial _{\mathfrak{b}_m}\,.
 \end{align}
 For the purpose of computing the scalar potential, it is convenient
 to compute the holomorphic Killing vectors
 $k^m,\,k^{\bar{m}}$. To this end we solve the equation:
 \begin{equation}
\delta_\alpha \Omega(z)^N=-\Omega(z)^M\,t_{\alpha
M}{}^N=k_\alpha^m\,\partial_m\Omega(z)+\ell_\alpha\,\Omega(z)\,,
 \end{equation}
and find:
\begin{equation}
k_{1_m}=-2\,z^m\,\partial_m\,\,;\,\,\,k_{2_m}=(1+(z^m)^2)\,\partial_m\,\,;\,\,\,k_{3_m}=(1-(z^m)^2)\,\partial_m\,.
\end{equation}
These are conveniently expressed in terms of a holomorphic
prepotential $\mathcal{P}_\alpha(z)$:
\begin{align}
\mathcal{P}_\alpha&=-\overline{V}^M\,t_{\alpha
M}{}^N\,\mathbb{C}_{NL}\,V^L\,,\nonumber\\
\mathcal{P}_{1_m}&=-i\,\frac{z^m+\bar{z}^m}{z^m-\bar{z}^m}\,,\,\,\mathcal{P}_{2_m}=i\,\frac{1+|z^m|^2}{z^m-\bar{z}^m}\,,\,\,\mathcal{P}_{2_m}=i\,\frac{1-|z^m|^2}{z^m-\bar{z}^m}\,,
\end{align}
the relation being:
\begin{equation}
k^{\bar{m}}_\alpha=-i\,g^{\bar{m}n}\,\partial_n\mathcal{P}_\alpha\,.
\end{equation}
\subsubsection{{\sc The gaugings with no Fayet Iliopoulos terms}}
 \label{noFIgauge}
 We first consider the case of no Fayet Iliopoulos terms, namely
 ($\Theta_M{}^a=0$). We can use the global symmetry $\mathrm{G}$  of the
 theory to simplify our analysis. Indeed the field equations and
 Bianchi identities are invariant if we $\mathrm{G}$-transform the
 field and embedding tensors at the same time. This is in particular
 true for the scalar potential $\mathcal{V}(\phi, \Theta)$:
 \begin{equation}
\forall g\in G\,\,\,:\,\,\,\,\mathcal{V}(\phi,
\Theta)=\mathcal{V}(g\star\phi, g\star\Theta)\,,\label{G-inv}
 \end{equation}
where $(g\star\phi)^r$ are the scalar fields obtained from $\phi^r$
by the action of the isometry $g$, and $(g\star\Theta)_M{}^\alpha$
is the $g$-transformed embedding tensor. Notice that we can have
other formal symmetries of the potential which are not in $\mathrm{U}_{SK}$.
Consider for instance the symplectic transformation:
\begin{equation}
\mathcal{S}={\rm
diag}(1,\varepsilon_m,1,\varepsilon_m)\,,\label{Sep}
\end{equation}
where $\varepsilon_m=\pm 1$,
$\varepsilon_1\varepsilon_2\varepsilon_3=1$. These transformations
correspond to the isometries $z^m\rightarrow \varepsilon_m\,z^m$,
which however do not preserve the physical domain defined by the
upper half plane for each complex coordinate: ${\rm Im}(z^m)>0$.
Therefore embedding tensors connected by such transformations are to
be regarded as physically inequivalent.
\par
 We have shown in sect. \ref{gaustu} that the embedding tensor, solution to the linear constraints and in the absence of Fayet Iliopoulos terms, is parameterized by two independent tensors $\xi^{(2)},\,\xi^{(3)}$ in the
 $\left(\frac{1}{2},\frac{1}{2},\frac{1}{2}\right)$ of $\mathrm{U}_{SK}$.
 These are then subject to the quadratic constraints that restrict the $\mathrm{U}_{SK}$-orbits of these two quantities.
 We can think of acting by means of $\mathrm{U}_{SK}$ on $\xi^{(2)}$, so as
 to make it the simplest possible. By virtue of eq. (\ref{G-inv}) this will not
 change the physics of the gauged model (vacua, spectra,
 interactions), but just make their analysis simpler.
 \par
Let us recall that the $\mathrm{U}_{SK}$-orbits of a single object, say
$\xi^{(2)M}$, in the
$\mathbf{W} = \left(\frac{1}{2},\frac{1}{2},\frac{1}{2}\right)$ representation
are described by a quartic invariant $\mathfrak{I}_4(\xi^{(2)})$, defined as:
\begin{equation}
\mathfrak{I}_4(\xi^{(2)})=-\frac{2}{3}\,t_{A\,M
N}\,t^A{}_{PQ}\,\xi^{(2)M}\xi^{(2)N}\xi^{(2)P}\xi^{(2)Q}\,.
\end{equation}
A very important observation is that by definition the $\mathbf{W}$ representation is that of the electro--magnetic-charges of a black-hole solution of ungauged supergravity. Hence the components of the $\xi^{(2)}$-tensor could be identified with the charges $\mathcal{Q}$ of such a black-hole and the classification of the orbits of $\mathrm{U}_{SK}$ in the representation $\mathbf{W}$ coincides with the classification of Black-Hole solutions. The quartic invariant is just the same that in the Black-Hole case determines the area of the horizon. Here we make the first contact with the profound relation that links the black-hole potentials with the gauging potentials.
The orbits in the $\left(\frac{1}{2},\frac{1}{2},\frac{1}{2}\right)$-representation are classified as follows\footnote{Strictly speaking, for all models in the sixth line of table \ref{homomodels}, there is a further fine structure (see  \cite{Borsten:2011ai}) in some of the orbits classified above which  depends on the $\mathrm{U}_{SK}$-invariant sign of the time-like component (denoted by $\mathcal{I}_2$) of the 3-vector $s^{x\,\alpha\beta} \xi^{(2)}_{\alpha\alpha_1\alpha_2}\xi^{(2)}_{\beta\beta_1\beta_2}\epsilon^{\alpha_1\beta_1}\epsilon^{\alpha_2\beta_2}$ (i.e. the $x=2$ component in our conventions). This further splitting in the STU model, however, is not relevant since yields isomorphic orbits.}:
\begin{itemize}
\item[i)] Regular, $\mathfrak{I}_4>0$, and there exists a $\mathbb{Z}_3$-centralizer;
\item[ii)]Regular, $\mathfrak{I}_4>0$, no $\mathbb{Z}_3$-centralizer;
\item[iii)]Regular, $\mathfrak{I}_4<0$;
\item[iv)]\emph{Light-like}, $\mathfrak{I}_4=0$, $\partial_M I_4\neq 0$;
\item[v)]\emph{Critical}, $\mathfrak{I}_4=0$, $\partial_M \mathfrak{I}_4= 0$, $t_A{}^{MN}\,\partial_M \partial_N\,I_4\neq 0$ ;
\item[vi)]\emph{Doubly critical}, $\mathfrak{I}_4=0$, $\partial_M \mathfrak{I}_4= 0$, $t_A{}^{MN}\,\partial_M \partial_N\,\mathfrak{I}_4=
0$ ,
\end{itemize}
where $\partial_M\equiv \partial/\partial \xi^{(2)M}$. The quadratic
constraints (\ref{2constr11}) restrict $\xi^{(2)}$ (and $\xi^{(3)}$)
to be either in the  \emph{critical} or in the
\emph{doubly-critical} orbit. Let us analyze the two cases
separately.
\paragraph{\sc $\xi^{(2)}$ Critical.}
The quadratic constraints imply $\xi^{(3)}=0$ and thus the embedding
tensor is parameterized by $\xi^{(1)}=\xi^{(2)}$, namely the diagonal
of the first two $\mathrm{SL}(2,\mathbb{R})$ groups in $\mathrm{G}_{SK}$. We
can choose  a representative of the orbit in the form:
\begin{equation}
\xi^{(2)}=g\,(0,1,c,0,0,0,0,0)\,.
\end{equation}
The scalar potential reads:
\begin{equation}
\mathcal{V}=\mathcal{V}_{gaugino,\,1}=
g^2\,e^{-\mathfrak{h}_1-\mathfrak{h}_2-\mathfrak{h}_3}\,\left((\mathfrak{b}_1+c\,\mathfrak{b}_2)^2+(e^{\mathfrak{h}_1}-c\,e^{\mathfrak{h}_2})^2\right)\,.
\end{equation}
The truncation to the dilatons ($\mathfrak{b}_m \, = \, 0$) is a consistent one:
\begin{equation}
\left.\frac{\partial \mathcal{V}}{\partial
\mathfrak{b}_m}\right\vert_{\mathfrak{b}_m=0}=0\,,
\end{equation}
and
\begin{equation}
\left.\mathcal{V}\right\vert_{\mathfrak{b}_m=0}=g^2\,\left(e^{-\frac{1}{2}(-\mathfrak{h}_1+\mathfrak{h}_2+\mathfrak{h}_3)}
-c\,e^{-\frac{1}{2}(\mathfrak{h}_1-\mathfrak{h}_2+\mathfrak{h}_3)}\right)^2\,.
\end{equation}
The above potential has an extremum if $c>0$, for
$e^{\mathfrak{h}_1}=c\,e^{\mathfrak{h}_2}$, while it is runaway if $c<0$. The
sign of $c$ is changed by a transformation of the kind (\ref{Sep})
with $\varepsilon_1=-\varepsilon_2=-\varepsilon_3=1$. For the reason
outlined above, in passing from a negative to a positive $c$, the
critical point of the potential moves to the unphysical domain
(${\rm Im}(z^2) < 0$). The gauging for $c=-1$ coincides with the one
considered  in \cite{mapietoine}, in the absence of Fayet Iliopoulos terms, with
potential (compare with eq.(\ref{Potentabel}):
\begin{equation}
\mathcal{V}_{CV}\, = \, \frac{e_0^2}{2\,{\rm
Im}(S)}\,\frac{P_2^+(y)}{P_2^-(y)}\,,
\end{equation}
where $P_2^\pm(y)=1-2 \,y_0\bar{y}_0\pm
2\,y_1\bar{y}_1+y^2\bar{y}^2$ and $y^2=y_0^2+y_1^2$. The two
potentials  are connected by the transformation relations between
the Calabi-Vesentini and the special coordinates spelled out in eq.(\ref{transformer})
and by setting $e_0=2\sqrt{2}\,g$.
\paragraph{\sc $\xi^{(2)}$ Doubly-Critical.}
We can choose  a representative of the orbit in the form:
\begin{equation}
\xi^{(2)}=g\,(1,0,0,0,0,0,0,0)\,.
\end{equation}
In this case $\xi^{(3)}$ is non-vanishing and has the form:
\begin{equation}
\xi^{(3)}=g'\,(1,0,0,0,0,0,0,0)\,.
\end{equation}
The gauging is electric ($\Theta^\Lambda=0$) and the gauge
generators $X_\Lambda=(X_0,X_m)$, $m=1,2,3$, satisfy the following
commutation relations:
\begin{equation}
[X_0,\,X_m]=M_m{}^n\,X_n,\,\,\,\,M_m{}^n={\rm
diag}(-2\,(g+g'),2\,g,\,2\,g')\,,
\end{equation}
all other commutators being zero. This gauging originates from a
Scherk-Schwarz reduction from $D=5$, in which the semisimple global
symmetry generator defining the reduction is the 2-parameter
combination $M_m{}^n$ of the $\mathfrak{so}(1,1)^2$ global symmetry
generators of the $D=5 $ parent theory.\par The scalar potential is
axion-independent and reads:
\begin{equation}
\mathcal{V}= (g^2+g
g'+g^{'2})\,e^{-\mathfrak{h}_1-\mathfrak{h}_2-\mathfrak{h}_3}\,.
\end{equation}
This potential is trivially integrable since it contains only one exponential of a single scalar field combination.
\subsubsection{\sc Adding ${\rm U}(1)$ Fayet Iliopoulos terms}
Let us now consider adding a component of the embedding tensor along
one generator of the ${\rm SO}(3)$ global symmetry group:
$\theta_M=\Theta_M{}^{a=1}$. The constraints on $\theta_M$ are
(\ref{2constr2}) and (\ref{2constr12}), which read:
\begin{equation}
\theta_M\,\mathbb{C}^{MP}\,X_{PN}{}^Q=0\,\,,\,\,\,X_{PN}{}^Q\,\theta_Q=0\,.
\end{equation}
while the constraints (\ref{2constr11}) on $\Theta_M{}^A$ are just
the same as before and induce the same restrictions on the orbits of
$\xi^{(2)},\,\xi^{(3)}$.  Clearly if $X_{PN}{}^Q=0$, namely
$\Theta_M{}^A=0$, no SK isometries are gauged and there are no
constraints on $\theta_M$. We shall consider this case
separately.\par The potential reads
\begin{align}
\mathcal{V}&= \mathcal{V}_{gaugino,\,1}+ \mathcal{V}_{gaugino,\,2}+
\mathcal{V}_{gravitino}\,,
\end{align}
where $\mathcal{V}_{gaugino,\,1}$ was constructed in the various
cases in the previous section, while:
\begin{align}
\mathcal{V}_{gaugino,\,2}+
\mathcal{V}_{gravitino}=\left(g^{\bar{m}n}\,\mathcal{D}_{\bar{m}}
\overline{V}^{M} \mathcal{D}_n\,V^{N}-3\,\overline{V}^{M}
V^{N}\right)\,\theta_M\,\theta_N\,, \label{salianka}
\end{align}
has just the form of an $\mathcal{N}=1$ potential generated by a superpotential:
\begin{equation}\label{superpatata}
    W_h \,=\, \theta_M \, \Omega_{SF}^M
\end{equation}
as discussed later in eq. (\ref{frattocchia}).
It is interesting to rewrite the above contribution to the potential
in terms of quantities which are familiar in  the context of  black
holes in supergravity. We use the property:
\begin{equation}
g^{\bar{m}n}\,\mathcal{D}_{\bar{m}} \overline{V}^{(M}
\mathcal{D}_n\,V^{N)}+\overline{V}^{(M}
V^{N)}=-\frac{1}{2}\,\mathcal{M}^{-1\,MN}\,,\label{contriV}
\end{equation}
where $\mathcal{M}_{MN}$ is the   symplectic, symmetric,
negative-definite matrix defined later in eq. (\ref{inversem4}) in terms of the $\mathcal{N}_{\Lambda\Sigma}(z,{\bar z})$ matrix which appears in the $D=4$ Lagrangian (See eq.(\ref{d4generlag})).
Let us now define
the complex quantity $Z=V^M\,\theta_M$. The FI contribution to the
scalar potential (\ref{contriV}) can be recast in the form:
\begin{equation}
\mathcal{V}_{gaugino,\,2}+
\mathcal{V}_{gravitino}=-\frac{1}{2}\,\theta_M\,\mathcal{M}^{-1\,MN}\,\theta_N-4\,|Z|^2=V_{BH}-4\,|Z|^2\,.
\label{madrileno}
\end{equation}
The first term has the same form as the (positive-definite) effective
potential for a static black hole with charges
$\mathcal{Q}^M=\mathbb{C}^{MN}\theta_N$, while the second one is the
squared modulus of the black hole \emph{central charge}. Notice that
we can also write
\begin{equation}
V_{BH}=-\frac{1}{2}\,\theta_M\,\mathcal{M}^{-1\,MN}\,\theta_N=|Z|^2+g^{\bar{m}n}\,D_{\bar{m}}
\overline{Z} D_n\,Z>0\,.
\end{equation}
Let us now study the full scalar potential in the relevant cases.
\paragraph{\sc $\xi^{(2)}$ Critical.}
In this case, choosing
\begin{equation}
\xi^{(2)}=g\,(0,1,c,0,0,0,0,0)\,.
\end{equation}
we find for $\theta_M$ the following general solution to the
quadratic constraints:
\begin{equation}
\theta_M=(0,\frac{f_1}{c},\,f_1,\,0,\,0,\,f_2,\,\frac{f_2}{c},\,0)\,,
\end{equation}
where $f_1,\,f_2$ are constants.
\par The scalar potential reads:
\begin{equation}
\label{guliashi}
\mathcal{V}=g^2\,e^{-\mathfrak{h}_1-\mathfrak{h}_2-\mathfrak{h}_3}\,\left((\mathfrak{b}_1+c\,\mathfrak{b}_2)^2+(e^{\mathfrak{h}_1}-c\,e^{\mathfrak{h}_2})^2\right)+
\frac{e^{-\mathfrak{h}_3}}{c}\,\left[(f_1+f_2\,\mathfrak{b}_3)^2+f_2^2\,e^{2\,\mathfrak{h}_3}\right]\,.
\end{equation}
The above potential (if $g>0$) has an extremum only for
$c<0,\,f_2>0$ and:
\begin{equation}
\mathfrak{h}_1=\mathfrak{h}_2+\log(-c),\,\mathfrak{h}_3=-\log\left(-\frac{2c
g}{f_2}\right)\,,\,\,\,y_3=-\frac{f_1}{f_2}\,,\,\,y_1=-c\,y_2\,.
\end{equation}
The potential at the extremum is
\begin{equation}
\mathcal{V}_0= 4\,g\,f_2>0\,,
\end{equation}
while the squared scalar mass matrix reads:
\begin{equation}
\left.(\partial_r\partial_s \mathcal{V}\,g^{st})\right\vert_0={\rm
diag}(2,2,1,1,0,0)\times \mathcal{V}_0\,.
\end{equation}
In this way we retrieve the stable dS vacuum of \cite{mapietoine}, discussed in Subsect. \ref{stabdesitter}, the two
parameters $f_1,\,f_2$ being related to $e_1$ and the de
Roo-Wagemann's angle.
\paragraph{\sc $\xi^{(2)}$ Doubly-Critical.}
In this case the constraints on $\theta_M$ impose:
\begin{align}
(g+g')\theta_1&=0\,,\,g\,\theta_2=0\,,\,g'\,\theta_3=0\,,\,g\,\theta^0=g'\,\theta^0=0\,,\,g\,\theta^1=g'\,\theta^1=0\,,\,
g\,\theta^2=g'\,\theta^2=0\,,\nonumber\\
g\,\theta^3&=g'\,\theta^3=0\,.
\end{align}
Under these conditions, unless $g=g'=0$, which is the case we shall
consider next, the FI contribution to the scalar potential vanishes.
\paragraph{\sc Case $\Theta_M{}^A=0$. Pure Fayet Iliopoulos gauging.}
In this case, we can act on $\theta_M$ by means of $\mathrm{U}_{SK}$ and
reduce it the theta vector to its canonical normal form:
\begin{equation}
\theta_M=(0,f_1,f_2,f_3,f^0,0,0,0)\,.
\end{equation}
The scalar potential reads:
\begin{align}
\mathcal{V}&= \mathcal{V}_{gaugino,\,2}+
\mathcal{V}_{gravitino}=-\sum_{m=1}^3\,e^{-\mathfrak{h}_m}\,\left(f_mf^0\,(\mathfrak{b}_m^2+e^{2\mathfrak{h}_m})+f_n
f_p\,\right)\,,
\end{align}
where $n\neq p\neq m$. The truncation to the dilatons is consistent
and we find:
\begin{equation}
\label{doppiocritFI}
\left.\mathcal{V}\right\vert_{\mathfrak{b}_m=0}=-\sum_{m=1}^3\,\left(f_mf^0\,e^{\mathfrak{h}_m}+f_n
f_p\,e^{-\mathfrak{h}_m}\right)\,,
\end{equation}
which is extremized with respect to the dilatons by setting
\begin{equation}
e^{2\mathfrak{h}_m}=\frac{f_n f_p}{f_mf^0}\,,
\end{equation}
and the potential at the extremum reads:
\begin{equation}
\mathcal{V}_0=-6\,\varepsilon\,\sqrt{f^0 f_1 f_2 f_3}<0\,,
\end{equation}
This extremum exists only if $f^0 f_1 f_2 f_3>0$. This implies that
$\theta_M$ should  be either in the orbit $i)$ ($\varepsilon=1$ in the above expression for $\mathcal{V}_0$) or in the orbit $ii)$ ($\varepsilon=-1$).   Using the
analogy between $\theta_M$ and black hole charges, these two orbits
correspond to BPS and non-BPS with $\mathfrak{I}_4>0$ black holes. The extremum
condition for $V_{BH}$ fixes the scalar fields at the horizon
according to the attractor behavior. Now the potential has an
additional term $-4\,|Z|^2$ which, however, for the orbits $i)$,
$ii)$, has the same extrema as $V_{BH}$ since its derivative with
respect to $z^m$ is $-4 \mathcal{D}_m Z\,\bar{Z}$ which vanishes for the $i)$
orbit since at the extremum of $V_{BH}$ (BPS black hole horizon)
$\mathcal{D}_m Z=0$, and for the $ii)$ orbit since at the extremum of $V_{BH}$
(black hole horizon) $Z=0$.
\par We conclude that in the ``BPS'' orbit $i)$ the extremum corresponds to an AdS-vacuum where the scalar mass
spectrum reads as follows:
\begin{equation}
\left.(\partial_r\partial_s \mathcal{V}\,g^{st})\right\vert_0={\rm
diag}\left(\frac{2}{3},\frac{2}{3},\frac{2}{3},\frac{2}{3},\frac{2}{3},\frac{2}{3}\right)\times
\mathcal{V}_0<0\,.
\end{equation}
These models have provided a useful supergravity framework where to study black hole solutions in anti-de Sitter spacetime \cite{adsblackholes}.\par
In the ``non-BPS'' orbit $ii)$ the potential has a de Sitter extremum which, however, is not stable, having takyonic directions:
\begin{equation}\left.(\partial_r\partial_s \mathcal{V}\,g^{st})\right\vert_0={\rm
diag}\left(-2,-2,2,2,2,2\right)\times
\mathcal{V}_0>0\,.
\end{equation}
We shall not consider this case in what follows.
\subsection{\sc Conclusions on the one-field cosmologies that can be derived from the gaugings of the $\mathcal{N}=2$ STU model}
Let us summarize the results of the above systematic discussion. From the gaugings of the $\mathcal{N}=2$ STU model one can obtain following dilatonic potentials:
\begin{description}
  \item[A)] \textbf{Critical Orbit without FI terms}. We have the potential:
  \begin{equation}\label{purto1}
{V}\, =\, g^2\,\left(e^{-\frac{1}{2}(-\mathfrak{h}_1+\mathfrak{h}_2+\mathfrak{h}_3)}
+ \,e^{-\frac{1}{2}(\mathfrak{h}_1-\mathfrak{h}_2+\mathfrak{h}_3)}\right)^2\,.
  \end{equation}
 In this case we have a consistent truncation to one dilaton by setting: $ \mathfrak{h}_1 \, = \, \mathfrak{h}_2 \, = \, \ell  \in \, \mathbb{R}$, since the derivatives of the potential with respect to $\mathfrak{h}_{1,2}$ vanish on such a line. The residual one dilaton potential is:
 \begin{equation}\label{cucco1}
   {V}\, =\, 4 \, g^2\, e^{- \mathfrak{h}_3}
 \end{equation}
 which upon use of the translation rule (\ref{babushka}) yields
 \begin{equation}\label{fulatto}
    \mathcal{V}\, =\, 12 \, g^2\, e^{- \frac{\varphi}{\sqrt{3}}}
 \end{equation}
 The above potential is trivially integrable, being a pure less than critical exponential.
  \item[B)] \textbf{Doubly Critical Orbit without FI terms}.We have the potential:
  \begin{equation}\label{purto2}
{V}\, =\, \mbox{const}\, e^{-\mathfrak{h}_1-\mathfrak{h}_2-\mathfrak{h}_3}
  \end{equation}
  Introducing the following field redefinitions:
  \begin{equation}\label{gurkina1}
    \phi_1 \, = \, \mathfrak{h}_1+\mathfrak{h}_2+\mathfrak{h}_3 \quad ; \quad  \phi_2 \, = \, \mathfrak{h}_2-\mathfrak{h}_3
    \quad ; \quad  \phi_3 \, = \, -2 \, \mathfrak{h}_1+\mathfrak{h}_2+\mathfrak{h}_3
  \end{equation}
 the kinetic term:
   \begin{equation}\label{kinesi1}
    \mbox{kin} \, = \, \frac{1}{2} \, \left (\dot{ \mathfrak{h}}_1+\dot{\mathfrak{h}}_2+\dot{\mathfrak{h}}_3 \right)
  \end{equation}
  is transformed into:
  \begin{equation}\label{kinesi2}
    \mbox{kin} \, = \, \frac{1}{6} \, \dot{ \phi}_1+\frac{1}{4} \, \dot{ \phi}_2+\frac{1}{12} \, \dot{ \phi}_3
  \end{equation}
  while the potential (\ref{purto2}) depends only on $\phi_1$. Hence we can consistently truncate to one field by setting
  $\phi_2=\phi_3 \, = \, \mbox{const}$ and upon use of the translation rule (\ref{babushka}) we obtain a trivially integrable over critical exponential potential:
  \begin{equation}\label{fulatto2}
    \mathcal{V}\, =\, \mbox{const} \, e^{- \sqrt{3} \, \varphi}
 \end{equation}
 \item[C)]  \textbf{Critical Orbit with FI terms}. This case leads to the potential (\ref{guliashi}) which, as we showed, reproduces the potential (\ref{potentissimo}) of the $\so(2,1)\times \uu(1)$ gauging extensively discussed in sect. \ref{stabdesitter}. Such a potential admits a stable de Sitter vacuum and a consistent one dilaton truncation to a model with a $cosh$ potential which is not integrable, since the intrinsic index $\omega$ does not much any one of the three integrable cases.
 \item[D)]  \textbf{Doubly Critical Orbit with FI terms}. Upon a constant shift of the dilatons in eq.(\ref{doppiocritFI}) this gauging leads to the following negative potential:
     \begin{equation}\label{papalone}
        V \, = \, - \, 2\,  \sqrt{f^0 \, f_1 \, f_2 \, f_3} \sum_{i=1}^3 \, \cosh\left[ \mathfrak{h}_i \right]
     \end{equation}
that has a stable anti de Sitter extremum. We have a consistent truncation to one-field by setting to zero any two of the three dilatons. Upon use of the translation rule (\ref{babushka}) we find the potential:
\begin{equation}\label{gugullo}
    \mathcal{V} \, = \, - \, \mbox{const}\left(2+\cosh\left[\frac{\varphi}{\sqrt{3}}\right]\right)
\end{equation}
which does not fit into any one of the integrable series of tables \ref{tab:families} and \ref{Sporadic}.
\end{description}
Hence apart from pure exponentials without critical points no integrable models can be fitted into any gauging of the  $\mathcal{N}=2$ STU model.
\section{\sc  $\mathcal{N}=1$ models with a superpotential}
Let us now turn to consider the case of $\mathcal{N}=1$ Supergravity coupled to Wess--Zumino multiplets \cite{cfgv}.
Following the notations of \cite{castdauriafre}, the general bosonic Lagrangian of this class of models is\footnote{Observe that here we consider only the graviton multiplet coupled to Wess-Zumino multiplets. There are no gauge multiplets and no $D$-terms. The embedding mechanisms discussed in \cite{secondosashapietro} is lost a priori from the beginning.}
\begin{equation}\label{n1sugra}
    \mathcal{L}^{\mathcal{N}=1}_{SUGRA} \, = \, \sqrt{-g} \, \left[ \mathcal{R}[g] \, +  \,2\, g^{HK}_{ij^\star} \, \partial_\mu z^i \, \partial^\mu {\bar z}^{j^\star} \, - \, 2\, V(z,{\bar z}) \,\right ]\ ,
\end{equation}
where the scalar metric is K\"ahler (the scalar manifold must be Hodge--K\"ahler)
\begin{eqnarray}
  g_{ij^\star} &=& \partial_i \, \partial_{j^\star} \, \mathcal{K} \\
  \mathcal{K}  &=& \overline{\mathcal{K}}   \, = \, \mbox{K\"ahler potential}
\end{eqnarray}
and the potential is
\begin{eqnarray}
    V & = & 4 \, e^2 \,\exp \left[ \mathcal{K} \right]\left ( g^{ij^\star}\, \mathcal{D}_i W_h(z)\, \mathcal{D}_{j^\star} \overline{W_h}({\bar z}) \, - \, 3 \,  |W_h(z)|^2 \right )\ ,
    \label{frullini}
\end{eqnarray}
where the superpotential $W_h(z)$ is a holomorphic function. Furthermore
\begin{eqnarray}
  \mathcal{D}_i \, W&=& \partial_i W \, + \,  \partial_i \mathcal{K} \, W \nonumber \\
  \mathcal{D}_{j^\star}\, { \overline{W}} &=& \partial_{j^\star} \overline{W} \, + \,
  \partial_{j^\star} \mathcal{K} \, \overline{W}\label{felucide}
\end{eqnarray}
are usually referred to as K\"ahler covariant derivatives. They arise since $W_h(z)$, rather than a function, is actually a holomorphic section of the line bundle $\mathcal{L} \rightarrow \mathcal{M}_{K} $ over the K\"ahler manifold whose first Chern class is the K\"ahler class, as required by the definition of Hodge--K\"ahler manifolds. In other words, $c_1\left( \mathcal{L}\right) \, = \, \left [ \mathrm{K} \right ]$, the latter being the K\"ahler two--form. The fiber metric on this line bundle is $h \, = \, \exp\left [ \mathcal{K}\right]$, so that a generic section $W(z,{\bar z})$
of $\mathcal{L}$ (not necessarily holomorphic) admits the invariant norm
\begin{equation}\label{invanorma}
    \parallel W \parallel ^2 \, \equiv \, W\, \overline{W} \, \exp\left [ \mathcal{K}\right]
\end{equation}
A generic gauge transformation of the line bundle takes the form
\begin{eqnarray}
  W^\prime(z,{\bar z}) &=& \exp\left [ \frac{1}{2} f(z)\right ] \times  W(z,{\bar z}) \nonumber\\
  {\overline{W}}^\prime(z,{\bar z}) &=& \exp\left [ \frac{1}{2} \overline{f(z)}\right ] \times  \overline{W}(z,{\bar z}) \, \label{frattocchia}
\end{eqnarray}
where $f(z)$ is a holomorphic complex function. Under the gauge transformation (\ref{frattocchia}), the fiber metric changes according to
\begin{equation}\label{coriandolo}
    \mathcal{K}^\prime(z,{\bar z}) \, = \, - \, \mathcal{K}^\prime(z,{\bar z}) + \mbox{Re} f(z)
\end{equation}
while the norm (\ref{invanorma}) stays invariant. It is important to stress that the same K\"ahler metric
$g_{ij^\star} \, = \, \partial_i \partial_{j^\star} \, \mathcal{K}^\prime$ would be obtained by the same token from $\mathcal{K}^\prime$.
All transition functions from one local trivialization of the line bundle to another one are of the form (\ref{frattocchia}) and
(\ref{coriandolo}), with an appropriate $f(z)$. The fiber metric introduces a canonical connection $\theta \, = \, h^{-1}\partial h$ leading to the  covariant derivatives (\ref{felucide}). In covariant notation, the potential (\ref{frullini}) takes the form
\begin{equation}\label{rucolafina}
    V \, = \, \, 4 \, e^2 \, \left( \parallel \mathcal{D} W \parallel^2 \, - \, 3 \, \parallel  W \parallel^2 \right)
\end{equation}
where by definition
\begin{eqnarray}
  \parallel \mathcal{D} W \parallel^2 &=& g^{ij^\star}\mathcal{D}_i \, W \, \mathcal{D}_{j^\star}\, \overline{W} \,\exp \left [ \mathcal{K} \right]\nonumber \\
  \, \parallel  W \parallel^2  &=&  W \,  { \overline{W}}\, \exp\left [ \mathcal{K}\right] \label{trifoglio}
\end{eqnarray}
Let us now consider the notion of covariantly holomorphic section, defined by the condition
\begin{equation}\label{condizia}
    \mathcal{D}_{j^\star}\,W \, = \,0
\end{equation}
From any covariantly holomorphic section, one can retrieve a holomorphic one by setting
\begin{equation}\label{olomorfina}
    W_{h}(z) \, = \, \exp\left [ - \,\frac{1}{2} \, \mathcal{K}\right] \,W \, \quad \Rightarrow \quad \partial_{j^\star} W_h \, = \, 0
\end{equation}
\par
By hypothesis the superpotential $W$ that appears in the potential (\ref{rucolafina}) is covariantly constant. The compact notation (\ref{rucolafina}) is very instructive since it stresses that the scalar potential results from the difference of two positive definite terms originating from two different contributions. The first contribution is the absolute square of the auxiliary fields appearing in the supersymmetry transformations of the spin $\frac{1}{2}$--fermions (the chiralinos belonging to Wess--Zumino multiplets), while the second is the square of the auxiliary field appearing in the supersymmetry transformation of the spin
$\frac{3}{2}$--gravitino. Indeed
\begin{eqnarray}
  \delta_{SUSY} \chi^i & = &  {\rm i} \, \partial_\mu z^i \, \gamma^\mu \, \epsilon^\bullet \, + \, \mathcal{H}^i \, \epsilon_\bullet \\
  \delta_{SUSY} \Psi_{\mu\bullet} &=& \mathcal{D}_\mu \, \epsilon_\bullet \, + \, S \, \gamma_\mu \, \epsilon^\bullet
  \label{ballaconilupi}
\end{eqnarray}
where $\epsilon^\bullet \, , \epsilon_\bullet$ denote the two chiral projections of the supersymmetry parameter and the scalar field dependent auxiliary fields are
\begin{eqnarray}
  S &=& {\rm i}\, e\, \sqrt{\parallel W\parallel^2} \, = \, {\rm i}\, e\, \sqrt{| W_h|^2} \,
  \exp\left[\frac{1}{2} \mathcal{K} \right] \nonumber\\
  \mathcal{H}^i &=& 2 \, e\, g^{ij^\star} \,  \, \mathcal{D}_{j^\star} W \, \exp\left[\frac{1}{2} \mathcal{K} \right]\label{auxiliary}
\end{eqnarray}
\par
This structure of the potential shows that any de Sitter vacuum characterized by a potential $V(z_0)$ that is positive at the extremum necessarily breaks supersymmetry since this implies that  the chiralino auxiliary fields are different from zero
in the vacuum $<\mathcal{H}^i> \, = \, \mathcal{H}^i(z_0) \, \ne \, 0$.

Let us also stress that the parameter $e$ appearing in the potential is just a dimensionful parameter which fixes the scale of all the masses generated by the gauging, \textit{i.e.} by the introduction of a superpotential.

\subsection{\sc  One--field models}
In this general framework the simplest possibility is a model with one scalar multiplet assigned to the homogeneous
K\"ahler manifold
\begin{equation}\label{tripini}
    \mathcal{M}_{K} \, = \, \frac{\mathrm{SU(1,1)}}{\mathrm{U(1)}}
\end{equation}
and K\"ahler potential
\begin{equation}\label{kelero1}
    \mathcal{K} \, = \, - \, \log \, \left[ (z \, - \, \bar{z})^q\right ]\ ,
\end{equation}
which leads to the K\"ahler metric
\begin{equation}\label{kelero2}
    g_{z\bar{z}} \, = \, - \, \frac{q}{(z \, - \, \bar{z})^2}\ ,
\end{equation}
where $q$ is an integer number. Its favorite value, $q=3$, corresponds to the $\mathcal{N}=1$ truncation of the $\mathcal{N}=2$ model $S^3$ that, on its turn arises from the $\mathrm{STU}$ model discussed in the previous section upon identification of the three scalar multiplets $S,T$ and $U$. Alternatively, the case $q=1$ corresponds to the $\mathcal{N}=1$  truncation of an $\mathcal{N}=2$ theory with vanishing Yukawa couplings. Because of their $\mathcal{N}=2$ origin, both instances of the familiar Poincar\'e Lobachevsky plane are not only Hodge-K\"ahler but actually \textit{special K\"ahler} manifolds.

In the notation of \cite{PietroSashaMarioBH1}, the holomorphic symplectic section governing this geometry is given
by the four--component vector
\begin{equation}\label{seziona}
    \Omega \, = \,\left\{-\sqrt{3}z^2,z^3,\sqrt{3} z,1\right\}\ ,
\end{equation}
which transforms in the spin $j\, = \, \frac{3}{2}$ of the $\mathrm{SL(2,\mathbb{R})} \sim \mathrm{SU(1,1)}$ group that happens to be
four--dimensional symplectic
\begin{equation}
\label{frilli}
\mathrm{SL(2,\mathbb{R})} \, \ni \,\left(\begin{array}{ll}
 a & b \\
 c & d
\end{array} \right) \, \Longrightarrow \, \left(
\begin{array}{llll}
 d a^2+2 b c a & -\sqrt{3} a^2
   c & -c b^2-2 a d b &
   -\sqrt{3} b^2 d \\
 -\sqrt{3} a^2 b & a^3 &
   \sqrt{3} a b^2 & b^3 \\
 -b c^2-2 a d c & \sqrt{3} a
   c^2 & a d^2+2 b c d &
   \sqrt{3} b d^2 \\
 -\sqrt{3} c^2 d & c^3 &
   \sqrt{3} c d^2 & d^3
\end{array}
\right) \, \in \, \mathrm{Sp(4,\mathbb{R})}
\end{equation}
where the preserved symplectic metric is
\begin{equation}\label{goriaci}
 \mathbb{C} \, = \,   \left(
\begin{array}{llll}
 0 & 0 & 1 & 0 \\
 0 & 0 & 0 & 1 \\
 -1 & 0 & 0 & 0 \\
 0 & -1 & 0 & 0
\end{array}
\right)
\end{equation}
According to the general set up of Special Geometry (for a recent review see \cite{pietroGR}), the K\"ahler potential (\ref{kelero1}) is retrieved letting
\begin{equation}\label{fagano}
    \mathcal{K}(z,{\bar z}) \, = \, - \log \left[ - {\rm i} \Omega \, \mathbb{C} \, \
    \overline{\Omega} \right ]
\end{equation}
Independently of the special structure that is essential for $\mathcal{N}=2$ supersymmetry, at the $\mathcal{N}=1$ level
one can consider general superpotentials that  are consistent with the Hodge--K\"ahler structure, provided they are holomorphic,
namely provided they can be expanded in a power series of the unique complex field $z$
\begin{equation}\label{supipoti}
    W_h(z) \, = \, \sum_{n \, \in \, \mathbb{N}} \, c_n \, \, z^n \ ,
\end{equation}
where the $c_n$  are complex coefficients. The sum over $n$ extends to a finite or infinite subset of the natural
numbers $\mathbb{N}$, while rational or irrational powers leading to cuts are excluded in order to obtain properly transforming
sections of the Hodge line bundle.

Notwithstanding this wider choice available at the $\mathcal{N}=1$ level, it is interesting to note that in discussing black--hole solutions of the corresponding $\mathcal{N}=2$ model one is lead to an effective sigma--model whose Lagrangian resembles closely the effective Lagrangian of the cosmological sigma--model and displays a potential that is also built in terms of a superpotential, although the latter is more restricted. The comparison between cosmological and black--hole constructions provides inspiring hints on the choice of appropriate superpotentials.
Let us briefly see how this works.
\subsection{\sc Cosmological versus black--hole potentials}
The common starting point for black--hole and cosmological solutions is the general form of the bosonic portion of the four--dimensional Supergravity, which takes the form (for a recent review see Chapter 8, Vol 2 in \cite{pietroGR} and all references therein)
\begin{eqnarray}
\mathcal{L}^{(4)} &=& \sqrt{|\mbox{det}\, g|}\left[R[g] - \frac{1}{2}
\partial_{ {\mu}}\phi^a\partial^{ {\mu}}\phi^b \mathfrak{g}_{ab}(\phi) \,
+ \,
2 \, \mbox{Im}\mathcal{N}_{\Lambda\Sigma}(\phi) \, F_{ {\mu} {\nu}}^\Lambda
F^{\Sigma| {\mu} {\nu}}\right. \nonumber\\
&&+\left. \, e^2 \, V(\phi) \, \right] \, + \,
\mbox{Re}\mathcal{N}_{\Lambda\Sigma}(\phi)\, F_{ {\mu} {\nu}}^\Lambda
F^{\Sigma}_{ {\rho} {\sigma}}\epsilon^{ {\mu} {\nu} {\rho} {\sigma}}\,  ,
\label{d4generlag}
\end{eqnarray}
where $F_{ {\mu} {\nu}}^\Lambda\equiv (\partial_{ {\mu}}A^\Lambda_{ {\nu}}-\partial_{ {\nu}}A^\Lambda_{ {\mu}})/2$ are the field strengths of the vector fields, $\phi^a$ denotes the collection of $n_{\mathrm{s}}$ scalar fields
parameterizing the scalar manifold $ \mathcal{M}_{scalar}^{D=4}$, with $\mathfrak{g}_{ab}(\phi)$ its metric and the field--dependent complex matrix $\mathcal{N}_{\Lambda\Sigma}(\phi)$ is fully determined by constraints imposed by duality symmetries. In addition, the scalar potential $V(\phi)$ is determined by the appropriate gauging procedures, while $e$ is the gauge coupling constant, which vanishes in ungauged supergravity.
\par
Although the discussion can be extended also to higher $\mathcal{N}$, for simplicity we focus on the $\mathcal{N}=2,1$ cases, where the real scalar fields are grouped in complex combinations $z^i$ and their kinetic term becomes
\begin{equation}\label{calerone}
    \frac{1}{2}
\partial_{ {\mu}}\phi^a\partial^{ {\mu}}\phi^b \mathfrak{g}_{ab}(\phi) \, \mapsto \, 2 \, g_{ij^\star}(z,{\bar z}) \, \partial_\mu z^i \,
\partial^{ {\mu}} \, {\bar z}^{j^\star}
\end{equation}
In the case of extremal black--hole solutions of ungauged Supergravity ($e=0$), the four--dimensional metric is of the form
\begin{equation}\label{metruzza}
    ds^2_{BH} \, = \, - \, \exp[U(\tau)] \,  dt^2  \, + \, \, \exp[- U(\tau)] \, dx^i \otimes dx^j \, \delta_{ij}
\end{equation}
where $\tau \, = \, - \,\left(\sum_{i=1}^3 \, x_i^2\right)^{\,-\,\frac{1}{2}}$ is the reciprocal of the radial distance, one is lead to the effective Euclidian $\sigma$--model (for a recent review see chapter Chapter 9, Volume Two in \cite{pietroGR} and all references therein)
\begin{eqnarray}
S_{BH} & \equiv & \int \, {\cal L}_{BH}(\tau) \, d\tau \quad \nonumber\\
 {\cal L}_{BH}(\tau ) & = & \frac{1}{4} \,\left( \frac{dU}{d\tau} \right)^2 +
 g_{ij^\star} \, \frac{dz^i}{d\tau} \,  \frac{dz^{j^\star}}{d\tau} + e^{U}
 \, V_{BH}(z, {\bar z} , \mathcal{Q})
 \label{effact}
\end{eqnarray}
The geodesic potential $V(z, {\bar z} , \mathcal{Q})$ is defined by
\begin{equation}\label{geopotentissimo}
    V_{BH}(z, {\bar z}, \mathcal{Q})\, = \, \frac{1}{4} \, \mathcal{Q}^t \, {\cal M}_4^{-1}\left( {\cal N}\right) \, \mathcal{Q} \ .
\end{equation}
Here $\mathcal{Q}$ is the vector of electric and magnetic charges of the hole,
which transforms in the same representation of the K\"ahler isometry group $\mathrm{G}$ as the symplectic section of Special Geometry.
In the $S^3$ case $G=\mathrm{SL(2,\mathbb{R})}$ and the four charges of the hole
\begin{equation}\label{grutinata}
    \mathcal{Q} \, = \, \left\{p_1,p_2,q_1,q_2\right\}
\end{equation}
transform by means of  the matrix (\ref{frilli}).
The $(\mathrm{2n+2})\times(\mathrm{2n+2}) $ matrix  ${\cal M}_4^{-1}$ appearing in eq.~(\ref{geopotentissimo}) is given in terms of the $(\mathrm{n+1})\times(\mathrm{n+1}) $  matrix $\mathcal{N}_{\Lambda\Sigma}(\phi)$ that appears in the $4D$ Lagrangian. In detail,
\begin{eqnarray}
\mathcal{M}_4^{-1} & = &
\left(\begin{array}{c|c}
{\mathrm{Im}}\mathcal{N}\,
+\, {\mathrm{Re}}\mathcal{N} \, { \mathrm{Im}}\mathcal{N}^{-1}\, {\mathrm{Re}}\mathcal{N} & \, -{\mathrm{Re}}\mathcal{N}\,{ \mathrm{Im}}\,\mathcal{N}^{-1}\\
\hline
-\, { \mathrm{Im}}\mathcal{N}^{-1}\,{\mathrm{Re}}\mathcal{N}  & { \mathrm{Im}}\mathcal{N}^{-1} \
\end{array}\right) \ , \label{inversem4}
\end{eqnarray}
where $n$ is the number of vector multiplets coupled to Supergravity.

Starting instead from the spatially flat cosmological metric
\begin{equation}\label{metruzzolla}
    ds^2_{Cosm} \, = \, - \, \exp[3A(t)] \,  dt^2  \, + \, \, \exp[2 A(t)] \, dx^i \otimes dx^j \, \delta_{ij}
\end{equation}
which, in the language of the preceding sections, corresponds to the gauge $\mathcal{B}=3A$,  one is led to the effective sigma model
\begin{eqnarray}
S_{Cosm} & \equiv & \int \, {\cal L}_{Cosm}(t) \, dt \quad \nonumber\\
 {\cal L}_{Cosm}(\tau ) & = & - \,\frac{3}{2} \,\left( \frac{dA}{dt} \right)^2 +
 g_{ij^\star} \, \frac{dz^i}{dt} \,  \frac{dz^{j^\star}}{dt} + e^{6 A}
 \, V_{Cosm}(z, {\bar z})
 \label{effactbis}
\end{eqnarray}
where $V_{Cosm}(z, {\bar z})\, = \, e^2 \, V(\phi)$ is the scalar potential produced by gauging that, in an $\mathcal{N}=1$ theory, or in an $\mathcal{N}=2$ one with only abelian gauge groups (Fayet--Iliopoulos terms), admits the representation in terms of a holomorphic superpotential recalled in eq.(\ref{frullini}).
The similarity between the cosmological and black--hole cases becomes striking if one recalls that the black--hole geodesic potential (\ref{geopotentissimo}) admits the alternative representation
\begin{eqnarray}\label{potenzialusgeodesicus}
V_{BH}(z, {\bar z}, \mathcal{Q})&= &   -\,\frac{1}{2} \,\left( \vert Z \vert ^2 +   g^{ij^\star} \mathcal{D}_i  Z \, \mathcal{D}_{j^\star} \bar{Z} \right)\ .
\end{eqnarray}
Here $Z$ denotes the field--dependent central charge of the supersymmetry algebra
\begin{equation}\label{centralcharge}
    Z \, \equiv \, \exp \left[ \frac{1}{2} \, \mathcal{K}(z,z)\right] \,   \mathcal{Q}^T \, \mathbb{C} \, \Omega (z)\ ,
\end{equation}
$\Omega (z)$ denotes the holomorphic symplectic section of special K\"ahler geometry (that of  eq.~(\ref{Omegabig})
for the $\mathrm{STU}$ model, or that of eq.~(\ref{seziona}) for the $S^3$ model) and $ \mathcal{K}(z,z)$ denotes the K\"ahler potential.
Introducing the black--hole holomorphic superpotential
\begin{equation}\label{governoladro}
    W_{BH}(z) \, \equiv \, \mathcal{Q}^T \, \mathbb{C} \, \Omega (z)
\end{equation}
eq.~(\ref{potenzialusgeodesicus}) for the geodesic potential can be recast in the form
\begin{eqnarray}\label{potenzialato}
V_{BH}(z, {\bar z}, \mathcal{Q})&= &   -\,\frac{1}{2} \,\exp \left[\mathcal{K}(z,z)\right] \,
\left(   g^{ij^\star} \mathcal{D}_i  W_{BH} \, \mathcal{D}_{j^\star} \bar{W}_{BH} + \vert W_{BH} \vert ^2 \right)
\end{eqnarray}
which is almost identical to eq.~(\ref{frullini}) yielding the cosmological potential, up to a crucial change of sign and coefficient. The coefficient $-3$ of the second term becomes $+1$, and in this fashion the black hole potential is strictly positive definite since it is the sum of two squares. Yet the entire discussion suggests that black--hole superpotentials, that are group theoretically classified by the available $\mathrm{G}$--orbits of charge vectors $\mathcal{Q}$, form a good class of superpotentials also for Gauged Supergravity models. Indeed we already saw, by means of the systematic analysis of the STU model,  that black--hole superpotentials encode and exhaust the available abelian gaugings for  $\mathcal{N}=2$ supergravity theories.
\subsection{\sc Cosmological Potentials from the $S^3$ model}

Relying on the preceding discussion, let us consider the abelian gaugings of the $S^3$ model provided by the superpotential
\begin{equation}
\label{chiridone}
    W_{\mathcal{Q}}(z) \, = \, \mathcal{Q} \mathbb{C} \Omega \, = \, -q_2 z^3+\sqrt{3} q_1
   z^2+\sqrt{3} p_1 z+p_2 \ ,
\end{equation}
which happens to be the most general third--order polynomial. Let us stress that in multi--field models based on larger special K\"ahler homogeneous manifolds $\mathrm{G/H}$, despite the existence of many coordinates $z_i$, the order of the superpotential will stay three since this is the polynomial order of the symplectic section for all such special geometries. Inserting (\ref{chiridone}) into
(\ref{frullini}) yields the four--parameter potential
\begin{equation}\label{4parami}
  V(z,{\bar z},\mathcal{Q}) \, = \,  -\frac{{\rm i} \left(2
   p_1^2+\left((z+{\bar z})
   q_1+2 \sqrt{3} z {\bar z}
   q_2\right) p_1+2 z
   {\bar z} q_1^2+p_2 \left(3
   (z+{\bar z}) q_2-2
   \sqrt{3}
   q_1\right)\right)}{z-{\bar z}}
\end{equation}
in which one can decompose $z$ into its real and imaginary parts according to
\begin{equation}\label{donnaiolo}
    z \, = \, {\rm i} \, e^\mathfrak{h} \, + \, \mathfrak{b}
\end{equation}
Not every choice of the charge vector $\mathcal{Q}$ allows for a consistent truncation to a vanishing axion, guaranteed by the condition
\begin{equation}\label{lordoftherings}
    \partial_\mathfrak{b} \,V(z,{\bar z},\mathcal{Q})|_{\mathfrak{b}=0} \, = \,0
\end{equation}
and yet there is a representative with such a property for every $\mathrm{SL(2,\mathbb{R})}$ orbit in the $j=\frac{3}{2}$ representation except for the largest one. Following the results of \cite{noinilpotenti} one can identify the orbits
\begin{enumerate}
  \item  The very small orbit with a parabolic stability group $\mathcal{O}_1 \, = \, \left\{p_1\to 0,p_2\to
   0,q_1\to 0,q_2\to \mathfrak{q}\right\}$
  \item The  small orbit with no stability group $\mathcal{O}_2 \, = \, \left\{p_1\to \sqrt{3}
   \mathfrak{p},p_2\to 0,q_1\to 0,q_2\to
   0\right\}$
  \item  The  large orbit with a $\mathbb{Z}_3$ stability group $\mathcal{O}_3 \, = \, \left\{p_1\to 0,p_2\to
   \mathfrak{p},q_1\to -\sqrt{3} \mathfrak{q},q_2\to
   0\right\}$ ($\mathfrak{p}\mathfrak{q}<0$ regular BPS in black hole constructions)
  \item The  large orbit with no stability group $\mathcal{O}_4 \, = \, \left\{p_1\to 0,p_2\to
   \mathfrak{p},q_1\to \sqrt{3} \mathfrak{q},q_2\to
   0\right\}$ ($\mathfrak{p}\mathfrak{q}>0$ regular non-BPS in black hole constructions)
   \item The very large orbit with no stability group $\left\{p_1\to p_1,p_2\to
   0,q_1\to q_1,q_2\to
   q_2\right\}$ \ ,
\end{enumerate}
and the following superpotentials and potentials:
\begin{description}
  \item[$\mathcal{O}_1$] The superpotential is purely cubic $W\, = \, -\mathfrak{q} \,z^3$ and the potential vanishes
  \begin{equation}\label{flatto}
    V \, = \, 0
  \end{equation}
  This is an instance of flat potentials \cite{noscale}. Namely supersymmetry is broken by the presence of non vanishing auxiliary fields, yet the vacuum energy is exactly zero and the ground state is Minkowski space.
  \item[$\mathcal{O}_2$] The superpotential is linear  $W\, = \, 3 \mathfrak{p} z$ and the consistent truncation to zero axion yields a pure exponential
  \begin{equation}\label{expoto}
    V \, = \, -3 e^{-\mathfrak{h}} \mathfrak{p}^2
  \end{equation}
  This potential is trivially integrable.
 \item[$\mathcal{O}_3$] The superpotential is quadratic  $W\, = \, \mathfrak{p}-3 \mathfrak{q} z^2$ and the consistent truncation to zero axion yields the following potential
  \begin{equation}\label{cossho}
    V \, = \, -3 e^{-\mathfrak{h}} \mathfrak{q} \left(\mathfrak{p}+e^{2 \mathfrak{h}} \,
   \mathfrak{q}\right)\, \simeq \, - 3 \mathfrak{q}^2 \, \cosh \hat{\mathfrak{h}}
   \end{equation}
   The last form of the potential can be always achieved by means of a constant shift of the scalar field $\mathfrak{h}\mapsto \mathfrak{h} + \mbox{const} $. In  this case the intrinsic index is:
   \begin{equation}\label{formidabile}
    \omega \, = \, \frac{1}{3}
   \end{equation}
   since the kinetic term of the $S^3$-model corresponds to $q=3$. It is different from the value $\omega = 1$ which is obtained from the non abelian $\so(1,2)$-gauging of the same model, yet it is still different from  either one of the integrable indices: $\omega \ne \sqrt{3}$ and $\omega \ne \frac{2}{\sqrt{3}}$.
   This result confirms what we already learned. Consistent one-field truncations of Gauged Supergravity easily yield cosmological models of the $cosh$-type,  yet non integrable ones. It is interesting to note that the $cosh$ case is in correspondence with the regular BPS black holes.
 \item[$\mathcal{O}_4$] The superpotential is quadratic  $W\, = \, \mathfrak{p}+3 \mathfrak{q} z^2$ but with a different relative sign between the constant and quadratic terms. The consistent truncation to zero axion yields the following potential
  \begin{equation}\label{cosshobis}
    V \, = \, 3 e^{-h} \mathfrak{p} \mathfrak{q}-3 e^h \mathfrak{q}^2\, \simeq \, - 3 \mathfrak{q}^2 \, \sinh \hat{\mathfrak{h}}
   \end{equation}
   As in the previous case the last form of the potential can always be achieved by means of a constant shift of the scalar field.  It is interesting to note that the $sinh$ case of potential  is in correspondence with the regular non-BPS black holes. Once again the index $\omega$ is not a critical one for integrability.
  \item[$\mathcal{O}_5$] In this case no consistent truncation to zero axion does exist.
\end{description}

\section{\sc The supersymmetric integrable model with one multiplet}
\label{integsusymodel}
The unique integrable model that so far we have been able to fit into the considered supersymmetric framework with just one multiplet (the K\"ahler metric is fixed once for all to the choice (\ref{tripini})   belongs to the series $I_2$ of table \ref{tab:families} and occurs for the under--critical value $\gamma = \frac{2}{3}$.  Before proceeding with the further analysis of this particular integrable model it is just appropriate to stress that
 in a couple of separate publications  Sagnotti and collaborators \cite{dks},\cite{dkps} have also shown that the phenomenon of climbing scalars, displayed by all of the integrable models we were able to classify, has the potential ability to explain the  oscillations in the low angular momentum part of the CMB  spectrum, apparently observed by PLANCK. In his recent talk given at the Dubna SQS2013 workshop, our coauthor Sagnotti has also shown a best fit to the PLANCK data for the low $\ell$ part of the spectrum, by using precisely the series of integrable potentials $I_2$\footnote{In comparing the following equation with the Table of paper \cite{primopapero}, please note the coefficient $\sqrt{3}$ appearing in the exponents that has been introduced to convert the unconventional normalization of the field $\varphi$ used there to the canonical normalization of the field $\phi$ used here.}
\begin{equation}\label{gammaserie}
    V(\phi) \, = \, a \, \exp\left[ 2\, \sqrt{3} \, \gamma \, \phi\right] + b \, \exp\left[  \sqrt{3} \, (\gamma +1)\, \phi\right]
\end{equation}
with the particularly nice value $\gamma \, = \, -\ft 76$ (see \cite{secondosashapietro} for details about the D-map insertion into supergravity). Here the different subcritcal value $\gamma \, = \, \frac 23$ is select by supergravity when we try to realize the integral model through a superpotential (F-embedding).
Indeed this potential can be  obtained from the $S^3$-model with a carefully calibrated and unique superpotential that   now we describe. We immediately anticipate that such a superpotential is not of the form discussed in the previous section and therefore strictly corresponds to an $\mathcal{N}=1$ theory and not to an abelian gauging of the $\mathcal{N}=2$ model. Technically, the difficulty met when trying to fit an integrable case into supergravity coupled just to one multiplet resides in the following. If the superpotential involves powers only up to the cubic order, as pertains to the construction via symplectic sections, the dilaton truncation can contain at most two types of exponentials, one positive and one negative, so that one can reach either $\cosh p \mathfrak{h}$ or $\sinh p\mathfrak{h}$ models. Yet the obtained index $p$ is always $1$, different  from the $p=3,2$ required by integrability. In order to get higher values of $p$, one would need higher powers $z^n$ in the superpotential, but as the degree of $W(z)$ increases one is confronted with new problems: one can generate higher exponentials $\exp\left [ p \mathfrak{h}\right]$ but only positive ones, while negative exponents are bounded from below, so that the list ends with $\exp\left [ - 3\, \mathfrak{h}\right]$. On the other hand, together with the highest positive exponential $\exp\left [ p_{max} \mathfrak{h}\right]$, also subleading ones  for $ 0< p < p_{max}$ appear and cannot be eliminated by a choice of coefficients in $W(z)$. As a result, the possible match with integrable models of the $cosh$, and $sinh$ type is ruled out, as the match with the sporadic potentials of table \ref{Sporadic}, all of which have the property of being symmetric in positive and negative exponentials. One is thus left with the two series $I_2$ and [9] of table \ref{tab:families}. The last is easily ruled out, since the exponents $6\gamma$ and $\frac{6}{\gamma}$ can be simultaneously integer only for $\gamma=1,2,3$, and no superpotential produces these values without producing other exponentials with intermediate subleading exponents. The hunting ground is thus restricted to the series $I_2$ of table \ref{tab:families} (the $cosh$ models have already been discussed), where one is to spot a combination of powers in $W(z)$ that gives rise to only two exponents in the potential, whose indices should be related by the very restricted relation defining the series.
A careful and systematic analysis led us to the unique solution provided by the following superpotential:
\begin{equation}\label{integsuppot}
    W_{int} \, = \, \lambda  z^4 \, +\, {\rm i} \, \kappa  z^3 \ ,
\end{equation}
where $\lambda$ and $\kappa$ are real constants. Performing the construction of the scalar potential one is led to
\begin{equation}\label{integSupPot}
   V_{int} (z,{\bar z})\, = \, \frac{z^2 {\bar z}^2 \lambda
   \left(3 z^2 \,\kappa +4 {\rm i}
   {\bar z} z^2 \lambda  -4 {\rm i}
   {\bar z}^2 \lambda  z+3
   {\bar z}^2 \kappa
   \right)}{3 (z-{\bar z})^2}\ ,
\end{equation}
To study the extrema of the above potential and for convenience in  the further development of the integration it is useful to change parametrization, reabsorbing the overall coupling constant $\lambda$ into a rescaling of the space-time coordinates and setting:
\begin{equation}\label{changeofvariable}
    \lambda \, = \,\frac{6}{\sqrt{5}} \quad ; \quad \kappa \, = \, \frac{2\omega}{\sqrt{5}}
\end{equation}
In this way the potential (\ref{integSupPot}) becomes
\begin{equation}
    V_{int} (z,{\bar z}) \, = \, \frac{12 z^2  {\bar z}^2 \left((4 {\rm i}
   {\bar z}+\omega ) z^2-4 {\rm i} {\bar z}^2
   z+{\bar z}^2 \omega \right)}{5
   (z-{\bar z})^2} \label{integSupPotBis}
\end{equation}
Next let us consider the derivative of the potential with respect to the complex field $z$:
\begin{eqnarray}
    \partial_z V_{int} & = & \frac{24 z {\bar z}^2 \left((4 {\rm i}
   {\bar z}+\omega ) z^3+2 {\bar z} (-5 {\rm i}
   {\bar z}-\omega ) z^2+6 {\rm i} {\bar z}^3
   z-{\bar z}^3 \omega \right)}{5
   (z-{\bar z})^3} \label{derivoz} \\
   & =& \frac{6}{5} b e^{-2 h} \left(b^2+e^{2
   h}\right) \left(3 \left(4 e^h-\omega
   \right) b^2+e^{2 h} \left(\omega +12
   e^h\right)\right) \label{realpartder}\\
   && + {\rm i} \, \left(-\frac{6}{5} e^{-3 h} \left(b^2+e^{2 h}\right)
  \left(\left(\omega -2 e^h\right) b^4+e^{2h} \left(8 e^h-\omega \right) b^2+2 e^{4 h}
  \left(\omega +5 e^h\right)\right) \right) \nonumber\\
  \label{imagpartder}
\end{eqnarray}
In eq.s (\ref{realpartder},\ref{imagpartder}) we have separated the real and imaginary parts of the potential derivative after replacing the field $z$ with its standard parametrization in terms of a dilaton and an axion:
\begin{equation}\label{fruttosio}
    z \, = \, {\rm i} \exp[h] \, + \, b
\end{equation}
In order to get a true extremum both the real and imaginary part of the derivative should vanish for appropriate values of $b$ and $h$. We begin with considering the zeros of the real part (\ref{realpartder}) in the axion $b$. It is immediately evident that there are three of them:
\begin{equation}\label{zerosReal}
    b \, = \, 0 \quad ; \quad b \, = \, \pm \, \frac{\sqrt{-e^{2 h} \omega -12 e^{3
   h}}}{\sqrt{3} \sqrt{4 e^h-\omega }}
\end{equation}
The first zero in (\ref{zerosReal}) is always available. The other two can occur only if $\omega <0$ is negative and
\begin{equation}\label{condiziata}
    -e^{2 h} \omega -12 e^{3
   h} >0
\end{equation}
If we choose the first zero (truncation of the axion) and we insert it into the imaginary part of the derivative we get:
\begin{equation}\label{allowatoprimo}
    -\frac{12}{5} e^{3 h} \omega -12 e^{4 h} \, = \, 0 \quad \Rightarrow \quad h \, = \,\left\{\begin{array}{ll}-\log \left(-\frac{5}{\omega }\right) & \mbox{if} \,\,\omega < 0 \\
    -\,\infty & \mbox{always} \end{array} \right.
\end{equation}
In the case the second and third zeros displayed in (\ref{zerosReal}) are permissible ($\omega < 0$), substituting their values in the imaginary part of the derivative (\ref{imagpartder}) we obtain the condition
\begin{equation}\label{nonallowato}
    -\frac{128 e^{3 h} \left(2 e^h-\omega \right)
   \omega ^3}{45 \left(4 e^h-\omega \right)^3}\, = \, 0 \quad \Rightarrow \quad h \, = \,- \infty
\end{equation}
Indeed for $\omega < 0$, no other solution of the above equation are available.
We conclude that if $\omega >0$ the only extremum of the potential is at $h \, = \, -\,\infty$ where the potential vanishes so that such an extremum corresponds to a \textit{Minkowski vacuum}. If $\omega <0 $ we have instead an additional extremum at:
\begin{equation}\label{puffoidato}
    z_0 \, = \, {\rm i} \, \frac{|\omega|}{5}
\end{equation}
where the potential takes the following negative value:
\begin{equation}\label{pastruffo}
    V_{int}(z_0) \, = \, \frac{6 \omega ^5}{15625} \, < \, 0
\end{equation}
Hence the extremum (\ref{puffoidato}) defines an anti de Sitter vacuum. We can wonder whether such an AdS vacuum is either supersymmetric or stable. The first possibility can be immediately ruled out by computing the derivative of the superpotential at the extremum:
\begin{equation}\label{dersupext}
    \partial_z W_{int}(z)|_{z=z_0} \, = \, -\frac{6 i \omega ^3}{125 \sqrt{5}} \, \ne \, 0
\end{equation}
Since $\partial_z W_{int}(z)$ does not vanish at the extremum, the auxiliary field of the \textit{chiralino} is different from zero and supersymmetry is broken. In order to investigate stability of the AdS vacuum we have to consider the Breitenlohner-Freedman bound \cite{bretelloneF} which, in the normalizations of \cite{castdauriafre} (see page 462 of Vol. I) is given by:
\begin{equation}\label{bretella}
    \lambda_i \, \ge \, \frac{3}{4} \, V_{int}(z_0)
\end{equation}
where by $\lambda_i$ we have denoted the Hessian of the potential $\partial_i\partial_j V_{int}$ calculated at the extremum
(\ref{puffoidato}). Using ${h,b}$ as field basis we immediately obtain:
\begin{equation}\label{balengone}
    \partial_i\partial_j V_{int}|_{z=z_0} \, = \,
    \left( \begin{array}{ll}
 -\frac{24 \omega ^5}{3125} & 0 \\
 0 & -\frac{84 \omega ^3}{625}
\end{array}
\right)
\end{equation}
from which the two eigenvalues are immediately read off and seen to be both positive. Hence the Breitenlohner-Freedman bound is certainly satisfied and we can conclude that for $\omega <0$ we have two vacua, a Minkowski vacuum at infinity and a \textit{stable, non supersymmetric $\mathrm{AdS}$ vacuum} at the extremum (\ref{puffoidato}). For $\omega >0$, instead we have only the Minkowski vacuum at infinity.
\subsection{\sc Truncation to zero axion}
\begin{figure}[!hbt]
\begin{center}
\iffigs
 \includegraphics[height=55mm]{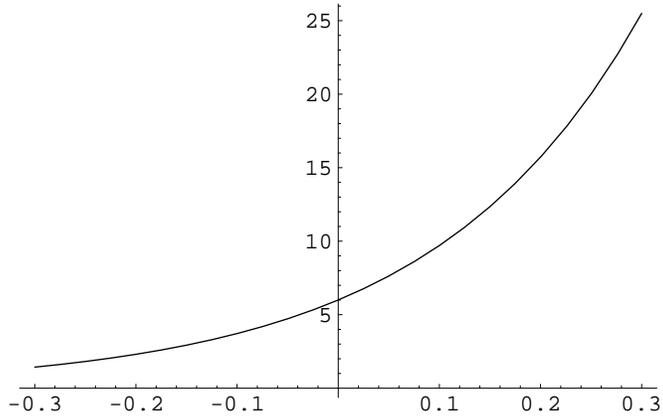}
 \vskip 2cm
 \includegraphics[height=55mm]{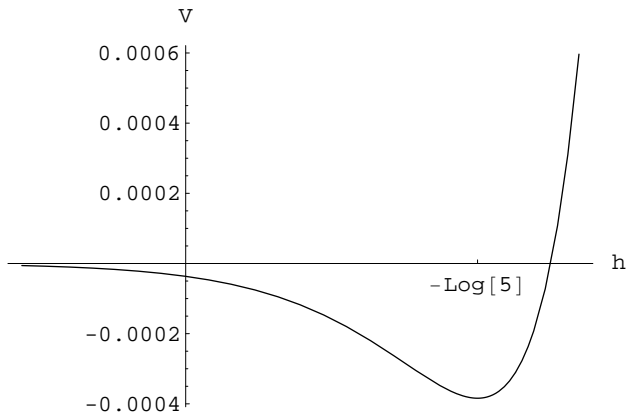}
\else
\end{center}
 \fi
\caption{\it
In this figure we display the behavior of the supersymmetric integrable potential \ref{farimboldobis} for the two choices $\omega =1$ (above) and $\omega = -1$ (below)
}
\label{zuzzolo}
 \iffigs
 \hskip 1cm \unitlength=1.0mm
 \end{center}
  \fi
\end{figure}

The potential (\ref{integSupPotBis}) of the considered supergravity model can be consistently truncated to a vanishing value of the axion $b$, since its derivative derivative with respect to $b$ vanishes at $b=0$. Imposing such a truncation from
(\ref{integSupPotBis}) we obtain the following  dilatonic potential
\begin{equation}\label{farimboldobis}
  V_{Int} \, = \, \frac{6}{5} e^{4 \mathfrak{h}} \left(\omega +4 e^\mathfrak{h}\right)
\end{equation}
which by means of the replacement (\ref{babushka}) is mapped into the case $\gamma=\frac{2}{3}$ of the series $I_2$ of integrable potentials listed in table \ref{tab:families}.
According to the previous analysis of extrema of the full theory, we see that,
depending on the sign of the parameter $\omega$, this potential is either monotonic or it has a minimum (see fig.\ref{zuzzolo})
The important thing to note is that when it exists, the extremum of the potential is always at a negative value of the potential. It corresponds to the stable $\mathrm{AdS}$ vacuum discussed in the previous subsection. As its well known the $\mathrm{AdS}$ has no parametrization in terms of spatially flat constant time slices. Hence if we assume, to begin with, a spatially flat ansatz for the metric, as we do in eq. (\ref{piatttosa}), no solution of the Friedman equations can stabilize the scalar field at the $\mathrm{AdS}$ extremum. Indeed the exact solutions produced by the available general integral show that the scalar field always flows to infinity at the beginning and end of cosmic time.
\subsection{\sc Explicit integration of the supersymmetric integrable model}
In order to integrate the field equation of this model, it is convenient to write down  the explicit form of the Lagrangian which has the following form
\begin{eqnarray}\label{frangente}
   \mathcal{L}_{int} & = & e^{3  {A}(t)- {\mathcal{B}}(t)}
   \left(-\frac{3}{2}
    {A}'(t)^2+\frac{3}{2}
   \mathfrak{h}'(t)^2-\frac{6}{5} e^{2  {\mathcal{B}}(t)+4
   \mathfrak{h}(t)} \left(\omega +4
   e^{h(t)}\right)\right)
\end{eqnarray}
and following the strategy outlined in \cite{primopapero}, one can move on to two new functions $U(\tau)$ and $V(\tau)$
\begin{eqnarray}
  A(\tau) &=& \frac{1}{5} \log (U(\tau ))+\log (V(\tau )) \nonumber\\
  B(\tau) &=& 2 \log (V(\tau ))-\frac{2}{5} \log (U(\tau ))\nonumber\\
  \mathfrak{h}(\tau) &=& \frac{1}{5} (\log (U(\tau ))-5 \log (V(\tau
   ))) \label{integtrasformazia}
\end{eqnarray}
Inserting the transformation (\ref{integtrasformazia}) into the Lagrangian (\ref{frangente}) this becomes
\begin{equation}\label{fiduciato}
\mathcal{L}_{int} \, = \,  -4 U(\tau )^{6/5}-\omega  V(\tau ) U(\tau
   )-U'(\tau ) V'(\tau )
\end{equation}
while the Hamiltonian constraint takes the form
    \begin{equation}\label{sfiducioso}
  \mathcal{H} \,  = \,   4 (U(\tau
   )^{6/5}+\omega \,V(\tau )
   U(\tau )-U'(\tau ) V'(\tau
   ) \, = \, 0
\end{equation}
The field equations associated with (\ref{fiduciato}) have the following triangular form:
\begin{equation}\label{equemozionica}
    \begin{array}{lcl}
 \omega  U(\tau )-U''(\tau ) & = & 0  \\
 \omega  V(\tau )-V''(\tau )+\frac{24}{5}
   \sqrt[5]{U(\tau )} &=& 0
   \end{array}
\end{equation}
and can be integrated by means of trigonometric or hyperbolic functions depending on the sign of $\omega$.
\subsection{\sc Trigonometric solutions in the potential with $AdS$ extremum}
If we pose $\omega\, = \, -\nu^2$ the first equation becomes the equation of the standard harmonic oscillator and we have:
{\small
\begin{eqnarray}
  U(\tau) &=& a \cos (\nu  \tau )+b \sin (\nu  \tau ) \label{armonium}\\
  V(\tau) &=& \left( 4 \cot \left(\nu  \tau +\tan
   ^{-1}\left(\frac{a}{b}\right)\right) \, \times \right.\nonumber \\
  &&\left. _2F_1\left(\frac{1}{2},\frac{9}{10};\frac{
   3}{2};\cos ^2\left(\nu  \tau +\tan
   ^{-1}\left(\frac{a}{b}\right)\right)
   \right) \sin ^2\left(\nu  \tau +\tan
   ^{-1}\left(\frac{a}{b}\right)\right)^{9/10
   } \left(b \cos (\nu  \tau )-a \sin (\nu  \tau
   )\right)\right. \nonumber\\
   && \left. +5 \left(\cos (\nu  \tau ) \left(c (a
   \cos (\nu  \tau )+b \sin (\nu  \tau
   ))^{4/5} \nu ^2+4 a\right)\right.\right.\nonumber\\
   &&\left. +\sin (\nu  \tau
   ) \left(d (a \cos (\nu  \tau )+b \sin (\nu
    \tau ))^{4/5} \nu ^2+4 b\right)\right) \times (5
   \nu ^2 (a \cos (\nu  \tau )+b \sin (\nu\tau ))^{-4/5})
\end{eqnarray}
}
where  $_2F_1$ denotes a hypergeometric function of the specified indices. The parameters $a,b,c,d$ are four integration constants, on which the Hamiltonian constraint imposes the condition
\begin{equation}\label{dariopisco}
    (b c + a d) \, = \, 0 \ .
   \end{equation}
We solve the constraint by setting $d=-\rho \, a$, $c=\rho \,b$. In this way we obtain an explicit general integral depending on three parameters $(a,b,\rho)$. The explicit form of the solution for the scale factor, for the $\exp[\mathcal{B}]$ function and for the scalar field $\mathfrak{h}(\tau)$ are given below.
{\small
\begin{eqnarray}
 a(\tau;a,b,\rho,\nu) &=& \frac{\mathfrak{J}(\tau;a,b,\rho,\nu)}{5 \nu ^2 (a \cos (\nu  \tau )+b \sin (\nu
   \tau ))^{3/5}}\nonumber \\
  \mathfrak{J}(\tau;a,b,\rho,\nu) &=& 5 \left(\sin (\nu  \tau ) \left(4 b-a \nu ^2
   \rho  (a \cos (\nu  \tau )+b \sin (\nu
   \tau ))^{4/5}\right)\right.\nonumber\\
   && \left.+ \cos (\nu  \tau )
   \left(b \rho  (a \cos (\nu  \tau )+b \sin
   (\nu  \tau ))^{4/5} \nu ^2+4
   a\right)\right)\nonumber \\
   && + 4 \cot \left(\nu  \tau +\tan
   ^{-1}\left(\frac{a}{b}\right)\right) \,
   _2F_1\left(\frac{1}{2},\frac{9}{10};\frac{3
   }{2};\cos ^2\left(\nu  \tau +\tan
   ^{-1}\left(\frac{a}{b}\right)\right)\right) \times \nonumber\\
   && (b \cos (\nu  \tau )-a \sin (\nu  \tau )) \sin
   ^2\left(\nu  \tau +\tan
   ^{-1}\left(\frac{a}{b}\right)\right)^{9/10}\nonumber\\
  \exp[\mathcal{B}(\tau;a,b,\rho,\nu)] &=& \frac{\left(\mathfrak{J}(\tau;a,b,\rho,\nu)\right)^2}{25 \nu ^4 (a \cos (\nu  \tau )+b \sin (\nu
   \tau ))^2}\\
 \mathfrak{h}(\tau;a,b,\rho,\nu) &=& \log\left[\frac{5 \nu ^2 (a \cos (\nu  \tau )+b \sin (\nu
   \tau ))^{7/5}}{ \mathfrak{J}(\tau;a,b,\rho,\nu)} \right] \label{grantrigsoluz}
\end{eqnarray}
}
From the explicit form of the solution the structure of its time development  is not immediately evident. Yet it is clear that  it must be periodic, since all addends are constructed in terms of trigonometric functions with the same frequency $\nu$.  Therefore we are lead to suspect that the scalar field will go to infinity and the scale factor to zero in a periodic fashion. In other words we expect solutions with a Big Bang and a Big Crunch. This expectation is sustained by the general arguments of paper \cite{primopapero}. Indeed the considered scalar potential has an absolute minimum, yet this minimum is at a negative value, so that in the phase portrait of the equivalent first system there is no fixed point and under these conditions the only possible solutions are blow-up solutions, physically corresponding to Big-Bang/Big Crunch universes.
\subsubsection{\sc Structure of the moduli space of the general integral}
In order to understand the actual form and the behavior of these type of solutions it is convenient  to investigate first the  physical interpretation of the three integration constants $a,b,\rho$ that we have introduced and reduce the general integral to a simpler canonical form.
\par
An \textit{a priori} observation valid for all the solutions of Friedman equations is that the effective parameter labeling such solutions is only one, two parameters being accounted for by the uninteresting overall scale of the scale--factor and  by the equally uninteresting possibility of shifting the parametric time $\tau$ by a constant. What has to be done case by case   is to  work out those combinations of the parameters that can be disposed of by the above mentioned symmetries and single out the unique meaningful deformation parameter.
\par
In the present case we begin by noting that all functions in the solution depend on $\tau$ only through the combination $\nu \tau$. Hence the frequency parameter $\nu$ can be reabsorbed by rescaling the parametric time:
\begin{equation}\label{reassorbo}
    \nu \tau \, = \, \tau^\prime.
\end{equation}
In other words, without loss of generality we can set $\nu=1$. Secondly let us consider the following rescaling of the solution parameters:
\begin{equation}\label{riscalaggio}
    a \, \mapsto \, \lambda \, a \quad ; \quad b \, \mapsto \, \lambda \, b \quad ; \quad \rho \, \mapsto \, \lambda^{-4/5} \, \rho
\end{equation}
under such a transformation we have:
\begin{eqnarray}
   a(\tau;\, \lambda \, a, \,\lambda \, b, \, \lambda^{-4/5} \, \rho, \, 1)  &=& \lambda^{2/5} \, a(\tau;\, a, \, b, \, \rho, \, 1) \nonumber \\
   \exp[\mathcal{B}(\tau;\, \lambda \, a, \,\lambda \, b, \, \lambda^{-4/5} \, \rho, \, 1)]  &=& \exp[\mathcal{B}(\tau;\, a, \, b, \, \rho, \, 1)]\nonumber \\
  \mathfrak{h}(\tau;\, \lambda \, a, \,\lambda \, b, \, \lambda^{-4/5} \, \rho, \, 1) &=& \mathfrak{h}(\tau;\, a, \, b, \, \rho, \, 1) \label{bamboccio}
\end{eqnarray}
From this we deduce that a suitable combination of the parameters $a,b,\rho$ is just the overall scale of the scale factor, as we announced. Using the symmetry of eq.(\ref{bamboccio}) we could for instance fix the gauge where either one of the three parameters $a,b,\rho$ is $1$. Yet we can do much better if we realize that the ratio $a/b$ actually amounts to a shift of the parametric time. Indeed by means of several analytic manipulations we can prove the following identities:
\begin{eqnarray}
   a\left(\tau + \arctan\left(\frac{b}{a}\right);\, a, \, b, \, \rho, \, 1\right)&=&  \frac{4}{5} \left(
   \sqrt{{b^2}+{a^2}} \cos (\tau
   )\right)^{2/5} \times \nonumber \\
   &&\left(\cos ^2(\tau )^{9/10} \,
   _2F_1\left(\frac{1}{2},\frac{9}{10};\frac{3}{2};\sin ^2(\tau )\right)\, \tan ^2(\tau )+5\right) \nonumber\\
   && -\rho \, \left(\sqrt{{b^2}+{a^2}}\, \cos (\tau )\right)^{6/5} \tan (\tau ) \nonumber\\
   \exp\left[\mathcal{B}\left(\tau + \arctan\left(\frac{b}{a}\right);\, a, \, b, \, \rho, \, 1\right)\right]&=& \frac{1}{25} \left(4 \cos ^2(\tau
   )^{9/10} \, _2F_1\left(\frac{1}{2},\frac{9}{10};\frac{3}{2};\sin ^2(\tau )\right)\, \tan ^2(\tau ) \right. \nonumber \\
   &&\left. -5 \rho  \left(\sqrt{{b^2}+{a^2}}\cos (\tau)\right)^{4/5} \tan (\tau)+20\right)^2 \nonumber\\
 \mathfrak{h}\left(\tau + \arctan\left(\frac{b}{a}\right);\, a, \, b, \, \rho, \, 1\right)&=& -\log \left(\frac{4}{5} \cos ^2(\tau
   )^{9/10} \, _2F_1\left(\frac{1}{2},\frac{9}{10};\frac{3}{2};\sin ^2(\tau )\right)\tan ^2(\tau )\right.\nonumber\\
   &&\left.-\rho  \left(\sqrt{{b^2}+{a^2}} \cos (\tau)\right)^{4/5} \tan (\tau )+4\right)\label{kukletta}
\end{eqnarray}
In this way we realize that after the shift $\tau \, \mapsto \, \tau \, + \, \arctan\left(\frac{b}{a}\right)$ the solution functions depend only on the two parameters $ \sqrt{{b^2}+{a^2}} $ and $\rho$. Furthermore the explicit result (\ref{kukletta}) suggests that we redefine these latter as follows:
\begin{equation}\label{rifriggo}
    \sqrt{{b^2}+{a^2}} \, = \, \Lambda^{\frac{5}{2}} \quad ; \quad \rho \, = \, \frac{Y}{\Lambda^2}
\end{equation}
so that we obtain:
\begin{eqnarray}
  a\left(\tau,\Lambda,Y\right) \, = \, \Lambda\,  \mathfrak{a}(\tau,Y) &=& \Lambda \,\left[ \frac{4}{5} \cos ^{\frac{2}{5}}(\tau )
   \left(\left(\cos \tau \right)^{9/5} \,
   _2F_1\left(\frac{1}{2},\frac{9}{10};
   \frac{3}{2};\sin ^2(\tau )\right)
   \tan ^2 (\tau)+5\right)\right.\nonumber\\
   &&\left.-Y \cos
   ^{\frac{1}{5}}(\tau ) \sin (\tau )\right]\nonumber\\
 \exp\left[\mathcal{B}\left(\tau ,Y \right)\right]&=& \frac{1}{25} \left(4 \cos ^2(\tau
   )^{9/10} \,  _2F_1\left(\frac{1}{2},\frac{9}{10};\frac{3}{2};\sin ^2(\tau )\right)\tan ^2(\tau )
   -\frac{5 Y \sin (\tau)}{\cos ^{\frac{1}{5}}(\tau )}+20\right)^2 \nonumber\\
  \mathfrak{h}\left(\tau,Y\right)&=& -\log \left(\frac{4}{5} \cos ^2(\tau
   )^{9/10} \,
   _2F_1\left(\frac{1}{2},\frac{9}{10};
   \frac{3}{2};\sin ^2(\tau )\right)
   \tan ^2(\tau )-\frac{Y \sin (\tau
   )}{\cos ^{\frac{1}{5}}(\tau
   )}+4\right) \nonumber\\
   \label{confettofalqui}
\end{eqnarray}
which puts into evidence the only relevant deformation parameter, namely $Y$.  We can get some understanding of the meaning of this latter by plotting the solution functions for various values of $Y$. We begin by analyzing the simplest and most symmetrical solution at $Y=0$.
\subsubsection{\sc The simplest trigonometric solution at $Y=0$}
\begin{figure}[!hbt]
\begin{center}
\iffigs
 \includegraphics[height=60mm]{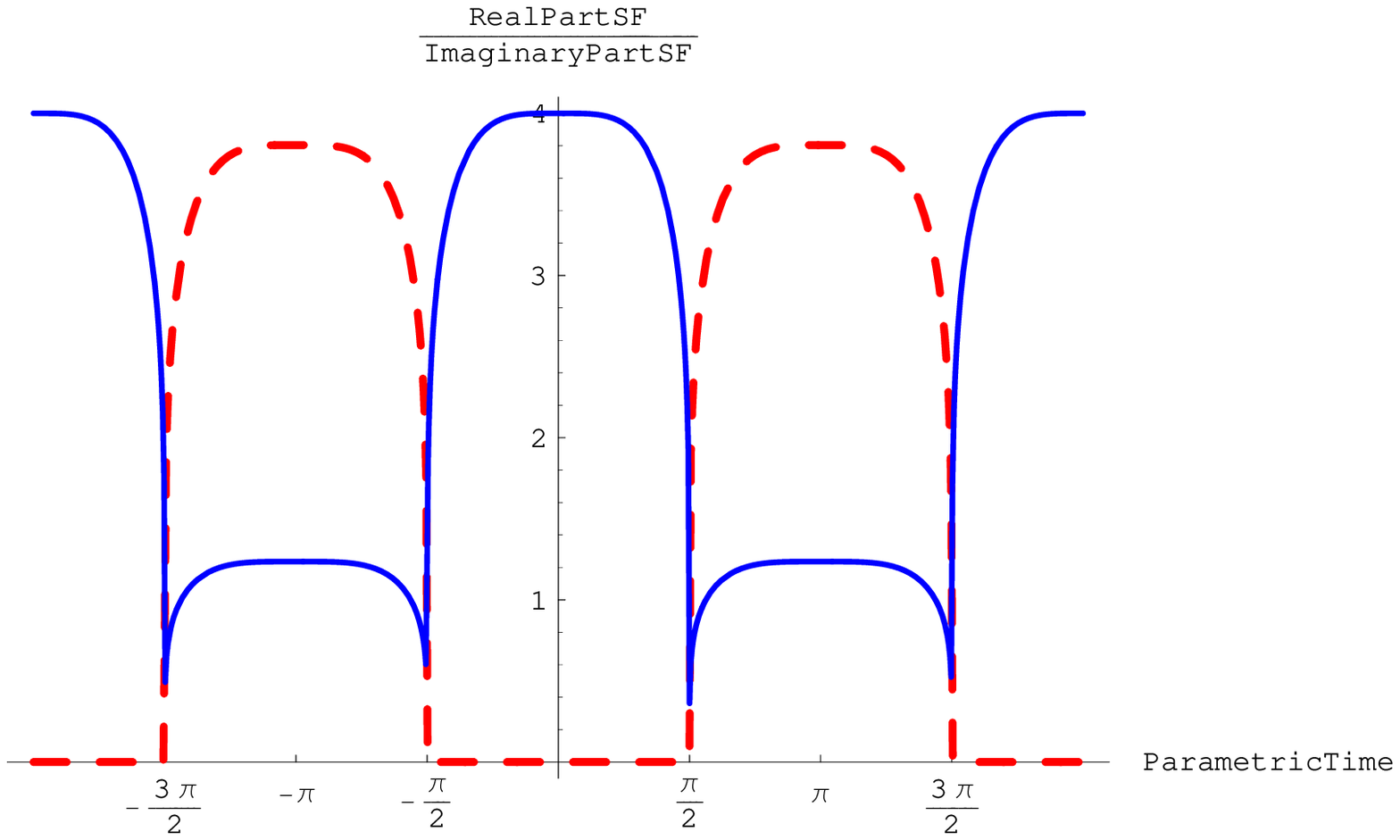}
 \includegraphics[height=60mm]{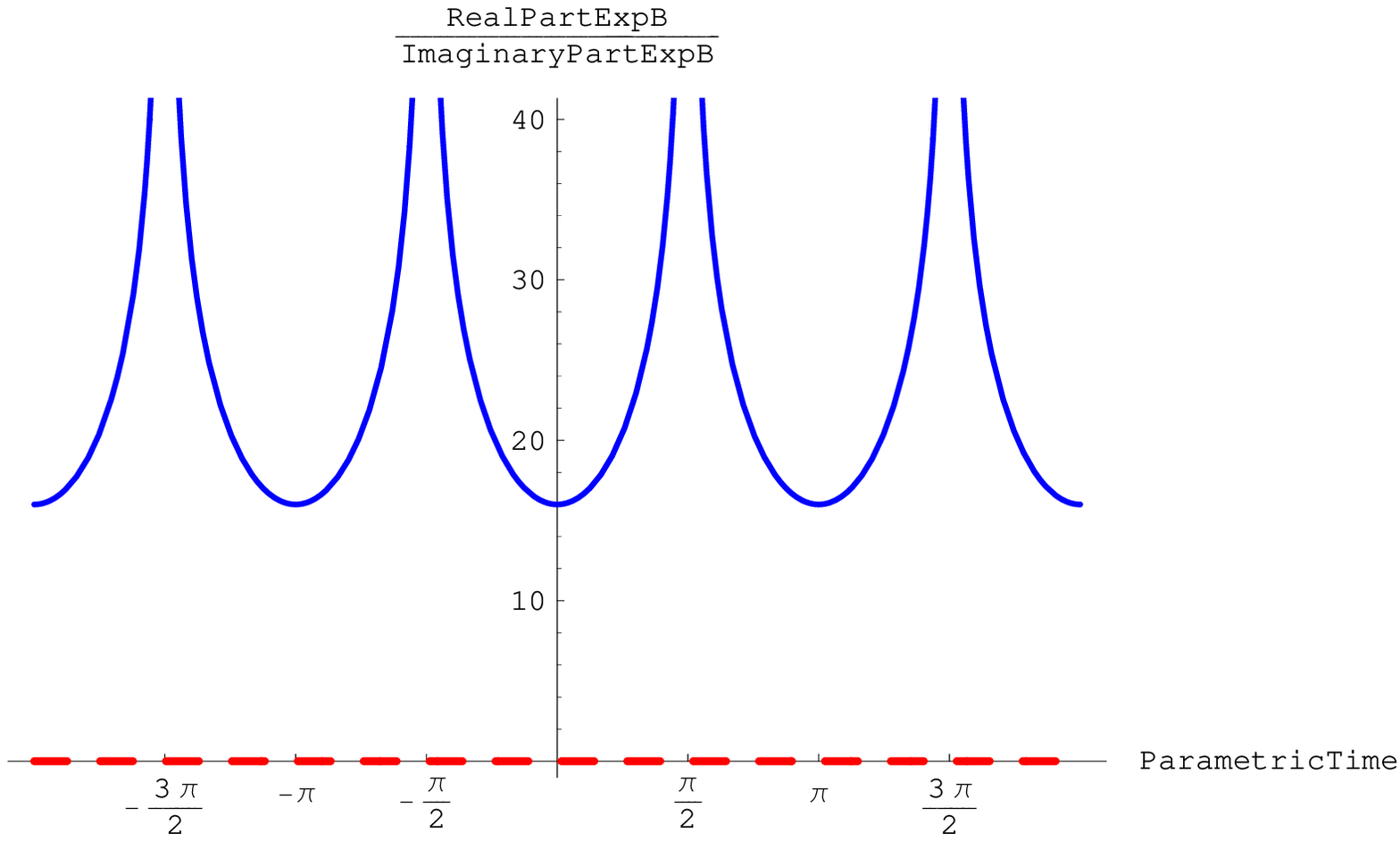}
 \includegraphics[height=60mm]{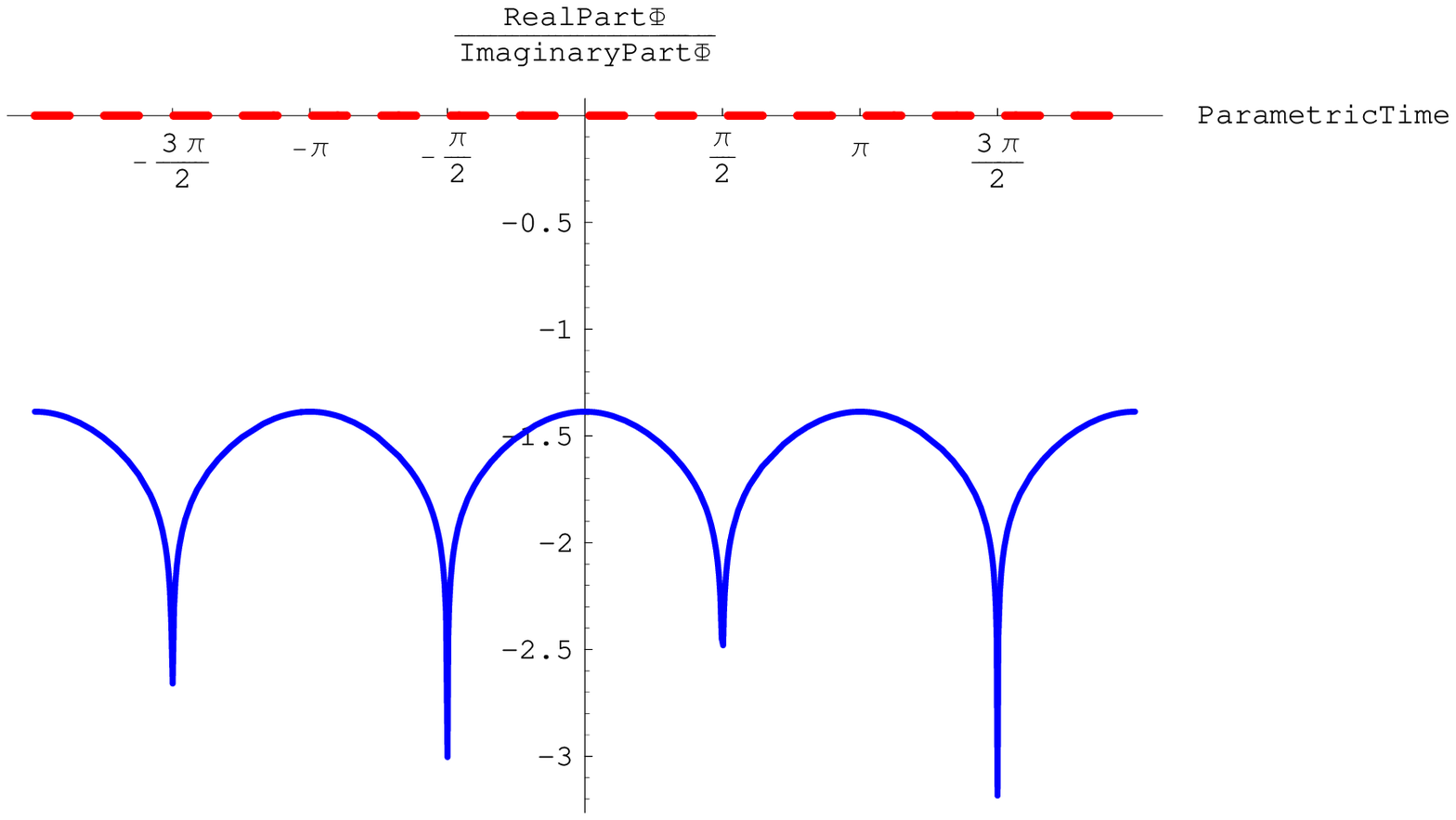}
\else
\end{center}
 \fi
\caption{\it
Plots of the real and imaginary parts of the scale factor, $\exp[B]$ factor and scalar field for the case of parameter $Y=0$. In the three diagrams the solid line represents the real part, while the dashed line represents the imaginary part. We note the periodicity of all the functions and the windows where the three of them are simultaneously real. Taking as basic reference window the interval $\left[-\frac{\pi}{2},\frac{\pi}{2}\right]$ we note the reflection symmetry of the plots with respect to the point $\tau =0$.}
\label{Y0triplots}
 \iffigs
 \hskip 1cm \unitlength=1.1mm
 \end{center}
  \fi
\end{figure}
\begin{figure}[!hbt]
\begin{center}
\iffigs
 \includegraphics[height=55mm]{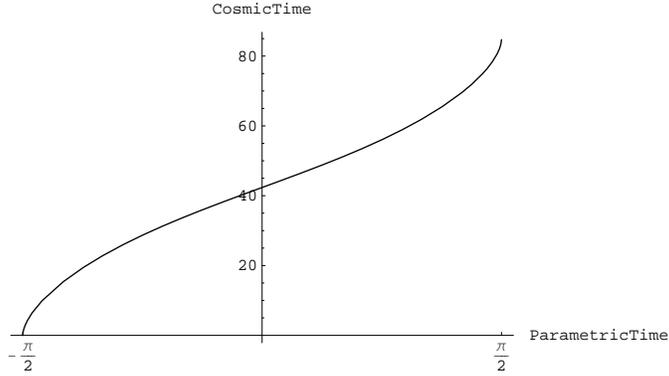}
\else
\end{center}
 \fi
\caption{\it
In this figure we display the behavior of the cosmic time with respect to the parametric time for the   trigonometric type of solution of the supersymmetric cosmological model with parameter $Y=0$.
}
\label{cosmicomica}
 \iffigs
 \hskip 1cm \unitlength=1.0mm
 \end{center}
  \fi
\end{figure}
\begin{figure}[!hbt]
\begin{center}
\iffigs
 \includegraphics[height=55mm]{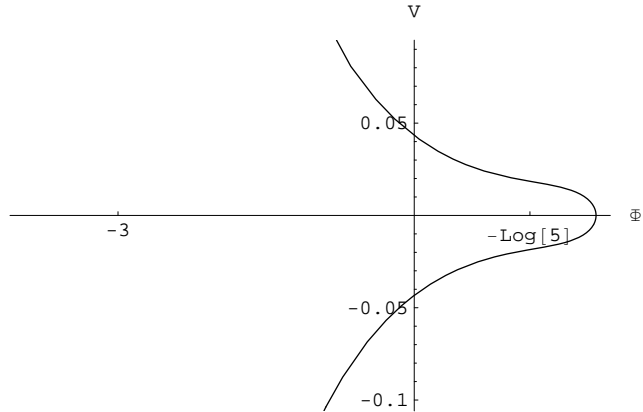}
\else
\end{center}
 \fi
\caption{\it
In this figure we display the phase portrait of the solution defined by $Y=0$. The axes are the scalar field $\Phi \equiv \mathfrak{h}$ and its derivative with respect tot the cosmic time $V \equiv \partial_{T_c} \mathfrak{h}$. The extremum of the potential is at $\Phi_0 \, = \, - \log[5]$. It is reached by the solution however with a non vanishing velocity. The field also reaches vanishing velocity yet not an extremum of the potential. Hence there is no fixed point and the trajectory is from infinity to infinity with no fixed point.
}
\label{facciaportratto}
 \iffigs
 \hskip 1cm \unitlength=1.0mm
 \end{center}
  \fi
\end{figure}
\begin{figure}[!hbt]
\begin{center}
\iffigs
 \includegraphics[height=60mm]{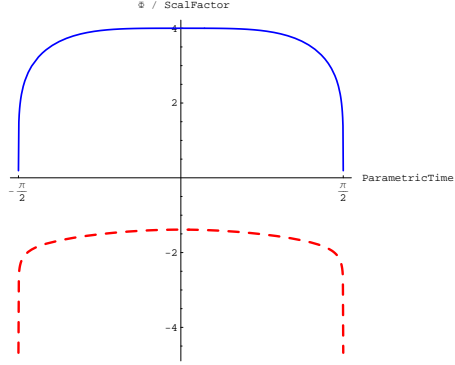}
\else
\end{center}
 \fi
\caption{\it
In this figure we display plots (in parametric time) of  the scale factor (solid line) and of the scalar field (dashed line) for the trigonometric type at $Y=0$. It is evident that in a finite time the Universe undergoes a Big Bang, a decelerated expansion and then a Big Crunch. At the same time the scalar field climbs from $-\infty$ to a maximum and then descends again to $-\infty$.
}
\label{cuginetto}
 \iffigs
 \hskip 1cm \unitlength=1.0mm
 \end{center}
  \fi
\end{figure}

In fig.\ref{Y0triplots} we display the behavior of the real and imaginary parts of three main functions composing the solution in the case $Y=0$. From these plots it is evident that a physical  solution of scalar-matter coupled gravity exists only in those windows where the three functions are simultaneously real, for instance in the interval $\left[-\frac{\pi}{2},\frac{\pi}{2}\right]$. In such an interval of the parametric time $\tau$ the scale factor goes from a zero to another zero so that the cosmic evolution should correspond to an Universe that starts with a Big Bang and finishes its life collapsing into a Big Crunch. To put such a conclusion on a firm ground we have actually to verify that both zeros of the scale factor do indeed correspond to a true space-like singularity and this can only be done by considering the intrinsic components of the curvature two-form showing that they all blow up to infinity in the initial and final point. This we will do shortly. First let us verify analytically the limit of the scale factor and of the scalar field in the initial and final point of the reality domain of the solution. We find:
\begin{eqnarray}
\label{apertivo}
  \lim_{\tau \rightarrow \pm \frac{\pi}{2}} \mathfrak{ a}(\tau;0) &=& 0 \nonumber\\
  \lim_{\tau \rightarrow \pm \frac{\pi}{2}}  \mathfrak{h}(\tau,0) &=& -\infty\nonumber\\
  \lim_{\tau \rightarrow \pm \frac{\pi}{2}} \exp[\mathcal{B}(\tau,0)]  &=& +\infty
\end{eqnarray}
This means that a life-cycle of this universe is contained in the following finite interval of parametric time $\left[ -\frac{\pi}{2}\, , \, \frac{\pi}{2} \right]$, which by the $\exp[\mathcal{B}(\tau;1,1,0,1)]$ function is monotonically mapped into a finite interval of Cosmic time. Indeed defining:
\begin{equation}\label{cosmicotempus}
    T_c(\tau) \, = \, \int_{-\frac{\pi}{2}}^{\tau} \,\exp[\mathcal{B}(t;1,1,0,1)] \, \mathrm{d}t
\end{equation}
we find:
\begin{equation}\label{corleonite}
    T_c(-\frac{\pi}{2}) \, = \, 0 \quad ; \quad T_c(\frac{\pi}{2}) \, = \, 84.7046
\end{equation}
the plot of $T_c(\tau)$ being displayed in fig.\ref{cosmicomica}
Due to these properties of the cosmic time function we do not loose any essential information by plotting the solution in parametric rather than in cosmic time. The essential difference between this case and the case of positive potentials with
positive extrema discussed in \cite{primopapero} is best appreciated by considering the phase-portrait of the this solution presented in fig.\ref{facciaportratto}
The absence of a fixed point implies the structure of a blow-up solution with a Big Bang and a Big Crunch which is displayed in fig.\ref{cuginetto}.
As we already emphasized the interpretation of Big Bang and Big Crunch is suggested by the plots, yet it has to be verified by an appropriate study of the curvature singularity.
\subsubsection{\sc The curvature two-form and its singularities}
Throughout this paper we  consider metrics of the form (\ref{piatttosa}). It is important to calculate the explicit general form of the curvature 2-form associated with such metrics. To this effect we introduce the vielbein:
\begin{equation}\label{vielbeinus}
    E^1 \, = \, \exp\left[\mathcal{B}(\tau)\right] \, d\tau \quad ; \quad E^i \, = \, \quad \exp\left[A(\tau)\right]\, dx^i  \quad (i\, = \, 2,3,4)
\end{equation}
and we obtain the following result for the matrix valued curvature two-form:
\begin{eqnarray}\label{curbura}
   & R^{AB} \equiv d\omega^{AB} \, + \, \omega^{AC} \, \wedge \, \omega^{DB} \, \eta_{CD} \, = \,  & \nonumber\\
   & \null & \nonumber\\
   & \left(
\begin{array}{c|c|c|c}
 0 & -\frac{E^1 \wedge E^2
   \left(\mathfrak{b}
   \mathfrak{a}''-\mathfrak{a}'
   \mathfrak{b}'\right)}{\mathfrak{a}
   \mathfrak{b}^3} &
   -\frac{E^1 \wedge E^3
   \left(\mathfrak{b}
   \mathfrak{a}''-\mathfrak{a}'
   \mathfrak{b}'\right)}{\mathfrak{a}
   \mathfrak{b}^3} &
   -\frac{E^1 \wedge E^4
   \left(\mathfrak{b}
   \mathfrak{a}''-\mathfrak{a}'
   \mathfrak{b}'\right)}{\mathfrak{a}
   \mathfrak{b}^3} \\
   \null & \null & \null & \null \\
   \hline
   \null & \null & \null & \null \\
 \frac{E^1 \wedge E^2
   \left(\mathfrak{b}
   \mathfrak{a}''-\mathfrak{a}'
   \mathfrak{b}'\right)}{\mathfrak{a}
   \mathfrak{b}^3} & 0 &
   -\frac{E^2 \wedge E^3
   (\mathfrak{a}')^2}{\mathfrak{a}^2 \mathfrak{b}^2} &
   -\frac{E^2)\wedge E^4)
   (\mathfrak{a}')^2}{\mathfrak{a}^2 \mathfrak{b}^2} \\
    \null & \null & \null & \null \\
   \hline
   \null & \null & \null & \null \\
 \frac{E^1 \wedge E^3
   \left(\mathfrak{b}
   \mathfrak{a}''-\mathfrak{a}'
   \mathfrak{b}'\right)}{\mathfrak{a}
   \mathfrak{b}^3} &
   \frac{E^2 \wedge E^3
   (\mathfrak{a}')^2}{\mathfrak{a}^2 \mathfrak{b}^2} & 0 &
   -\frac{E^3 \wedge E^4
   (\mathfrak{a}')^2}{\mathfrak{a}^2 \mathfrak{b}^2} \\
    \null & \null & \null & \null \\
   \hline
   \null & \null & \null & \null \\
 \frac{E^1 \wedge E^4
   \left(\mathfrak{b}
   \mathfrak{a}''-\mathfrak{a}'
   \mathfrak{b}'\right)}{\mathfrak{a}
   \mathfrak{b}^3} &
   \frac{E^2 \wedge E^4
   (\mathfrak{a}')^2}{\mathfrak{a}^2 \mathfrak{b}^2} &
   \frac{E^3 \wedge E^4
   (\mathfrak{a}')^2}{\mathfrak{a}^2 \mathfrak{b}^2} & 0
\end{array}
\right) &
\end{eqnarray}
having denoted by $\omega^{AB}$ the Levi-Civita spin connection defined by:
\begin{equation}\label{levicivita}
    dE^A \, + \, \omega^{AB} \, \wedge \, E^{C} \, \eta_{BC} \, = \,0
\end{equation}
and having introduced the following notation:
\begin{equation}\label{franzusko}
    \mathfrak{a} \, \equiv \, \exp[A(\tau)] \, = \,  a(\tau) \quad ; \quad \mathfrak{b} \, \equiv \, \exp[\mathcal{B}(\tau)]
\end{equation}
If  the functions $\mathfrak{a},\mathfrak{b}$ are specialized to the  form (\ref{confettofalqui}), we obtain some rather formidable, yet fully  explicit analytic expressions that correspond to the intrinsic  components of the Riemann tensor for the solution under consideration. We can calculate the limit of the curvature two-form when $\tau$ approaches a zero of the scale-factor. In the case $Y=0$, the only zeros are at $\tau\, = \, \pm \frac{\pi}{2}$ and we find:
\begin{equation}\label{limitus}
\lim_{\tau \rightarrow \pm \frac{\pi}{2}} \, R^{AB} \, = \,    \left(
\begin{array}{llll}
 0 & \infty  E^1\wedge E^2 & \infty
   E^1\wedge E^3 & \infty  E^1\wedge
   E^4 \\
 -\infty  E^1\wedge E^2 & 0 & -\infty
    E^2\wedge E^3 & -\infty
   E^2\wedge E^4 \\
 -\infty  E^1\wedge E^3 & \infty
   E^2\wedge E^3 & 0 & -\infty
   E^3\wedge E^4 \\
 -\infty  E^1\wedge E^4 & \infty
   E^2\wedge E^4 & \infty  E^3\wedge
   E^4 & 0
\end{array}
\right)
\end{equation}
Hence the Riemann tensor diverges in all directions and both the initial and final zero of the scale factor correspond to true singularities confirming their interpretation as the Big Bang and Big Crunch points.
Actually we can make the statement even more precise. In the case of the $Y=0$ solution we can calculate the asymptotic expansion of the curvature components in the neighborhood of the two singularities and we find:
\begin{equation}\label{divergendo}
R^{AB} \, \stackrel{\tau \rightarrow \pm \pi /2}{\approx }\,  \frac{1}{(\mp \frac{\pi}{2}+\tau)^{6/5}}\,
 \frac{25 \Gamma \left(\frac{3}{5}\right)^4}{8 \pi ^2
   \Gamma \left(\frac{1}{10}\right)^4} \, \left(
\begin{array}{llll}
 0 & E^1\wedge E^2 & E^1\wedge E^3
   & E^1\wedge E^4 \\
 -E^1\wedge E^2 & 0 & -\frac{1}{2}
   E^2\wedge E^3 & -\frac{1}{2}
   E^2\wedge E^4 \\
 -E^1\wedge E^3 & \frac{1}{2}
   E^2\wedge E^3 & 0 & -\frac{1}{2}
   E^3\wedge E^4 \\
 -E^1\wedge E^4 & \frac{1}{2}
   E^2\wedge E^4 & \frac{1}{2}
   E^3\wedge E^4 & 0
\end{array}
\right)
\end{equation}
Hence all the components of the intrinsic curvature tensor have the same  degree of divergence which is identically at the Big Bang and at the Big Crunch. This reflects the already noted $\mathbb{Z}_2$ symmetry of the solution.
\subsubsection{\sc $Y$-deformed solutions}
\begin{figure}[!hbt]
\begin{center}
\iffigs
 \includegraphics[height=60mm]{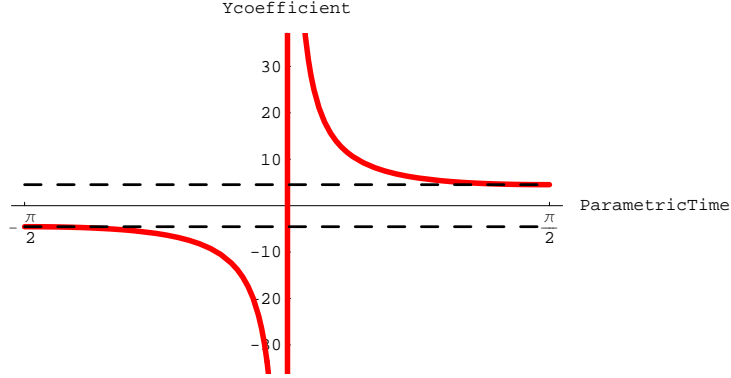}
\else
\end{center}
 \fi
\caption{\it
Plot of the function $\mathfrak{f}(\tau)$. A zero of the scale factor occurs when $\mathfrak{f}(\tau_0) = Y$. Hence for all those values of $Y$ that are never attained by the function $\mathfrak{f}(\tau)$ in the range $\left [ -\frac{\pi}{2} , \frac{\pi}{2} \right ]$ there is no early Big Crunch. The two straight asymptotic lines are at the values $\pm Y_0 \, = \, \pm \frac{4 \sqrt{\pi } \Gamma
   \left(\frac{11}{10}\right)}{\Gamma
   \left(\frac{3}{5}\right)}$
}
\label{ziotta}
 \iffigs
 \hskip 1cm \unitlength=1.0mm
 \end{center}
  \fi
\end{figure}

\begin{figure}[!hbt]
\begin{center}
\iffigs
 \includegraphics[height=55mm]{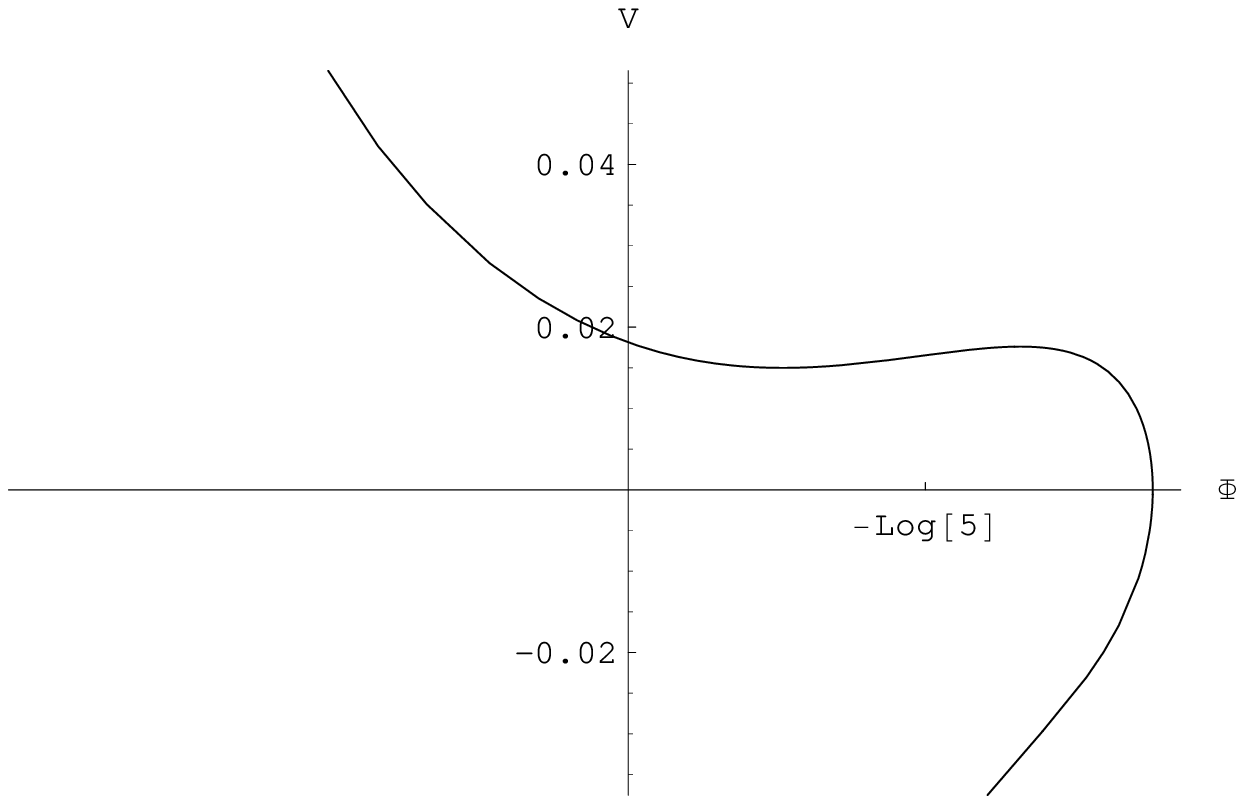}
\else
\end{center}
 \fi
\caption{\it
In this figure we display the phase portrait of the solution defined by $Y=1$. The axes are the scalar field $\Phi \equiv \mathfrak{h}$ and its derivative with respect tot the cosmic time $V \equiv \partial_{T_c} \mathfrak{h}$. The extremum of the potential is at $\Phi_0 \, = \, - \log[5]$. It is reached by the solution however with a non vanishing velocity. The field also reaches vanishing velocity yet not at the extremum of the potential. Hence there is no fixed point and the trajectory is  infinite with no fixed point. The symmetric shape of the $Y=0$ trajectory is  lost.
}
\label{facciaportratto2}
 \iffigs
 \hskip 1cm \unitlength=1.0mm
 \end{center}
  \fi
\end{figure}

\begin{figure}[!hbt]
\begin{center}
\iffigs
 \includegraphics[height=60mm]{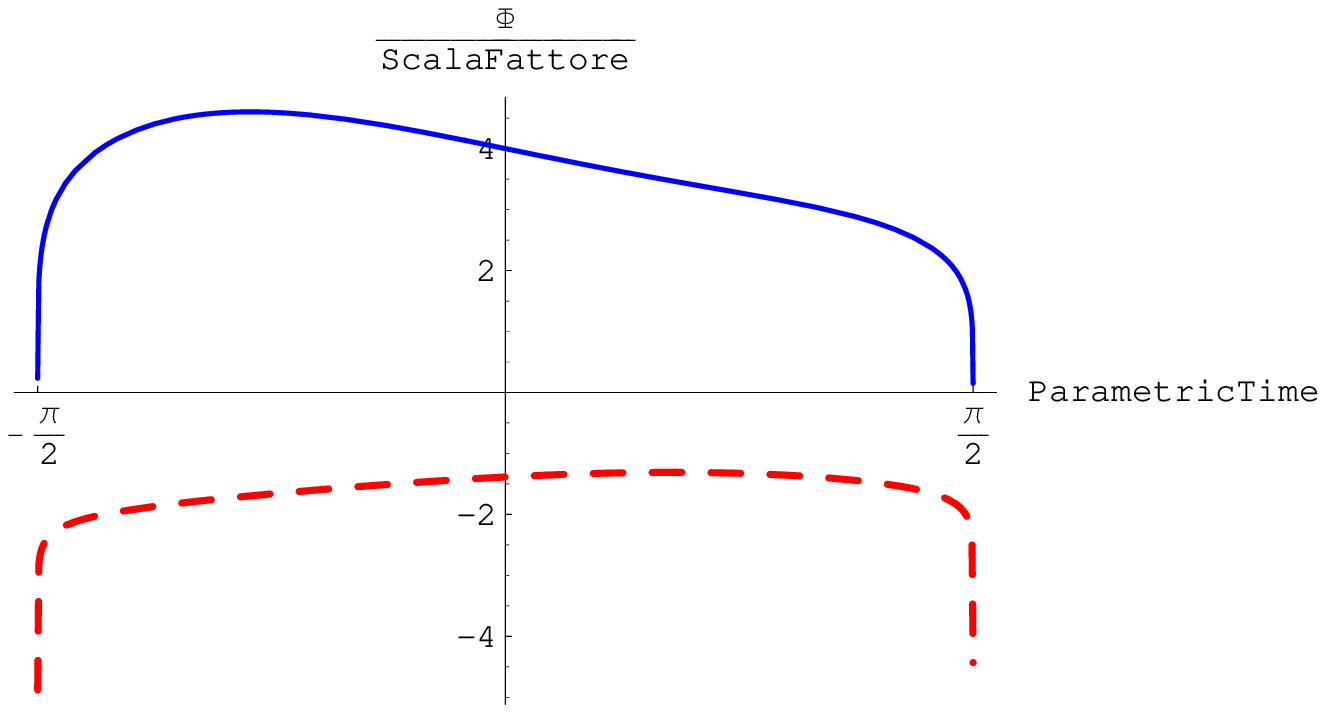}
\else
\end{center}
 \fi
\caption{\it
In this figure we display plots (in parametric time) of  the scale factor (solid line) and of the scalar field (dashed line) for the trigonometric type of solutions at $Y=1$. It is evident that in a finite time the Universe undergoes a Big Bang, a decelerated expansion and then a Big Crunch. At the same time the scalar field climbs from $-\infty$ to a maximum and then descends again to $-\infty$.
}
\label{nipotino}
 \iffigs
 \hskip 1cm \unitlength=1.0mm
 \end{center}
  \fi
\end{figure}

\begin{figure}[!hbt]
\begin{center}
\iffigs
 \includegraphics[height=50mm]{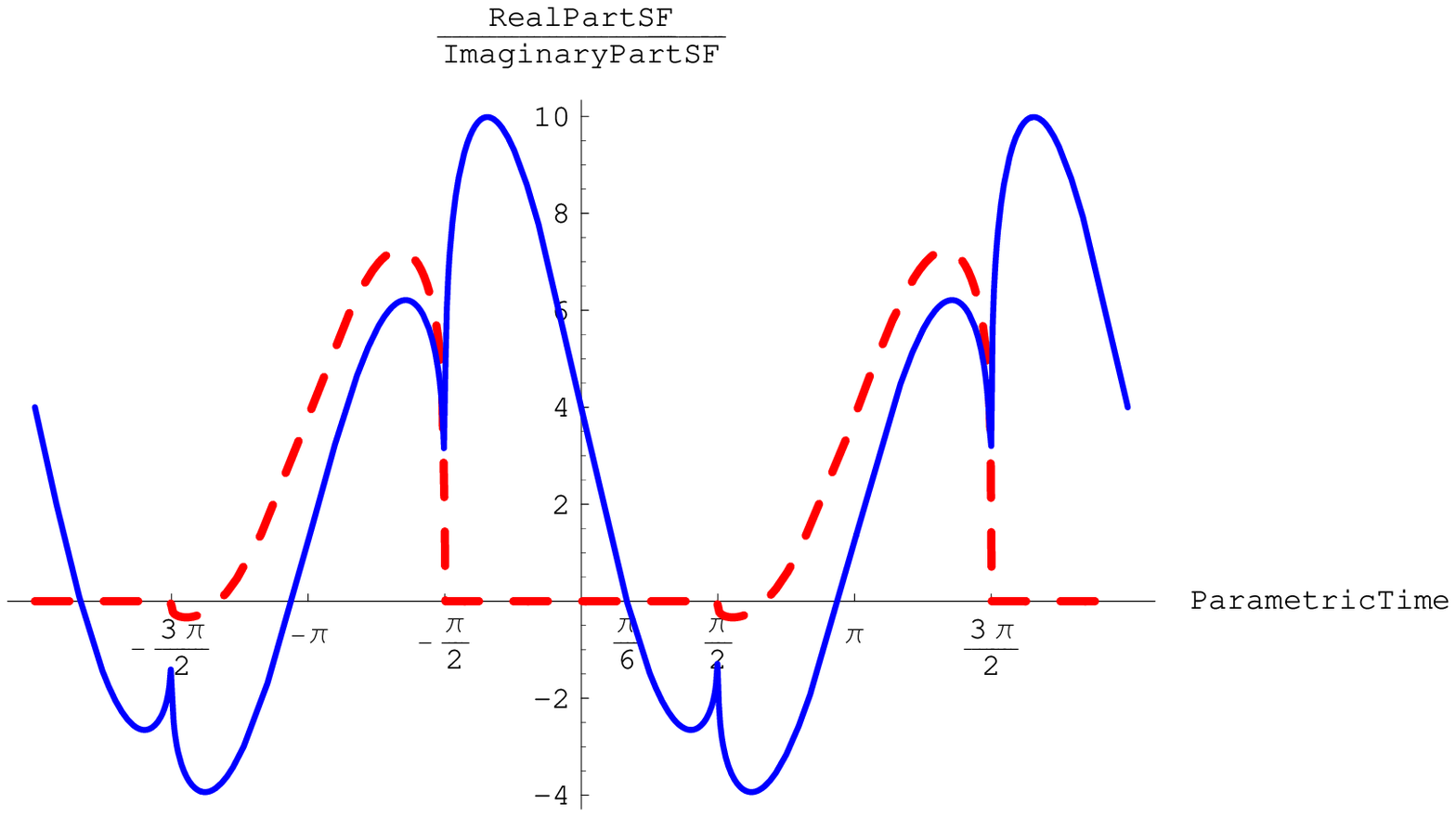}
 \includegraphics[height=50mm]{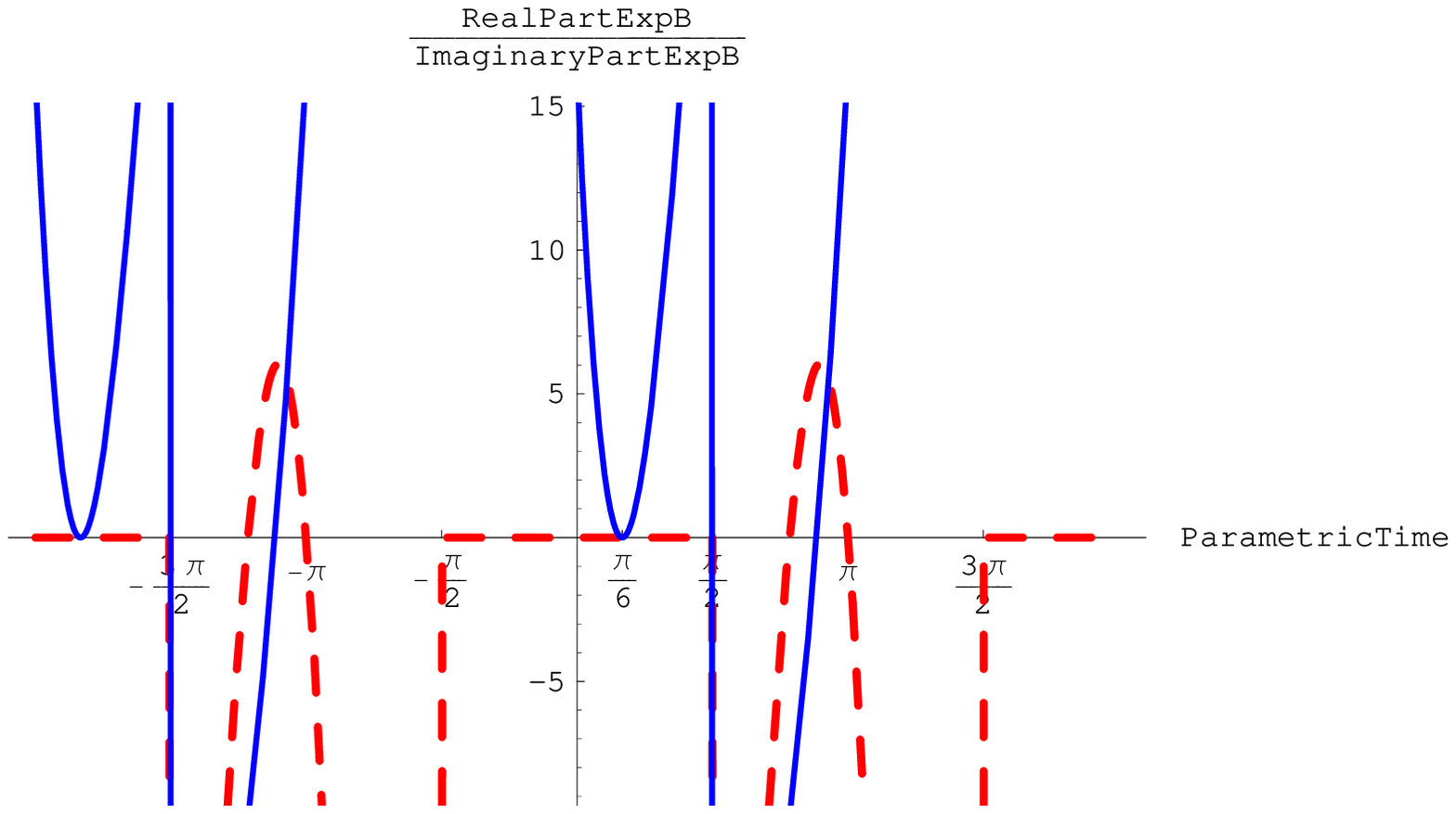}
 \includegraphics[height=50mm]{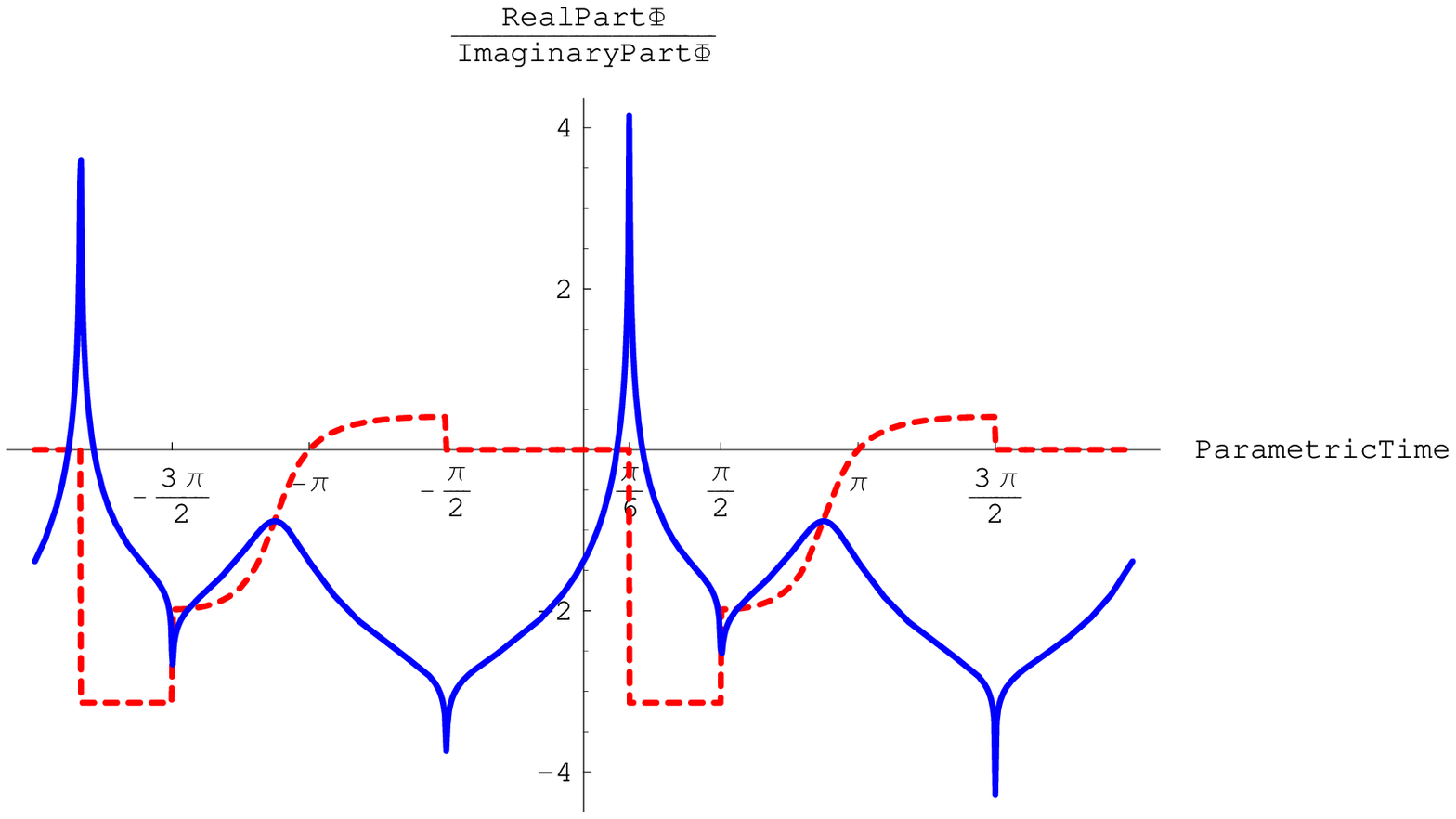}
\else
\end{center}
 \fi
\caption{\it
Plots of the real and imaginary parts of the scale factor, of the $\exp[\mathcal{B}]$-factor and of the scalar field for the case of parameter $Y=Y_\bullet$, which is overcritical $Y_\bullet > Y_0$.  In the three diagrams the solid line represents the real part, while the dashed line represents the imaginary part.  The interval  in which the three functions are simultaneously real is now reduced to  $\left[-\frac{\pi}{2},\frac{\pi}{3}\right]$. In the real range the scalar fields climbs from $-\infty$ to $+\infty$. }
\label{Ybultriplots}
 \iffigs
 \hskip 1cm \unitlength=1.1mm
 \end{center}
  \fi
\end{figure}

\begin{figure}[!hbt]
\begin{center}
\iffigs
 \includegraphics[height=60mm]{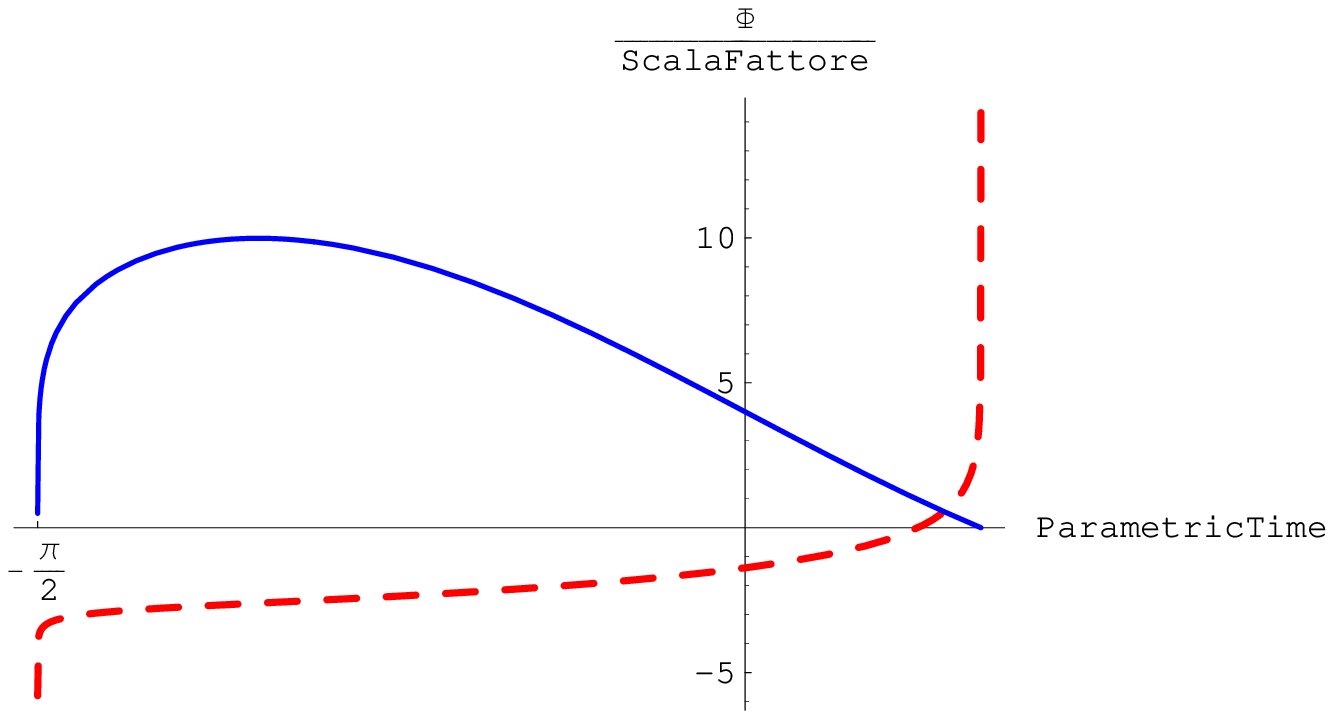}
\else
\end{center}
 \fi
\caption{\it
In this figure we display plots (in parametric time) of  the scale factor (solid line) and of the scalar field (dashed line) for the supercritical trigonometric type of solutions at $Y=Y_\bullet$. It is evident that in the finite parametric time interval $\left[-\frac{\pi}{2},\frac{\pi}{3}\right]$ the Universe undergoes a Big Bang, a decelerated expansion and then a Big Crunch. At the same time the scalar field climbs from $-\infty$ to   $+\infty$.
}
\label{giunglatroops}
 \iffigs
 \hskip 1cm \unitlength=1.0mm
 \end{center}
  \fi
\end{figure}

\begin{figure}[!hbt]
\begin{center}
\iffigs
 \includegraphics[height=55mm]{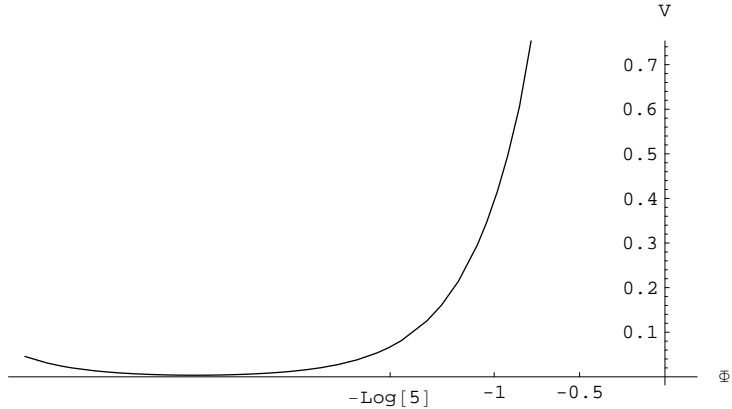}
\else
\end{center}
 \fi
\caption{\it
In this figure we display the phase portrait of the solution defined by the supercritical value $Y=Y_\bullet$. The axes are the scalar field $\Phi \equiv \mathfrak{h}$ and its derivative with respect tot the cosmic time $V \equiv \partial_{T_c} \mathfrak{h}$. The extremum of the potential is at $\Phi_0 \, = \, - \log[5]$. It is reached by the solution however with a non vanishing velocity. The field also reaches vanishing velocity yet differently from the previous case before rather than after the extremum. This allows for the continuous climbing of the field up to $+\infty$.
}
\label{facciaportratto3}
 \iffigs
 \hskip 1cm \unitlength=1.0mm
 \end{center}
  \fi
\end{figure}

We established that the actual moduli space of the trigonometric solutions is provided by the deformation parameter $Y$. It is  interesting to explore the quality of the solutions that this latter  parameterizes. The first fundamental question is whether all solutions have a Big Bang and a Big Crunch or other behaviors are possible. Periodicity of the solution functions guarantees that, in any case, the scale factor has zeros at $\tau = \pm \frac{\pi}{2} + n \times \pi $ for $n\in \mathbb{Z}$, yet there is also another possibility which has to be taken into account:  an additional zero might  or might not occur
in the interval $\left [ -\frac{\pi}{2} , \frac{\pi}{2} \right ]$. This depends on the value of $Y$.
Given the form (\ref{confettofalqui}) of the solution, a zero of the scale factor can occur at a value $\tau_0$ which satisfies the equation:
\begin{equation}\label{goliardo}
    Y \, = \, \frac{4}{5} \cos
   ^{\frac{1}{5}}\left(\tau _0\right)
   \csc \left(\tau _0\right) \left(\cos
   ^2\left(\tau _0\right)^{9/10} \,
   _2F_1\left(\frac{1}{2},\frac{9}{10};
   \frac{3}{2};\sin ^2\left(\tau
   _0\right)\right) \tan ^2\left(\tau
   _0\right)+5\right) \, \equiv \, \mathfrak{f}(\tau_0)
\end{equation}
The plot of the function $\mathfrak{f}(\tau)$ defined above is displayed in fig.\ref{ziotta}.
We see that for
\begin{equation}\label{gomorroida}
    |Y| \, \le \, Y_0 \, \equiv \, \frac{4 \sqrt{\pi } \Gamma
   \left(\frac{11}{10}\right)}{\Gamma
   \left(\frac{3}{5}\right)}
\end{equation}
the candidate Big Bang and Big Crunch are at $\tau \, = \, \pm \frac{\pi}{2}$, while for $|Y| \, > \, Y_0$ the candidate Big Bang is at $\tau \, = \, - \frac{\pi}{2}$, while the Big Crunch occurs earlier at:
\begin{equation}\label{georgy}
    \tau_0 \, = \, \mathfrak{f}^{-1}(Y)
\end{equation}
It is reasonable to expect a significantly different structure of the solution in the two cases.
\paragraph{$Y\ne0$ but less than critical.}
\begin{figure}[!hbt]
\begin{center}
\iffigs
 \includegraphics[height=50mm]{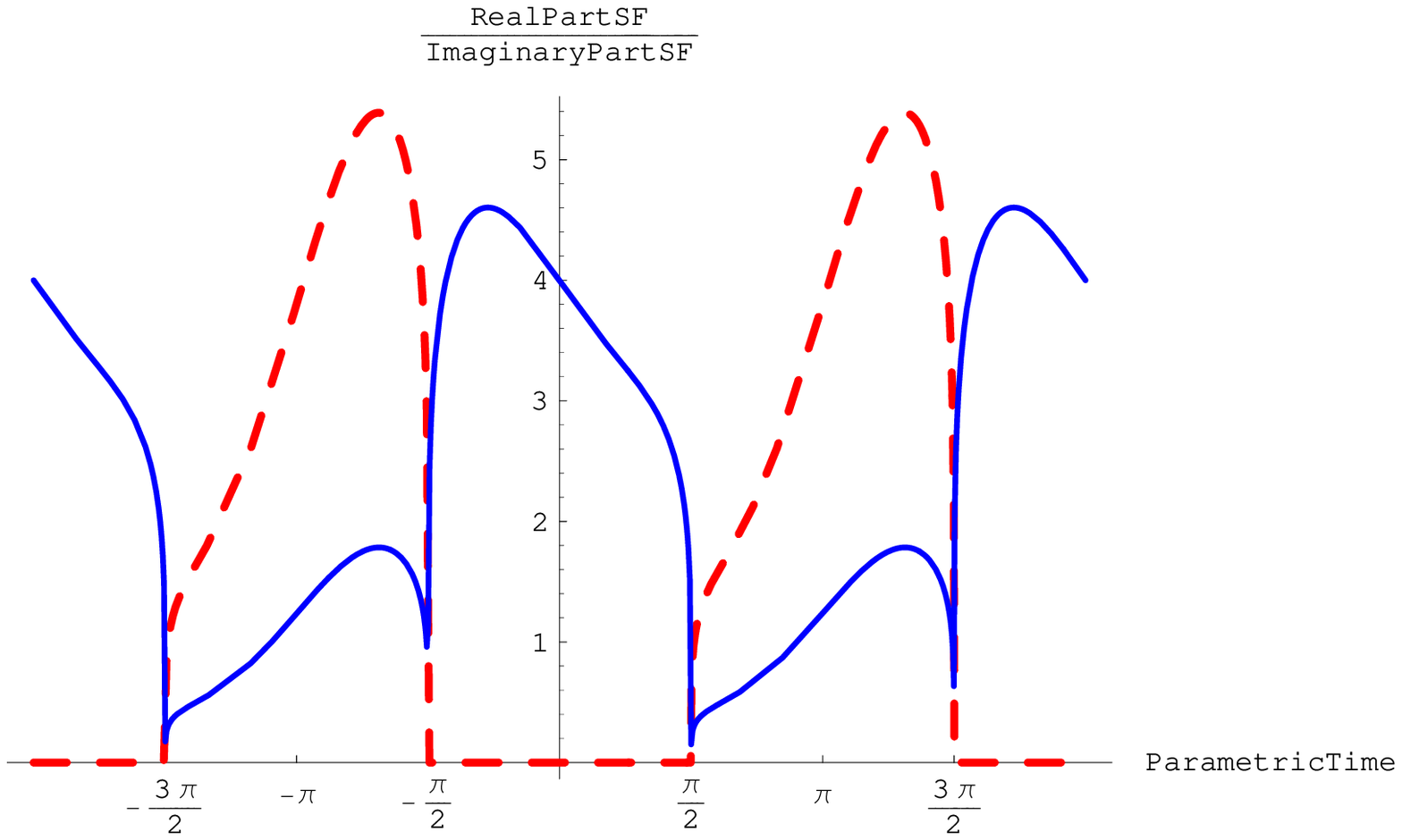}
 \includegraphics[height=50mm]{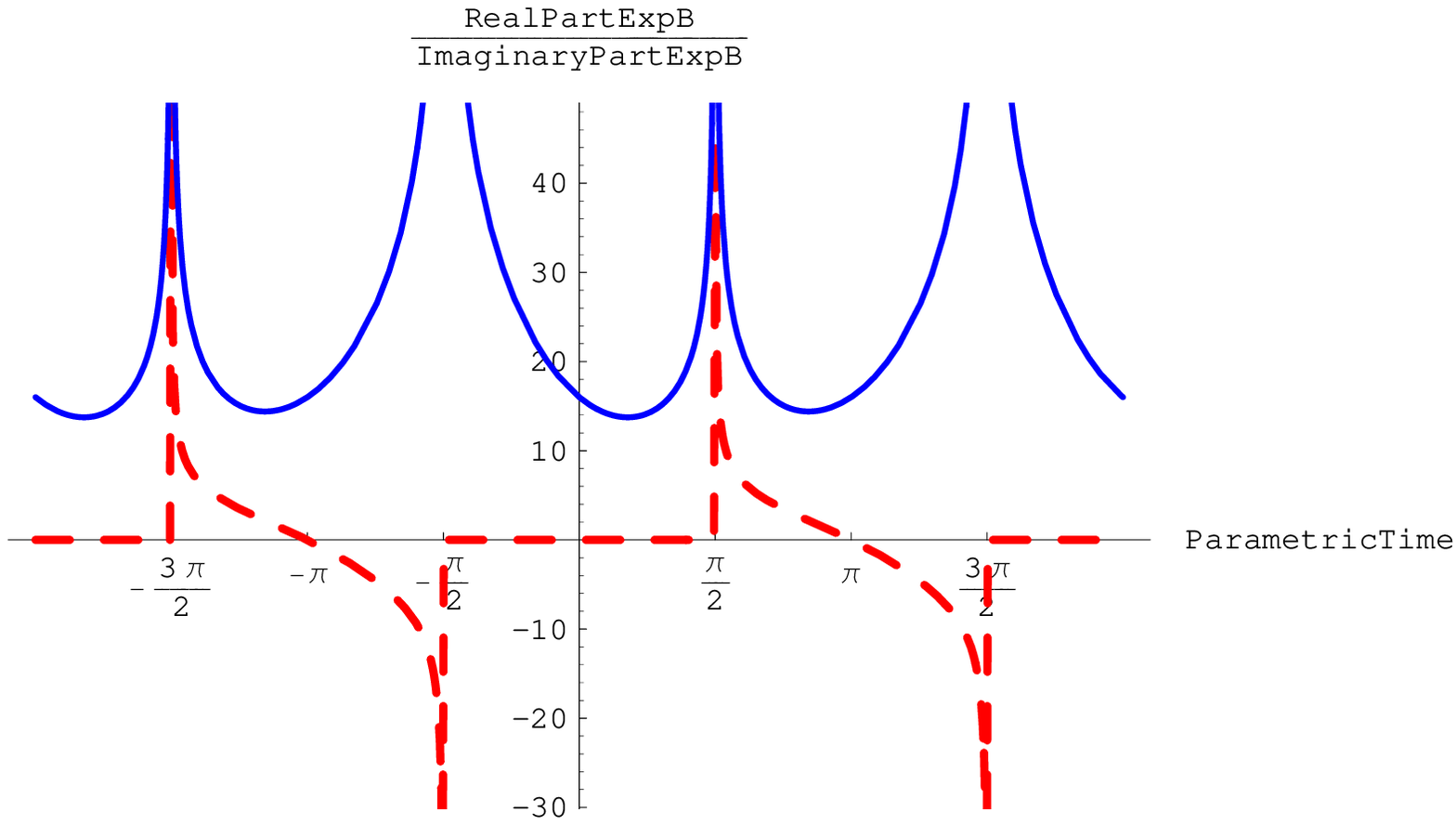}
 \includegraphics[height=50mm]{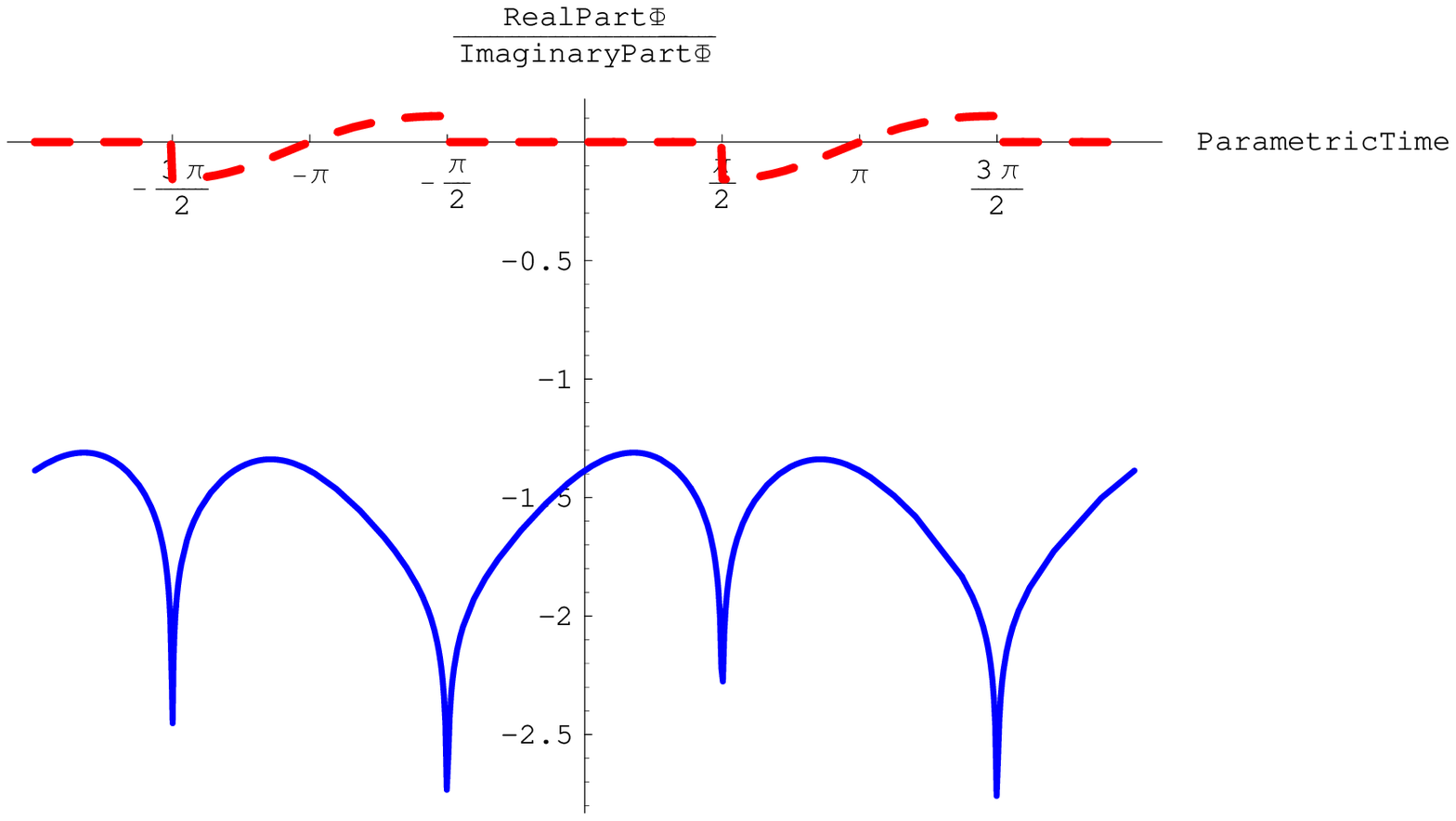}
\else
\end{center}
 \fi
\caption{\it
Plots of the real and imaginary parts of the scale factor, of the $\exp[\mathcal{B}]$-factor and of the scalar field for the case of parameter $Y=1$, which is non zero but smaller than the critical value $Y_0$  In the three diagrams the solid line represents the real part, while the dashed line represents the imaginary part.  The interval  in which the three function are simultaneously real is still $\left[-\frac{\pi}{2},\frac{\pi}{2}\right]$ as in the $Y=0$ case. Yet the shape of the plots is no longer symmetric and also the $\exp[B]$ factor and the scalar field start developing imaginary parts that  instead are identically zero over the full range of $\tau$ when $Y=0$. }
\label{Y1triplots}
 \iffigs
 \hskip 1cm \unitlength=1.1mm
 \end{center}
  \fi
\end{figure}
In fig.\ref{Y1triplots} we display the plot of the real and imaginary parts for the three functions composing the solutions for the less than critical case $Y=1$. As we see the shape of the plots is no longer symmetric and imaginary parts are developed also by the scalar field and by the $\exp[\mathcal{B}]$-factor,  yet the candidate Big Crunch occurs once again at the parametric time $\tau =\frac{\pi}{2}$. Furthermore the scalar field, after climbing to some finite value, drops again to $-\infty$ at the end of the life cycle of this Universe. The no longer symmetric phase-portrait of this solution is displayed in fig.\ref{facciaportratto2}.
The verification that the zeros of the scale factor are indeed singularities is done by inspecting the divergences of the curvature two-form  in $\pm \frac{\pi}{2}$. In this case it is much more difficult to calculate the coefficients of the asymptotic expansion, yet is sufficiently easy to determine the divergence order of the various components. We find:
\begin{equation}
R^{AB} \, \stackrel{\tau \rightarrow \pm \pi /2}{\approx }\,  \left(
\begin{array}{llll}
 0 & \mathcal{O}\left(\frac{1}{\left(\tau \mp \frac{\pi }{2}\right)^{6/5}}\right)
   & \mathcal{O}\left(\frac{1}{\left(\tau \mp \frac{\pi }{2}\right)^{6/5}}\right)
   & \mathcal{O}\left(\frac{1}{\left(\tau \mp \frac{\pi }{2}\right)^{6/5}}\right)
   \\
 \mathcal{O}\left(\frac{1}{\left(\tau \mp \frac{\pi }{2}\right)^{6/5}}\right) & 0
   & \mathcal{O}\left(\frac{1}{\left(\tau \mp \frac{\pi }{2}\right)^{14/5}}\right)
   & \mathcal{O}\left(\frac{1}{\left(\tau \mp \frac{\pi }{2}\right)^{14/5}}\right)
   \\
 \mathcal{O}\left(\frac{1}{\left(\tau \mp \frac{\pi }{2}\right)^{6/5}}\right) &
   \mathcal{O}\left(\frac{1}{\left(\tau \mp \frac{\pi }{2}\right)^{14/5}}\right) &
   0 & \mathcal{O}\left(\frac{1}{\left(\tau \mp \frac{\pi
   }{2}\right)^{14/5}}\right) \\
 \mathcal{O}\left(\frac{1}{\left(\tau \mp \frac{\pi }{2}\right)^{6/5}}\right) &
   \mathcal{O}\left(\frac{1}{\left(\tau \mp \frac{\pi }{2}\right)^{14/5}}\right) &
   \mathcal{O}\left(\frac{1}{\left(\tau \mp \frac{\pi }{2}\right)^{14/5}}\right) &
   0
\end{array}
\right) \label{divergendo2}
\end{equation}
At $Y\ne 0$, differently from the $Y=0$ case  there are two different velocities of approach to infinity for the curvature components. Half of them go faster and half of them go slower. Yet the relevant point is that all of them blow up and certify that we are in presence of a true singularity, both at the beginning and at the end of time.
The overall shape of the solution for the scalar field and for the scale factor is displayed in fig.\ref{nipotino}.
\paragraph{Overcritical $Y>Y_0$}. When the parameter $Y$ is over critical we have a new zero of the scale factor which occurs at some $\tau_0$ in the fundamental interval $\left[-\frac{\pi}{2},\frac{\pi}{2}\right]$. A practical way to deal with this type of solutions is to invert the procedure and use $\tau_0$ as parameter by setting $Y=\mathfrak{f}(\tau_0)$. As an illustrative example we choose $\tau_0 \, = \, \frac{\pi}{6}$ and we get:
\begin{equation}\label{ybullo}
 Y_\bullet \, = \,  \mathfrak{f}\left(\frac{\pi}{6}\right) \, = \,  4 \times 2^{4/5} \sqrt[10]{3}+\frac{2}{5} \,
   _2F_1\left(\frac{1}{2},\frac{9}{10};
   \frac{3}{2};\frac{1}{4}\right)
\end{equation}
The behavior of the real and imaginary parts of the three functions composing the solution for $Y=Y_\bullet$ is displayed in fig.\ref{Ybultriplots}.
The new character of the solution is immediately evident from such plots. The earlier zero of the scale factor is in correspondence with a divergence of the scalar field that now climbs from $-\infty$ to $+\infty$ as it is displayed in fig.\ref{giunglatroops}
The structure  of the phase-portrait changes significantly with respect to the subcritical cases and it is displayed in fig.\ref{facciaportratto3}
At the Big Bang point $\tau \, = \, -\frac{\pi}{2}$ the curvature components diverge just in the same way as in eq.(\ref{divergendo2}) namely the fastest approach to infinity is $\mathcal{O}\left( \frac{1}{(\tau+\pi/2)^{14/5}}\right)$. Instead at the new Big Crunch point $\tau=\pi/3$ the fastest diverging components of the curvature tensor have a much stronger singularity, namely they diverge as $\mathcal{O}\left( \frac{1}{(\tau+\pi/2)^{7}}\right)$. This further shows the clear-cut separation between less than critical and over critical solutions of the trigonometric type.
\par
Apart from this finer structure the above detailed analysis has explicitly demonstrated the main point which we want to stress since it is  somehow new in General Relativity. Notwithstanding the spatial flatness of the metric and notwithstanding the positive asymptotic behavior of the potential $V(\mathfrak{h})$ that goes to $+\infty$ for large values of the scalar field $\mathfrak{h}$, the presence of  a negative extremum of $V(\mathfrak{h})$, (does not matter whether maximum or minimum) always implies a collapse of the Universe at a finite value of the cosmic or parametric time.
\par
The Big Crunch collapse is the typical destiny of a closed Universe with positive spatial curvature. Therefore one is naturally led to inquiry whether such Universes as those discussed above have just the same causal structure as a closed Universe. To give an answer to such a question we consider the Particle and Event Horizons.
\subsubsection{\sc Particle and Event Horizons}
\label{parteventhoriz}
\begin{figure}[!hbt]
\begin{center}
\iffigs
 \includegraphics[height=50mm]{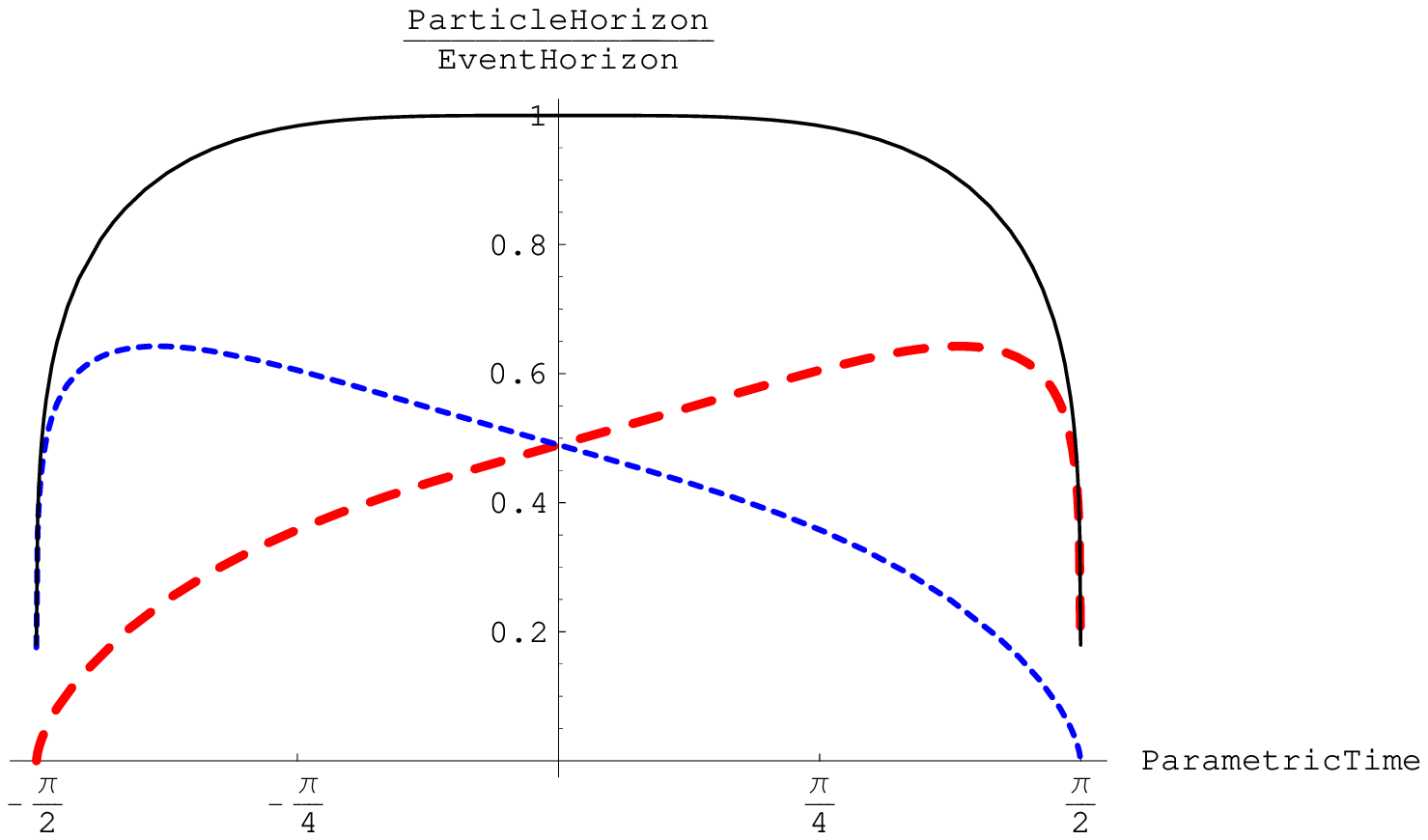}
 \includegraphics[height=50mm]{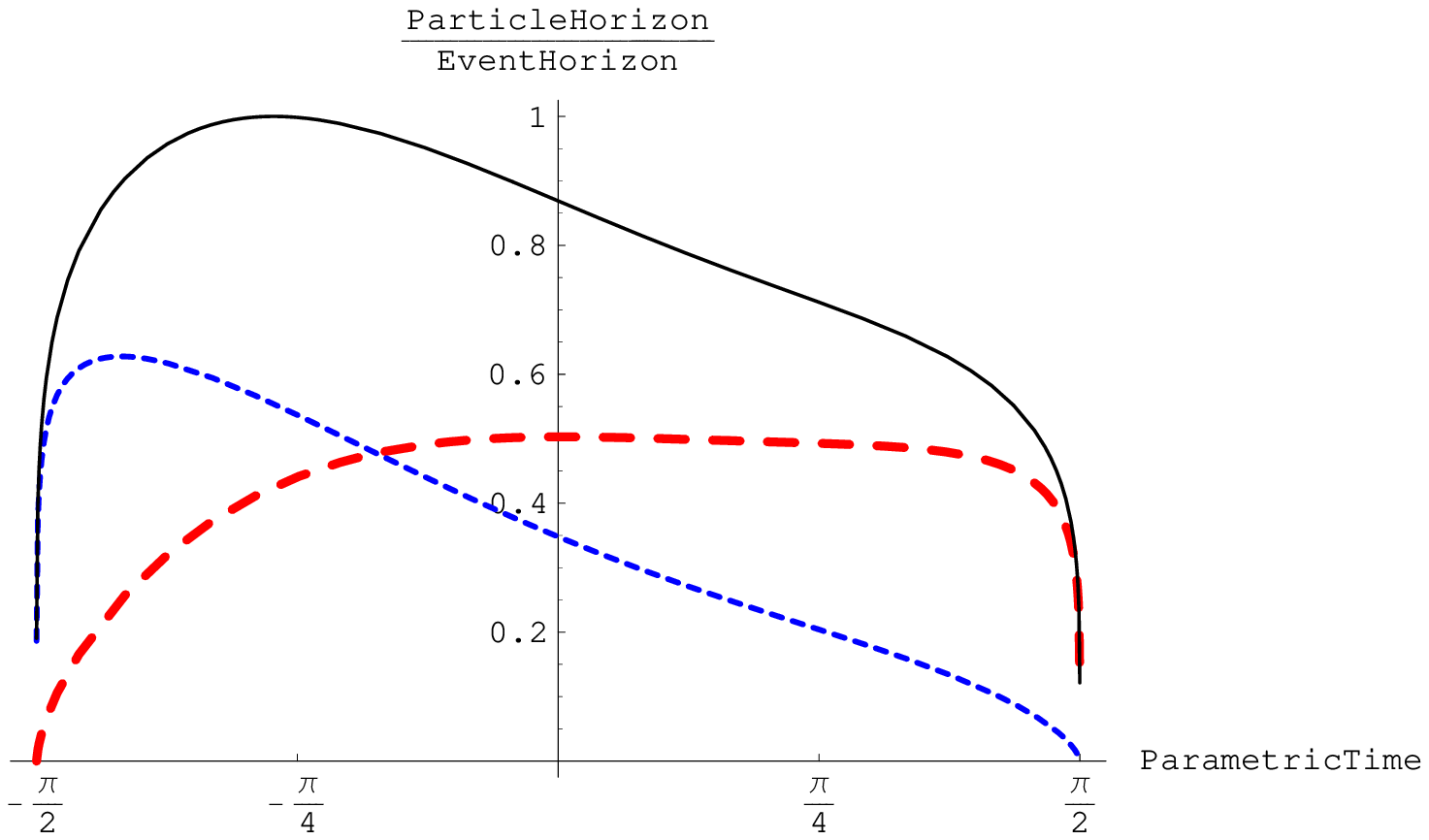}
 \includegraphics[height=50mm]{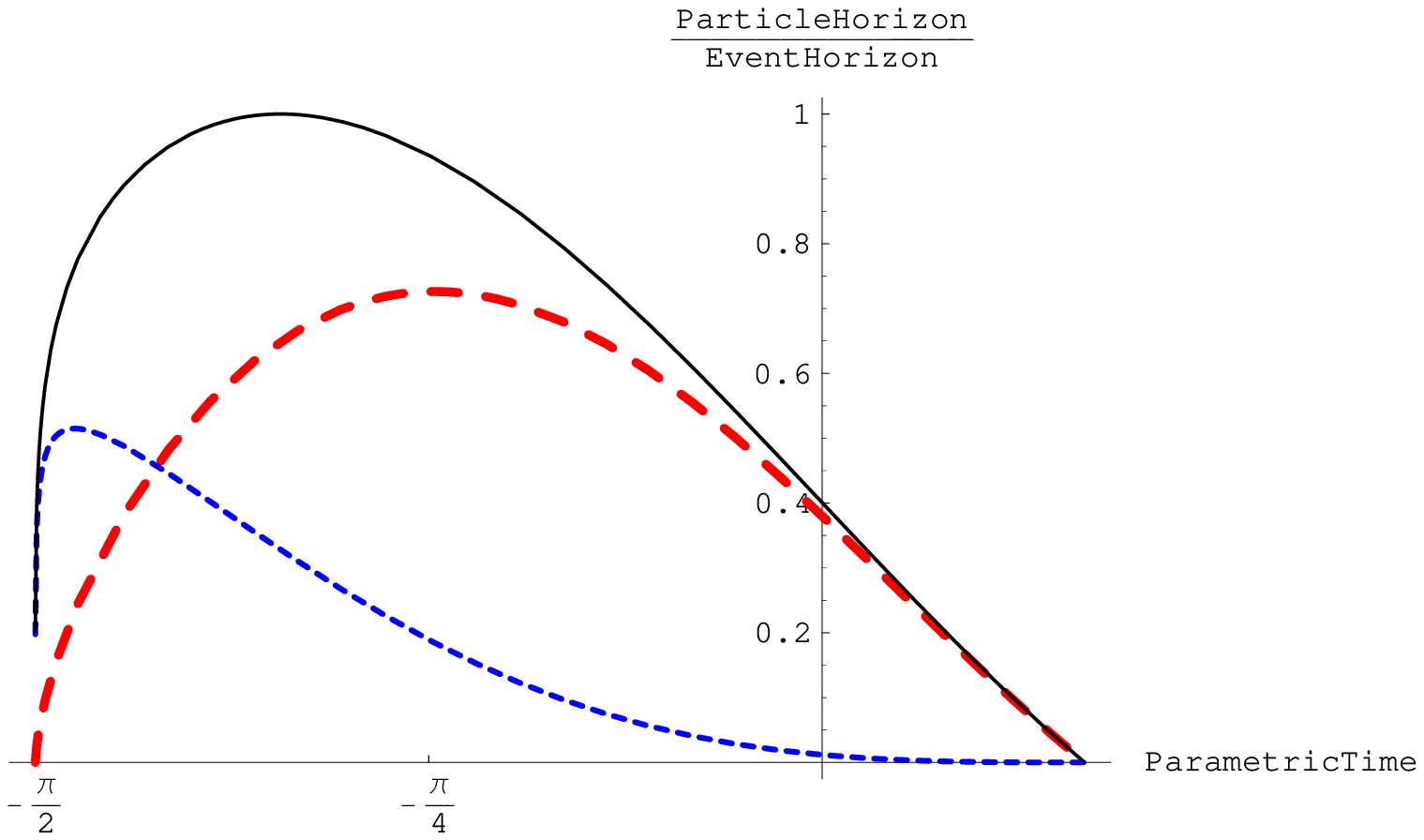}
\else
\end{center}
 \fi
\caption{\it
The three plots in this figure  respectively refer to the solutions $Y=0$, $Y=1$ and $Y=Y_\bullet$.  In each plot the solid line represents the Scale Factor, the dashed line with longer dashes represents the Particle Horizon, while the dashed line with shorter and denser dashes represents the Event Horizon. In all cases the Event Horizon goes to zero faster than the Particle Horizon and invisible portion of the Universe become visible to the same observer at later times.}
\label{partucla}
 \iffigs
 \hskip 1cm \unitlength=1.1mm
 \end{center}
  \fi
\end{figure}

Two important concept in Cosmology are those of Particle and Event Horizons. Given a metric of the form (\ref{piatttosa}) let us rewrite it in polar coordinates:
\begin{equation}\label{polaris}
    ds^2 \, = \, \exp\left[\mathcal{B}(\tau) \right] \, d\tau^2 - \mathfrak{a}^2(\tau) \, \left( dr^2 + r^2 d\Omega^2\right)
\end{equation}
where, as usual, $d\Omega^2$ denotes the metric on a two-sphere, and let us consider the radial light-like geodesics defined by the equation:
\begin{equation}\label{bomboladigas}
   0 \, = \, \exp\left[2\mathcal{B}(\tau) \right] \, d\tau^2 - \mathfrak{a}^2(\tau) \,  dr^2  \quad \rightarrow \quad \int_{0}^R \, dr \, = \, \int_{- \frac{\pi}{2}}^T \, d\tau \, \frac{\exp\left[\mathcal{B}(\tau) \right] }{\mathfrak{a}(\tau)}
\end{equation}
From (\ref{bomboladigas}) it follows that at any parametric time $T$ after the Big Bang, the remotest radial coordinate  from which we can receive a signal  is given by:
\begin{equation}\label{curlandia}
    R(T) \, = \,\int_{- \frac{\pi}{2}}^T \, d\tau \, \frac{\exp\left[\mathcal{B}(\tau) \right] }{\mathfrak{a}(\tau)}
\end{equation}
and in any case the maximal physical value of such a radial coordinate is given by:
\begin{equation}\label{curlandiabis}
    r_{max} \, = \,\int_{- \frac{\pi}{2}}^{T_{max}} \, d\tau \, \frac{\exp\left[\mathcal{B}(\tau) \right] }{\mathfrak{a}(\tau)}
\end{equation}
where $T_{max}$ is the Big Crunch parametric time. It is therefore convenient to measure radial coordinates $r$ in fractions of this maximal one and measure the scale factor in fraction of the maximal one attained during time evolution:
\begin{equation}\label{massimino}
    a_{max} \, \equiv \, \mathfrak{a}(\hat{\tau}) \quad ; \quad \mbox{where} \quad \partial_\tau \mathfrak{a}(\tau)|_{\tau = \hat{\tau}} \, = \, 0
\end{equation}
 Setting:
\begin{equation}\label{gugulini}
   {\bar{R}}(T) \, = \, \frac{R(T)}{r_{max}} \quad ; \quad \bar{\mathfrak{a}}(\tau) \, = \, \frac{\mathfrak{a}(\tau)}{a_{max}}
\end{equation}
we conclude that the fastest distance from which an observer can receive a signal at any instant of time is:
\begin{equation}\label{patacchio}
    \mathcal{P}(T) \, = \, \frac{\mathfrak{a}(T)}{a_{max} \, r_{max} } \, \int_{- \frac{\pi}{2}}^T \, d\tau \, \frac{\exp\left[\mathcal{B}(\tau) \right] }{\mathfrak{a}(\tau)}
\end{equation}
By definition this distance is the \textit{Particle Horizon} and defines the portion of Space that is visible by any Observer living at time $T$.
 On the other hand  the \textit{Event Horizon} is the boundary of the Physical Space from which no signal will ever reach an Observer living at time $T$ at any time of his future.  In full analogy with equation (\ref{patacchio}) the Event Horizon is defined by:
\begin{equation}\label{spazzolone}
    \mathcal{E}(T) \, = \, \frac{\mathfrak{a}(T)}{a_{max} \, r_{max} } \, \int_{T}^{T_{max}} \, d\tau \, \frac{\exp\left[\mathcal{B}(\tau) \right] }{\mathfrak{a}(\tau)}
\end{equation}
It is well known, (see for instance \cite{pietroGR}) that in a matter dominated, closed Universe the Particle Horizon and the Event Horizon exactly coincide. This means that in such a Universe, the portion of space that is invisible to an observer living at time $T$ will remain invisible to him also at all later times. Furthermore in such a Universe the Particle/Event Horizon contracts to zero exactly at the moment when the Universe reaches its maximum extension. An observer living at that time is completely blind and will stay blind all the rest of his life.
In the Universes we have considered in this section things go quite differently since the Particle and the Event Horizon do not coincide and actually have a somehow opposite behavior. Plots of the Particle and Event horizon are shown in figure \ref{partucla} for the three solutions of trigonometric type we have been considering.
In all cases the Particle Horizon does not coincide with the Event Horizon. At the beginning the latter is larger than the former, which means that there is a portion of the invisible Universe which will reveal itself to the a given observer in his future. Then the Particle Horizon grows and the Event Horizon rapidly decreases. This means that as time goes on larger and larger becomes the visible Universe but also larger and larger the portion that will never reveal itself to an observer living at that time. In all cases before the end of its life the entire Universe becomes visible to an observer living at that  time. This happens at relatively early times for supercritical solutions with $Y> Y_0$.
\par
This rather intricate structure is quite different from that of a Closed Universe with positive spatial curvature. Contrary to generally accepted lore, these solutions of Einstein Klein Gordon equations show that spatial flatness of the Universe should not lead us to automatically exclude the possibility of a final collapse into a singularity. We think that this is an important warning and for this reason we analyzed the case in depth. A second motivation for our detailed analysis was the confirmation of the climbing, descending mechanism that strictly correlates the space-time singularities  with the divergences of the scalar field that can only start and end up at infinity. Without a positive extremum the scalar field cannot stop at a fixed point and space-time has no other option than exploding and then collapsing.
\par
In the next section we consider the much simpler and rather smoothly behaved hyperbolic solutions that occur when the potential has no extremum.
\subsection{\sc Hyperbolic solutions in the run away potential without finite extrema}
When we choose $\omega=\nu^2$ the potential has no minimum, the solution of eq.s(\ref{equemozionica}) drastically simplifies and it is provided in terms of exponential functions.
\par
Explicitly we obtain:
{\small
\begin{eqnarray}
\label{HyperUVsol}
  U(\tau) &=& a \, e^{\nu  \tau } +b \, e^{-\nu
   \tau } \\
  V(\tau) &=& \left(e^{-\nu  \tau }
   \left(\left(e^{2 \nu  \tau }
   a+b\right) \left(c e^{2 \nu
   \tau } \nu ^2+d \nu ^2-4 e^{\nu
   \tau } \sqrt[5]{e^{\nu  \tau }
   a+b e^{-\nu  \tau
   }}\right)\right.\right.\nonumber\\
   &&\left.\left.-e^{\nu  \tau }
   \sqrt[5]{e^{\nu  \tau } a+b
   e^{-\nu  \tau }} \left(b-a e^{2
   \nu  \tau }\right)
   \left(\frac{e^{2 \nu  \tau }
   a}{b}+1\right)^{4/5} \,
   _2F_1\left(\frac{2}{5},\frac{4}{
   5};\frac{7}{5};-\frac{a e^{2 \nu
    \tau
   }}{b}\right)\right)\right)\times\nonumber\\
   && \left(e^{2
   \nu  \tau } a+b\right) \nu ^2
\end{eqnarray}
}
where $a,b,c,d$ are integration constants. Once inserted in the formula (\ref{integtrasformazia}) for the physical fields
the solution (\ref{HyperUVsol}) produces a solution of the original equation upon implementation of the same constraint
(\ref{dariopisco}) as in the trigonometric case that we can solve with the same position, namely  by setting $d=-\rho \, a$, $c=\rho \,b$.
The final form of the solution depending on three parameters is the following one:
\begin{eqnarray}
  \mathfrak{P}(\tau;a,b,\rho,\nu) &=& e^{-\nu  \tau } \sqrt[5]{e^{\nu
   \tau } a+b e^{-\nu  \tau }}
   \left(\left(e^{2 \nu  \tau }
   a+b\right) \left(b e^{2 \nu
   \tau } \rho  \nu ^2-a \rho  \nu^2 \right.\right.\nonumber\\
   &&\left.\left. -4 e^{\nu  \tau }
   \sqrt[5]{e^{\nu  \tau } a+b
   e^{-\nu  \tau }}\right)-e^{\nu
   \tau } \sqrt[5]{e^{\nu  \tau }
   a+b e^{-\nu  \tau }} \left(b-a
   e^{2 \nu  \tau }\right)
   \left(\frac{e^{2 \nu  \tau }
   a}{b}+1\right)^{4/5} \, \times \right.\nonumber\\
   &&\left.
   _2F_1\left(\frac{2}{5},\frac{4}{
   5};\frac{7}{5};-\frac{a e^{2 \nu
    \tau }}{b}\right)\right) \nonumber\\
  a(\tau;a,b,\rho,\nu) &=& \frac{\mathfrak{P}(\tau;a,b,\rho,\nu)}{\left(e^{2 \nu  \tau } a+b\right)
   \nu ^2} \nonumber \\
   \end{eqnarray}
   \begin{eqnarray}
  \exp\left[\mathcal{B}\right](\tau;a,b,\rho,\nu) &=& \frac{\left(\mathfrak{P}(\tau;a,b,\rho,\nu)\right)^2}{\left(e^{\nu  \tau } a+b e^{-\nu
   \tau }\right)^{4/5} \left(e^{2
   \nu  \tau } a+b\right)^2 \nu ^4} \nonumber\\
  \mathfrak{h}(\tau;a,b,\rho,\nu) &=& \log \left[\frac{\left(e^{\nu  \tau } a+b e^{-\nu
   \tau }\right)^{2/5} \left(e^{2
   \nu  \tau } a+b\right) \nu ^2}{\mathfrak{P}(\tau;a,b,\rho,\nu)} \right]
\end{eqnarray}
\subsubsection{\sc The simplest hyperbolic solution}
The simplest solution of the hyperbolic type is obtained for the choice $a=0$, $c=0$, $b=1$, $\rho=1$, since in this case the hypergeometric function disappears and we simply get:
\begin{eqnarray}
\label{finocchius}
  \mathbf{a}(t,\nu) &\equiv & a\left(t-\frac{5 \log \left(\frac{\nu
   ^2}{5}\right)}{6 \nu };0,1,1,\nu\right) \, = \, \frac{5^{2/3} e^{-\frac{2 t
   \nu }{5}}
   \left(-1+e^{\frac{6 t \nu
   }{5}}\right)}{\nu ^{4/3}} \nonumber \\
  \exp\left[\mathbf{B}(t,\nu) \right] &\equiv& \exp\left[\mathcal{B}\left(t-\frac{5 \log \left(\frac{\nu
   ^2}{5}\right)}{6 \nu };0,1,1,\nu\right)\right] \, = \,  \frac{25 \left(-1+e^{\frac{6 t
   \nu }{5}}\right)^2}{\nu ^4}\nonumber\\
  \mathbf{h}(t,\nu) &\equiv & \mathfrak{h}\left(t-\frac{5 \log \left(\frac{\nu
   ^2}{5}\right)}{6 \nu };a,b,\rho,\nu\right) \, = \, \log \left(\frac{1}{5
   \left(-1+e^{\frac{6 t \nu
   }{5}}\right)}\right)+2 \log (\nu )
\end{eqnarray}
The shift in the parametric time variable $\tau \rightarrow t -\frac{5 \log \left(\frac{\nu^2}{5}\right)}{6 \nu }$ has been specifically arranged in such a way that $t=0$ is a zero of the scale factor, namely corresponds to the Big Bang. Furthermore, in this case, which involves only elementary transcendental functions, the relation between parametric and cosmic time can be explicitly evaluated. We have:
\begin{equation}\label{radiboga}
    T_c(t) \, \equiv \, \int_0^t \, dx \,\exp\left[\mathbf{B}(x,\nu) \right] \, = \, \frac{25 t}{\nu ^4}-\frac{125 e^{\frac{6 t \nu
   }{5}}}{3 \nu ^5}+\frac{125 e^{\frac{12 t \nu
   }{5}}}{12 \nu ^5}+\frac{125}{4 \nu ^5}
\end{equation}
This allows for a simple evaluation of the asymptotic behavior of both the scale factor and the scalar field for asymptotic very late and very early times. We calculate the limit:
\begin{eqnarray}\label{gordiano}
    \lim_{t\,\to \, \infty} \, \frac{\log[\mathbf{a}(t,\nu)]}{\log[T_c(t)]} & = & \frac{1}{3}
\end{eqnarray}
This means that at late times, independently from the parameter $\nu$ the scale factor behaves like the cubic root of the cosmic time.
\begin{equation}
   \mathbf{a}(T_c,\nu) \, \stackrel{T_c \to \infty}{\simeq} \, \mbox{const} \times T_c^{\frac{1}{3}} \nonumber \\
  \label{asintotus1}
\end{equation}
This corresponds to an equation of state of type \ref{equatastata} with $w=1$. In view of eq.s(\ref{patatefritte}) this means that at late times the predominant contribution to the energy density is the kinetic one, the potential energy being negligible.  Such a conclusion can be matched with the information on the asymptotic behavior of the scalar field for late times. This latter can be worked in the following way. As $t \to \infty$ (for $\nu >0$) we have:
\begin{equation}\label{ciaccius1}
    T_c \, \stackrel{t \to \infty}{\simeq} \,\frac{125 e^{\frac{12 t \nu
   }{5}}}{12 \nu ^5}
\end{equation}
while for the scalar field we get:
\begin{equation}\label{ciaccius2}
\mathbf{h}(t,\nu)    \, \stackrel{t \to \infty}{\simeq} \, -\frac{6 t \nu }{5}
\end{equation}
Combining the two results we get:
\begin{equation}\label{ciaccius3}
 \mathbf{h} \, \stackrel{T_c \to \infty}{\simeq} \, - \frac{1}{2} \, \log[T_c]
\end{equation}
namely, the scalar field goes logarithmically to $-\infty$ when plotted against cosmic time. Obviously the value of the potential (\ref{farimboldobis}) at $\mathfrak{h} \, = \, -\infty$ is zero and this explains the asymptotic dominance  of the kinetic energy.
\par
\begin{figure}[!hbt]
\begin{center}
\iffigs
 \includegraphics[height=60mm]{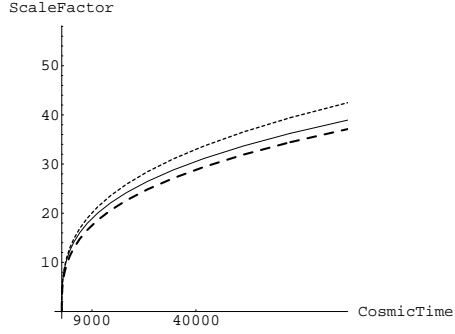}
 \vskip 2cm
 \includegraphics[height=60mm]{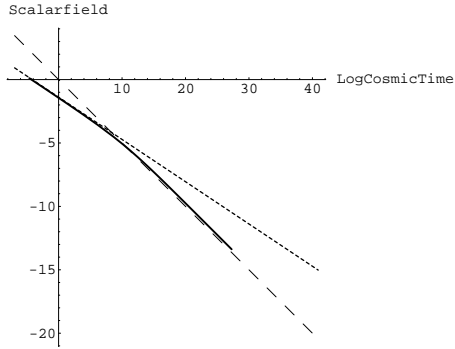}
\else
\end{center}
 \fi
\caption{\it
Here we present the behavior of the scale factor and of the scalar field for the simplest of the hyperbolic type solutions ($\omega = \nu^2 $) of the cosmological model based on the potential of
eq.(\ref{farimboldobis}).
The analytic form of the solution is given in
eq.(\ref{finocchius}).
For the plot we have chosen $\nu = \frac{1}{4}$. In the first graph, describing the scale factor, the solid line is the actual solution while the dashed curves are of the form $\alpha_{1,2} \, T_c^{\frac 13}$ with two different coefficient $\alpha_1 = \frac{3^{2/3}}{{10}^{1/3}}$ and $\alpha_2 = {\frac{3}{5}}^{1/3}$. The first curve is tangential to the solution at $T_c\to 0$ while the second is tangential to the solution at $T_c\to \infty$. The same style of presentation is adopted in the second picture. Here we plot the scalar field against the logarithm of the cosmic time. The two dashed straight lines represent the curves $-\frac{1}{3}\,  \log \left[ T_c \right] $, and $-\frac{1}{2}\, \log \left[ T_c \right]$. The first is tangential to the solution at $T_c \to 0$, the second is tangential to the solution at $T_c\to \infty$. }
\label{soluziasemplice}
 \iffigs
 \hskip 1cm \unitlength=1.1mm
 \end{center}
  \fi
\end{figure}
\begin{figure}[!hbt]
\begin{center}
\iffigs
 \includegraphics[height=40mm]{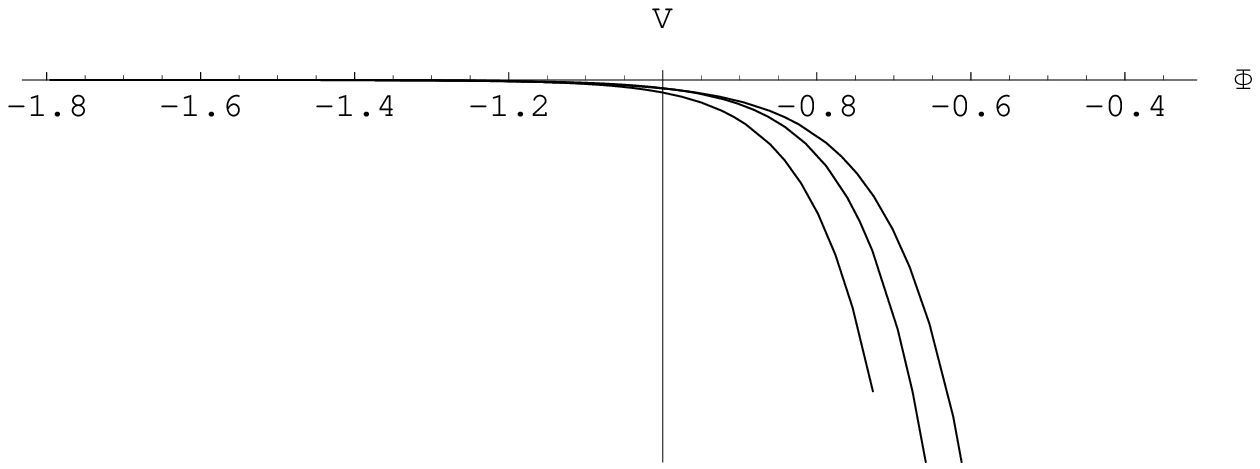}
 \vskip 2cm
 \includegraphics[height=40mm]{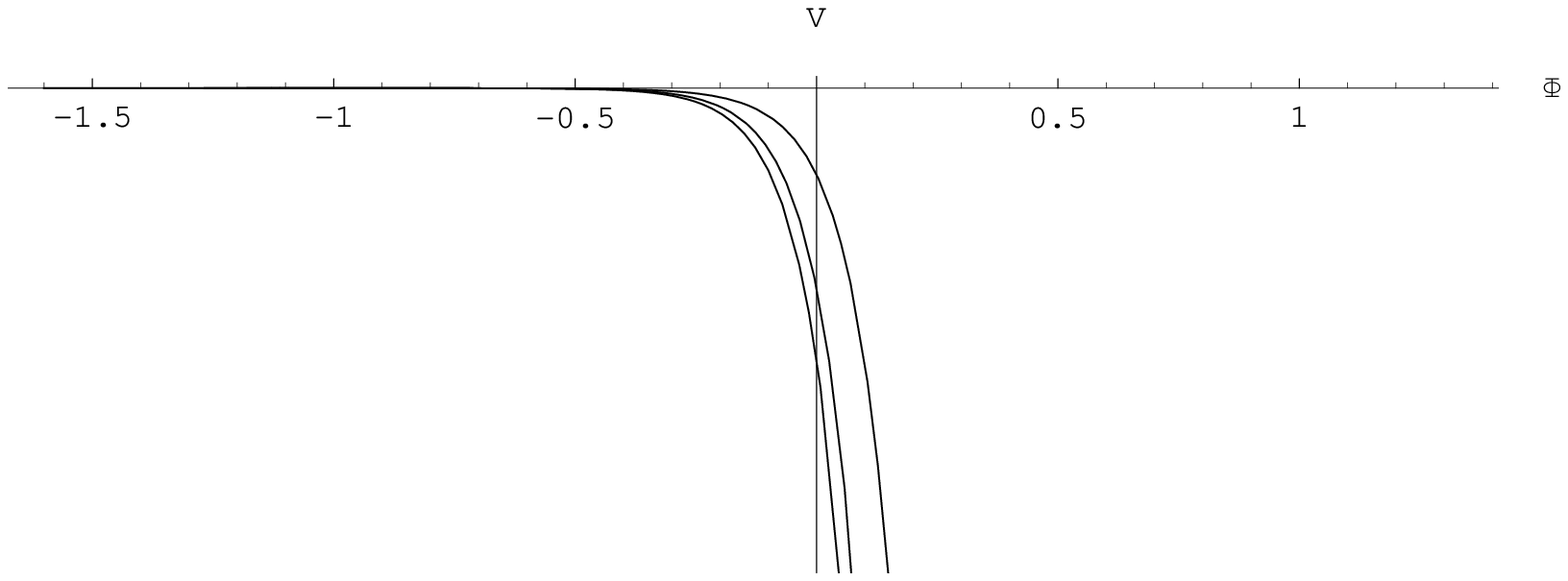}
\else
\end{center}
 \fi
\caption{\it
 In this picture we present some trajectories making up the phase portrait for the solutions of hyperbolic type presented in this section. In the upper figure we use only the simple solution involving elementary functions. In the lower figure we utilize also solutions involving the hypergeometric term. The quality of the portraits is just the same. }
\label{curbettis}
 \iffigs
 \hskip 1cm \unitlength=1.1mm
 \end{center}
  \fi
  \end{figure}

To work out the behavior at very early times it is more complicated, yet we can predict it by inspecting the behavior of the energy density and of the pressure. Inserting the form of the solution and of the potential in eq.(\ref{patatefritte}) we obtain the parametric time behavior of the energy density and of the pressure \footnote{Note that to obtain this result we have calculated the derivative of the scalar field with respect to the cosmic time and not with respect to the parametric time}:
\begin{eqnarray}
\label{quaquastata}
  \rho &=& \frac{3 \nu ^8 \left(-4 \nu ^2+2 e^{\frac{6 t \nu
   }{5}} \left(2 \nu ^2+5\right)+e^{\frac{12 t \nu
   }{5}} \left(3 \nu ^2-5\right)-5\right)}{15625
   \left(-1+e^{\frac{6 t \nu }{5}}\right)^6} \\
  p &=& \frac{3 \nu ^8 \left(4 \nu ^2-2 e^{\frac{6 t \nu
   }{5}} \left(2 \nu ^2+5\right)+e^{\frac{12 t \nu
   }{5}} \left(3 \nu ^2+5\right)+5\right)}{15625
   \left(-1+e^{\frac{6 t \nu }{5}}\right)^6}
\end{eqnarray}
Expanding both functions in power series  for $t\sim 0$ we obtain:
\begin{eqnarray}
  \rho & \stackrel{t \to 0}{\sim} & \frac{\nu ^4}{1728 t^6}-\frac{\nu ^5}{2592
   t^5}+O\left(\frac{1}{t^4}\right) \\
  p & \stackrel{t \to 0}{\sim} & \frac{\nu ^4}{1728 t^6}-\frac{13 \nu ^5}{12960
   t^5}+O\left(\frac{1}{t^4}\right)
\end{eqnarray}
 Both the pressure and the energy density diverge as $1/t^6$ plus subleading singularities; the identity of the coefficient in the leading pole of both expansions implies that also at very early times the effective equation of state is
 \begin{equation}\label{equastata}
    p \, = \, \rho \quad \Leftrightarrow \quad w \, = \, 1
 \end{equation}
 which implies the following behavior for the scale factor:
 \begin{equation}
   \mathbf{a}(T_c,\nu) \, \stackrel{T_c \to 0}{\simeq} \, \mbox{const} \times T_c^{\frac{1}{3}} \nonumber \\
  \label{asintotus0}
\end{equation}
With same technique we can also work out the asymptotic behavior of the scalar field in the origin of time:
\begin{equation}\label{ciaccius5}
 \mathbf{h} \, \stackrel{T_c \to 0}{\simeq} \, - \frac{1}{3} \, \log[T_c]
\end{equation}
In fig.\ref{soluziasemplice} we present the plots of an explicit example of such solutions.
Finally in fig.\ref{curbettis} we present the phase portrait for this type of solutions of the hyperbolic type. We compare the phase portrait of the simple solutions we have analyzed above with the phase portrait of the generic solutions that involve also the hypergeometric term. The quality of the picture is essentially the same, yet there is an  important critical difference concerning the asymptotic behavior.
\subsubsection{\sc Hyperbolic solutions displaying  the second asymptotic behavior for late times}
According to the analysis of \cite{primopapero} and of the previous literature \cite{dks,lm,exponential_pot}, when the potential is of the exponential type considered in this paper, namely
\begin{equation}
 \mathcal{V}(\varphi)\ =  \ \Sigma_{k=1}^{n}\,{\cal V}_{0k}\,e^{2\,\gamma_k\,\varphi}\ , \quad \gamma_{k} \ > \ \gamma_{k+1}\
\label{SingAn2}
\end{equation}
there are two possible different asymptotic behaviors of the scale factor in the vicinity of a Big Bang or of a Big Crunch. One behavior is universal and it is the one  we have met in the previous example of the simplest hyperbolic solution, namely:
\begin{equation}\label{gorko}
    a(T_c) \, \sim \, T_c^{\frac{1}{3}} \quad \Leftrightarrow \quad w\, = \, 1 \, \quad \mbox{kinetic asymptopia}
\end{equation}
The \textit{universal asymtopia} is the only one available both at the beginning and at the end of time when the solution for the scalar field is \textit{climbing}. As already stressed it corresponds to a complete dominance of the kinetic energy of the scalar field with respect to its potential one. On the other hand if the scalar is descending there are a priori two possible asymptopia available: in addition to the universal kinetic one (\ref{gorko}), there is also the following additional one:
\begin{equation}\label{babele}
      a(T_c) \, \sim \, T_c^{\frac{1}{3\,\gamma_{dom}^2}} \quad \Leftrightarrow \quad w\, = \, 2 \, \gamma_{dom}^2 -1 \, \quad \mbox{potential asymptopia}
\end{equation}
where $\gamma_{dom}$ is the coefficient of the dominant exponential appearing in the potential, once this latter is written in its normal form (\ref{SingAn2}) by means of the replacement (\ref{babushka}). For descending scalars that tend to $-\infty$, the dominant exponential is that with the smaller positive $\gamma_k$. In our case $\gamma_{dom} \, = \, \frac{2}{3}$ so that the second asymptopia available in the case of descending solutions, is:
\begin{equation}\label{babelebis}
      a(T_c) \, \sim \, T_c^{\frac{3}{4}} \quad \Leftrightarrow \quad w\, = \, - \, \frac{1}{9} \quad \mbox{potential asymptopia}
\end{equation}
The simplest asymptotic solution described in the previous section does not take advantage of the second possibility for its  asymptotic behavior: both at very early and at very late times the scale factors goes as $T_c^{\frac{1}{3}}$. This is no longer the case for the solutions with parameter $a\ne0$ where the contribution from the hypergeometric function is switched on. Indeed we have verified that for all such solutions the scale factor goes as $T_c^{\frac{1}{3}}$ near the initial singularity but diverges as $T_c^{\frac{3}{4}}$ for late times.
\begin{figure}[!hbt]
\begin{center}
\iffigs
 \includegraphics[height=40mm]{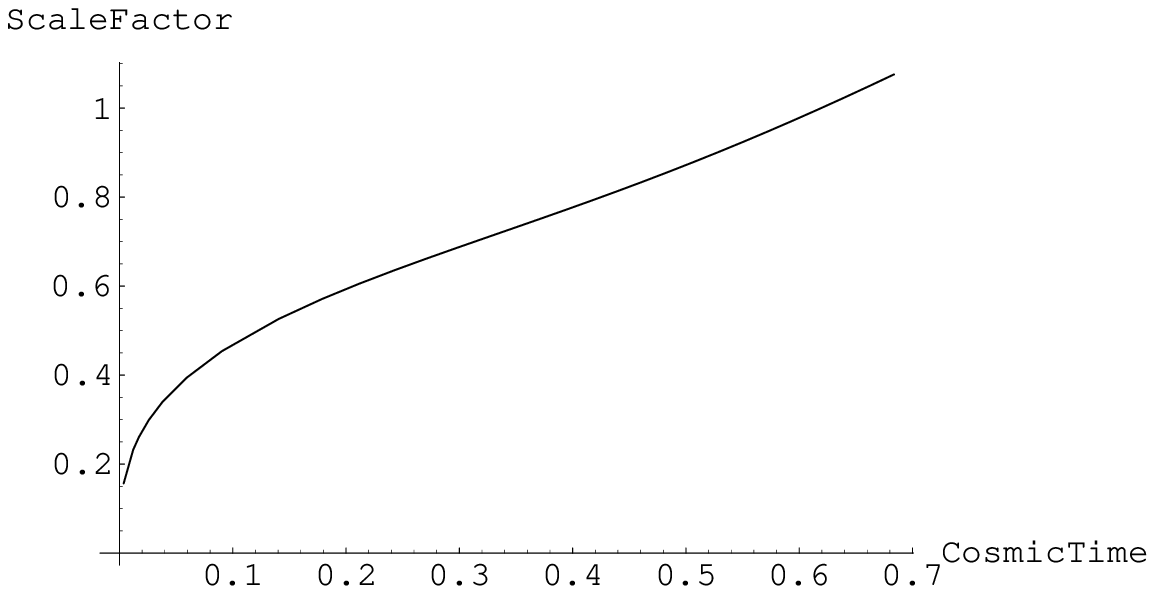}
 \includegraphics[height=40mm]{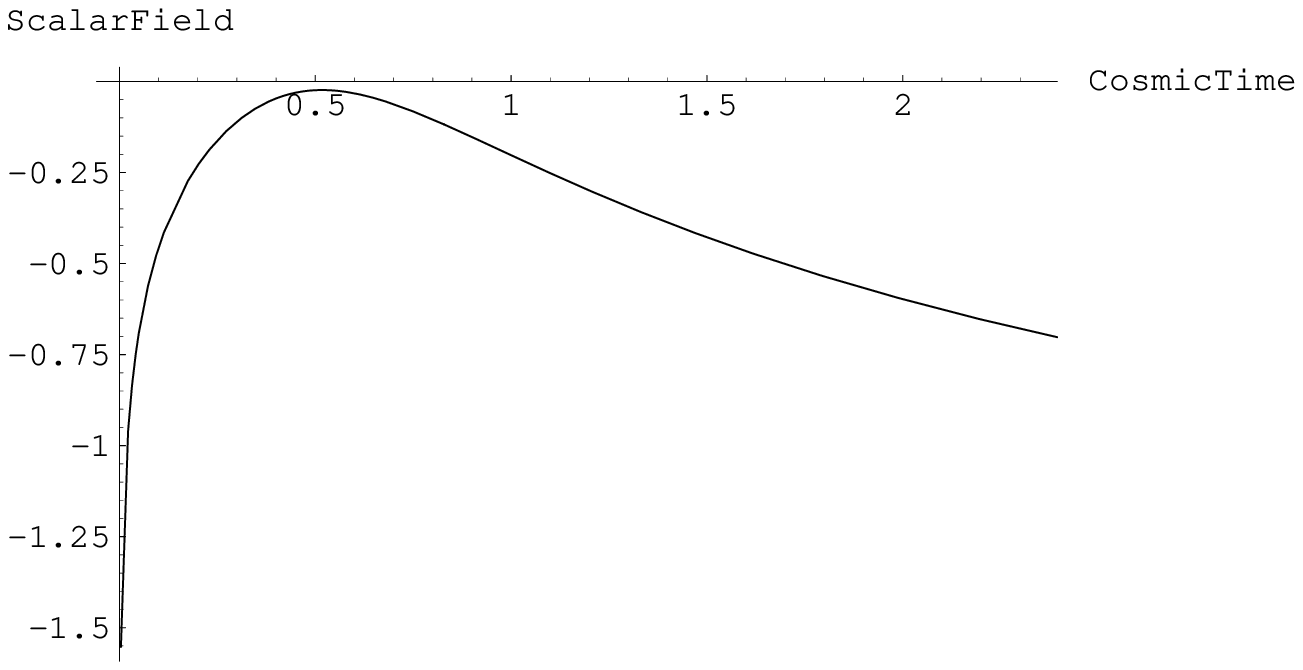}
 \vskip 2cm
 \includegraphics[height=40mm]{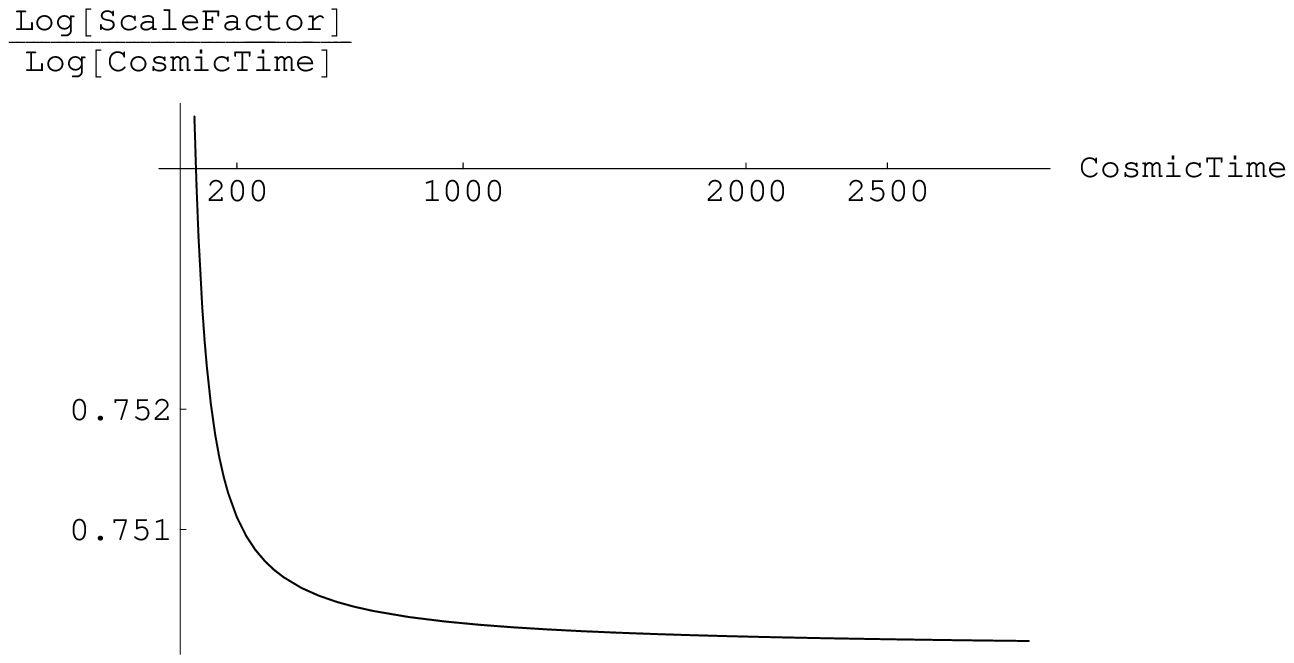}
\else
\end{center}
 \fi
\caption{\it
 In this picture we present the plots of the scale factor (first plot) and of the scalar field (second plot) in the hyperbolic solution with parameters $a= -1,b=\rho=\nu=1$.  The asymptotic behavior for late cosmic times of the scale factor is visualized by the third plot of the logarithm of the function divided by the logarithm of its argument. It is quite evident that this ratio goes rapidly to $0.75 \, = \, \frac{3}{4}$. }
\label{scappavia}
 \iffigs
 \hskip 1cm \unitlength=1.1mm
 \end{center}
  \fi
  \end{figure}
 An example of such a behavior is provided in the plots of fig.\ref{scappavia}. The effective equation of state at late times corresponds to a negative pressure although to a very weak one $w \, = \, -\frac{1}{9}$. This negative pressure is provided by a small dominance of the potential energy over the kinetic one and causes an indefinite expansion of the universe slightly stronger than that of a matter dominated Universe ($T_c^{\frac{2}{3}}$) yet very far from an exponential one.
\section{\sc Analysis of the $\cosh$-model}
\label{zerlina}
We come finally to one of the main questions posed in this paper, namely how much valuable are the exact solutions of integrable one-field cosmologies as simulations or approximations of the solutions of the physical cosmological models, in particular of those provided by consistent one-field truncations of supergravity theories. These latter, as we have emphasized, are mostly not integrable, as we have already seen and as we are going to show further in the last section.
In other words the present section is devoted to assess the viability of the \textit{minimalist approach}.
\par
Assuming that cosmological models derived from supergravity are not integrable, we still would like to ascertain whether  integrable potentials that are similar to actual physical ones have solutions that  simulate in a reasonable way the solutions of the supergravity derived models.
\par
In this contest a primary example is provided by the $cosh$ model  already introduced in previous sections and defined by the Lagrangian (\ref{lagrucconata})
that depends on the three parameters  $q$, $ p$ and $\mu^2$.
As we already pointed out, the family of $\cosh$--models defined by the Lagrangian (\ref{lagrucconata}) with the potential
\begin{equation}\label{gorrona}
    V(\mathfrak{h}) \, = \, - \, \mu^2 \, \cosh \left[ p\, \mathfrak{h}\right] \quad ; \quad \mu^2 \, = \,\mbox{either positive  or negative}
\end{equation}
includes two integrable cases  when $\frac{p}{\sqrt{3 q}}=1$ or $\frac{p}{\sqrt{3 q}}=\frac{2}{3}$. At the same time the case $p=1$, $q=1$, which by no means is integrable, corresponds to the consistent one-field truncation of the $\mathrm{STU}$ model. Furthermore, other instances of the same Lagrangian are expected to appear in the consistent truncation of other Gauged Supergravity models.
\par
Hence the model (\ref{lagrucconata}) is a perfect testing ground for the questions we have posed.
\par
An important conclusion that was reached in section \cite{primopapero} is that the qualitative behavior of solutions is dictated by the type of critical points possessed by the equivalent first order dynamical system that, on its turn is just dictated by the properties of extrema of the potential. Hence the first question that arises in connection with our model (\ref{lagrucconata}) is whether its critical points are always of the same type or fall into different classes depending on the parameters $p,q$.
\par
To this effect we begin by summarizing some of the results of sections\cite{primopapero} in a  language  less mathematically oriented and closer to the  jargon of  the supergravity community . Furthermore in such a summary we utilize the customary physical normalizations of Friedman equations, recalled in eq.s(\ref{fridmano}), rather than the normalizations and notations of \cite{primopapero} that are less familiar to cosmologists.
\subsection{\sc Summary of the mathematical results on the structure of Friedman equations and on the qualitative description of their solutions}
\label{sec:qualitativesummary}
Choosing the standard gauge ${\cal B} \, = \, 0 $
and considering the standard form (\ref{fridmano}) of Friedman equations,  the main crucial observation that was put forward in \cite{primopapero},  is that the logarithm of the scale factor $A(t)$ is a cyclic variable since it appears only through the Hubble function, namely covered by a derivative.
The next crucial observation was that the second order differential system (\ref{fridmano}) can be rewritten
in two different ways as a system of first order ordinary differential equations for two variables. These rewritings  were named \textit{irreducible subsystems} and it was advocated that each of them, when solved, generates solutions of the initial second order system (\ref{fridmano}). Adopting such a language allowed for the use of some powerful theorems that can predict the general qualitative behavior of the solutions of Friedman equations once the potential $V(\phi)$ is specified.
\par
The two subsystems are hereby rewritten in the standard notation of General Relativity and Cosmology:
\paragraph{\sc Subsystem I}. The first subsystem uses as independent variables the scalar field $\phi(t)$ and its time derivative $\dot{\phi}(t)$, which is renamed $\mathrm{v}(t)$. Hence one writes:
\begin{eqnarray}
&& \dot{\phi} \ = \ \mathrm{v} \ ,\nonumber \\
&& \dot{\mathrm{v}} \ = \, - \, 3 \, \sigma \, \mathrm{v} \, \sqrt{\frac{1}{3} \, \mathrm{v}^{\,2} \ + \ \frac{2}{3}\, {V}(\phi)}\ - \
{V}^{\,\prime}(\phi)\ ,
\label{MODE3}
\end{eqnarray}
where $\sigma \, = \, \pm 1$ takes into account the two branches of the square root in solving the quadratic equation for the Hubble function. One has to \textit{exclude} those branches of the solutions of  (\ref{MODE3}) that satisfy the following conditions:
\begin{eqnarray}
&& \frac{1}{3} \, \mathrm{v}^{\,2} \ + \ \frac{2}{3}\, {V}(\phi)\, = \, 0 \quad if \quad
{V}(\phi) \, \neq \, 0 \ ,\nonumber \\
&& \frac{1}{3} \, \mathrm{v}^{\,2} \ + \ \frac{2}{3}\, {V}(\phi) \, < \, 0
\label{MODE3a}
\end{eqnarray}
The remaining solutions are named \textit{admissible}.
When  an admissible  solution of  eqs. (\ref{MODE3}) is given, namely when the functions $\phi(t)$ and $\mathrm{v}(t)$ have been determined, the Hubble function $H(t)$ is immediately obtained,
\begin{eqnarray}
H(t) \ = \, \sigma \, \sqrt{\frac{1}{3} \, \mathrm{v}^{\,2}(t) \ + \ \frac{2}{3}\, {V}(\phi(t))}\
\label{MODE5}
\end{eqnarray}
and, by means of a further integration, one obtains also the scale factor:
\begin{equation}\label{corinzio}
    a(t) \, = \, \exp \left[ \int \, H(t) \, dt \,\right]
\end{equation}
\paragraph{\sc Subsystem II.} The second subsystem uses as independent variables the scalar field and the Hubble function. So doing we are lead to the following first order equations:
\begin{eqnarray}
&& \dot{\phi} \, = \, \sigma \, \left(3 \, H^{\,2} \ - \ 2\, {V}(\phi)\right)^{\frac{1}{2}}\quad ; \quad \sigma = \pm 1 \nonumber \\
&& \dot{H} \ = \ -\, \left(3 \, H^{\,2} \ - \ 2\, {V}(\phi)\right)\
\label{MODE4}
\end{eqnarray}
As in the  first instance of irreducible subsystem also here we have to exclude unadmissible branches of solutions to eq.s (\ref{MODE4}), namely those   that satisfy the following conditions:
\begin{eqnarray}
&& 3 \, H^{\,2} \ - \ 2\, {V}(\phi) \, = \, 0 \quad if \quad
{V}^{\,\prime}(\phi) \ \neq \ 0 \ , \nonumber\\
&& 3 \, H^{\,2} \ - \ 2\, {V}(\phi) \, = \, 0 \, < \, 0
\label{MODE4a}
\end{eqnarray}
It is easily verified that all equations of the original Friedman system (\ref{fridmano})
follow from either one  of the subsystems (\ref{MODE3}) and (\ref{MODE4}).
\par
In mathematical language, both subsystems (\ref{MODE3}) and (\ref{MODE4}) are nonlinear autonomous first-order ordinary differential equations over a two--dimensional Euclidean plane, namely either $\mathbb{R}^2 \, \ni \, (\phi, \, \mathrm{v})$ or $\mathbb{R}^2 \, \ni \,(\phi, \, H)$.
\par
The mathematical theory of planar dynamical systems is highly developed  and allows for a qualitative analysis of both the  local and the global behavior of their \textit{phase portraits}, namely of their  trajectories (also named orbits). According to such a theory a generic planar system is very regular: it admits  only a  few different types of trajectories and limit sets. Explicitly we can have:
\begin{description}
  \item[a)] periodic orbits, named also cycles,
  \item[b)] heteroclinic orbits that connect two different critical points of the system
  \item[c)] homoclinic orbits that start from a critical point and return to it at the end of time
  \item[d)] trajectories that connect  the point at infinity of $\mathbb{R}^2$ with a fixed point.
\end{description}
As a result no planar dynamical system can  be chaotic. This property distinguishes planar systems very strongly  from dynamical systems in dimensions higher than two, where various chaotic regimes are generically allowed.
\par
From these considerations it follows that one-field cosmologies are not chaotic and one can obtain a qualitative understanding of their solutions. In the integrable case the solutions can also be worked out analytically.
The relevant point is that such analytical solutions can be taken as trustable models of the behavior of the solutions also for entire classes of potentials whose integrable representatives occur only at very special values of their parameters.
\subsubsection{\sc Subsystem I : qualitative analysis}
To illustrate in a concrete manner these general ideas we choose to work with the subsystem I.
Fixed points of the subsystem  (\ref{MODE3}) are defined by the following equations
\begin{eqnarray}
\mathrm{v}_0 \, = \, 0\ , \quad {V}^{\,\prime}(\phi_0)\, = \, 0
\label{MODE6}
\end{eqnarray}
and are admissible if they satisfy the condition:
\begin{eqnarray}
{V}(\phi_0)\, \geq \, 0
\label{MODE6a}
\end{eqnarray}
In plain physical words a fixed point of this dynamical system is just a vacuum solution of scalar  coupled  gravity, namely a constant configuration of the scalar field that is an extremum of the potential. At the same time the space-time metric is either the Minkowski metric if $V(\phi_0) =0$ or the de Sitter metric if $V(\phi_0) = \Lambda >0$.
\par
From the dynamical system point of view, if the condition (\ref{MODE6a}) is not fulfilled, then the subsystem does not possesses fixed points at all, i.e. all its phase space points are regular. Without fixed points a nonlinear system admits only monotonic solutions, which can also blow up in a finite time. From the physical point of view an extremum of the potential at a negative value of the potential corresponds to an anti de Sitter space, yet anti de Sitter, differently from de Sitter admits no representation in terms of flat constant time slices, which is our initial assumption. Hence the only consequence can be a blowing up solution with a Big Bang followed by a  Big Crunch.
\paragraph{\sc Linearization of the first order system in a neighborhood of a fixed point}
Let us consider the linearization of the first order system in a neighborhood of the fixed point by setting:
\begin{eqnarray}
  \phi&=& \phi_0 + \Delta\phi \nonumber\\
  \mathrm{v} &=& \Delta \mathrm{v} \label{fluttuo}
\end{eqnarray}
To first order in the deviations we obtain
\begin{eqnarray}
&& \Delta\dot{\phi} \ = \ \Delta\mathrm{v} \ ,\nonumber \\
&& \Delta\dot{ \mathrm{v}} \ = \ -\, \sigma \, \sqrt{ 6 \,{V}(\phi_0)}\, \Delta \mathrm{v}\ - \, {V}^{\,\prime \prime}(\phi_0) \, \Delta\phi \ + \ h.o.t.\ ,
\label{MODE7}
\end{eqnarray}
where the abbreviation h.o.t. means higher  order terms.
\par
As explained in \cite{primopapero}, the eigenvalues of the linearization matrix:
\begin{equation}
\mathcal{M} \, = \, \left[\begin{array}{cc} 0 & 1 \\ - V^{\prime\prime}_0  & - \, \sigma\, \sqrt{6 \, V_0}\\
\end{array}\right]\label{somber}
\end{equation}
namely:
\begin{eqnarray}
 \lambda_{\pm} \ = \, \frac{1}{\sqrt{2}}\,\left(-\,\sigma \, \sqrt{3 \, V_0}\, \pm \,  \sqrt{3 \, V_0\,  - \,  2 \, V^{\prime \prime}_0 } \right)
\label{MODE8}
\end{eqnarray}
characterize the type of the corresponding critical points and consequently define the phase portrait of the linearization. The main theorem that allows to predict the qualitative behavior of the solution states the following.
In the case the fixed point is hyperbolic, namely both eigenvalues have a non vanishing real part, (i.e. $\mathrm{Re}\,(\lambda_{\pm})\,\neq\,0$) the phase portraits of the nonlinear system and of its linearization are diffeomorphic  in a  finite neighborhood of the hyperbolic fixed point.
Hence the analysis of the linearization gives a valuable information about the phase portrait of the original nonlinear system we are interested in.
\par
We have in particular the following classification of non degenerate fixed points for which, by definition, the linearization matrix has no zero eigenvalue:
\paragraph{\sc Classification of fixed point types}
\begin{description}
  \item[a) \textbf{Saddle }] When the two real eigenvalues have opposite sign $\lambda_+ >0, \,\lambda_- < 0$ or
  $\lambda_+ <0, \,\lambda_- > 0$.
  \item[b) \textbf{Node }] When the two real eigenvalues have the same sign $\lambda_+ >0, \,\lambda_- > 0$ or
  $\lambda_+ <0, \,\lambda_- < 0$.
  \item[c) \textbf{Improper Node}] When the two eigenvalues coincide $\lambda_+ \, = \, \lambda_-$, yet the linearization matrix (\ref{somber}) is not diagonal.
   \item[d) \textbf{Degenerate Node}] When  the linearization matrix (\ref{somber}) is proportional to identity.
  \item[e)\textbf{Focus }] When the two eigenvalues have non vanishing both the  imaginary part and the real part and are complex conjugate to each other $\lambda_\pm \, = \, x \, \pm {\rm i} \, y$.
    \item[f) \textbf{Center } ]When the two eigenvalues are purely imaginary and conjugate to each other.
\end{description}
For each of these fixed point types the trajectories have a distinct behavior that we are going to illustrate by means of our concrete example, namely the \textit{Cosh Model} analysed in the present section.
\par
Furthermore we should recall the result  that the subsystem (\ref{MODE3}) has no periodic trajectories according to Dulac's criterion since:
\begin{eqnarray}
\frac{\partial \dot{\phi}}{\partial \phi} \ + \
\frac{\partial \dot{\mathrm{v}}}{\partial \mathrm{v}} \ \equiv \ -\,2\,\sigma \, \frac{\mathrm{v}^{\,2} \ + \  {V}(\phi)}{\sqrt{\frac{1}{3}\,v^{\,2} \ + \ \frac{2}{3}\, {V}(\phi)}}
\label{MODE3per}
\end{eqnarray}
does not change  sign over the whole two-dimensional plane. Thus, we are led to the conclusion that this subsystem can have only fixed points (i.e. vacuum solutions) as well as heteroclinic/homoclinic orbits,  orbits connecting  infinity with a fixed point or orbits connecting infinity with infinity in the case of fixed points of the saddle type.
As we know  the case $p=1$, $q=1$ is the one which appears in the non-abelian gauging of the $\mathrm{STU}$ model and the case $p=1$, $q=3$ can be obtained in the $S^3$ model by means of an abelian gauging.
\par
Although the $\mathcal{N}=2$ case is not integrable, yet it belongs to same subclass (Node) as the integrable case $\frac{p}{\sqrt{3 q}}=\frac{2}{3}$. Hence we can probably learn about its behavior from an analysis of the  integrable case close to it.
\par
The natural question which we have posed, namely  \textit{how much do the solutions of the physically relevant $\cosh$-models depart from the exact solutions of the  integrable members of the same family} can now be partially answered.   As we just stressed, the physically relevant $\mathcal{N}=2$ case is of the \textit{Node type} so that any integrable model with the same type of fixed point would just capture all the features of the physical $\mathrm{STU}$ model. The other integrable case $\frac{p}{\sqrt{3 q}}=1$ is instead of the \textit{Focus type}, therefore it has a little bit more of structure with respect to the $\mathrm{STU}$-model. Any other supersymmetric one-field model with a \textit{Cosh potential} of the focus type, although not integrable might be well described by the $\frac{p}{\sqrt{3 q}}=1$ integrable member of the family (\ref{lagrucconata}).
\subsection{\sc Normal Form  of the Cosh-model}
In this spirit let us first show how the \textit{Cosh model} can be put into a normal form, displaying a unique  parameter $\omega$ whose value will determine the type of fixed point and, for two special choices, yield two integrable models.
\par
To this effect we introduce the following rescaled fields and variables:
\begin{equation}\label{cambiovario}
 \mathfrak{h}[t] \, = \, \frac{\phi(\tau)}{\sqrt{q}} \quad ; \quad t \, = \, \frac{\sqrt{2} \tau }{\mu } \quad ; \quad A(t) \, = \, A(\tau) \quad ; \quad \omega \, \equiv \, \frac{p}{\sqrt{q}}
\end{equation}
In terms of these new items, the effective Lagrangian (\ref{lagrucconata}) becomes
\begin{equation}\label{borragine}
 \mathcal{L}  \, = \,  e^{3 A(\tau) \, - \, \mathcal{B}(\tau)}\left\{-\frac{3}{2} A'(\tau
   )^2+\frac{1}{2} \phi'(\tau
   )^2\, \mp \, 2 e^{2\mathcal{B}(\tau)} \, \cosh [ \omega \phi(\tau)
   ]\right\} \ ,
\end{equation}
where the sign choice distinguishes two drastically different systems. The first choice yields a positive definite potential with an absolute minimum that allows for a stable de Sitter vacuum, while the second yields a potential unbounded from below with an absolute maximum.
If we choose the gauge $\mathcal{B}=0$, the field equations of this system, including the hamiltonian constraint can be written as the following three Friedman equations:
\begin{equation}\label{friemano}
    \begin{array}{lcl}
 \frac{a'(\tau )^2}{a(\tau
   )^2}-\frac{1}{3}
   \phi'(\tau )^2\, \mp \,\frac{4}{3}
   \cosh [\omega  \phi(\tau
   )] & = & 0\\
   \null & \null & \null \\
 \frac{2}{3} \phi'(\tau
   )^2\, \mp \, \frac{4}{3} \cosh [\omega
   \phi(\tau
   )]+\frac{a''(\tau )}{a(\tau )} & = & 0\\
    \null & \null & \null \\
 \pm \, 2 \omega  \sinh [\omega
   \phi(\tau )]+\frac{3
   a'(\tau ) \phi'(\tau
   )}{a(\tau )}+\phi''(\tau ) & = & 0 \\
\end{array}
\end{equation}
where $a(\tau)\, = \, \exp[A(\tau)]$.
\subsection{\sc The general integral for the case $\omega = \sqrt{3}$.}
In the integrable case $\omega\, = \,\sqrt{3}$, by means of the integrating transformation described in \cite{primopapero} we obtain the following general solution of eq.s (\ref{friemano}) depending on three parameters, the scale $\lambda$ and the two angles $\psi$ and $\theta$, which applies to the case of the positive potential (upper choice in eq.(\ref{borragine})):
\begin{eqnarray}
  a_+(\tau) &=& \sqrt[3]{\left(\lambda  \cos (\psi )
   \cosh \left(\sqrt{3} \tau
   \right)+\lambda  \sinh
   \left(\sqrt{3} \tau
   \right)\right)^2-\lambda ^2 \cos
   ^2\left(\theta -\sqrt{3} \tau
   \right) \sin ^2(\psi )} \\
  \phi_+(\tau) &=& \frac{1}{\sqrt{3}}\log \left[\frac{\cos (\psi )
   \cosh \left(\sqrt{3} \tau
   \right)-\cos \left(\theta
   -\sqrt{3} \tau \right) \sin (\psi
   )+\sinh \left(\sqrt{3} \tau
   \right)}{\cos (\psi ) \cosh
   \left(\sqrt{3} \tau \right)+\cos
   \left(\theta -\sqrt{3} \tau
   \right) \sin (\psi )+\sinh
   \left(\sqrt{3} \tau
   \right)}\right] \label{soluziapiu}
\end{eqnarray}
For the negative potential (lower choice in eq.(\ref{borragine})) we find instead:
\begin{eqnarray}
  a_-(\tau) &=& \sqrt[3]{\lambda ^2 \left(\cosh
   ^2(\psi ) \sinh ^2\left(\theta
   -\sqrt{3} \tau \right)-\left(\sin
   \left(\sqrt{3} \tau \right)-\cos
   \left(\sqrt{3} \tau \right) \sinh
   (\psi )\right)^2\right)} \\
\phi_-(\tau)&=& \frac{1}{\sqrt{3}} \, \log \left[ \frac{\sin
   \left(\sqrt{3} \tau \right)+\cosh
   (\psi ) \sinh \left(\theta
   -\sqrt{3} \tau \right)-\cos
   \left(\sqrt{3} \tau \right) \sinh
   (\psi )}{\sin \left(\sqrt{3} \tau
   \right)-\cosh (\psi ) \sinh
   \left(\theta -\sqrt{3} \tau
   \right)-\cos \left(\sqrt{3} \tau
   \right) \sinh (\psi
   )}\right]\label{soluziameno}
\end{eqnarray}
One important observation is the following. In the case of the positive potential, by choosing the parameters $\psi=0,\theta=0$ we obtain the very simple solution:
\begin{equation}\label{carriome}
    a_0(\tau) \, = \, \exp\left[ {\frac{2 \tau }{\sqrt{3}}} \right] \, \lambda
   ^{2/3} \quad ; \quad \phi\, = \, 0 \quad \Rightarrow \quad \mathfrak{h}\, = \, 0
\end{equation}
This is the de Sitter solution where the scalar field is stationary at its absolute minimum and the scale factor grows exponentially.
Such a solution is ruled out in the case of the negative potential which does not allow for any static scalar field solution.
\paragraph{\sc The second hamiltonian structure}
We have explicitly integrated the integrable cosmological model and we have obtained its general integral. We can address the question why is it integrable? The answer is that it admits not just one rather two functionally independent conserved hamiltonians. Examining their structure is worth doing since it provides hints about the underlying properties of the field theory that might be responsible for the emergence
of integrability at the cosmological level. Consider then the Lagrangian and the hamiltonian for the model under consideration:
\begin{eqnarray}
    \mathcal{L}_{0} & = & \exp[3 A ] \,  \left\{ \frac{q}{2} \, \dot{\mathfrak{h}}^2 \, - \, \frac{3}{2} \, \dot{A}^2 \, - \, \mu^2 \cosh \left [ 3 \, \mathfrak{h} \right ] \right\}\nonumber\\
    \mathcal{H}_{0} & = & \exp[3 A ] \,  \left\{ \frac{q}{2} \, \dot{\mathfrak{h}}^2 \, - \, \frac{3}{2} \, \dot{A}^2 \, + \, \underbrace{\mu^2 \cosh \left [ 3 \, \mathfrak{h} \right]}_{V_0(\mathfrak{h})} \right\}
\end{eqnarray}
By means of direct evaluation we can easily check that the following two functionals:
\begin{eqnarray}
  \mathcal{H}_{1} &=& -\frac{1}{2} e^{3 A }
   \left(2 \mu ^2 \cosh
   ^2\left(\frac{3
   \mathfrak{h} }{2}\right)+3 \dot{A} ^2
   \cosh ^2\left(\frac{3
   \mathfrak{h} }{2}\right)+3 \sinh
   ^2\left(\frac{3
   \mathfrak{h} }{2}\right) \dot{\mathfrak{h}} ^2+3
   \sinh (3 \mathfrak{h} ) \dot{A}
   \dot{\mathfrak{h}} \right) \nonumber\\
   \null & \null &\null \label{hamma1}\\
  \mathcal{H}_{2} &=& \frac{1}{2} e^{3 A }
   \left(-2 \mu ^2 \sinh
   ^2\left(\frac{3
   \mathfrak{h} }{2}\right)+3 \dot{A} ^2
   \sinh ^2\left(\frac{3
   \mathfrak{h} }{2}\right)+3 \cosh
   ^2\left(\frac{3
   \mathfrak{h} }{2}\right) \dot{\mathfrak{h}} ^2+3
   \sinh (3 \mathfrak{h} ) \dot{A}
   \dot{\mathfrak{h}} \right) \nonumber\\
   \label{hamma2}
\end{eqnarray}
satisfy the following conditions:
\begin{eqnarray}
  \mathcal{H}_{1}+\mathcal{H}_{2} &=& \mathcal{H}_{0} \\
  \frac{\mathrm{d}}{\mathrm{d}t} \mathcal{H}_{1,2}&=& 0 \quad \mbox{Upon use of field equations from Lagrangian}
\end{eqnarray}
Hence $\mathcal{H}_{1,2} $ are the two conserved hamiltonians that guarantee the integrability of the system. As we know, the actual solution of Friedman equations is obtained  by enforcing also the constraint $\mathcal{H}_{0}\, = \,0$ so that for our general solution
(\ref{soluziapiu}-\ref{soluziameno}) we have: $\mathcal{H}_{1}\,=\,- \,\mathcal{H}_{2}$.
\subsection{\sc The general integral in the case $\omega = \frac{2}{\sqrt{3}}$}
In this case as in the previous one the solution can be obtained by means of the same substitution  described \cite{primopapero}, yet with respect to the previous case there is one relevant difference. In this case the gauge $\mathcal{B}=0$ cannot be   chosen
and there is a difference between the cosmic time $t_c$ and the parametric time $\tau$. The form of the space-time metric is the following:
\begin{equation}\label{spaziailtempone}
    ds^2 \, = \, - \, \exp\left [ - 2 A(\tau) \right ] \, \mathrm{d}\tau^2  \, + \, \exp\left [  2 A(\tau) \right ] \, \mathrm{d}\vec{x}^2
\end{equation}
corresponding to the gauge $\mathcal{B}(\tau) \, = \, - A(\tau)$.
In principle the general integral depends on three integration constants, but in this case one of them can be immediately reabsorbed into a shift of the parametric time and thus we are left  only with two relevant constants.
\par
At the end of the computations the general integral can be written in a very suggestive and elegant form in terms of the four roots of a quartic polynomial. Precisely we have:
\begin{eqnarray}
  A(\tau)&=& \log \left[\frac{2 \sqrt[4]{\left(\tau
   -\lambda _1\right)
   \left(\tau -\lambda
   _2\right) \left(\tau
   -\lambda _3\right)
   \left(\tau -\lambda
   _4\right)}}{\sqrt{3}}\right] \label{AfattoCosh2} \\
  \phi(\tau) &=& \frac{1}{4} \sqrt{3} \log
   \left(\frac{\left(\tau
   -\lambda _3\right)
   \left(\tau -\lambda
   _4\right)}{\left(\tau
   -\lambda _1\right)
   \left(\tau -\lambda
   _2\right)}\right) \\
  \mathcal{B}(\tau) &=& -A(\tau) \label{soluzionna}
\end{eqnarray}
One important caveat, however, is the following. Let us name
\begin{equation}\label{maxmin}
    \mbox{Re} \lambda_{min}   \quad ; \quad \mbox{Re} \lambda_{max}
\end{equation}
the smallest and the largest of the real parts of the four roots. The functions (\ref{soluzionna}) provide an exact solution of the Friedman equations under two conditions:
\begin{description}
  \item[A)]  The parametric time $\tau$ is either in the range $\mbox{Re} \lambda_{max} \, \le  \tau \, \le +\infty$ or in the range
  $\mbox{Re} \lambda_{min} \, \ge  \tau \, \ge -\infty$
  \item[B)]  The four roots satisfy the following constraint:
  \begin{equation}
2 \lambda _1 \lambda _2-\lambda _3
   \lambda _2-\lambda _4 \lambda
   _2-\lambda _1 \lambda _3-\lambda _1
   \lambda _4+2 \lambda _3 \lambda _4  \, = \, 0 \label{custretto1}
\end{equation}
\end{description}
When $ \tau$ is in the range $\mbox{Re} \lambda_{min} \, \le  \tau \, \ge \, \mbox{Re} \lambda_{max}$, the expressions (\ref{soluzionna}) do not satisfy Friedman equations.
 \par
Solving (\ref{custretto1}) explicitly and replacing  such solution back into the expression (\ref{soluzionna}) of the fields we obtain the general integral depending on three parameters. As we already stressed one of these three parameters can always be reabsorbed by means of a shift of the  parametric time coordinate $\tau$, as it is evident from the form (\ref{soluzionna}). A convenient way of taking into account these gauge fixing is provided by solving the constraint (\ref{custretto1}) in terms of two real parameters $(\alpha,\beta)$, as it follows:
\begin{eqnarray}
  \lambda_1 &=& \alpha -\sqrt{2} \sqrt{\alpha ^2-\beta } \nonumber\\
  \lambda_2 &=& \alpha +\sqrt{2} \sqrt{\alpha ^2-\beta }\nonumber\\
  \lambda_3 &=& -\alpha -\sqrt{2} \sqrt{\alpha ^2+\beta } \nonumber\\
  \lambda_4 &=& \sqrt{2} \sqrt{\alpha ^2+\beta }-\alpha \label{fraticello}
\end{eqnarray}
which greatly facilitates the discussion since depending on whether $|\beta| < \alpha^2$ or $|\beta| >  \alpha^2$, we either have four real roots and hence four zeros of the scale factor or two real roots and two complex conjugate ones. If $|\beta| = \alpha^2$ we have only real roots but three of them coincide and the fourth is different. Finally if both $\alpha$ and $\beta$ vanish the four roots coincide and the corresponding solution degenerates into the de Sitter solution. Let us substitute explicitly the values (\ref{fraticello}) into (\ref{soluzionna}) and obtain the general integral in the form:
\begin{eqnarray}
  a(\tau,\alpha,\beta) \, \equiv \, \exp\left[A(\tau,\alpha,\beta)\right] &=&\frac{2 \sqrt[4]{\alpha ^4-6 \tau ^2
   \alpha ^2+8 \beta  \tau  \alpha +\tau
   ^4-4 \beta ^2}}{\sqrt{3}} \label{afattoCosh2}\\
  \phi(\tau,\alpha,\beta) &=& \frac{1}{4} \sqrt{3} \log
   \left(\frac{\alpha ^2-2 \tau  \alpha
   -\tau ^2+2 \beta }{\alpha ^2+2 \tau
   \alpha -\tau ^2-2 \beta }\right)\label{scalfattoCosh2}\\
  \exp\left[B(\tau,\alpha,\beta)\right] &=& \frac{\sqrt{3}}{2 \sqrt[4]{\alpha ^4-6 \tau ^2
   \alpha ^2+8 \beta  \tau  \alpha +\tau
   ^4-4 \beta ^2}} \label{BfattoCosh2}
\end{eqnarray}
which will be very useful in the discussion of the of the properties of solutions in the vicinity of the fixed point which for this case happen to be of the Node-type.
\subsection{\sc Discussion of the fixed points}
Inserting the potential $2\, \cosh\left( \omega \phi\right)$ into the formulae (\ref{somber}) and (\ref{MODE8})
for the linearization matrix and for its eigenvalues eigenvalues we immediately obtain:
\begin{equation}\label{mygoodness}
   \mathcal{M} \, = \, \left[ \begin{array}{cc}
                                0 & 1 \\
                                -2 \, \omega^2 & \sqrt{12}
                              \end{array}
   \right] \quad  \Rightarrow \quad \lambda_\pm \, = \, - \sqrt{3} \pm \sqrt{3\, - \, 2 \omega^2}
\end{equation}
From this we learn that for $0 \, < \, \omega \, < \, \sqrt{\frac{3}{2}}$ the fixed point at $\phi=0$ is of the \textit{Node type}.
For $\omega = \sqrt{\frac{3}{2}}$ the fixed point is exactly of the \textit{improper node type}, while for $\omega > \sqrt{\frac{3}{2}}$ it is always of the focus type. What this implies for the behavior of the scalar field we will shortly see.
\par
One important point to stress concerns the initial conditions that have always got to be of the Big Bang type (initial singularity). The exact solutions of the integrable case are very instructive in this respect. Fixing the Big Bang initial condition $a(0)=0$ at a finite reference time ($\tau = 0$) from (\ref{soluziapiu}) we obtain a relation between the angle $\psi$ and the angle $\theta$
\begin{equation}\label{bigbango}
    \cos (\psi)
      -  \cos^2\left(\theta\right) \sin ^2(\psi ) \, = \, 0
\end{equation}
which inserted back into the formula (\ref{soluziapiu}) for the scalar field implies $\phi(0) \, = \, \infty$.
This is not a peculiarity of the integrable model, rather it is a general fact. The zeros at a finite time of the scale factor are always in correspondence with a singularity of the scalar field. Hence the only initial condition that can be fixed independently at the Big Bang is the initial velocity of the field $\dot{\phi}(0)$.
\par
Emerging from the initial singularity at $\pm \infty$ the scalar field can flow to its fixed point value, namely zero (if the potential is positive) and the way it does so depends on the fixed-point type. However it can also happen that before reaching the fixed point the scalar field goes again to $\pm \infty$. In this case we have a blow-up solution where, at a finite time the scale factor goes again to zero and we have a Big Crunch. This  happens only in the case the  extremal point of the  potential (either minimum or maximum) is negative. If the extremal point of the potential is positive we always have an asymptotic de Sitter destiny of the considered Universe.
\subsection{\sc Behavior of solutions in the neighborhood of a Node critical point}
\begin{figure}[!hbt]
\begin{center}
\iffigs
 \includegraphics[height=70mm]{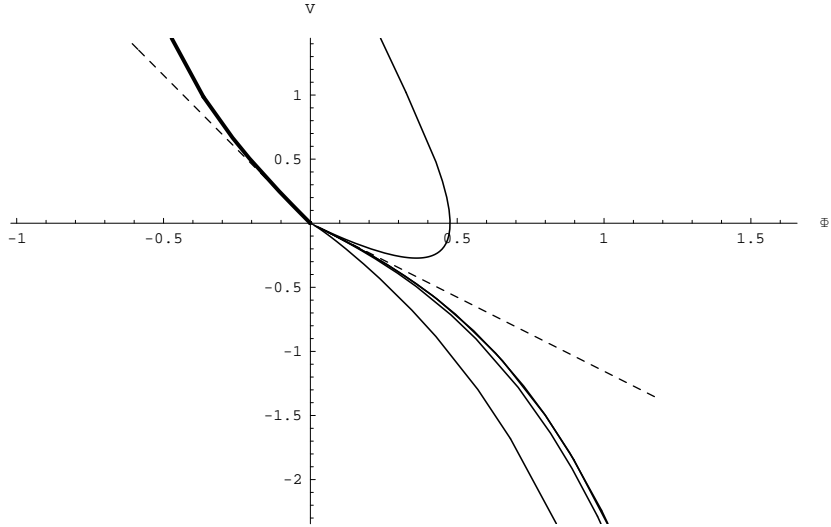}
\else
\end{center}
 \fi
\caption{\it
In this figure are shown the $(\phi,\mathrm{v})$ trajectories corresponding to six solutions of the $\omega \, = \frac{2}{\sqrt{3}}$ Cosh-model with parameters $(\alpha,\beta)$ = $(0,\frac 1 2)$, $(2,0)$, $(5,0)$, $(3,-3)$, $(2,5)$, $(1,-18)$. The case $(0,\frac 1 2)$ corresponds to the separatrix which approches the fixed point along the tangent vector $\{ -\frac{\sqrt{3}}{4} , 1\}$. All the other solutions approach the fixed point along the tangent vector $\{ -\frac{\sqrt{3}}{2} , 1\}$. The two dashed lines represent these two tangent vectors.
\label{separanonsepara}}
\iffigs
 \hskip 1cm \unitlength=1.0mm
\end{center}
\fi\end{figure}

The best way to discuss the quality of solutions is by means of the so called phase portrait of the dynamical system, where we plot the trajectories of the solution in the plane $\{\phi\, , \, v\}$. As we have already emphasized there are solutions that go to the fixed point $\{\phi\, , \, v\}= \{0,0\}$ and solutions that never reach it. The solutions that reach the fixed point have a universal type of behavior in its neighborhood which we now describe for the case of the Node. We illustrate such a behavior by means of the exact solutions of the integrable case $\omega = \frac{2}{\sqrt{3}}$. The two eigenvectors corresponding to the two eigenvalues  in eq.(\ref{mygoodness}) are:
\begin{equation}\label{eigenfunctions}
    \mathbf{v}_{\pm} \, = \,\left\{
\begin{array}{rcl}
 \{\frac{\sqrt{3-2 \omega ^2}-\sqrt{3}}{2
   \omega ^2} &,& 1 \} \\
   \null&\null&\null \\
 \{-\frac{\sqrt{3-2 \omega ^2}+\sqrt{3}}{2
   \omega ^2} &,& 1\}
\end{array}\right. \quad \stackrel{\omega = \frac{2}{\sqrt{3}}}{\Longrightarrow} \quad \left\{
\begin{array}{rcl}
\{ -\frac{\sqrt{3}}{4} &,& 1\} \\
 \{-\frac{\sqrt{3}}{2} &,& 1\}
\end{array}
\right.
\end{equation}
According to theory all solutions of the differential system approach the fixed point along the eigenvector vector $\{-\frac{\sqrt{3-2 \omega ^2}+\sqrt{3}}{2 \omega ^2} , 1\} \, \Rightarrow \, \{-\frac{\sqrt{3}}{2} , 1\}$ corresponding to the eigenvalue of largest absolute value, except a unique, exceptional one, named the \textbf{separatrix} that approaches the fixed point along the other eigenvector  $\{ -\frac{\sqrt{3}}{4} , 1\}$. This can be checked analytically computing a limit. Let us define the function:
\begin{equation}\label{tangentus}
    T(\tau,\alpha,\beta) \, = \, \frac{\partial_\tau \phi(\tau,\alpha,\beta) \, a(\tau\,\alpha,\beta)}{\phi(\tau,\alpha,\beta)} \, =\, \frac{\partial_{t_c} \phi(t_c)}{\phi(t_c)}
\end{equation}
which, due to the metric (\ref{spaziailtempone}), represents the ratio of the logarithmic derivative of the scalar field with respect to the cosmic time. By explicit calculation we find:
\begin{eqnarray}
  \lim_{\tau \rightarrow \pm \infty} \, T(\tau,\alpha,\beta) &=& \mp\frac{2}{\sqrt{3}} \\
  \lim_{\tau \rightarrow \pm \infty} \, T(\tau,\alpha,0) &=& \mp\frac{2}{\sqrt{3}} \\
  \lim_{\tau \rightarrow \pm \infty} \, T(\tau,0,\beta) &=& \mp\frac{4}{\sqrt{3}}
\end{eqnarray}
We conclude that the separatrix solution is given by the choice  $\alpha=0$, all the other solutions being regular. An inspiring view of the phase portrait is given in fig.\ref{separanonsepara}
\subsection{\sc Analysis of the separatrix solution}
\begin{figure}[!hbt]
\begin{center}
\iffigs
 \includegraphics[height=55mm]{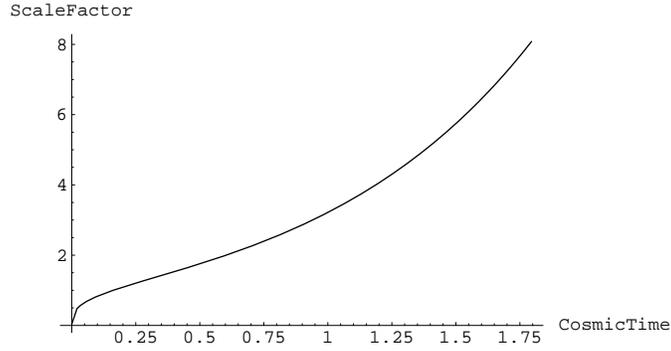}
 \vskip 2cm
 \includegraphics[height=55mm]{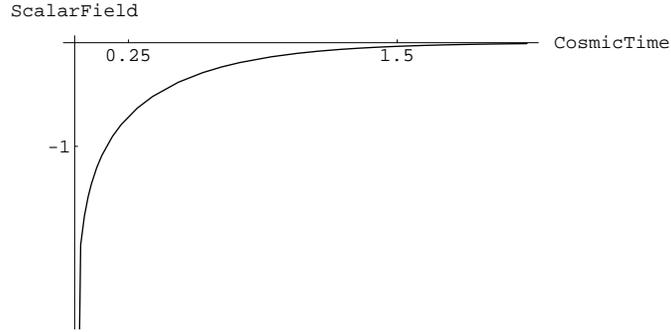}
\else
\end{center}
 \fi
\caption{\it
Behavior of the scale factor and of the scalar field in the case of the separatrix solution. As one can see we have a climbing scalar that from minus infinity goes asymptotically to its extremum value $\phi = 0$, while the scale factor has an asymptotic exponential behavior as in most of the other regular solutions that go the fixed point, the specialty of this solution is visible only in the phase portrait.
}
\label{separando}
 \iffigs
 \hskip 1cm \unitlength=1.0mm
 \end{center}
  \fi
\end{figure}

Because its exceptional status, the separatrix solution is worth to be analysed in some detail.
Explicitly we have:
\begin{eqnarray}
  a(\tau) &=& \frac{2 \sqrt[4]{\tau
   ^4-1}}{\sqrt{3}} \\
  \phi(\tau) &=&\frac{1}{4} \sqrt{3} \log
   \left(\frac{\tau ^2-1}{\tau
   ^2+1}\right) \\
  e^{B} &=& \frac{\sqrt{3}}{2 \sqrt[4]{\tau
   ^4-1}}
\end{eqnarray}
The roots of $a(\tau) $ are $\pm {\rm i}$ and $\pm 1$. Hence the solution exists and is real only for $|\tau| >1$. So we have two real identical branches of the solution in the range $[-\infty, -1]$ and in the range $[1, +\infty]$. The cosmic time can be explicitly integrated and admits  the following expression in terms of a generalized hypergeometric function:
\begin{equation}\label{cosmictimeSeparo}
\int_{1}^{T}\, \frac{\sqrt{3}}{2 \sqrt[4]{\tau
   ^4-1}} \, \mathrm{d}\tau   \, = \,  \frac{1}{2} \sqrt{3}
   \left(-\frac{\,
   _3F_2\left(1,1,\frac{5}{4};
   2,2;\frac{1}{T^4}\right)}{1
   6 T^4}+\log (T)+\frac{1}{8}
   (-\pi +\log (64))\right)
\end{equation}
The behavior of the scale factor and of the scalar field for the separatrix solution are displayed in fig.\ref{separando}
\subsection{\sc Analysis of a solution with four real roots}
\begin{figure}[!hbt]
\begin{center}
\iffigs
 \includegraphics[height=70mm]{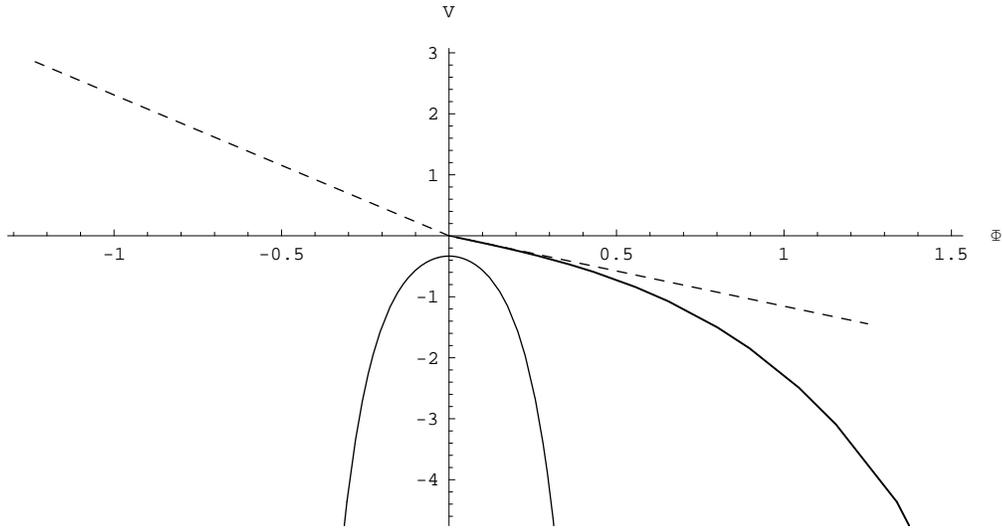}
\else
\end{center}
 \fi
\caption{\it In this figure we display the two trajectories in $(\phi,\mathrm{v})$-plane corresponding to the solution of parameter $(  \alpha,\beta)=(2,0)$. One branch connects infinity with infinity and never reaches the fixed point. This branch is actually fake since in this range of the parametric time Friedman equations are not satisfied!  The other physical branch which satisfies Friedaman equation, connects infinity with the fixed point which is reached along the universal tangent vector of all solutions except the separatrix.
}
\label{duebraccine}
 \iffigs
 \hskip 1cm \unitlength=1.0mm
 \end{center}
  \fi
\end{figure}

\begin{figure}[!hbt]
\begin{center}
\iffigs
 \includegraphics[height=55mm]{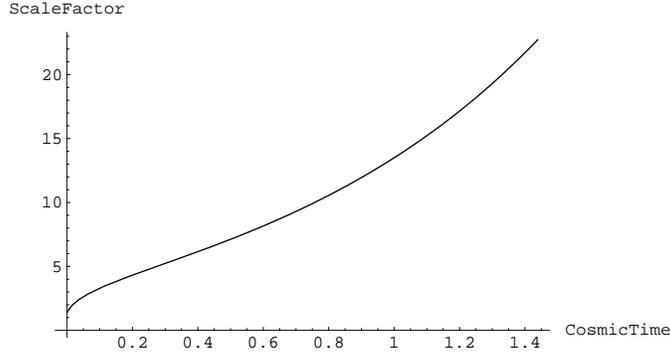}
 \vskip 2cm
 \includegraphics[height=55mm]{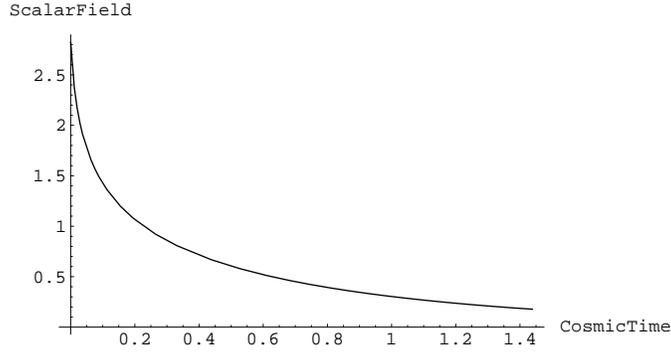}
\else
\end{center}
 \fi
\caption{\it
In this figure we display the behavior of the scale factor and of the scalar field for the asymptotic branch of the solution $(\alpha,\beta) = (2,0)$. The scalar descends from $+\infty$ to $0$ in an infinite  time. In a parallel way the scale factor  approaches an asymptotically exponential behavior.
}
\label{turoldo}
 \iffigs
 \hskip 1cm \unitlength=1.0mm
 \end{center}
  \fi
\end{figure}

Next we analyze the exact solution with parameters $(\alpha , \beta) = (2,0)$.
In this case the solution has the following form:
\begin{eqnarray}
\label{duebraccia}
  a(\tau)&=& \frac{2 \sqrt[4]{\tau ^4-24 \tau
   ^2+16}}{\sqrt{3}}\\
  \phi(\tau) &=& \frac{1}{4} \sqrt{3} \log
   \left(\frac{\tau ^2+4 \tau -4}{\tau
   ^2-4 \tau -4}\right) \\
  e^B &=& \frac{\sqrt{3}}{2 \sqrt[4]{\tau ^4-24 \tau
   ^2+16}}
\end{eqnarray}
and the scale factor admit four real roots, namely:
\begin{equation}\label{ruttoni}
 \lambda_{1,2,3,4} \, = \,    \left\{2+2 \sqrt{2},-2+2 \sqrt{2},2-2
   \sqrt{2},-2-2 \sqrt{2}\right\}
\end{equation}
The scalar factor is real in the intervals $[-\infty, \lambda_4]$, $[\lambda_3, \lambda_2]$ and $[\lambda_1, +\infty]$. In the two identical branches (it suffices to change $\tau \leftrightarrow - \tau$) $[-\infty, \lambda_4]$ and $[\lambda_1, +\infty]$, the solution reaches the fixed point and it is asymptotically de Sitter (exponential increase of the scale factor). On the other hand in the branch $[\lambda_3, \lambda_2]$, we might think that we have a solution that begins with a Big Bang and ends up with Big Crunch (the Universe collapses) in a finite amount of cosmic time.  Yet this branch of the functions (\ref{duebraccia}) simply does not satisfy Friedman equations and such an embarrassing solution does not exist!  The two phase trajectory composing the solution and the fake  solution are displayed in fig.\ref{duebraccine}
The physical branch of this solution which is connected with the fixed point and has an asymptotically de Sitter behavior is displayed in fig.\ref{turoldo}
\subsection{\sc Numerical simulations}
In this section our goal  is to explore  the phase portrait and the behavior of the solutions of Friedman equations for a few different values of $\omega$ which corresponds to different fixed point types. We consider the following four cases:
\begin{equation}\label{perdinci}
    \omega \, = \, \underbrace{1}_{\mbox{Node}} \, , \, \underbrace{\frac{2}{\sqrt{3}}}_{\mbox{Node \& Integrable}} \, , \,
    \underbrace{\sqrt{3}}_{\mbox{focus \& integrable}} \, , \, \underbrace{3}_{\mbox{focus}}
\end{equation}
and we make a comparison of their behavior by solving numerically the Friedman equations with the same initial conditions in the four cases. We cannot choose exactly $a(0)\,=\,0$ since this corresponds to a singularity, so we just choose $a(0)$ quite small and $\phi(0)$ quite large. A precise way of choosing the initial conditions to be applied to all four cases can be provided  by the analytic solution determined by the integrable cases. This time we use the solution eq.(\ref{soluziapiu}) of the integral model $\omega \, = \, \sqrt{3}$  characterized by parameters:
\begin{equation}\label{gonzallo}
    \lambda \, = \, 1 \quad ; \quad \theta \, = \, \pi \quad ; \quad \xi \, = \, \frac{\pi}{6}
\end{equation}
we obtain:
\begin{equation}\label{gordolino}
    a(0) \, = \,  2^{-1/3} \quad ; \quad \phi(0) \, = \, \frac{\log
   \left(\frac{1+\sqrt{3}}{-1+\sqrt{3}}\right)}{\sqrt{3}} \quad ;\quad \dot{\phi}(0) \, = \, -2
\end{equation}
which will be the initial values for the integration programme in all four cases (\ref{perdinci}).
The result is provided in a series of figures.
\par
\begin{figure}[!hbt]
\begin{center}
\iffigs
 \includegraphics[height=50mm]{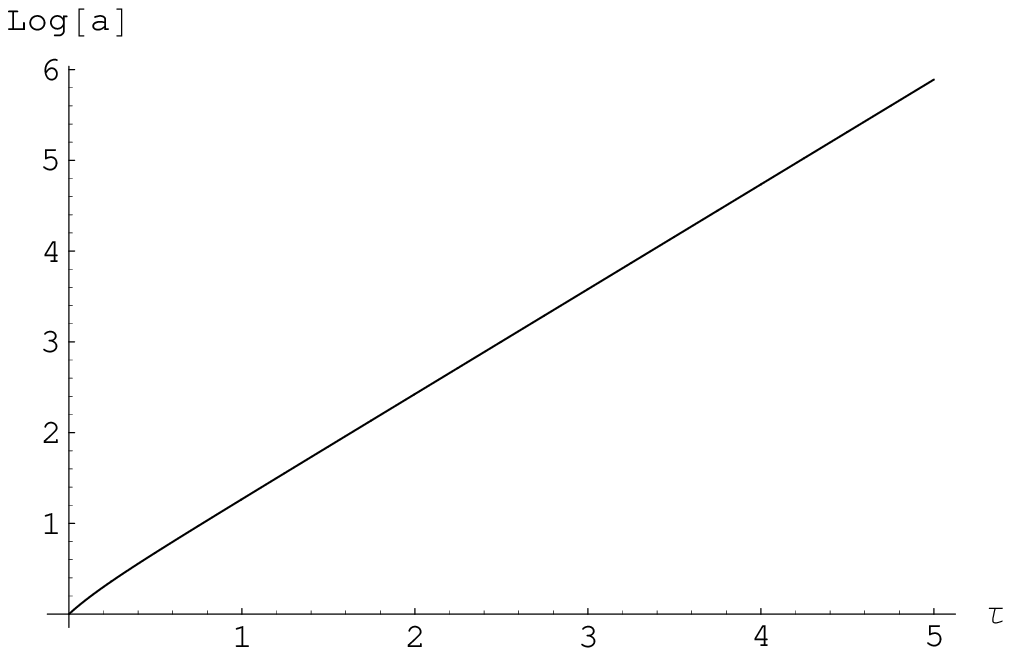}
 \vskip 2cm
 \includegraphics[height=50mm]{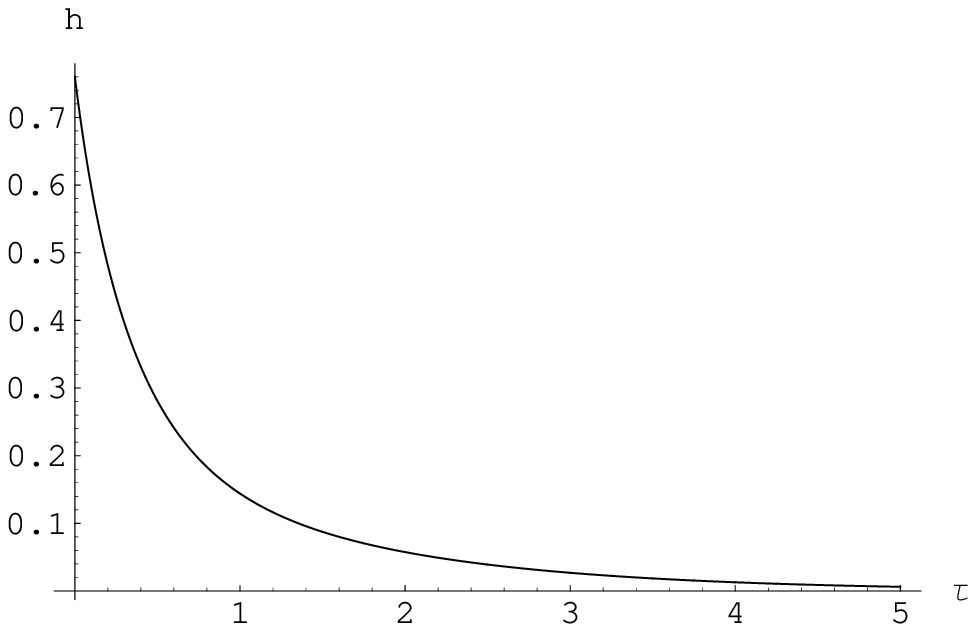}
\else
\end{center}
 \fi
\caption{\it
Setting the initial conditions eq.(\ref{gordolino}), in this figure we display the evolution of the scale factor $a(\tau)$ and of the scalar field $\phi(\tau)$ for the case $\omega = 1$, where the fixed point is of the node type. As we see we just have a descending scalar that goes smoothly to the fixed value with no oscillations.
}
\label{psiconano1a}
 \iffigs
 \hskip 1cm \unitlength=1.0mm
 \end{center}
  \fi
\end{figure}
\begin{figure}[!hbt]
\begin{center}
\iffigs
 \includegraphics[height=60mm]{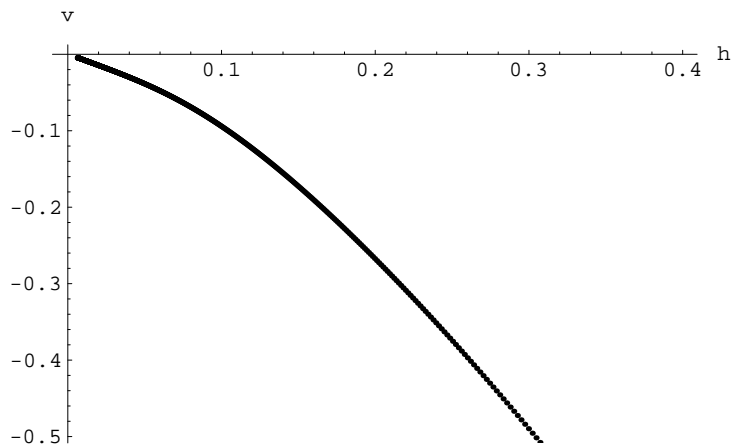}
\else
\end{center}
 \fi
\caption{\it
In this figure we display the trajectory determined by the  initial conditions of eq.(\ref{gordolino})  in the phase space $(\phi,\mathrm{v})$ for the case $\omega \, = \, 1$. This trajectory goes from infinity to the fixed point without winding around it (node) and following the direction fixed by the linearization matrix
}
\label{psiconano1b}
 \iffigs
 \hskip 1cm \unitlength=1.0mm
 \end{center}
  \fi
\end{figure}
\begin{figure}[!hbt]
\begin{center}
\iffigs
 \includegraphics[height=50mm]{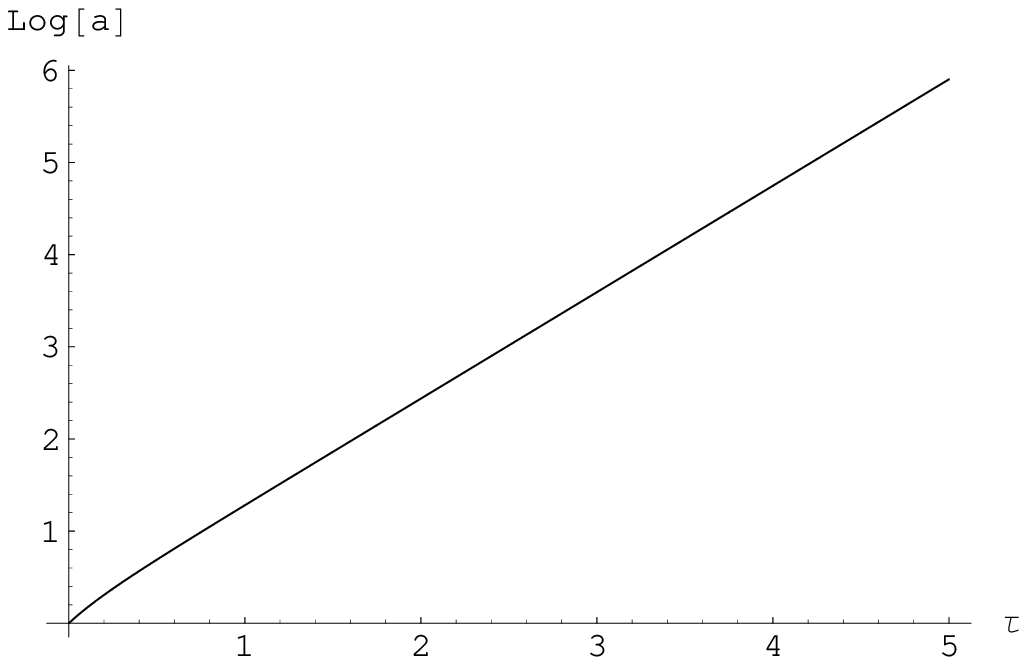}
 \vskip 2cm
 \includegraphics[height=50mm]{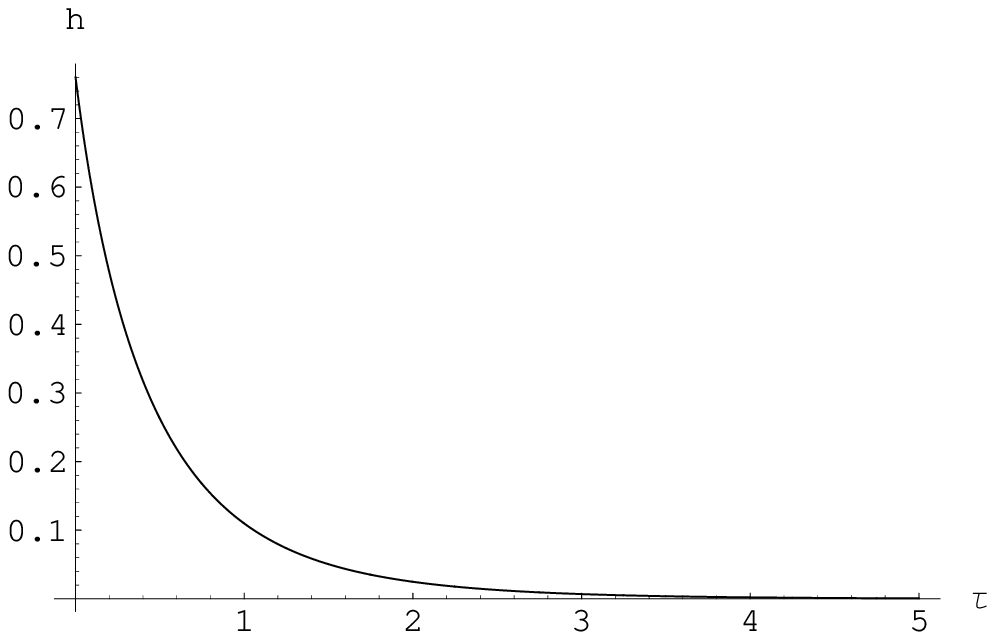}
\else
\end{center}
 \fi
\caption{\it
Setting the initial conditions eq.(\ref{gordolino}), in this figure we display the evolution of the scale factor $a(\tau)$ and of the scalar field $\phi(\tau)$ for the case $\omega = \frac{2}{\sqrt{3}}$, which is actually integrable. Its analytic behavior was extensively analysed before.
}
\label{psiconano2a}
 \iffigs
 \hskip 1cm \unitlength=1.0mm
 \end{center}
  \fi
\end{figure}
\begin{figure}[!hbt]
\begin{center}
\iffigs
 \includegraphics[height=60mm]{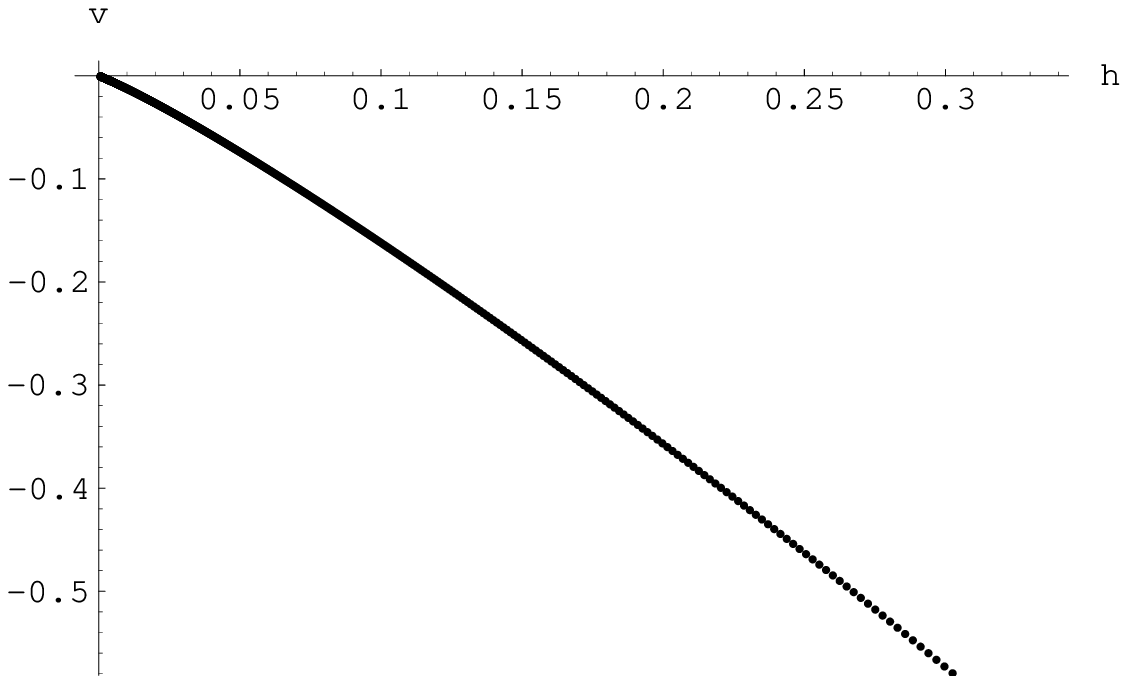}
\else
\end{center}
 \fi
\caption{\it
In this figure we display the trajectory determined by the  initial conditions of eq.(\ref{gordolino})  in the phase space $(\phi,\mathrm{v})$ for the case $\omega \, = \, \frac{2}{\sqrt{3}}$. This trajectory goes from infinity to the fixed point without winding around it: Node.
}
\label{psiconano2b}
 \iffigs
 \hskip 1cm \unitlength=1.0mm
 \end{center}
  \fi
\end{figure}
\begin{figure}[!hbt]
\begin{center}
\iffigs
 \includegraphics[height=50mm]{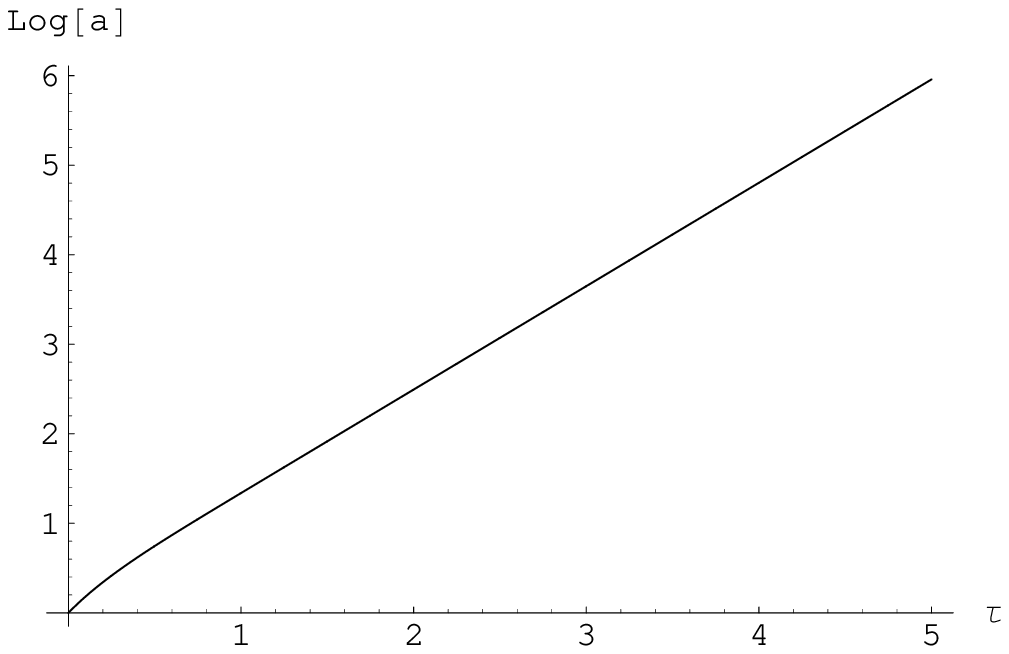}
 \vskip 2cm
 \includegraphics[height=50mm]{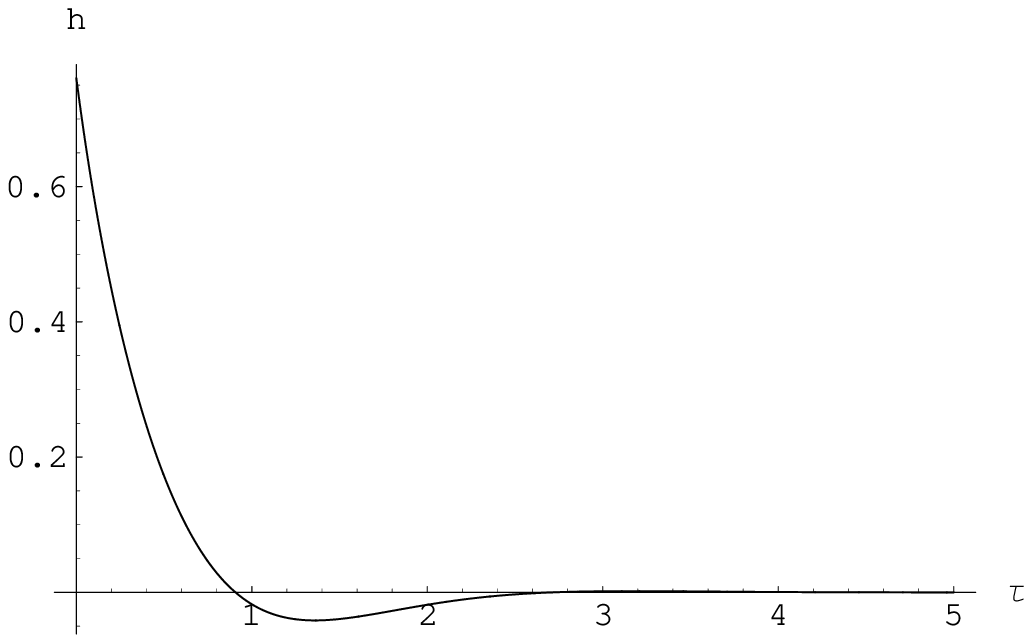}
\else
\end{center}
 \fi
\caption{\it
Setting the initial conditions eq.(\ref{gordolino}), in this figure we display the evolution of the scale factor $a(\tau)$ and of the scalar field $\phi(\tau)$ for the case $\omega = \sqrt{3}$, where the fixed point is of the focus type. This is an integrable case for which we posses the analytical solutions As we see we have a descending scalar that goes to a negative valued minimum and then climbs again  to its fixed point value  at zero.
}
\label{psiconano3a}
 \iffigs
 \hskip 1cm \unitlength=1.0mm
 \end{center}
  \fi
\end{figure}
\begin{figure}[!hbt]
\begin{center}
\iffigs
 \includegraphics[height=60mm]{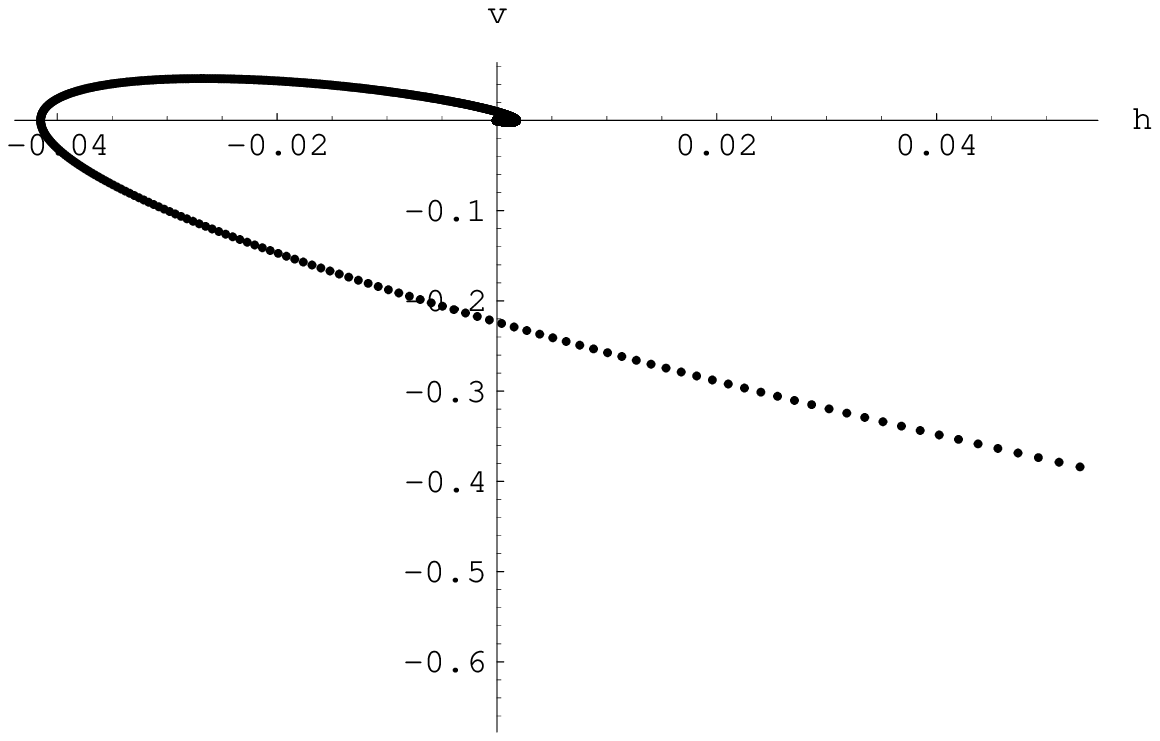}
\else
\end{center}
 \fi
\caption{\it
In this figure we display the trajectory determined by the  initial conditions of eq.(\ref{gordolino})  in the phase space $( \phi,\mathrm{v})$ for the case $\omega \, = \, \sqrt{3}$. This trajectory goes from infinity to its fixed point  winding a little bit around it (focus).
}
\label{psiconano3b}
 \iffigs
 \hskip 1cm \unitlength=1.0mm
 \end{center}
  \fi
\end{figure}
\begin{figure}[!hbt]
\begin{center}
\iffigs
 \includegraphics[height=50mm]{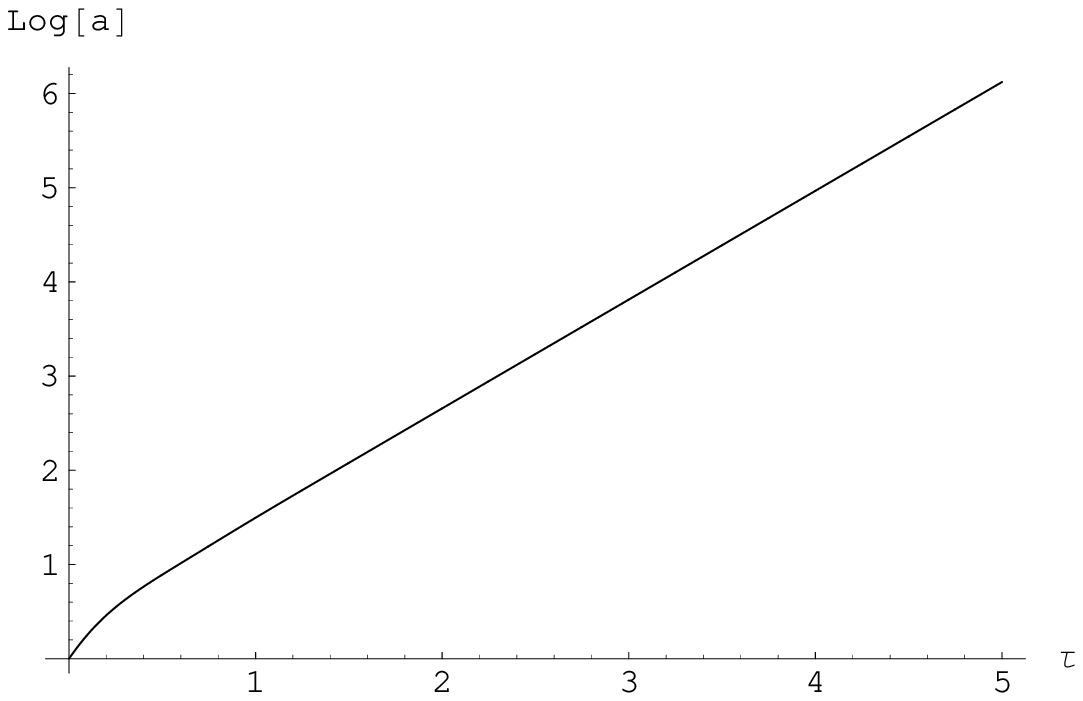}
 \vskip 2cm
 \includegraphics[height=50mm]{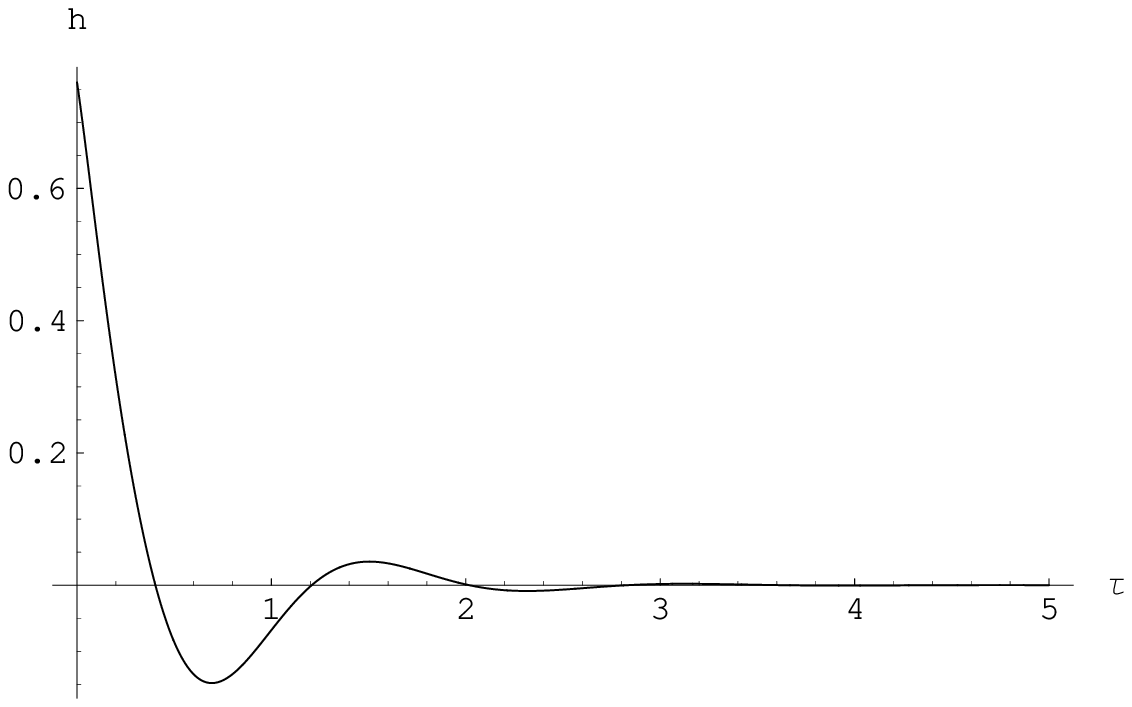}
\else
\end{center}
 \fi
\caption{\it
Setting the initial conditions eq.(\ref{gordolino}), in this figure we display the evolution of the scale factor $a(\tau)$ and of the scalar field $\phi(\tau)$ for the case $\omega = 3$, where the fixed point is of the focus type. As we see we have a descending scalar that goes to a negative valued passing through various oscillations.
}
\label{psiconano4a}
 \iffigs
 \hskip 1cm \unitlength=1.0mm
 \end{center}
  \fi
\end{figure}
\begin{figure}[!hbt]
\begin{center}
\iffigs
 \includegraphics[height=60mm]{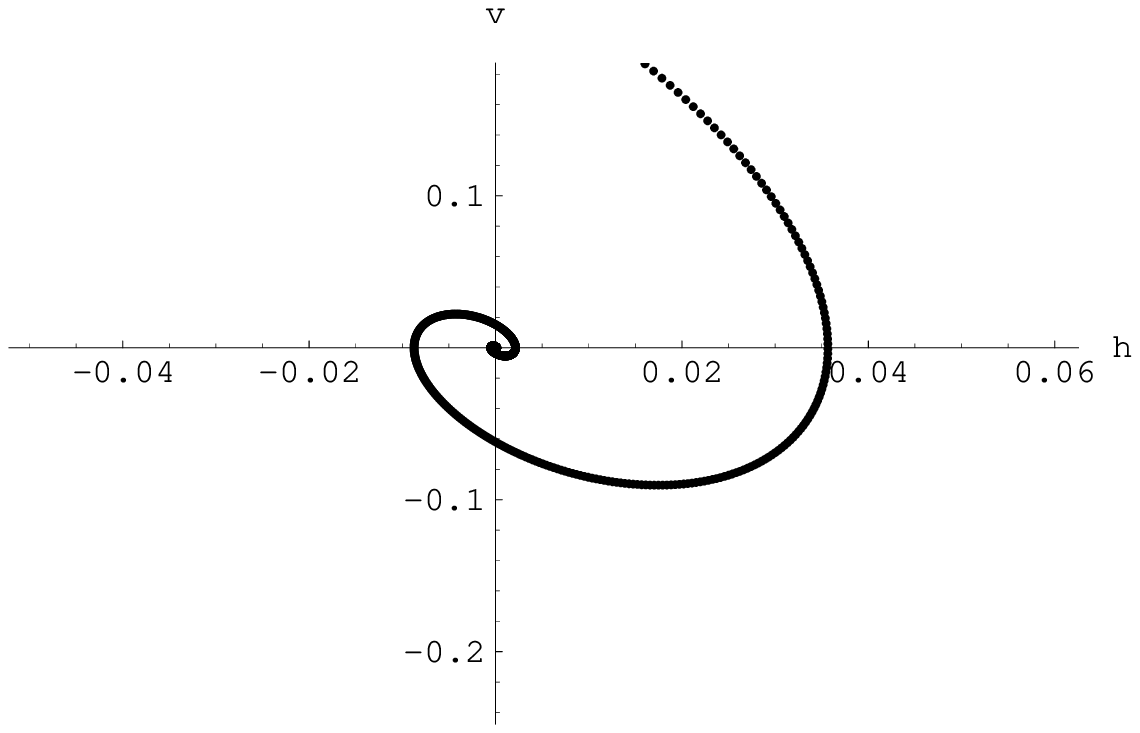}
\else
\end{center}
 \fi
\caption{\it
In this figure we display the trajectory determined by the  initial conditions of eq.(\ref{gordolino})  in the phase space $(h=\phi,\mathrm{v})$ for the case $\omega \, = \, 3$. This trajectory goes from infinity to its fixed point  winding few times  around it (focus).
}
\label{psiconano4b}
 \iffigs
 \hskip 1cm \unitlength=1.0mm
 \end{center}
  \fi
\end{figure}

We begin with the node case $\omega\, = \, 1$ whose behavior is displayed in fig.s \ref{psiconano1a} and \ref{psiconano1b}
.
We continue  with the integrable case $\omega\, = \, \frac{2}{\sqrt{3}}$ whose behavior is displayed in fig.s \ref{psiconano2a} and \ref{psiconano2b}.
Next  we consider the integrable case $\omega\, = \, \sqrt{3}$ whose behavior, already of the focus type, is displayed in fig.s \ref{psiconano3a} and \ref{psiconano3b}.
Finally  we consider the  case $\omega\, = \, 3$ which is not integrable, yet shares with the integrable case s$\omega= \sqrt{3}$ the quality of its fixed point, namely  the focus type. The behavior is displayed in fig.(\ref{psiconano4a})
and in fig.\ref{psiconano4b}.
The conclusion of this comparison is that what we can learn from the integrable models is the behavior of solutions that share the same type of fixed point.  Also  the asymptotic behavior for very late times of both the scalar field and of the scale factor is captured by the integrable case and is shared by all other members of the family.  The structure of the scalar field behavior at finite times is rather strongly dependent from the value of $\omega$. For sufficiently large $\omega$ we get the focus case and the scalar oscillates around the fixed point.    The larger is $\omega$ the more the trajectory winds around the fixed point. This winding corresponds to oscillations of the scalar field which in cosmology are potentially interesting since they might be at the heart of the reheating mechanism after inflation.
\par
\begin{figure}[!hbt]
\begin{center}
\iffigs
 \includegraphics[height=50mm]{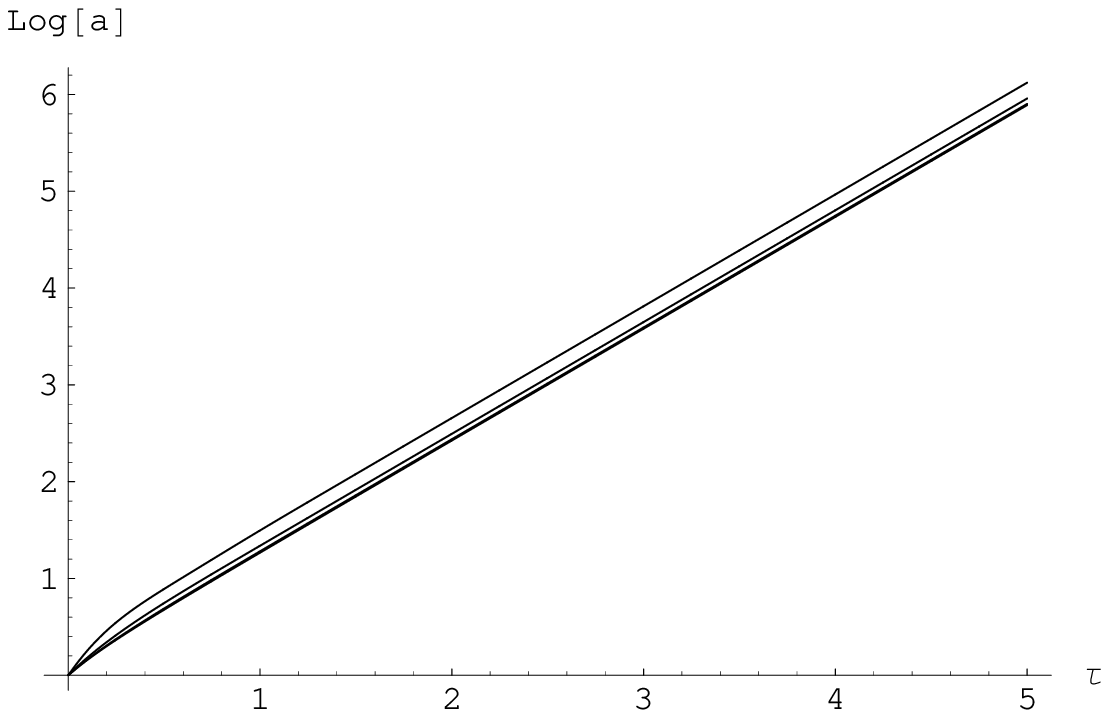}
 \vskip 2cm
 \includegraphics[height=50mm]{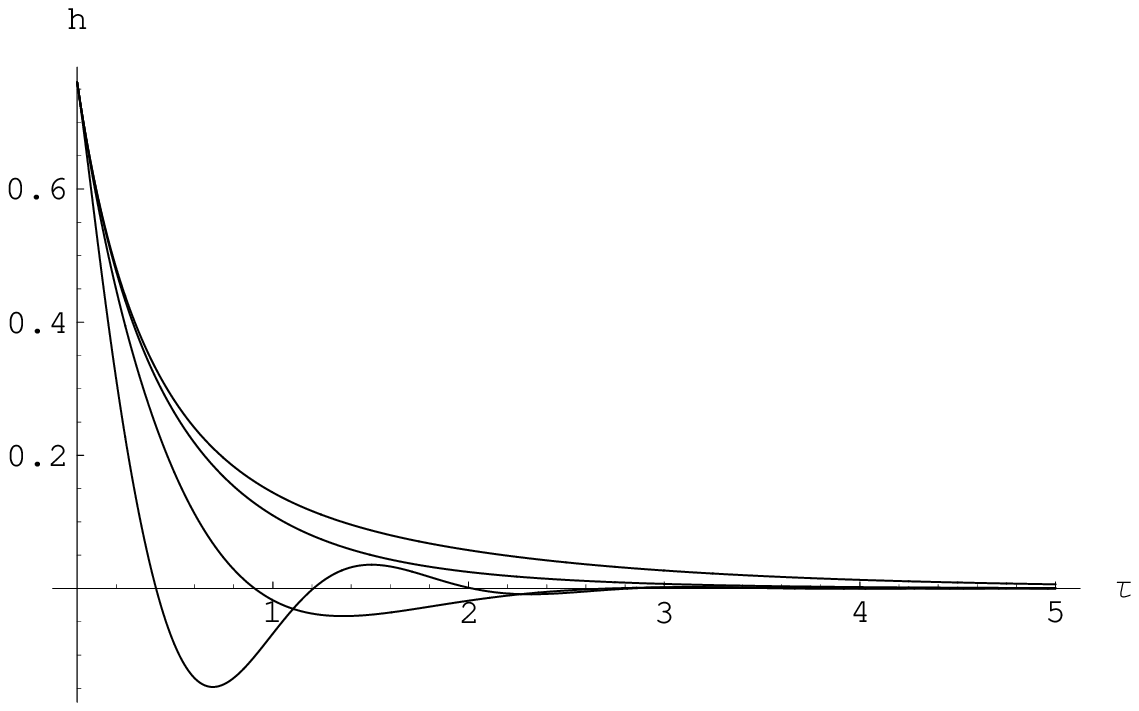}
\else
\end{center}
 \fi
\caption{\it
In this figure we display in the same plot the behavior of the scale factor and of the scalar field starting with the same initial condition but following the equations for the different values  of $\omega$ considered in eq.(\ref{perdinci})
}
\label{buonanno}
 \iffigs
 \hskip 1cm \unitlength=1.0mm
 \end{center}
  \fi
\end{figure}

The structure of possible trajectories  is summarized in fig.\ref{buonanno} that plots together the behavior of the scalar field for the various considered values of $\omega$ and so does for the scale factors
\section{\sc A brief scan of $\mathcal{N}=1$ superpotentials from Flux Compactifications}
\label{fluxscan}
A lot of progress has been made since the year 2000 in the context of flux compactifications of string theory with the aim of obtaining four-dimensional effective theories with phenomenologically desirable features, among which, after the discovery of the Universe  late-time acceleration,  with high relevance, ranks the need to find de Sitter vacua.
\par
A vast literature in this field deals with the search of \textit{flux backgrounds} that are compatible with minimal $\mathcal{N}=1$ supersymmetry in $D=4$ and relies on the mechanism, firstly discovered in \cite{gkp} of inducing an effective $\mathcal{N}=1$ superpotential from fluxes. This token has been extensively utilized in those compactifications that give rise to an $\mathrm{STU}$-model as low energy description \cite{Derendinger:2005ph,Derendinger:2004kf,Denef:2007pq,Kachru:2002he,Kachru:2004jr}.
The final outcome of these procedures, which take their motivation in orbifold and orientifold compactifications \cite{orientifolds}, is just an explicit expression for the superpotential $W(S,T,U)$, which can be used to calculate the scalar potential, whose extrema and consistent one-field truncations  can be studied in a systematic way. The marvelous thing of this description is that each coefficient in the development of $W(S,T,U)$ in the constituent fields, $S,T,U$ has a direct interpretation in terms of fluxes and, in an appropriate basis, it admits only quantized integer values.
\par
Recently a series of papers has appeared \cite{Dibitetto:2012xd,Dibitetto:2011gm,Danielsson:2012by} aiming at a systematic charting of the landscape of these superpotentials and of the extrema of their corresponding scalar potentials.
\par
In the present section we briefly consider such a landscape with the aim to single out consistent one-dilaton truncations and work out the corresponding potential $\mathcal{V}(\varphi)$, to be compared with our list of integrable ones.  As a result we obtain several examples of one field potentials, all falling into the general family of linear combinations of exponentials that we consider, yet very seldom   satisfying the severe relations on the exponents and the coefficients  that are required for integrability. Although in this run we identified only one  new integrable model,  the lesson that we  learn from it is that, by considering truncations of multi-field supergravities, the variety of possible outcomes is significantly enlarged. Indeed  what matters are the intrinsic indices $\omega_i \, = \, \, p_i/\sqrt{3\,q}$, the numbers $p_i$ being the exponent coefficients of the  field $\mathfrak{h}$ that is kept in the truncation, while $q$ is the coefficient of the kinetic term of the latter. In view of this, when $\mathfrak{h}$ happens to be a linear combination of several other dilatons, the coefficients of the linear combination play a role, both in generating a variety of $p_i$.s and in giving rise to non standard $q$.s, the final result being difficult to be predicted a priori.
On the other hand, the coefficients of the appropriate linear combination that can constitute a consistent truncation, are
searched for by diagonalizing the mass matrix of the theory in the vicinity of an extremum. Indeed the mass eigenstates around an exact vacuum of the theory constitute a natural basis where some fields can be put consistently to zero with the exception of one which survives.
\par
We plan to use the landscape chartered by Dibitetto et al as a mean to illustrate the above ideas.
\subsection{\sc The $\mathrm{STU}$ playing ground}
Originally, in orbifold compactifications on $\mathrm{T}^6/\mathbb{Z}_2\times \mathbb{Z}_2$, one arrives at seven complex moduli fields:
\begin{equation}\label{dottorbalordo}
    S, \, T_1, \,T_2, \,T_3, \, U_1, \,U_2, \,U_3
\end{equation}
the first being related to the original dilaton and Kalb-Ramond field of $10$-dimensional supergravity, the remaining six being the appropriate complexifications of the six radii of $\mathrm{T}^6$. Embedding these fields into $\mathcal{N}=4$ supergravity, which is a necessary intermediate step when supersymmetry is halved by the orbifold projection, Dibitetto et al. \cite{Dibitetto:2011gm} have been able to reduce the playing ground for the search of flux induced superpotentials to three fields:
\begin{equation}\label{STUmolti}
    S \quad ; \quad T \, = \, T_1 =T_2=T_3 \quad ; \quad U \, = \, U_1 =U_2=U_3
\end{equation}
This is done by consistently truncating $\mathcal{N}=4$ supergravity to the singlets with respect to an appropriately chosen global $\mathrm{SO(3)}$ symmetry group. This truncation breaks supersymmetry to $\mathcal{N}=1$. As a result of the identifications in eq.(\ref{STUmolti}) the K\"ahler potential for the residual fields takes the following form:
\begin{equation}\label{carlettopoto}
   \mathcal{K} \, = \, - \, \log[-i\,(S-\bar{S})]\, - \, \log[i\,(T-\bar{T})^3]\, - \, \log[ i\,(U-\bar{U})^3]
\end{equation}
Next the authors of \cite{Dibitetto:2011gm} have identified a list of $32$ monomials out of which the superpotential can be constructed as a linear combination. Assigning a real coefficient $\lambda_{1,\dots,16}$ to the even powers and an imaginary coefficient ${\rm i} \mu_{1,\dots,16}$ to the odd ones, one guarantees a priori that the truncation to zero axions for all the three fields is consistent. With this proviso, the most general $\mathcal{N}=1$ superpotential considered by Dibitetto et al is the following one:
\begin{eqnarray}
  W_{gen} &=& \lambda _1+U^2 \lambda _2+S U \lambda
   _3+S U^3 \lambda _4+T U \lambda _5+T
   U^3 \lambda _6+S T \lambda _7+S T
   U^2 \lambda _8 \nonumber\\
   && +T^3 U^3 \lambda
   _9+T^3 U \lambda _{10}+S T^3 U^2
   \lambda _{11}+S T^3 \lambda
   _{12}\nonumber\\
   && +T^2 U^2 \lambda _{13}+T^2
   \lambda _{14}+S T^2 U^3 \lambda
   _{15}+S T^2 U \lambda _{16}\nonumber\\
   && +{\rm i} U \mu
   _1+{\rm i} U^3 \mu _2+{\rm i} S \mu _3+{\rm i} S U^2
   \mu _4+i T \mu _5+i T U^2 \mu _6+{\rm i} S
   T U \mu _7+{\rm i} S T U^3 \mu _8\nonumber\\
   && +{\rm i} T^3
   U^2 \mu _9+{\rm i} T^3 \mu _{10}+{\rm i} S T^3
   U^3 \mu _{11}+{\rm i} S T^3 U \mu _{12}\nonumber\\
   &&+{\rm i}
   T^2 U^3 \mu _{13}+{\rm i} T^2 U \mu
   _{14}+{\rm i} S T^2 U^2 \mu _{15}+{\rm i} S T^2
   \mu _{16}
\end{eqnarray}
The interesting point is that each of the $\lambda$.s and each of the $\mu$.s has a precise interpretation in terms of $10$-dimensional fluxes of various type.
\par
A completely general approach to the study of vacua and possible one-field truncations would consist of the following precise algorithm:
\begin{center}
\textbf{Truncation Charting Algorithm from Flux Superpotentials (TCAFS)}
\end{center}
 \begin{enumerate}
   \item Calculate from $W_{gen}$ the scalar potential $V_{gen}\left(\lambda,\mu,h_{1,2,3},b_{1,2,3}\right)$ depending on three dilatons, three axions and 32 real parameters $\{\lambda,\mu\}$.
   \item Consistently truncate this potential to zero axions $\hat{V}_{gen}\left(\lambda,\mu,h_{1,2,3}\right)=V_{gen}\left(\lambda,\mu,h_{1,2,3},0\right)$.
   \item Calculate the three derivatives of the potential with respect to the remaining dilatons $\partial_{h_i} \hat{V}$  and impose that they are zero at $h_{1,2,3}\, = \,0$:
       \begin{equation}\label{estremadura}
        \partial_{h_i} \hat{V}\left(\lambda,\mu,h_{1,2,3}\right)|_{h_{1,2,3}=0} \, = \, 0
       \end{equation}
     These conditions impose that the base point of the manifold $S=T=U={\rm i}$ should be an extremum of the potential. This choice corresponds to no loss of generality, since the dilatons are defined up to a translation and any extremum can be mapped into the reference point  $S=T=U={\rm i}$ at the prize of  rescaling some of the coefficients $\{\lambda,\mu\}$. Hence, as long as we keep $\{\lambda,\mu\}$ general we do not loose anything by deciding a priori where the extremum should be located. In this way eq.s(\ref{estremadura}) become a set of three algebraic equations of higher order for the coefficients $\{\lambda,\mu\}$.
   \item Solve, if possible, the algebraic equations (\ref{estremadura}). In principle this results into a set of $m$ solutions:
       \begin{equation}\label{solutini}
        \lambda_i \, = \, \lambda_i^{(\alpha)} \quad ; \quad \mu_i \, = \, \mu_i^{(\alpha)} \quad ; \quad \alpha\, = \, 1,\dots,m
       \end{equation}
   \item Replace one by one the solutions (\ref{solutini}) into $\hat{V}_{gen}\left(\lambda,\mu,h_{1,2,3}\right)$, obtaining $m$ potentials of the three dilaton fields:
       \begin{equation}\label{pistacchi}
        V^{(\alpha)}(h_1,h_2,h_3) \, \equiv \, \hat{V}_{gen}\left(\lambda^{(\alpha)},\mu^{(\alpha)},h_{1,2,3}\right) \quad ; \quad \alpha\, = \, 1,\dots,m
       \end{equation}
       which, by construction, have an extremum in $h_{1,2,3} \, = \,0$. Verify whether each extremum corresponds to Minkowski ($V^{(\alpha)}(0) \, = \, 0$), de Sitter ($V^{(\alpha)}(0) \, > \, 0$) or anti de Sitter ($V^{(\alpha)}(0) \, < \, 0$) space.
   \item For each potential $V^{(\alpha)}(h_1,h_2,h_3)$ calculate the mass-matrix in the extremum:
   \begin{equation}\label{massamatrata}
    M_{ij}^{(\alpha)} \, = \, \partial_i\partial_j V^{(\alpha)}(h_1,h_2,h_3)|_{h_{1,2,3}=0} \quad ; \quad \alpha\, = \, 1,\dots,m
   \end{equation}
   and the corresponding eigenvalues $\Lambda_{I}^{(\alpha)}$ ($I=1,2,3$) and eigenvectors $\vec{v}_I^{(\alpha)}$.
   \item From the eigenvalues $\Lambda_{I}^{(\alpha)}$ ($I=1,2,3$) we learn about stability or instability of the corresponding vacuum. By means of the corresponding eigenvectors introduce a new basis of three fields well-adapted to the potential $V^{(\alpha)}$:
       \begin{equation}\label{gomoide}
        \phi_I^{(\alpha)} \, \equiv \, \vec{v}_I^{(\alpha)} \cdot \vec{h} \quad ; \quad \alpha\, = \, 1,\dots,m \quad I\, =\, 1,2,3
       \end{equation}
    \item Transform the potential and the kinetic term to the new well adapted basis and inspect if truncation to any of the $\phi_{1,2,3}$, by setting to zero the other two is consistent.
    \item In case of positive answer to the previous question calculate  the effective coefficient $q$ in the kinetic term and by means of the transformation (\ref{babushka}) produce a potential $\mathcal{V}(\varphi)$ to be compared with the list of integrable ones.
 \end{enumerate}
 The problem with the above algorithm is simply computational. Using all of the 32 terms in the superpotential and truncating to zero axions we are left with a three-dilaton potential that contains 480 terms and the three algebraic equations  (\ref{estremadura}) in  32 unknowns are too large to be solved by standard codes in MATHEMATICA. Some strategy to reduce the parameter space has to be found. What we were able to do with ease was just to test the TCAFS on some reduced space suggested by the special superpotentials reviewed in \cite{Dibitetto:2011gm}. We did not assume the coefficients presented in that paper, we simply restricted the parameter space to that spanned by the monomials included in each of these superpotentials and by running the TACFS algorithm on such parameter space we retrieved exactly the same results presented in  \cite{Dibitetto:2011gm}.  For all these extrema we have also calculated the mass-matrix and we have found some consistent one field truncations for which we could determine the corresponding one-field potential. As anticipated they all fall in the family of linear combination of exponentials and although none coincides with an integrable one we start seeing new powers and new structures that in the one-field constructions were absent.
 \par
 Finally with some ingenuity we were able to derive new interesting instances of superpotentials that lead to interesting one-field truncations. In one case we obtained a new instance of a supersymmetric integrable cosmological model.
 \subsection{\sc Locally Geometric Flux induced superpotentials}
 In \cite{Dibitetto:2011gm}, the authors consider a particular superpotential that is denominated \textit{locally geometric} since its origin is claimed to arise from a combination of geometric type IIB fluxes with non geometric ones, the resulting composition still admitting a \textit{locally geometric} description. Leaving aside the discussion of its ten dimensional origin in type IIB or type IIA compactifications the above mentioned superpotential has the following form:
 \begin{equation}\label{partus}
  W_{locgeo} \, = \,  \lambda _1+S U^3 \lambda _4+T U \lambda _5+S T U^2 \lambda _8
 \end{equation}
and corresponds to a truncation of the $32$ dimensional parameter space to a four-dimensional one spanned by $\{\lambda_1,\lambda_4 ,
\lambda_5,\lambda_8\}$. Using the standard parametrization of the fields:
\begin{equation}\label{fischiuto}
   S \, = \, {\rm i} \, \exp[h_1] + b_1 \quad ; \quad T \, = \, {\rm i} \, \exp[h_2] + b_2 \quad ; \quad  U \, = \, {\rm i} \, \exp[h_3] + b_3
\end{equation}
and implementing the first two steps of the TCAFS we obtain the following potential:
\begin{eqnarray}\label{Vlocgeo}
    V & = &\frac{1}{64} e^{-h_1-3 h_2-3 h_3}
   \lambda _1^2-\frac{1}{32} e^{-3 h_2}
   \lambda _1 \lambda _4+\frac{1}{64}
   e^{h_1-3 h_2+3 h_3} \lambda
   _4^2+\frac{1}{32} e^{-2 h_2+h_3}
   \lambda _4 \lambda _5 \nonumber\\
   &&-\frac{1}{192}
   e^{-h_1-h_2-h_3} \lambda
   _5^2-\frac{1}{32} e^{-2 h_2-h_3}
   \lambda _1 \lambda _8+\frac{1}{32}
   e^{-h_2} \lambda _5 \lambda
   _8-\frac{1}{192} e^{h_1-h_2+h_3}
   \lambda _8^2
\end{eqnarray}
Implementing next the steps 3 and 4 of the TCAFS we obtain five non vanishing solutions for the $\lambda$-coefficients that can be displayed by writing the corresponding superpotentials:
\begin{eqnarray}
  W_{(2)}^{locgeo} &=& 1+3\, T U+3\, S T U^2-S U^3 \label{theo2}\\
   W_{(3a)}^{locgeo}&=& 5-9 \,T U+3\, S T U^2-S U^3\label{theo3a}\\
   W_{(Mink)}^{locgeo} &=& 1+S U^3 \label{theoMink}\\
   W_{(1)}^{locgeo} &=&-1-3 \,T U+3\, S T U^2-S U^3\label{theo1}\\
   W_{(3b)}^{locgeo} &=&-\frac{1}{3}-T U+3\, S T
   U^2+\frac{5 \,S U^3}{3}\label{theo3b}
\end{eqnarray}
The names given to these solutions are taken from the nomenclature utilized in table 5 of \cite{Dibitetto:2011gm} since the superpotentials we found exactly correspond to those  considered there, up to a multiplicative overall constant in the last case.
The values in the extremum  of the corresponding scalar potentials are:
\begin{equation}\label{governolo}
    \left\{\underbrace{V_{(2)}^{locgeo}(\vec{0})}_{\mathrm{dS}}, \, \underbrace{V_{(3a)}^{locgeo}(\vec{0})}_{\mathrm{AdS}}, \, \underbrace{V_{(Mink)}^{locgeo}(\vec{0})}_{\mathrm{Mink}}, \, \underbrace{V_{(1)}^{locgeo}(\vec{0})}_{\mathrm{AdS}}, \, \underbrace{V_{(3b)}^{locgeo}(\vec{0}) }_{\mathrm{AdS}}\right\} \, = \, \left\{\frac{1}{16},-\frac{15}
   {16},0,-\frac{3}{16},-\frac{5}{48}\right\}
\end{equation}
Hence we conclude that we have one  de Sitter vacuum, one Minkowski vacuum and three anti de Sitter vacua. This is just the same result found by the authors of \cite{Dibitetto:2011gm}.  The corresponding dilaton potentials that give rise to such vacua at their extremum have the following explicit form:
\begin{eqnarray}
  V_{(2)}^{locgeo}(\vec{h})&=&-\left( -\frac{1}{32} e^{-3
   h_2}-\frac{9
   e^{-h_2}}{32}-\frac{1}{64}
   e^{-h_1-3 h_2-3
   h_3}+\frac{3}{32} e^{-2
   h_2-h_3}+\frac{3}{64}
   e^{-h_1-h_2-h_3}\right)\nonumber\\
   &&\left. +\frac{3}{3
   2} e^{-2
   h_2+h_3}+\frac{3}{64}
   e^{h_1-h_2+h_3}-\frac{1}{64
   } e^{h_1-3 h_2+3 h_3}\right)\label{theo2V} \\
  V_{(3a)}^{locgeo}(\vec{h}) &=& -\left(-\frac{5}{32} e^{-3
   h_2}+\frac{27
   e^{-h_2}}{32}-\frac{25}{64}
   e^{-h_1-3 h_2-3
   h_3}+\frac{15}{32} e^{-2
   h_2-h_3}\right.\nonumber\\
   &&\left.+\frac{27}{64}
   e^{-h_1-h_2-h_3}-\frac{9}{3
   2} e^{h_3-2
   h_2}+\frac{3}{64}
   e^{h_1-h_2+h_3}-\frac{1}{64
   } e^{h_1-3 h_2+3 h_3}\right) \label{theo3aV}\\
V_{(Mink)}^{locgeo}(\vec{h}) &=&- \left(\frac{1}{32} e^{-3
   h_2}-\frac{1}{64} e^{-h_1-3
   h_2-3 h_3}-\frac{1}{64}
   e^{h_1-3 h_2+3 h_3} \right)\\
  V_{(1)}^{locgeo}(\vec{h})  &=&-\left( \frac{1}{32} e^{-3
   h_2}+\frac{9
   e^{-h_2}}{32}-\frac{1}{64}
   e^{-h_1-3 h_2-3
   h_3}-\frac{3}{32} e^{-2
   h_2-h_3}\right)\nonumber\\
   &&\left.+\frac{3}{64}
   e^{-h_1-h_2-h_3}-\frac{3}{3
   2} e^{h_3-2
   h_2}+\frac{3}{64}
   e^{h_1-h_2+h_3}-\frac{1}{64
   } e^{h_1-3 h_2+3 h_3}\right)\label{theo1V}\\
  V_{(3b)}^{locgeo}(\vec{h}) &=&-\left( -\frac{5}{288} e^{-3
   h_2}+\frac{3
   e^{-h_2}}{32}-\frac{1}{576}
   e^{-h_1-3 h_2-3
   h_3}-\frac{1}{32} e^{-2
   h_2-h_3}\right.\nonumber\\
   &&\left. +\frac{1}{192}
   e^{-h_1-h_2-h_3}+\frac{5}{9
   6} e^{h_3-2
   h_2}+\frac{3}{64}
   e^{h_1-h_2+h_3}-\frac{25}{5
   76} e^{h_1-3 h_2+3 h_3}\right)\label{theo3bV}
\end{eqnarray}
We continue the development of the TCAFS algorithm, case by case.
\subsubsection{\sc The $\mathrm{dS}$ potential}
Calculating the mass matrix in the extremum of the potential $ V_{(2)}^{locgeo}(\vec{h})$ we obtain:
\begin{equation}\label{massmat1}
    M_{mass} \, = \, \left(
\begin{array}{lll}
 - \frac{1}{16} & 0 & 0 \\
 0 & 0 & 0 \\
 0 & 0 & 0
\end{array}
\right)
\end{equation}
Hence we have one negative and two null eigenvalues which means that this $dS$ vacuum is unstable. Since the mass matrix is diagonal the charge eigenstates, namely the fields coincide with the mass eigenstates and we can explore if there are consistent truncations. By direct evalutation of the derivatives we find that there are two consistent one-field truncations:
\begin{description}
  \item[A-truncation.]  $h_1 \, = \, h_2 \, = \, 0$. In this case the residual potential is:
  $$V\, = \, \frac{1}{16} e^{-3 h_3} \left(1-3 e^{h_3}+3 e^{2 h_3}\right)$$
  Since the kinetic term of $h_3$ has a factor $q=3$, by means of the substitution (\ref{babushka}), we obtain:
   \begin{equation}\label{potentus1}
    \mathcal{V}(\varphi) \, = \, \frac{e^{-\varphi}}{16}-\frac{3}{16} e^{-2\varphi /3}+\frac{3 e^{-\varphi /3}}{16}
   \end{equation}
   which is not any of the integrable potentials but belongs to the same class.
     \item[B-truncation.] $h_2 \, = \, h_3 \, = \, 0$. In this case the residual potential is:
     $$V\, = \,- \, \frac{1}{32}\left(-4+e^{-h_1}+e^{h_1}\right)$$
     Since the kinetic term of the $h_1$ field has a factor $q=1$, by means of the substitution (\ref{babushka}), we obtain:
     \begin{equation}\label{potentus2}
     \mathcal{V}(\varphi) \, = \,   \frac{1}{16}\left(2-\text{Cosh}\left[\frac{\varphi}{\sqrt{3}}\right]\right)
     \end{equation}
     which also belongs to the class of exponential potentials here considered but does not fit into any integrable series or sporadic case.
\end{description}
\subsubsection{\sc The $\mathrm{AdS}$ potential 3a}
Calculating the mass matrix in the extremum of the potential $ V_{(3a)}^{locgeo}(\vec{h})$ we obtain:
\begin{equation}\label{massmat2}
    M_{mass} \, = \,  - \, \left(
\begin{array}{lll}
 \frac{1}{16} & -\frac{3}{4} &
   -\frac{3}{4} \\
 -\frac{3}{4} & -3 &
   -\frac{3}{2} \\
 -\frac{3}{4} & -\frac{3}{2} &
   -3
\end{array}
\right)
\end{equation}
The eigenvalues of this mass-matrix are:
\begin{equation}\label{eigevallla2}
\mathrm{Eigenval} \, = \,     \left\{- \underbrace{\frac{1}{32} \left(-71-\sqrt{6481}\right )}_{> \, 0},\underbrace{\frac{3}{2}}_{> \, 0},\underbrace{- \frac{1}{32}\left(-71+\sqrt{6481}\right)}_{< \, 0}\right\}
\end{equation}
showing that this anti de Sitter vacuum is stable since all the eigenvalues satisfy the Breitenlohner-Freedman bound $\lambda_i \, > \, -\, \frac{45}{48}$ . The corresponding mass eigenstates are the following fields
\begin{equation}\label{basato}
    \left \{\phi_1,\, \phi_2,\, \phi_3 \right \} \, = \, \left\{\frac{1}{24}
   \left(-73+\sqrt{6481}\right
   )
   h_1+h_2+h_3,-h_2+h_3,-\frac
   {1}{24}
   \left(73+\sqrt{6481}\right)
   h_1+h_2+h_3\right\}
\end{equation}
Calculating the derivatives we verify that there is no consistent truncation of this potential.
\subsubsection{\sc The $\mathrm{AdS}$ potential 3b}
Calculating the mass matrix in the extremum of the potential $ V_{(3b)}^{locgeo}(\vec{h})$ we obtain:
\begin{equation}\label{massmat3}
    M_{mass} \, = \, - \,  \left(
\begin{array}{lll}
 \frac{1}{144} & \frac{1}{12}
   & -\frac{1}{12} \\
 \frac{1}{12} & -\frac{1}{3} &
   \frac{1}{6} \\
 -\frac{1}{12} & \frac{1}{6} &
   -\frac{1}{3}
\end{array}
\right)
\end{equation}
The eigenvalues of this mass-matrix are:
\begin{equation}\label{eigevallla3}
\mathrm{Eigenval} \, = \,     \left\{\underbrace{\frac{1}{288} \left(71+\sqrt{6481}\right )}_{> \, 0},\underbrace{\frac{1}{6}}_{> \, 0},\underbrace{\frac{1}{288}\left(71-\sqrt{6481}\right)}_{< \, 0}\right\}
\end{equation}
showing that also also this anti de Sitter vacuum is stable since all eigenvalues satisfy the Breitenlohner Freedman bound $\lambda_i \, > \, - \, \frac{5}{64}$. The corresponding mass eigenstates are the following fields
\begin{equation}\label{basatobis}
    \left \{\phi_1,\, \phi_2,\, \phi_3 \right \} \, = \, \left\{\frac{1}{24}
   \left(-73+\sqrt{6481}\right
   )
   h_1-h_2+h_3,h_2+h_3,-\frac{
   1}{24}
   \left(73+\sqrt{6481}\right)
   h_1-h_2+h_3\right\}
\end{equation}
Calculating the derivatives we verify that there is no consistent truncation of this potential.
\subsubsection{\sc The $\mathrm{AdS}$ potential 1}
Calculating the mass matrix in the extremum of the potential $ V_{(1)}^{locgeo}(\vec{h})$ we obtain:
\begin{equation}\label{massmat4}
    M_{mass} \, = \, - \,  \left(
\begin{array}{lll}
 \frac{1}{16} & 0 & 0 \\
 0 & -\frac{3}{8} & 0 \\
 0 & 0 & -\frac{3}{8}
\end{array}
\right)
\end{equation}
which is diagonal and the eigenvalues are immediately read off:
\begin{equation}\label{eigevallla4}
\mathrm{Eigenval} \, = \,     \left\{\underbrace{- \frac{1}{16}}_{<\, 0},\underbrace{\frac{3}{8}}_{< \, 0},\underbrace{\frac{3}{8}}_{> \, 0}\right\}
\end{equation}
showing that also this anti de Sitter vacuum is stable. Indeed also in this case the Breitenlohner-Freedman bound is satisfied $\lambda_i \, > \, - \, \frac{9}{64}$. The  mass eigenstates coincide in this case with the charge eigenstates and we have two consistent one-field truncations:
\begin{description}
  \item[A-truncation.]  $h_1 \, = \, h_3 \, = \, 0$. In this case the residual potential is:
  $$V\, = \, \frac{3}{16} e^{-2 h_2}-\frac{3 e^{-h_2}}{8}$$
  Since the kinetic term of $h_2$ has a factor $q=3$, by means of the substitution (\ref{babushka}), we obtain:
   \begin{equation}\label{potentus4}
    \mathcal{V}(\varphi) \, = \, \frac{3}{16} e^{-2 \varphi /3}-\frac{3 e^{-\varphi /3}}{8}
   \end{equation}
   which is not any of the integrable potentials but belongs to the same class.
     \item[B-truncation.] $h_2 \, = \, h_3 \, = \, 0$. In this case the residual potential is:
     $$V\, = \, - \, \frac{1}{32}
   \left(4+e^{-h_1}+e^{h_1}\right)$$
     Since the kinetic term of the $h_1$ field has a factor $q=1$, by means of the substitution (\ref{babushka}), we obtain:
     \begin{equation}\label{potentus5}
     \mathcal{V}(\varphi) \, = \,   - \, \frac{1}{16} \left(\cosh\left(\frac{\varphi}{\sqrt{3}}\right)+2\right)
     \end{equation}
     which also belongs to the class of exponential potentials here considered but does not fit into any integrable series or sporadic case.
\end{description}
\subsection{\sc Another more complex example}
Inspired by the superpotential presented in eq.(5.12) of \cite{Dibitetto:2011gm} we have also considered the following extension of the superpotential (\ref{partus}):
\begin{eqnarray}\label{gonzaldino}
    \hat{W}^{locgeo} & = &  \lambda _1+S U^3 \lambda _4+TU \lambda _5+S T U^2\lambda _8 \nonumber\\
    &&+T^3 U^3 \lambda_9+S T^3 \lambda _{12}+T^2U^2 \lambda _{13}+S T^2 U\lambda _{16}
\end{eqnarray}
which leads to a potential with 30 terms and and 26 different type of exponentials. Finding all the roots of the equations that determine the existence of an extremum turned out to be to difficult, yet apart from the already known solution of the previous sections we were able to find by trial and error another solution corresponding to the following superpotential which depends on the overall parameter $\lambda_{4}$:
\begin{equation}\label{fasciofascio}
\hat{W}_0 \, = \,     -T U\lambda _4-S T^2 U \lambda _4-2 S TU^2 \lambda _4
+2 T^2 U^2 \lambda_4+S U^3 \lambda _4+T^3 U^3 \lambda_4
\end{equation}
The corresponding scalar potential, which for brevity we do not write, can be consistently  truncated to the dilatons by setting all the axions to zero and  by construction it has an extremum in $S=T=U={\rm i}$ where it takes the positive value $\frac{\lambda _4^2}{12}$. Hence, choosing the superpotential (\ref{fasciofascio}) we find an $\mathrm{dS}$ vacuum. Calculating the mass matrix in this extremum we find:
\begin{equation}\label{matamat5}
    M \, = \, - \, \frac{\lambda_4^2}{24} \, \left(
\begin{array}{lll}
 1 & 3 & 0 \\
 3 & -1 & 0 \\
 0 & 0 & -4
\end{array}
\right)
\end{equation}
It is convenient to choose  $\lambda_4=\sqrt{24}$ and in this way the the eigenvalues of $M$ have the following simple form:
\begin{equation}\label{eigati}
    \mbox{Eigenvalues}[M] \, \equiv \, \Lambda_i \, = \, \left\{4,\sqrt{10},\, - \,\sqrt{10}\right\}
\end{equation}
The presence of a negative one among the eigenvalues (\ref{eigati}) shows that the constructed $\mathrm{dS}$-vacuum is unstable.
The  eigenstates corresponding to the above eigenvalues are the following fields:
\begin{equation}\label{autostatti}
    \left\{\phi_1 , \, \phi_2 , \, \phi_3 , \, \right\} \, = \, \left\{h_3,\left(\frac{1}{3}-\frac{\sqrt{10}}{3}\right)
   h_1+h_2,\left(\frac{1}{3}+\frac{\sqrt{10}}{3}\right) h_1+h_2\right\}
\end{equation}
By calculating the derivatives we find that setting $\phi_2\,=\,\phi_3\,=\, 0$ is a consistent truncation. The surviving potential has the form:
\begin{equation}\label{formidino}
V \, = \,     1+\frac{e^{-\phi _1}}{2}+2 e^{\phi_1}-3 e^{2 \phi _1}+\frac{3 e^{3 \phi _1}}{2}
\end{equation}
while the kinetic term of the field $\phi_1$ has $q=3$. By means of the transformation (\ref{babushka}) the potential (\ref{formidino}) is mapped into:
\begin{equation}\label{cassandra}
    \mathcal{V}(\varphi) \, = \, 1+\frac{e^{-\varphi /3}}{2}+2 e^{\varphi /3}-3 e^{2 \varphi /3}+\frac{3 e^{\varphi }}{2}
\end{equation}
which is a combination of four different exponentials but does not fit into any of the integrable cases listed by us in tables \ref{tab:families} and \ref{Sporadic}.
\par
We might find still more examples, yet we think that those provided already illustrate the variety of one-field multi exponential potentials one can  obtain by consistent truncations of Gauged Supergravity. In this large variety identifying, if any, combinations that perfectly match one of the integrable cases is quite difficult, in want of some strategy able to orient a priori our choices, yet with some art we were able to single out at least one of them.
\subsection{\sc A new integrable model embedded into supergravity}
Working in a reduced parameter space that was determined with some inspired guessing we found the following very simple superpotential:
\begin{equation}\label{gordingo}
    W_{integ} \, = \, \left(i T^3+1\right) \left(S U^3-1\right)
\end{equation}
which  inserted into the formula for the scalar potential, upon consistent truncation to no axions produces the following dilatonic potential:
\begin{eqnarray}
 V_{dil}(\vec{h}) &=& \frac{5}{32}+\frac{1}{32} e^{-3 h_2}+\frac{e^{3
   h_2}}{32}-\frac{1}{64} e^{-h_1-3 h_3}+\frac{1}{64}
   e^{-h_1-3 h_2-3 h_3}\nonumber\\
   &&+\frac{1}{64} e^{-h_1+3 h_2-3
   h_3}-\frac{1}{64} e^{h_1+3 h_3}+\frac{1}{64} e^{h_1-3
   h_2+3 h_3}+\frac{1}{64} e^{h_1+3 h_2+3 h_3} \label{dilatus}
\end{eqnarray}
Performing the following field redefinition:
\begin{equation}\label{rotaziosca}
    h_1\to \sqrt{3} \phi _2 \quad , \quad h_2\to -\frac{\phi
   _1}{\sqrt{3}}\quad , \quad h_3\to \phi _3
\end{equation}
which is a rotation that preserves the form of the dilaton kinetic term:
\begin{equation}\label{furbacchione}
    \frac{1}{2} \dot{h}_1^2 + \frac{3}{2} \dot{h}_2^2 + \frac{3}{2} \dot{h}_3^2 \, \rightarrow \, \frac{1}{2} \dot{\phi}_1^2 + \frac{3}{2} \dot{\phi}_2^2 + \frac{3}{2} \dot{\phi}_3^2
\end{equation}
 the dilaton potential (\ref{dilatus}) transforms into
\begin{eqnarray}
  V_{dil}(\vec{\phi}) &=& \frac{5}{32}+\frac{1}{32} e^{-\sqrt{3} \phi
   _1}+\frac{1}{32} e^{\sqrt{3} \phi _1}-\frac{1}{64}
   e^{-\sqrt{3} \phi _2-3 \phi _3}+\frac{1}{64}
   e^{-\sqrt{3} \phi _1-\sqrt{3} \phi _2-3 \phi
   _3}\nonumber\\
   &&+\frac{1}{64} e^{\sqrt{3} \phi _1-\sqrt{3} \phi
   _2-3 \phi _3}-\frac{1}{64} e^{\sqrt{3} \phi _2+3 \phi
   _3}+\frac{1}{64} e^{-\sqrt{3} \phi _1+\sqrt{3} \phi
   _2+3 \phi _3}+\frac{1}{64} e^{\sqrt{3} \phi
   _1+\sqrt{3} \phi _2+3 \phi _3}\nonumber\\
   &&\label{baruffa}
  \end{eqnarray}
  The above potential has a de Sitter extremum at $\phi_{1,2,3}\, = \, 0$:
  \begin{equation}\label{dSstabulo}
    \left.\frac{\partial}{\partial \phi_{1,2,3}} \, V_{dil}(\vec{\phi})\right|_{\phi_{1,2,3}\, = \, 0} \, = \, 0 \quad ; \quad
    \left. V_{dil}(\vec{\phi})\right|_{\phi_{1,2,3}\, = \, 0}\, = \, \frac{1}{4} \, > \, 0
  \end{equation}
This dS vacuum  is stable since the mass matrix:
  \begin{equation}\label{massamatriciotta}
  \mbox{Mass}^2 \, \equiv \, \left.\frac{\partial^2}{\partial \phi_{i}\partial \phi_{j}} \, V_{dil}(\vec{\phi})\right|_{\phi_{1,2,3}\, = \, 0} \, = \, \left(
\begin{array}{lll}
 \frac{3}{8} & 0 & 0 \\
 0 & \frac{3}{32} & \frac{3 \sqrt{3}}{32} \\
 0 & \frac{3 \sqrt{3}}{32} & \frac{9}{32}
\end{array}
\right)
  \end{equation}
 has two positive and one null eigenvalue:
 \begin{equation}\label{eigenvaluti}
    \mbox{Eigenvalues} \, \mbox{Mass}^2 \, = \, \left\{\frac{3}{8},\frac{3}{8},0\right\}
 \end{equation}
 The corresponding mass eigenstates are the following fields:
 \begin{equation}\label{caratto}
    \left \{ \omega_1 \, ,\, \omega_2 \, ,\,\omega_3 \right \} \, = \, \left\{\frac{\phi _2}{\sqrt{3}}+\phi _3,\phi _1,\phi
   _3-\sqrt{3} \phi _2\right\}
 \end{equation}
 and transformed to the $\omega_i$ basis the dilatonic potential (\ref{baruffa}) becomes:
 \begin{eqnarray}
   V_{dil}(\vec{\omega})\ &=& \frac{5}{32}-\frac{1}{64} e^{-3 \omega _1}-\frac{e^{3
   \omega _1}}{64}+\frac{1}{32} e^{-\sqrt{3} \omega
   _2}+\frac{1}{32} e^{\sqrt{3} \omega _2}+\frac{1}{64}
   e^{-3 \omega _1-\sqrt{3} \omega _2}\nonumber\\
   &&+\frac{1}{64} e^{3
   \omega _1-\sqrt{3} \omega _2}+\frac{1}{64}
   e^{\sqrt{3} \omega _2-3 \omega _1}+\frac{1}{64} e^{3
   \omega _1+\sqrt{3} \omega _2}\label{finocchio}
 \end{eqnarray}
 which is explicitly independent from the field $\omega_3$, (the massless field) and can be consistently truncated to either one of the two massive modes $\omega_{1,2}$. In terms of the mass-eigenstates the kinetic term has the following form:
 \begin{equation}\label{kineticoso}
    \mbox{kin} \, = \, \frac{9 \dot{\omega} _1^2}{8}+\frac{\dot{\omega} _2^2}{2}+\frac{3
   \dot{\omega} _3^2}{8}
 \end{equation}
 In view of this and using the translation rule (\ref{babushka}), the two potentials that we obtain from the two consistent truncations are:
 \begin{eqnarray}
  3  \left. V_{dil}(\vec{\omega})\right|_{\omega_{1}\, = \, 0 \, ;\, \omega_2 \, = \, \frac{\varphi}{\sqrt{3}}}&=& \frac{3}{8} (\cosh (\varphi )+1)\label{integpotenz1} \\
  3  \left. V_{dil}(\vec{\omega})\right|_{\omega_{1}\, = \, \frac{2 \, \varphi}{3\sqrt{3}} \, ;\, \omega_2 \, = \, 0} &=& \frac{3}{32} \left(\cosh \left(\frac{2 \varphi
   }{\sqrt{3}}\right)+7\right) \label{nonintegpotenz2}
 \end{eqnarray}
The potential (\ref{integpotenz1}) fits into the integrable series $I_1$ of table \ref{tab:families} with $C_{11}=C_{22}=C_{12}$, while the potential (\ref{nonintegpotenz2}) associated with the second consistent truncation does not fit into any integrable series.
 \begin{figure}[!hbt]
\begin{center}
\iffigs
 \includegraphics[height=70mm]{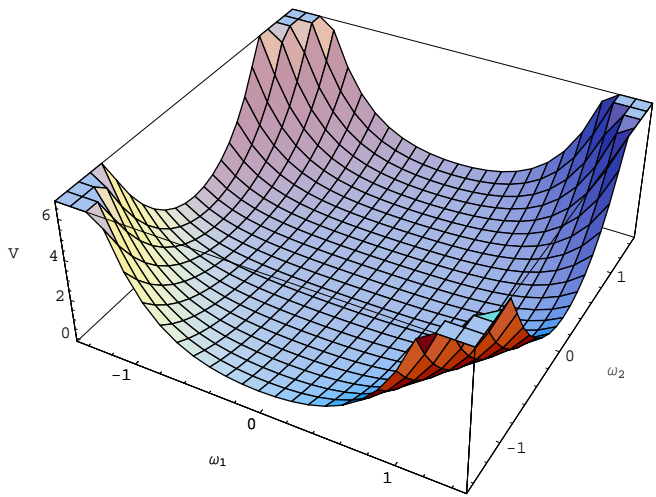}
\else
\end{center}
 \fi
\caption{\it
Structure of the two field supergravity potential which hosts an integrable $\cosh[\varphi]$ model. The integrable potential is cut out by the intersection with the plane $\omega_1=0$.}
\label{bellopotente}
 \iffigs
 \hskip 1cm \unitlength=1.1mm
 \end{center}
  \fi
\end{figure}
\par
The lesson that we learn from this example is very much illuminating in order to appreciate the significance and the role of integrable models in the framework of supergravity. Leaving aside the axions, that should be in any case taken into account, but that we assume to be stabilized at their vanishing values, a generic solution of the supergravity field equations would involve a scalar field moving, in the current example, on the two-dimensional surface of the potential displayed in fig.\ref{bellopotente}. Certainly the two field model is not integrable since it admits non-integrable reductions so that deriving generic solutions cannot be attained. Yet the existence of an integrable one-field reduction implies that we can work out some special exact solution of the multi-field theory by using the integrability of a particular reduction. Conceptually this means that the emphasis on general integrals usually attached to integrability is to be dismissed in this case. The general integral of the reduction is in any case a set of particular solutions of the complete physical theory which does not capture the full extensions of initial conditions. The correct attitude is to consider the whole machinery of integrability as an algorithm to construct particular solutions of Einstein equations that might be more or less relevant depending  on their properties.
\section{\sc  Summary and Conclusions}\label{sec:conclusion}
In the present paper we have addressed two questions
\begin{description}
  \item[A)] Whether any of the integrable one-field cosmological models classified in \cite{primopapero} can be fitted into Gauged Supergravity (in the $\mathcal{N}=1$ case we have restricted ourselves to F-term supergravitites, where the contribution of the vector multiplets to the scalar potential is negligible) based on constant curvature scalar manifolds $\mathrm{G/H}$ as consistent truncations to one-field of appropriate multi-field models.
  \item[B)] Whether the solutions of integrable one-field cosmological models can be used as a handy simulation of the behavior of unknown exact cosmological solutions of Friedman equations
\end{description}
Both questions have received a positive answer.
\par
As for question A) we have shown that the embedding of integrable one-field cosmologies into Gauged Supergravity is rather difficult but not impossible. Indeed we were able to identify two independent examples of integrable potentials that can be embedded into $\mathcal{N}=1$ supergravity, gauged by means of suitable and particularly nice superpotentials:
\begin{description}
  \item[Model One)] The first integrable supersymmetric model corresponds to $\mathcal{N}=1$ supergravity coupled to a single Wess-Zumino multiplet with the K\"ahler Geometry of $\frac{\mathrm{SU(1,1)}}{\mathrm{U(1)}}$ and the following quartic superpotential:
      \begin{equation}\label{superpot1}
        W_{int1}(z) \, = \, \frac{2}{\sqrt{5}} \left (3\, z^4 \, + \, {\rm i} \, \omega \, z^3\right)
      \end{equation}
    and gives rise, after truncation to the dilaton, to the integrable potential of series $I_2$ in table \ref{tab:families} with $\gamma \, = \, \frac{2}{3}$.
  \item [Model Two)] It is obtained within the STU-model with a pure $\mathcal{N}=1$ K\"ahler structure (the special K\"ahler structure is violated) and a very specific superpotential:
      \begin{equation}\label{superpot2}
    W_{int2}(S,T,U) \, = \, \left(i T^3+1\right) \left(S U^3-1\right)
\end{equation}
whose interpretation within flux compactification is an interesting issue to be  pursued further. A consistent truncation of this $\mathcal{N}=1$ model yields the integrable potential $I_1$ of table \ref{tab:families}.
\end{description}
An important additional result of the present paper is the complete classification of all possible $\mathcal{N}=2$ gaugings of the STU model that was presented in section \ref{STUgauginghi}. This classification  was performed within the embedding tensor formalism by means of which we reduced the enumeration of non-abelian gaugings  to the enumeration of admissible $\mathrm{G}$-orbits in the $\mathbf{W}$-representation, the same which black-hole charges are assigned to. In each admissible orbit we have the choice of switching on the Fayet Iliopoulos terms or keeping them zero.  This yields two different gaugings for each admissible orbit. Finally one can consider pure abelian  gaugings that are once again in correspondence with the $\mathrm{G}$-orbits in the $\mathbf{W}$-representation.
This classification provided two results. On one hand we verified that the only (stable) de Sitter vacuum is the one that was obtained several years ago in \cite{mapietoine}. On the other hand we might conclude that no integrable model can be embedded in any of these gauged models.
\par
Provisionally it follows that the  very few examples of integrable cosmologies admitting a supergravity embedding are found within the $\mathcal{N}=1$ framework with F-term gauging. On the other hand, utilizing the D-terms and the axial symmetric K\"ahlerian manifolds that are in the image of the D-map, infinite series of integrable cosmological models can be embedded into $\mathcal{N}=1$ Supergravity \cite{secondosashapietro}.
We plan to pursue further the classification of all the gaugings for the $\mathcal{N}=2$ models of table \ref{homomodels} in order to ascertain whether these conclusion hold true in all cases or whether there are new integrable truncations \cite{terzopapero}.
\par
As for question B) we have addressed  it concretely in the case of the $cosh$-model which emerges in many one-field truncations but it is integrable only for a few distinct values of the parameters. We have shown that if the non-integrable and the integrable considered models share the same type of fixed point (for instance node), then the solutions of the integrable case capture all the features of the solutions to the non integrable model and are actually numerically very close to them. Such a demonstration is forcibly only qualitative and can be just appreciated by looking at the plots. A precise algorithm to estimate the error is so far absent.
\par
The detailed analysis, presented in section \ref{integsusymodel}, of the space of solutions to the supersymmetric integrable \textit{Model One} revealed a new interesting phenomenon that, to the best of our knowledge, was so far undiscovered in General Relativity. As we stressed in paper \cite{primopapero}, whenever the scalar potential has an extremum at negative values the solutions of Friedman equation describe a Universe that, notwithstanding its spatial flatness, ends its life into a Big Crunch like a closed Universe with positive spatial curvature. This is already an interesting novelty but an even more striking one was discovered in our analysis of sect. \ref{parteventhoriz}. The causal structure of these spatially flat collapsing universes is significantly different from that of the closed universe, since here the particle and event horizons do not coincide and have an interesting evolution during the universe life cycle. We think that the possible cosmological implications of this mechanism should be attentively considered.
\vskip 2cm
\section*{Acknowledgments}
The authors wish to thank A. Sagnotti for enlightening discussions and critical reading of the manuscript. \\
The work of A.S. was supported in part by the RFBR Grants No. 11-02-01335-a, No. 13-02-91330-NNIO-a and No. 13-02-90602-Arm-a.

\end{document}